\documentclass[11pt,a4paper,twoside,openright]{book}              

\makeatletter
\g@addto@macro\bfseries{\boldmath}
\makeatother

\usepackage[Lenny]{fncychap}

\usepackage{epsfig,graphicx}
\usepackage[english,german]{babel} 
\usepackage[utf8]{inputenc} 
\usepackage{amssymb} 
\usepackage{amsmath} 
\usepackage{xspace} 
\usepackage{url} 
\usepackage{braket} 
\usepackage{slashed} 
\usepackage{multirow} 
\usepackage{bbold} 
\usepackage{dsfont} 
\usepackage{mathrsfs}
\usepackage[usenames,dvipsnames,svgnames,table]{xcolor}  
\usepackage{marginnote} 
\usepackage{minitoc} 
\dominitoc 
\usepackage{epigraph} 
\usepackage{pdfpages} 

\usepackage{fancyhdr}
\makeatletter
\renewcommand\chaptermark[1]{%
  \markboth{\MakeUppercase{%
    \ifnum \c@secnumdepth >\m@ne
      \if@mainmatter
        \@chapapp\ \thechapter. \ %
      \fi
    \fi
    #1}}{}%
}
\renewcommand\@mkboth[2]{\markboth{#1}{}}
\makeatother
\usepackage[sort&compress,numbers]{natbib}
\usepackage{doi}
\usepackage{hyperref} 

\usepackage[capitalise]{cleveref}
\usepackage[usenames,dvipsnames,svgnames,table]{xcolor}  

\hypersetup{%
    colorlinks=true, linktocpage=true, pdfstartpage=3, pdfstartview=FitV,%
    breaklinks=true, pdfpagemode=UseNone, pageanchor=true, pdfpagemode=UseOutlines,%
    plainpages=false, bookmarksnumbered, bookmarksopen=true, bookmarksopenlevel=1,%
    hypertexnames=true, pdfhighlight=/O,
    urlcolor=Cerulean, linkcolor=Cerulean, citecolor=Cerulean, 
}

\usepackage{booktabs} 
\usepackage[font=footnotesize,labelfont=bf]{caption}

\usepackage{relsize}
\def\babar{\mbox{\slshape B\kern-0.1em{\smaller A}\kern-0.1emB\kern-0.1em{\smaller A\kern-0.2em R}}\xspace}

\definecolor{crimson}{RGB}{220,20,60}

\newcommand{\cpt}{$\chi$PT\xspace}
\newcommand{\lcpt}{$\ell N_c\chi\textrm{PT}$\xspace}

\newcommand{\albl}{$a^{\textrm{HLbL}}_{\mu}$\xspace}
\newcommand{\alblp}{$a^{\textrm{HLbL;}P}_{\mu}$\xspace}
\newcommand{\PtoLL}{$P\rightarrow\bar{\ell}\ell$\xspace}

\title{\bf A theoretical study of meson transition form factors}    
\author{Pablo~Sanchez-Puertas}              
\date{\today}                           

\begin{document}                        

\frontmatter                            

\begin{titlepage}
\begin{center}

\vspace*{1cm}
{\Huge\textbf{A theoretical study of meson transition form factors}}
        
\vspace{1.5cm}
{\Large{  \textit{
Dissertation \\
zur Erlangung des Grades\\
,,Doktor der Naturwissenschaften''
}}}

\vspace{1.5cm}
{
\Large
am Fachbereich Physik, Mathematik und Informatik\\
der Johannes Gutenberg-Universit\"at Mainz
}

\vspace{3cm}
\includegraphics[width=8cm]{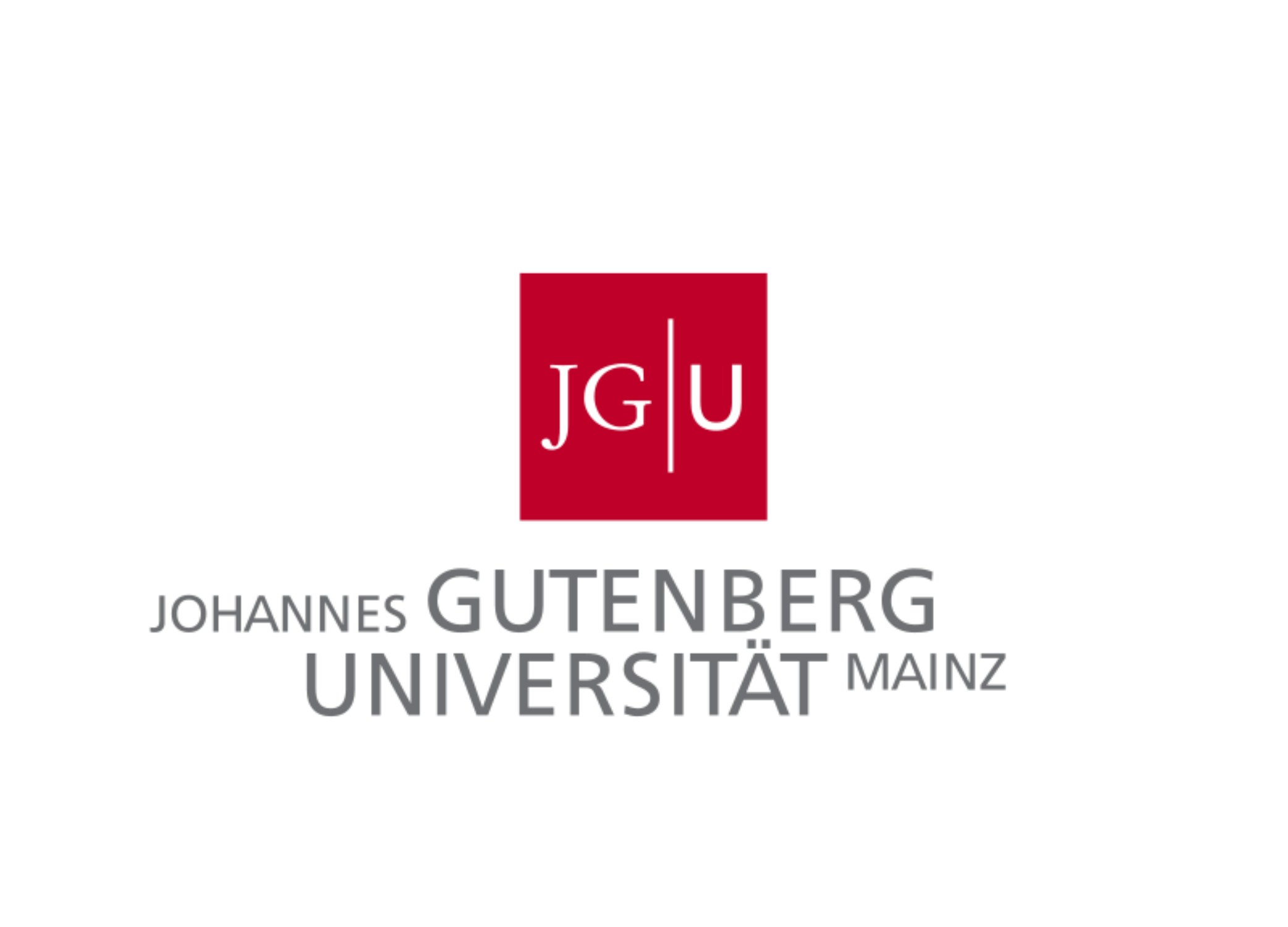}

\vfill

{\Large
{\textbf{Pablo Sanchez-Puertas}}\\
geboren in Granada (Spanien)\\
\vspace{1cm}
Mainz, 2016
}
        
\end{center}
\end{titlepage}


{\thispagestyle{empty}

\null\vfill

%
%

%
%
%



\cleardoublepage\thispagestyle{empty}

\noindent {\Huge{Abstract}}

\vfill

\noindent This thesis studies the lightest pseudoscalar mesons, $\pi^0$, $\eta$, and $\eta'$, through their transition form factors. 
Describing the underlying structure of hadrons is still a challenging problem in theoretical physics. 
These form factors, which can be experimentally measured, provide valuable information on the pseudoscalar 
meson inner structure and are of fundamental interest for describing their elementary interactions. Obtaining a precise 
description for these form factors has become a pressing subject given their role in one of the finest tests of our understanding
of particle physics: the anomalous magnetic moment of the muon. The foreseen experimental precision for this observable challenges 
the available theoretical descriptions so far.

Still, incorporating the available experimental information into a theoretical framework becomes increasingly difficult  
as the experimental precision improves, challenging simplified frameworks.
In this work, we propose to use the framework of Pad\'e theory in order to precisely describe these form factors in the space-like region, which provides a well-founded 
mathematically-based and data-driven approach for this task.

The first part of our study is devoted to extract the parameters relevant to our approach using the available single-virtual space-like data. 
The accuracy of the method, beyond that of previous approaches, has been later confirmed in experiments performed in the low-energy 
time-like region for the $\eta$ and $\eta'$ cases. To give consideration to these new results, we incorporated the corresponding data 
into our analysis.
The extension of the formalism to the most general double-virtual case is subsequently discussed, which requires the introduction, 
for the first time in this context, of Canterbury approximants, the bivariate version of Pad\'e approximants.

As a direct application of our results, the $\eta-\eta'$ mixing parameters have been extracted from the single-virtual transition form factors. 
The employed method provides an alternative to the traditional ones, obtaining competitive results while minimizing modeling errors.

Besides, our double-virtual description is employed for describing the rare decays of the pseudoscalar mesons into a lepton pair. 
The latter process offers an opportunity to test the doubly virtual pseudoscalar mesons transition form factors 
as well as an opportunity to discuss possible new physics contributions in light of the present discrepancies.

Finally, our approach is used to obtain a precise calculation for the pseudoscalar-pole contribution to the hadronic light-by-light piece of the  
anomalous magnetic moment of the muon. This includes, for the first time, a systematic error and meets the required precision in foreseen experiments.



\cleardoublepage\thispagestyle{empty}
\selectlanguage{german}%

\noindent {\Huge{Zusammenfassung}}

\vfill

\noindent Die vorliegende Dissertation befasst sich mit dem Studium der leichtesten pseudoskalaren Mesonen $\pi^0$, $\eta$, and $\eta'$ via 
deren {\"U}bergansformfaktoren. Eine Beschreibung der zugrunde liegenden Struktur der Hadronen stellt in der theoretischen Physik immer 
noch eine Herausforderung dar. Diese Formfaktoren, die experimentell bestimmt werden k{\"o}nnen, stellen eine wichtige Informationsquelle 
\"uber die innere Struktur pseudoskalarer Mesonen dar und sind von grundlegendem Interesse f\"ur die Beschreibung ihrer elementaren 
Wechselwirkungen. Der Erhalt einer pr\"azisen Beschreibung f\"ur diese Formfaktoren ist, mit Blick auf ihre Rolle in einem der genauesten Tests 
unseres Verst\"andnisses der Teilchenphysik: dem anomalen magnetischen Moment des Myons, zu einem dringlichen Thema geworden. 

Die Einarbeitung der verf\"ugbaren, experimentell ermittelten, Informationen in einen theoretischen Rahmen wird nach wie vor mit zunehmender 
Genauigkeit der Experimente schwieriger, was vereinfachte Modelle auf die Probe stellt. Im Rahmen dieser Arbeit schlagen wir vor, sich der 
Pad\'e-Approximation, ein sowohl mathematisch als auch auf Daten basierender und somit wohlbegr\"undeter Zugang zu diesem Problem ist, zu 
bedienen, um diese Formfaktoren in raumartigen Bereichen pr\"azise beschreiben zu k\"onnen. 

Im ersten Teil unserer Betrachtungen widmen wir uns unter Ausnutzung von Messwerten raumartiger Prozesse mit einem virtuellen Photon, der 
Extraktion der f\"ur unseren Zugang relevanten Parameter. Die Genauigkeit dieser Methode, die \"uber bisherige Versuche hinausgeht, wurde 
sp\"ater durch Experimente die im raumartigen Niedrigenergiesektor f\"ur die F\"alle von $\eta$ und $\eta'$ durchgef\"uhrt wurden, best\"atigt. 
In der Folge wird 
die Ausweitung des Formalismus auf den allgemeinsten Fall zweier virtueller Photonen diskutiert, was die in diesem Kontext erstmalige Einf\"uhrung 
der Canterbury-Approximation, der zweidimensionalen Pad\'e-Approximation, erfordert. 

Als eine direkte Anwendung unserer Ergebnisse, wurden die Parameter der Mischung $\eta-\eta'$ aus dem \"Ubergangsformfaktor eines virtuellen 
Photons ermittelt. Die verwendete Methode bietet eine Alternative zu traditionell verwendeten, wobei wir konkurrenzf\"ahige Ergebnisse erhalten 
und zugleich modellbezogene Fehler minimieren.

Zudem wird unsere Beschreibung von Prozessen mit zwei virtuellen Photonen auf die Beschreibung der seltenen Zerf\"alle eines pseudoskalaren 
Mesons in ein Leptonen-Paar angewendet. Der genannte Prozess bietet die Gelegenheit die \"Ubergansformfaktoren pseudoskalarer Mesonen f\"ur 
zwei virtuelle Photonen zu testen. 

Schlussendlich wird unsere Vorgehensweise dazu verwendet, eine genaue Berechnung f\"ur den Beitrag des pseudoskalaren Pols zum Anteil der 
hadronischen Licht-Licht-Streuung des anomalen magnetischen Moments des Myons zu erhalten. In dieser mitinbegriffen ist erstmalig ein systematischer 
Fehler und sie entspricht der f\"ur 
Experimente geforderten, ben\"otigten Genauigkeit.

\selectlanguage{english}%


\newenvironment{acknowledgements}%
{\cleardoublepage\thispagestyle{empty}\null\vfill
\begin{center}\textbf{\textit{\LARGE Acknowledgements}}\end{center}}%
{\vfill\null}
\begin{acknowledgements}

I would like to express my gratitude to Pere Masjuan for his guidance during these four years of doctoral 
studies. His door was always open when looking for help and I have profited from his knowledge in physics 
in countless and interesting discussions. This work would not have been possible without his guidance 
and great enthusiasm. 
The realization of this thesis would have not been possible either without the help of Marc 
Vanderhaeghen, to whom I thank for his support, reading and interesting comments concerning this manuscript.

During my time at Mainz, I could profit as well from the help and comments from many people at the 
Nuclear Physics department; a very special thanks goes to Tobias Beranek, Mikhail Gorchtein, Nikolay Kivel  and 
Oleksandr Tomalak, for many and valuable discussions and for sharing their wisdom with me. 
Besides, I am particularly indebted to my office mates, Patricia 
Bickert and Nico Klein for the hours stolen in trying to understand the basic concepts of chiral perturbation 
theory and to Hans Christian Lange for helping with the German translation. 
In addition, and besides Mainz, I was lucky to have a pleasant collaboration with R.~Escribano, without whom 
one chapter of this thesis could not have been possible, and his PhD student Sergi Gonzalez-Sol\'is, whom 
which I was luck to share my office and many discussions.

Finally, the greatest acknowledgement goes to my wife, for her support, love and   
encouragment, and for reading this manuscript since the early times.

\end{acknowledgements}

    \setcounter{tocdepth}{1}   
    \setcounter{minitocdepth}{1}
\tableofcontents                        
\chapter{Preface}
\label{chap:intro}

The past decades of fundamental research in particle physics have established the standard model (SM) of particle physics as the microscopic 
theory of fundamental interactions, encompassing the strong, weak and electromagnetic forces in a $SU(3)_c \times SU(2)_L \times U(1)_Y$ gauge 
theory\footnote{See Refs.~\cite{Peskin:1995ev,Donoghue} for a detailed review.}. 
Even if its formulation fits in a few lines, it provides the most successful theory ever formulated in the history of particle physics, 
and still stands in a good shape after thorough tests over the years ---some of them standing to astonishing precision.
\\

However, the SM as it is, was known to provide an incomplete description of nature even before its last piece remaining, the Higgs boson, was discovered in 2012 at the  
LHC experiment~\cite{Aad:2012tfa,Chatrchyan:2012xdj}. Firstly, the SM fails to incorporate Einstein's theory of general 
relativity ---quantizing gravity is still a fundamental problem in theoretical physics. Secondly, there is great evidence that the ordinary matter which is described within the SM cannot 
explain the rotational galaxy curves, which seems to require the existence of dark matter ---actually, the SM only describes the visible matter, which corresponds with around $5\%$ 
of the energy content of the universe, whereas dark matter~\cite{ArkaniHamed:2008qn} would account for $26\%$~\cite{Adam:2015rua,Ade:2015xua}. 
At larger cosmological scales, it is hard to explain the 
observed curvature of the universe without the presence of dark energy~\cite{Li:2011sd} ---accounting for the remaining $69\%$ energy content of the universe--- 
for which the vacuum energy is a possible candidate; the SM however provides a number which is far too large as compared to the observational requirements. Furthermore, the 
SM is not able to describe baryogenesis ---the CP violation within the SM is not large enough--- nor inflation. Asides, the SM does not contain a mass term for the neutrinos, 
which is required to explain neutrino oscillations; the origin of neutrino mass and whether neutrinos are Dirac or Majorana particles is still an open question. 
There exist in addition theoretical reasons for which the SM is thought to be just the low-energy manifestation of an ultraviolet (UV) completion including, at the very least, 
gravity ---known to be important at energies around the Planck scale  $\Lambda_{\textrm{Planck}}\sim10^{19}~\textrm{GeV}$. 
Furthermore, one of the most renowned issues has to deal with  the so called hierarchy problem~\cite{Martin:1997ns}. This is related to the large radiative corrections which 
the Higgs mass receives. These make natural to expect a mass close to the next scale of new-physics ---say $\Lambda_{\textrm{Planck}}$--- in contrast with 
its (now) well known mass $m_{H}=125.09(24)~\textrm{GeV}$~\cite{Agashe:2014kda}. Furthermore, our current SM knowledge suggests that the SM vacuum is only metastable. 
Besides, the SM does not explain the large hierarchies of masses and the 
existence of three family generations. All in all, an evidence strong enough to stimulate speculations on different 
kind of physics scenarios beyond the standard model (BSM).
\\

It has been a while right now since the first supersymmetric~\cite{Martin:1997ns} and technicolor~\cite{Weinberg:1979bn,Susskind:1978ms} models appeared as a plausible realization of 
nature providing an UV completion of the SM. Since then, they have been intensively searched for at the most energetic colliders of their time, such as 
SPS@CERN, SLC@SLAC, LEP@CERN,  Tevatron@Fermilab and, at this moment, LHC@CERN. 
Each of these new theories share the SM as their low-energy limit, but differ in their additional (heavier) particle spectra.  
Positive observations at colliders would immediately shed light into the uttermost structure of particle physics and represents the main tool to search for new  physics. 
The negative results for these searches so far has led to the development of some alternative models such as Little(st) 
Higgs~\cite{ArkaniHamed:2002pa,ArkaniHamed:2002qy}, and extra dimensions~\cite{ArkaniHamed:1998rs}, 
which so far have not been found either. This situation in the world of particle physics has led to the envisioning of new powerful colliders which would produce such hypothesized 
heavy particles. Nevertheless, it may be that such particles are too heavy to be produced at any envisaged collider so far, which poses a distressing scenario 
for the community. \\

Fortunately, quantum field theory provides alternative approaches to look for even heavier physics at lower energies. The quantum vacuum, with its fluctuations, offer us the 
opportunity to test the effects of these new particles at lower energies. Diagrammatically, this occurs through loop processes where heavy particles are virtually created and 
destroyed. From the modern point of view of effective field theories, this can be easily understood as a consequence of integrating-out the heavy fields, which produces 
additional higher-dimensional operators non-present in the SM. An outstanding example are those operators which drive 
the proton decay ---a process which is forbidden in the SM due to the accidental baryon number conservation. Actually, the stringent bounds for the ---so far unobserved--- proton 
decay, provide strong limits to the scales of Grand Unified theories 
as high as $10^{15}~\textrm{GeV}$, 
which is inconceivable to test at any collider. Additional examples of these tiny contributions appear as well 
within the SM, such as in flavor physics, where the effective contribution from the charm and top quark in $\bar{K}^0\!-\!K^0$~\cite{Gaillard:1974hs} 
and $\bar{B}^0\!-\!B^0$~\cite{Ellis:1977uk,Prentice:1987ap} mixing, respectively, already predicted the order of the charm quark mass as well as a heavy top mass prior 
to their discoveries. Even preciser estimates were obtained both for the top and the Higgs masses based on 
electroweak 
precision observables~\cite{ALEPH:2005ab} before they were discovered. 
The list of processes potentially sensitive to new physics effects is long, especially in flavor  
physics. Still at far lower energies, there is a world-famous observable which, given its experimental precision, plays an important 
role in looking for new physics and constraining BSM theories: the anomalous magnetic moment of the muon $(g_{\mu}-2)$~\cite{Jegerlehner:2009ry}. 
%
The latter is related to the magnetic dipole moment of the muon $\boldsymbol{\mu}_m$ governing its interaction with a 
classical magnetic field $\boldsymbol{B}$ through the Hamiltonian
\begin{equation}
  \mathcal{H} = -\boldsymbol{\mu}_m \cdot \boldsymbol{B}(\boldsymbol{x}), \quad \boldsymbol{\mu}_m = g_{\mu}\left(\frac{e\mathcal{Q}}{2m_{\mu}}\right) \boldsymbol{S}, 
\end{equation}
with $\boldsymbol{S}(\mathcal{Q})$ the muon spin(charge) and $g_{\mu}$ its gyromagnetic ratio, which classical value can be predicted from Dirac theory, obtaining $g_{\mu}=2$. 
Such quantity receives however quantum corrections ---arising within the SM of particle physics--- implying deviations from the $(g_{\mu}-2)=0$ value. Similarly, new kind 
of physics would produce additional 
corrections to this observable. Therefore, a very precise measurement of $(g_{\mu}-2)$ would allow to test BSM physics provided we are able to calculate 
the SM contributions to this observable, including quantum electrodynamics (QED), quantum chromodynamics (QCD) and electroweak (EW) contributions, 
to the astonishing precision to which  $a_{\mu}\equiv (g_{\mu}-2)/2$ is measured~\cite{Bennett:2006fi},
\begin{equation}
 a_{\mu}^{\textrm{exp}}  = 116592091(63) \times 10^{-11}.
\end{equation}
The current theoretical estimation for the SM contribution reads\footnote{See \cref{chap:gm2} for detailed numbers and references.} 
\begin{equation}
  a_{\mu}^{\textrm{th}}  = 116591815(57) \times 10^{-11},
\end{equation}
and leads to a $3.2\sigma$ discrepancy among theory and experiment. This has motivated speculations on BSM physics contributing to this observable. For this reason, 
two new experiments have been projected at Fermilab~\cite{LeeRoberts:2011zz} and J-PARC~\cite{Mibe:2010zz} aiming for an improved precision 
$\delta a_{\mu} = 16\times 10^{-11}$ in order to sort out the 
nature of the discrepancy ---note that even a negative result in the search for new physics effects would provide then a valuable constraint on BSM theories. 
However, this effort will be in vain unless a 
similar theoretical improvement is achieved, the current limiting factor being the SM hadronic corrections, among which, the leading order  
hadronic vacuum polarization (HVP) and hadronic light-by-light (HLbL) contributions dominate. 
Improving the current errors represents however an extremely difficult task, as these calculations involve non-perturbative hadronic physics 
which cannot be obtained from a first principles calculation ---the exception is of course lattice QCD which nevertheless requires some advances, 
specially for the HLbL, in order to improve current theoretical estimates. \\

The main objective of this thesis is to improve on a particularly large contribution dominating the HLbL 
---the pseudoscalar-pole contribution--- at the precision required for the new projected $(g_{\mu}-2)$ experiments. 
To this end, it is necessary to carefully describe the pseudoscalar meson interactions with two virtual photons. These 
are encoded in their transition form factors (TFFs), that must be described as precisely and 
model-independent as possible ---including an accurately defined error---  in order to achieve a precise and reliable result.  
Their study concerns the first part of this thesis. 
To this end, the methodology of Pad\'e approximants and multivariate extensions relevant for the double-virtual TFFs description are considered.
\\

Closely related to the HLbL, we address the calculation of the rare $P\to\bar{\ell}\ell$ decays, where $P=\pi^0,\eta,\eta'$ and $\ell=e,\mu$. These processes, showing a similar 
dependence on the pseudoscalar TFFs as the HLbL, not only offer a valuable check on our TFF description, but represent the only source of experimental information on the 
double-virtual TFF up to day ---describing the double-virtual TFF behavior is very important in order to achieve a precise HLbL determination. 
Beyond that, the large suppression of these processes within the SM offers an opportunity to search for possible new physics effects in these decays. 
In the light of the present experimental discrepancies, we carefully describe and discuss them together with their implications in $(g_{\mu}-2)$.
\\


Despite of the theoretical relevance and hype for new physics searches, there are still some interesting questions within the SM that need to be addressed
 ---the QCD spectrum among others. Whereas R.~L.~Jaffe affirmed that ``the absence of exotics is one of the most obvious features of QCD''~\cite{Jaffe:2004ph}, which 
can be supported from large-$N_c$ arguments, this is not clear at all as the recent discovery of a new plethora of the so-called $XYZ$ exotic states~\cite{Godfrey:2008nc}, 
or even the possible pentaquark states in the charm quark sector~\cite{Aaij:2015tga} shows. Even more elusive is the question of gluonium states ---purely gluonic quarkless bound states--- for which 
several candidates exist. Given their quantum numbers, it is possible that some gluonium admixture exists in the $\eta'$. 
Elucidating the $\eta$ and $\eta'$ structure has been a very interesting and controversial topic, which relevance is not only theoretical but phenomenological, 
as it enters a number of heavy mesons decays. The electromagnetic interactions, encoded in their TFFs, offer a probe to test the internal $\eta$ and $\eta'$ structure which we use  
in order to obtain a new determination for the $\eta-\eta'$ mixing parameters.\\

\section*{Outline}

The thesis is structured as follows: the fundamental concepts on QCD and TFFs employed in this thesis are presented in \cref{chap:QCD} along with the theory 
of Pad\'e approximants, which we adopt to describe the pseudoscalar TFFs. 
In \cref{chap:data}, we use the available data for the $\eta$ and $\eta'$ in order to extract the required low-energy parameters (LEPs). 
The excellent prediction that the method provides for the 
low-energy time-like region ---based on space-like data and proving the power of the approach--- is discussed and later incorporated into our analysis. 
This allows for an improvement in our LEPs extraction and provides a single description for the whole space-like and low-energy time-like regions. In \cref{chap:CA}, 
the generalization of Pad\'e approximants to the bivariate case is introduced for the first time in this context and carefully discussed, thus providing a framework to reproduce 
the most general doubly-virtual TFF. In \cref{chap:mixing}, we discuss as a first application from our outcome an alternative extraction for the $\eta-\eta'$ mixing 
parameters, which overcomes some problematics of previous approaches and incorporates subleading large-$N_c$ and chiral corrections. In \cref{chap:PLL}, we 
discuss a first application based on our TFF parameterization: the calculation of $P\to\bar{\ell}\ell$ decays, which are of interest given current 
experimental discrepancies. Our method improves upon previously existing VMD-based models, specially for the $\eta$ and $\eta'$. As a closure, 
a careful discussion on possible new-physics effects is presented. Finally, in \cref{chap:gm2}, we use our approach to calculate the pseudoscalar-pole contribution 
to the hadronic light-by-light $(g_{\mu}-2)$ contribution. For the first time, a systematic method properly implementing the theoretical constraints,  
not only for the $\pi^0$, but for the $\eta$ and $\eta'$ mesons and including a systematic error is achieved. Besides, the resulting calculation succeeds in obtaining a 
theoretical error in accordance to that which is foreseen in future $(g_{\mu}-2)$ experiments, which is the main goal of this thesis.

\mainmatter                             

\chapter{Quantum Chromodynamics and related concepts}
\label{chap:QCD}
\minitoc

\section{Introduction}

In this chapter, we introduce the essential concepts of Quantum Chromodynamics (QCD) that will be required along this thesis.
First, we introduce QCD, the quantum-field theory (QFT) of the strong interactions. 
We discuss then one of its central properties, asymptotic freedom. This feature, allowing to perform a perturbative expansion at high-energies, 
forbids at the same time a similar application at low energies. 
For this reason, we introduce chiral perturbation theory ($\chi$PT), the effective field theory of QCD at low energies, 
which is our best tool to describe the physics of pions ($\pi$), kaons ($K$) and eta ($\eta$) mesons at low-energies, providing the 
relevant framework for discussions in this thesis.
None of the previous descriptions are able to describe the intermediate energy region at around $1$~GeV though. A successful framework 
providing some insight in this intermediate energy regime, encompassing both the chiral expansion and perturbative QCD limits, 
is the limit of large number of colors, large-$N_c$. 
We argue that the success of resonant approaches inspired from such limit may be connected to the mathematical theory of Pad\'e approximants 
(PAs), which is subsequently introduced. Finally, we briefly describe the pseudoscalar transition form factors.

\section{Quantum Chromodynamics}

QCD is the microscopic theory describing the strong interactions in terms of quarks and gluons. 
The former are the matter building blocks of the theory, whereas the latter represent the force carriers.  
It consists of a Yang-Mills $SU(3)_c$ ---$c$ standing for color--- theory which Lagrangian is 
given as\footnote{Section based in Refs.~\cite{Peskin:1995ev,PDG:qcd}.}
\begin{equation}
\label{eq:QCDlag}
\mathcal{L}_{\textrm{QCD}} = \sum_f  \overline{q}_f (i\slashed{D} - m_q) q_f -\frac{1}{4}G^c_{\mu\nu}G^{c,\mu\nu},
\end{equation}
where $q_f$ represent the quark spinor fields transforming under the fundamental $SU(3)_c$ representation; as such, they are said to come in $N_c=3$ colors. 
Quarks come in addition in $n_f=6$ different species or flavors $f$, up $(u)$, down $(d)$, strange $(s)$, charm $(c)$, bottom $(b)$ and top $(t)$, with different 
masses $m_q$ spanning over five orders of magnitude. The symbol  $\slashed{D}=\gamma^{\mu}D_{\mu}$, with $\gamma^{\mu}$ the Dirac matrices (see \cref{app:conv}) and $D_{\mu}$ 
the covariant derivative
\begin{equation}
D_{\mu} = \partial_{\mu} -ig_s A_{\mu}^c t^c,
\end{equation}
with $g_s$ the strong coupling constant, $t^c=\lambda^c/2$ the fundamental representation group generators and $A_{\mu}^c$ the $N_c^2-1=8$ 
gluon fields transforming in the adjoint $SU(3)_c$ representation. Finally, the $G^c_{\mu\nu}$ term stands for the field strength 
tensor\footnote{The structure constants $f^{abc}$ are defined from $[t^a,t^b]=if^{abc}t^c$.}
\begin{equation}
 G^{c}_{\mu\nu}=\partial_{\mu}A_{\nu}^a  -  \partial_{\nu}A^a_{\mu} + g_s f^{abc}A^b_{\mu}A^c_{\nu}.
\end{equation}


The central property promoting QCD as the theory of the strong interactions is asymptotic freedom. 
In any QFT, renormalization effects lead to a non-constant coupling which is said to run with the energy. Such dependence is described 
with the help of the renormalization group (RG) equations for the coupling constant $\alpha_s = g_s^2/(4\pi)$~\cite{PDG:qcd}, 
\begin{equation}
\label{eq:asympf}
\mu^2 \frac{d \alpha_s}{d\mu^2} = \beta(\alpha_s) = -\alpha_s\left(\beta_0\frac{\alpha_s}{4\pi}  + \beta_1\left(\frac{\alpha_s}{4\pi}\right)^2 + \beta_2\left(\frac{\alpha_s}{4\pi}\right)^3 +  ... \right),
\end{equation}
where $\beta_0=(\frac{11N_c}{3} - \frac{2n_f}{3})$, being $n_f$ the number of active flavors; additional $\beta_{1,2,...}$ terms can be found, up to $\beta_4$, in 
Refs.~\cite{PDG:qcd,Baikov:2016tgj}. The remarkable property in \cref{eq:asympf} is the overall negative sign for $\beta_0>0$, i.e. for $n_f<\frac{11}{2}N_c<17$, 
deserving a Nobel prize in 2004 to D.~J.~Gross, H.~D.~Politzer and F.~Wilczek\footnote{Interesting enough, at order $\alpha_s$, \cref{eq:asympf} leads to the solution 
$\alpha_s(\mu) = 2\pi/(\beta_0\ln(\mu/\Lambda_{\textrm{QCD}}))$, which defines an intrinsic (certainly non-perturbative) scale $\Lambda_{\textrm{QCD}}$.}.
This sign implies the decreasing of the strong coupling constant at high energies ---asymptotic freedom--- and allows for an easy and standard perturbative expansion in terms 
of quarks and gluons degrees of freedom. This property will be used in \cref{sec:tffpqcd} to derive the high energy behavior for the pseudoscalar transition form 
factors (TFFs). In contrast, at low energies $\alpha_s$ increases, leading to a non-perturbative behavior and a strong-coupling regime, which is thought to be responsible 
for confinement, this is, the fact that free quarks and gluons are not observed in nature; instead, they bind together to form color-singlet states known as hadrons ---the 
pions and proton among them. 
It must be emphasized at this point that confinement cannot be strictly explained on the basis of \cref{eq:asympf}, which is based on a perturbative calculation. 
Indeed, describing confinement represents a still unsolved major theoretical challenge in mathematical physics as formulated for instance by the Clay Math institute~\cite{ClayInst}. 
Describing QCD at low-energies therefore represents a formidable task. So far, a first principles calculation based on \cref{eq:QCDlag} has only been achieved through 
Lattice QCD~\cite{PDG:lqcd}, an expensive computational numerical method based on the ideas from K.~Wilson~\cite{Wilson:1974sk} consisting  in a four dimensional 
euclidean space-time discretization of the QCD action. Additional, Dyson-Schwinger equations provides for a continuum non-perturbative approach to quantum field theories, 
which have been solved within some further approximation schemes. 
However, even if lattice calculations have shown a tremendous progress in the recent years, not all type of observables are at present accessible in lattice QCD. 
Furthermore, they are extremely costly and require some guidance 
when performing the required extrapolations. A viable and successful analytic approach comes by the hand of \cpt, the low-energy effective field theory of QCD.

\section{Low energy QCD: \cpt}
\label{sec:cptintro}

At the Lagrangian level, \cref{eq:QCDlag} is invariant by construction under Lorentz and local $SU(3)_c$ transformations. 
\cref{eq:QCDlag} is invariant too under the discrete charge conjugation $(C)$, parity $(P)$, and time reversal $(T)$ transformations. 
In addition, there exists on top an almost-exact accidental symmetry which is not obvious 
or explicit in the construction, this is, the chiral symmetry; using the left-handed $P_L=\frac{1-\gamma_5}{2}$ and right-handed $P_R=\frac{1+\gamma_5}{2}$ projectors, 
the QCD Lagrangian may be written as\footnote{Most of the notations and concepts in this section are taken from Ref.~\cite{Scherer:2012xha}.} 
\begin{equation}
\mathcal{L}_{\textrm{QCD}} =    i\overline{q}_{L}\slashed{D}q_{L}  +   i\overline{q}_{R}\slashed{D}q_{R}   -
  \overline{q}_{L}\mathcal{M} q_{R}  -\overline{q}_{R}\mathcal{M} q_{L}   -\frac{1}{4}G^c_{\mu\nu}G^{c,\mu\nu},
\end{equation}
where $\mathcal{M}=\operatorname{diag}(m_u,m_d,m_s,m_c,m_b,m_t)$ and $q_{L(R)} = P_{L(R)}q$ with $q=(u,d,s,c,b,t)^T$. If the 
quark masses were left apart, the Lagrangian would be symmetric as well under the chiral global\footnote{Global means that, unlike in gauge theories, 
$U_{L,R}\neq U_{L,R}(x)$, i.e., the transformation does not depend on the space-time coordinate.} transformations $q_{L(R)}\to U_{L(R)}q_{L(R)}$, where 
$U_L\neq U_R$ represents a unitary matrix in flavor space, mixing then different flavors. This is, QCD does not distinguish among chiral quark flavors.
Whereas the massless quark limit would represent a bad approximation for the heavy ($c$, $b$, $t$) quarks, this is not the case for the light 
($u$, $d$, $s$) ones; the fact that the light hadrons are much heavier than the light quark masses points that 
the light quark masses should have little, if anything, to do with the mechanism conferring light hadrons their masses. The origin of the latter should be traced back to 
confinement, and is responsible for generating most of the visible particle masses in the universe. Chiral symmetry should be therefore a good approximation for the light-quarks sector. 

Consequently, at the low energies where the heavy quarks do not play a role, we should find an approximate $U(3)_L \times U(3)_R$ symmetry. 
Through the use of Noether theorem, this would imply a set of $18$ conserved currents and associated charges. These are conveniently 
expressed in terms of the vector and axial currents 
\begin{equation}
\label{eq:Icurrents}
J_{\mu}^a=L_{\mu}^a+R_{\mu}^a = \overline{q}\gamma_{\mu}\frac{\lambda^a}{2} q  \qquad  \qquad   J_{5\mu}^a=R_{\mu}^a - L_{\mu}^a = \overline{q}\gamma_{\mu}\gamma_5\frac{\lambda^a}{2} q,
\end{equation}
where $\lambda^a/2$ are the group generators. There is an octet corresponding to the eight $\lambda^a$ Gell-Mann matrices and a singlet which, for later convenience, we define 
as $\lambda^{0}=\sqrt{2/3}~\mathds{1}_{3\times3}$. The symmetry group may be rewritten then as $U(3)_L \times U(3)_R = U(1)_V\times U(1)_A\times SU(3)_V \times SU(3)_A$.
However, the previous symmetry group holds only at the classical level; quantum corrections break the axial $U(1)_A$ symmetry. Precisely, the axial current divergence is given 
as~\cite{Peskin:1995ev} 
\begin{equation}
\label{eq:anom}
\partial^{\mu}J_{5\mu}^a = \left\{ \mathcal{P}^a,\mathcal{M}\right\} -\frac{g_s^2 }{16\pi^2} \epsilon^{\alpha\beta\mu\nu}G^b_{\alpha\beta}G^c_{\mu\nu} \operatorname{tr}\left( \frac{\lambda^a}{2} t^b t^c \right),   
\end{equation}
where the pseudoscalar current $\mathcal{P}^a=\overline{q}i\gamma_5\frac{\lambda^a}{2}q$ has been used, $t^{b,c}$ are the $SU(3)_c$ generators associated to the strong 
interactions and $\lambda^a/2$ those associated to the chiral transformations. For $SU(3)_{A}$, the associated generators are traceless matrices in flavor space, producing a 
vanishing trace for the rightmost term; this contrasts with the (flavor singlet) $U(1)_A$ transformations, which generator is proportional to the unit matrix in flavor 
space\footnote{Note that $\operatorname{tr}(t^at^b)=(1/2)\delta^{ab}$.}. Consequently, the singlet axial current is not 
conserved even in the chiral limit of vanishing quark masses $\mathcal{M}=0$; it is called therefore an anomalous symmetry. 

All in all, at the quantum level we should have an approximate $U(1)_V\times SU(3)_V\times SU(3)_A$ symmetry. The $U(1)_V$ symmetry is related to the baryon number 
conservation in the SM and is as important as to forbid the proton decay. The $SU(3)_V$ symmetry would imply the existence of degenerate-mass flavor multiplets 
in the hadronic spectrum, whereas the $SU(3)_A$ would imply analogous multiplets with opposite parity. 
However, the latter is not realized in nature: degenerate opposite parity multiplets are not found, indicating that the axial symmetry is spontaneously broken. 
This is thought to be related to the fact that, whereas the QCD Lagrangian is invariant under these 
transformations, the vacuum of the theory is not ---the complex structure of the QCD vacuum is thought to be the ultimate responsible for the spontaneous breaking of the chiral symmetry. 
An important consequence of this feature comes by the hand of Goldstone's theorem. Goldstone's theorem dictates that, whenever a global symmetry is spontaneously broken, massless 
goldstone bosons with the quantum number of the broken generators appear. In nature, it seems that the symmetry breaking pattern is 
$U(1)_V\times SU(3)_V\times SU(3)_A \to U(1)_V\times SU(3)_V$ and 8 pseudoscalar Goldstone bosons should appear in correspondence with the 8 broken $SU(3)_A$ generators. 
In nature, there are no massless particles to which such hypothetical states could be associated to. There exists however, an octet of pseudoscalar particles much lighter than the  
standard mesons. These are the $\pi$'s, $K$'s and $\eta$ mesons. It is believed that in the chiral limit $m_{u,d,s}\to 0$ such particles would correspond to the massless Goldstone 
bosons of the spontaneously broken chiral symmetry. In the real world, the explicit symmetry breaking by the quark masses is thought to give masses to these mesons, which are dubbed 
as pseudo-Goldstone bosons. 

Before proceeding to describe \cpt, it is worth to take a brief detour anticipating some of the consequences of the large-$N_c$ limit of QCD. Particularly, we are interested in 
the $U(1)_A$ axial anomaly. As we will comment in \cref{sec:largeN}, 't Hooft showed that in the large-$N_c$ limit, the strong coupling constant 
should be replaced as $g_s\to \tilde{g}_s/\sqrt{N_c}$, where $\tilde{g}_s$ is to be fixed as $N_c\to \infty$~\cite{'tHooft:1973jz}. Consequently, in the chiral limit 
($\mathcal{M}\to0$), \cref{eq:anom} reads
\begin{equation}
\partial^{\mu}J_{5\mu}^a = -\frac{\tilde{g}_s^2 }{16\pi^2N_c} \epsilon^{\alpha\beta\mu\nu}G^b_{\alpha\beta}G^c_{\mu\nu} \operatorname{tr}\left( \frac{\lambda^a}{2} t^b t^c \right) 
\xrightarrow{N_c\to\infty} 0,   
\end{equation}
and the singlet axial current is conserved too as $N_c\to\infty$. In such limit, the $U(1)_A$ anomalous symmetry would be recovered, and the spontaneously breaking of the chiral 
symmetry would come with an additional Goldstone boson, the $\eta'$. Consequently, considering $N_c$ as a parameter large enough, the $\eta'$ could be 
incorporated to the \cpt Lagrangian in a combined chiral and large-$N_c$ expansion, which is known as large-$N_c$ chiral perturbation theory ($\ell N_c\chi\textrm{PT}$). 
More formal arguments for the vanishing $\eta'$ mass in the large-$N_c$ chiral limit can be found in Ref.~\cite{Witten:1979vv}.

\subsection{The chiral expansion}

In the chiral limit of QCD, we believe in the existence of 8 masless Goldstone bosons associated to the breaking of the chiral 
symmetry ---9 if the large-$N_c$ limit is considered. Above this, there is a mass gap below the intrinsic scale that 
is generated in QCD, call it $\Lambda_{\chi}$, 
where the full zoo of hadronic particles appears. 
Quantitatively, this spectrum starts around $0.5$~GeV for scalar mesons, around $0.8$~GeV for vector mesons and around $1$~GeV for baryons. 
This situation calls for an effective field theory description of QCD at low-energies, in which the heavy hadrons above $\Lambda_{\textrm{QCD}}$ are 
integrated out from the theory, which is effectively described in terms of the relevant degrees of freedom, the Goldstone bosons. 
The effect of the physics above $\Lambda_{\chi}$ are encoded in a plethora of terms appearing in the effective Lagrangian 
---actually, as many of them as the underlying symmetries allow to include. 
Of course, writing down the most general effective Lagrangian allowed by the assumed symmetry principles of the 
theory represents a formidable ---if not impossible--- task, as it contains an infinite number of terms. The second ingredient for constructing a useful effective field theory 
is the presence of an expansion parameter, according to which only a finite number of terms is required in order to achieve a 
prescribed precision. For effective field theories of spontaneously broken symmetries, this is an expansion in terms of small momenta
$p^2/\Lambda_{\chi}^2$. In the real world, the small quark masses are non-zero, giving mass to the pseudo-Goldstone 
bosons. Still, these are much smaller than $\Lambda_{\chi}$, which allows to systematically incorporate additional terms 
accounting for the explicit symmetry breaking as an expansion in terms of $m_q/\Lambda_{\chi}$ 
---\cpt is therefore an effective field theory description of QCD in terms of small momenta and quark 
masses.

The theoretical framework to describe such theories was initiated 
by Weinberg~\cite{Weinberg:1966fm}, Coleman, Wess and Zumino~\cite{Coleman:1969sm} and in collaboration with Callan in~\cite{Callan:1969sn}.  
It generally implies that the Goldstone boson fields, $\phi(x)$, transform non-linearly upon the symmetry group; they are described then in terms of the $U(x)$ matrix
\begin{equation}
U(x) = \operatorname{exp}\left( \frac{i\phi(x)}{F} \right) = 1 +i\frac{\phi(x)}{F} + ... 
\end{equation}
with $F$ a parameter required to obtain a dimensionless argument 
and $\phi(x)$ the matrix associated to the Goldstone bosons
\begin{equation}  
\label{eq:cptG}
\phi(x)  \! =  \! \sum_{a=1}^8 \phi^a\lambda^a  \! = \! 
\left(
   \begin{array}{ccc}
      \pi^0 \! + \! \frac{1}{\sqrt{3}}\eta   &   \sqrt{2}\pi^+   &   \sqrt{2}K^+ \\
      \sqrt{2}\pi^-     & -\pi^0\!   + \! \frac{1}{\sqrt{3}}\eta   & \sqrt{2}K^0 \\
      \sqrt{2}K^-       & \sqrt{2}\bar{K}^0   & -\frac{2}{\sqrt{3}}\eta \\
   \end{array}
\right),
\end{equation}
which serves as a building block of the theory. In this way, one can write the most general Lagrangian according to the powers of momentum $p^n$ ---what is equivalent, the 
number of derivatives $\partial^n$--- and powers of the quark masses (accounting that $m_q\sim p^2$). Due to Lorentz 
invariance, derivatives appears in even numbers, $2n$, leading to the decomposition
\begin{equation}
\label{eq:cptcount}
 \mathcal{L} =  \mathcal{L}_2 + \mathcal{L}_4 + \mathcal{L}_6 + ... + \mathcal{L}_{2n} + ... \ .
\end{equation}
In addition, any of the pieces displayed above produces an infinite number of contributions ---Feynman diagrams--- to some given specific process. Consequently, an additional scheme 
classifying these pieces according to their relevance is required. This is achieved using Weinberg's power counting~\cite{Weinberg:1978kz}, which assigns a chiral dimension $D$ to every 
amplitude $\mathcal{M}$ (see \cref{app:conv}) arising from a particular diagram according to its properties upon momenta, $p$, and quark masses, $m_q$, scaling 
\begin{equation}
\label{eq:scaling}
\mathcal{M}(p_i,m_q) \to \mathcal{M}(t p_i, t^2m_q) = t^D\mathcal{M}(p_i,m_q).
\end{equation} 
The final result is given, in four space-time dimensions, in terms of the number of internal pseudo-Goldstone boson propagators, $N_I$, number of loops, $N_L$, and the number of 
vertices $N_{2k}$ from $\mathcal{L}_{2k}$ (see \cref{eq:cptcount}) as 
\begin{equation}
  D = 4N_L - 2N_I + \sum_{k=1}^{\infty} 2kN_{2k}.
\end{equation}

\subsection{Leading order Lagrangian}

The most general Lagrangian at leading order, $\mathcal{L}_2$, reads~\cite{Gasser:1983yg,Gasser:1984gg}
\begin{equation}
\mathcal{L}_2 = \frac{F^2}{4} \operatorname{tr}\left( D_{\mu}UD^{\mu}U^{\dagger}  \right)  +   \frac{F^2}{4} \operatorname{tr}\left( \chi U^{\dagger} + U\chi^{\dagger} \right),  
\end{equation} 
where $F$ is known as the pion decay constant in the chiral limit due to its relation at LO with the $\pi^{\pm}$ decay. 
The covariant derivative is defined as 
\begin{equation}
\label{eq:extvecax}
D_{\mu}U = \partial_{\mu}U - ir_{\mu}U + iUl_{\mu} = \partial_{\mu}U - i\left[ v_{\mu} , U \right]  - i\left\{ a_{\mu} , U \right\} 
\end{equation}
and allows to couple the pseudo-Goldstone bosons to external left ($l_{\mu}$) and right ($r_{\mu}$) handed ---alternatively vector ($v_{\mu}$) and axial ($a_{\mu}$)--- 
currents. 
Finally $\chi = 2B(s+ip)$, where $B$ is related to the quark condensate $\langle \bar{q}{q} \rangle_0$ in the chiral limit  and $s(p)$ are the 
external (pseudo)scalar currents\footnote{The elements $v_{\mu},a_{\mu},s,p$ are defined in terms of generating functional external currents  
$\mathcal{L}_{\textrm{ext}} = v_{\mu}^a \bar{q}\gamma^{\mu}\frac{\lambda^a}{2} q  +  a_{\mu}^a \bar{q}\gamma^{\mu}\gamma_5\frac{\lambda^a}{2} q   
-s^a\bar{q}\lambda^a q + p^a\bar{q}i\gamma_5 \lambda^a q \equiv\bar{q}\gamma^{\mu}( v_{\mu} +\gamma_5 a_{\mu})q - \bar{q}(s-i\gamma_5p){q} $ . }. 
This allows to introduce the finite quark masses 
effects via $s\to \mathcal{M}=\operatorname{diag}(m_u,m_d,m_s)$. 

The most general Lagrangian construction at the next order, $\mathcal{L}_4$, was discussed in the 
seminal papers from Gasser and Leutwyler~\cite{Gasser:1983yg,Gasser:1984gg}.

Finally, the large-$N_c$ limit allows to include the $\eta'$ as a ninth degree of freedom, giving birth to \lcpt, a low-energy description 
of QCD in terms of small momenta, quark masses and the large number of colors. In this framework, the expansion parameters are 
$p^2\!\sim\! m_q\!\sim\! N_c^{-1}\!\sim\! \mathcal{O}(\delta)$ and the expansion reads
\begin{equation}
\label{eq:lcptcount}
 \mathcal{L} =  \mathcal{L}^{(0)} + \mathcal{L}^{(1)} + \mathcal{L}^{(2)} + ... + \mathcal{L}^{(\delta)} + ... \ .
\end{equation}
In addition, the $N_c$ scaling has to be incorporated to \cref{eq:scaling}. As a result, it can be obtained among others that $F\sim\mathcal{O}(N_c^{1/2})$, or 
that loop processes as well as additional flavor traces are $N_c$-suppressed in this framework. 
The leading order Lagrangian is given as~\cite{Kaiser:2000gs}
%
\begin{equation}
\label{eq:lcptLO}
\mathcal{L}^{(0)} = \frac{F^2}{4} \operatorname{tr}\left( D_{\mu}UD^{\mu}U^{\dagger} \right) + \frac{F^2}{4} \operatorname{tr}\left( \chi U^{\dagger} + U\chi^{\dagger} \right) 
      -\frac{1}{2}\tau \left( \psi + \theta \right)^2,
\end{equation} 
where \cref{eq:cptG} is to be replaced by 
\begin{equation}  
\phi(x)  \! =  \! \sum_{a=0}^8 \phi^a\lambda^a  \! = \! 
\left(
   \begin{array}{ccc}
      \pi^0 \! + \! \frac{1}{\sqrt{3}}\eta +\frac{F}{3}\psi   &   \sqrt{2}\pi^+   &   \sqrt{2}K^+ \\
      \sqrt{2}\pi^-     & -\pi^0\!   + \! \frac{1}{\sqrt{3}}\eta +\frac{F}{3}\psi  & \sqrt{2}K^0 \\
      \sqrt{2}K^-       & \sqrt{2}\bar{K}^0   & -\frac{2}{\sqrt{3}}\eta  +\frac{F}{3}\psi \\
   \end{array}
\right)
\end{equation}
with $\lambda^0=\sqrt{2/3}~\mathds{1}_{3\times3}$ and $\psi\equiv \frac{\sqrt{6}}{F}\phi^0$, being $\phi^0$ the field to be related to the singlet 
Goldstone boson in the chiral large-$N_c$ limit. The $\tau$ term in \cref{eq:lcptLO} is 
connected with the $\bra{0} T\{ \omega(x)\omega(0) \} \ket{0}$ two-point function in the pure gluonic theory\footnote{The winding number density is defined as 
$\omega = -\frac{g_s^2}{32\pi^2}\epsilon^{\mu\nu\rho\sigma}G^c_{\mu\nu}G^c_{\rho\sigma} = -\frac{\alpha_s}{4\pi}G^c\tilde{G}^c$ with the dual tensor 
$\tilde{G}^{c,\mu\nu} = \frac{1}{2}\epsilon^{\mu\nu\rho\sigma}G^c_{\rho\sigma}$.}and $\theta$, the vacuum angle~\cite{Kaiser:2000gs}, represents an external 
current ---similar to the $2Bs$ term in $\chi$.
\\


\subsection{The effective Wess-Zumino-Witten action}

The Lagrangians described above ---including the higher order in \cref{eq:cptcount,eq:lcptcount}--- can be shown to be invariant under $\phi\to-\phi$ transformations 
if no external currents are considered, meaning that they always contain interactions with an even number of pseudo-Goldstone bosons.
This remains the case even if vector currents are included in the formalism. The preceding Lagrangians cannot describe the $\pi^0\to\gamma\gamma$ and related decays. 
The $\pi^0\to\gamma\gamma$ decay has indeed been a fascinating process in the history of particle physics, the underlying mechanism driving this decay remaining a mystery  
until the independent discovery of the anomalies ---the breaking of classical symmetries in QFT--- in 1969 by Adler~\cite{Adler:1969gk} and Bell-Jackiw~\cite{Bell:1969ts}  
(ABJ anomaly). The ABJ anomaly can be used then to predict the $\pi^0\to\gamma\gamma$ decay in the chiral limit of QCD ---see for instance~\cite{Goity:2002nn}. 
The systematic incorporation of anomalies 
into chiral Lagrangians is accomplished by the use of the Wess-Zumino-Witten (WZW) action~\cite{Wess:1971yu,Witten:1983tw}, which introduces additional terms involving 
an odd number of Goldstone bosons as well as terms such as $\phi\gamma\gamma, \phi^3\gamma$, etc (see Ref.~\cite{Scherer:2012xha,Kaiser:2000gs}). 
For our case of interest, we refer to the leading term inducing $P\to\gamma\gamma$ decays~\cite{Leutwyler:1997yr}
%
\begin{equation}
\label{eq:cptwzw}
\mathcal{L}_{\textrm{WZW}} = \frac{N_c\alpha}{8\pi}\epsilon^{\mu\nu\rho\sigma}F_{\mu\nu}F_{\rho\sigma} \operatorname{tr}\left( \mathcal{Q}^2\phi \right)
\end{equation}
which is valid both, for \cpt and \lcpt, where it appears at order $\mathcal{L}_4$ and $\mathcal{L}^{(1)}$, respectively. In the expression above, 
$\mathcal{Q}=\operatorname{diag}(2/3,-1/3,-1/3)$ is the charge operator and $\phi = \lambda^a\phi^a$.
\\

\subsection{Basic leading order results: masses and decay constants}

As an example, we outline here the LO results for the pseudoscalar masses and decay constants in \lcpt. 
From the Lagrangian \cref{eq:lcptLO}, and taking $\chi\to2B\mathcal{M}$, we obtain for the kinetic terms at LO
\begin{align}
\mathcal{L}^{(0)}_{\textrm{kin}} =& \  \partial_{\mu}\pi^+\partial^{\mu}\pi^-  - 2B\hat{m} \pi^+\pi^- + \frac{1}{2}\left(  \partial_{\mu}\pi^0\partial^{\mu}\pi^0  - 2B\hat{m} \pi^0\pi^0\right)   \nonumber\\
&+ \partial_{\mu}K^+\partial^{\mu}K^-  +  \partial_{\mu}\bar{K}^0\partial^{\mu}K^0   -  B(\hat{m}+m_s)\left( K^+K^- + \bar{K}^0K^0 \right)\nonumber\\
&+   \frac{1}{2}\left( \partial_{\mu}\eta_8\partial^{\mu}\eta_8   - B\left( \tfrac{2\hat{m}+4m_s}{3} \right) \eta_8\eta_8 \right) 
 - \frac{1}{2} (\eta_8\eta_0 + \eta_0\eta_8) \tfrac{2\sqrt{2}}{3}(\hat{m}-m_s) \nonumber\\
& +  \frac{1}{2}\left(  \partial_{\mu}\eta_0\partial^{\mu}\eta_0 - B\left( \tfrac{4\hat{m}+2m_s}{3}\right)\eta_0\eta_0 -\tfrac{6\tau}{F^2}\eta_0\eta_0  \right).  \label{eq:cptmass} 
\end{align}
In the expression above, the isospin-symmetric limit $m_u=m_d\equiv \hat{m}$ has been used. \cref{eq:cptmass} allows to identify the pions and kaons 
masses at LO
\begin{equation}
\label{eq:toppmass}
  m_{\pi^{\pm}}^2 = m_{\pi^0}^2 \equiv \mathring{M}_{\pi}^2  = 2B\hat{m}   \qquad     m_{K^{\pm}}^2 = m_{K^0}^2   \equiv \mathring{M}_{K}^2 = B(\hat{m}+m_s).
\end{equation}
The $\eta$ and $\eta'$ masses require additional work since the terms from the third line in \cref{eq:cptmass} are non-diagonal, leading to the $\eta-\eta'$ mixing. 
This will be discussed in more detail in \cref{chap:mixing}. For the moment, let us note that in standard \cpt $\eta=\eta_8$, which receives mass from the 
quarks alone. The singlet component $\eta_0$ acquires a large topological mass $M_{\tau}^2=6\tau/F^2$ absent in the octet terms.

Finally, we define the pseudoscalar decay constants, which are of major interest for discussing the $\eta-\eta'$ mixing in \cref{chap:mixing} 
as well as for calculating 
new physics contributions to $P\to\bar{\ell}\ell$ decays in \cref{chap:PLL}, where $P=\pi^0,\eta,\eta'$. 
The pseudoscalar decay constants are defined in terms of the matrix elements of the pseudoscalars with the axial current
\begin{equation}
\label{eq:Fp}
\bra{0} J_{5\mu}^a \ket{P(p)} \equiv ip_{\mu}F_P^a, \qquad  J_{5\mu}^a = \bar{q}\gamma_{\mu}\gamma_5\frac{\lambda^a}{2}q.
\end{equation}
They can be obtained at LO from \cref{eq:lcptLO} taking an external axial current $a_{\mu}\equiv a_{\mu}^a\frac{\lambda^{a}}{2}$, see \cref{eq:extvecax}. The relevant term reads
\begin{equation}
-F \operatorname{tr} \left( \partial^{\mu}\phi~a_{\mu}  \right) = -\frac{F}{2} \operatorname{tr} \left( \partial^{\mu}\phi \lambda^a   \right)a_{\mu}^a .
\end{equation}
Identifying $\lambda^a$ with the relevant $SU(3)$ matrix, i.e., $\lambda^3$ for the $\pi^0$, one obtains in \cpt that 
$F_{\pi^{\pm}}=F_{\pi^0}=F_{K^{\pm}}=F_{K^0} = F_{\eta} \equiv F$. In \lcpt, the $\eta-\eta'$ mixing makes this picture more complicated for the $\eta$ and $\eta'$ mesons.

The success obtained in \cpt at higher orders (state of the art is $\mathcal{O}(p^6)$) in predicting different observables shows a good performance 
of the theory, which is to day our best tool to produce analytical calculations for low-energy hadronic physics. 
Still, the theory is not expected to be valid above some scale, often defined as $\Lambda_{\chi} \equiv 4\pi F$, which is below the pQCD applicability range. 
For a particular process, the natural scale at which one can expect a poor performance is given by the closest relevant hadronic resonance which 
has not been included in the theory as an active degree of freedom.
Unfortunately, this avoids to match the theory with pQCD.

\section{Closing the gap: large-$N_c$ QCD}
\label{sec:largeN}

Describing all the QCD phenomenology with its great complexity represents a challenging task. 
The complex analytic structure which QCD requires ---think about reproducing all nuclear physics as a part--- makes an analytic description 
nonviable. Consequently, so far, only perturbative expansions have reached success in analytically describing particular sectors of QCD, but the lack of an apparent perturbative parameter 
of the theory at all scales avoids the whole QCD description within a single framework.
However, 't Hooft pointed out that there might be such a candidate for an expansion parameter in QCD, this is, the limit of large number of colors, 
large $N_c$~\cite{'tHooft:1973jz}. Its phenomenological success and the fact that it is the only framework justifying some known features of QCD, such as Regge theory or the OZI rule 
among others, makes this approximation to QCD very useful even if so far it only produces a qualitative picture of QCD rather than a quantitative one\footnote{This introduction 
is mainly based on Refs.~\cite{Witten:1979kh,Lebed:1998st,Manohar:1998xv}.}.\\

The large-$N_c$ limit of QCD is based on the combinatorics $SU(N_c)$ group factors arising in diagrammatic calculations. Recall for instance the RG equation 
for the strong coupling constant $\alpha_s$ in \cref{eq:asympf}. There, $N_c$ plays a relevant role in the leading coefficient for the $\beta$-function 
$\beta_0= \frac{1}{3}\left(11N_c - 2n_f \right)$. In the large-$N_c$ limit, the first part dominates. 
Actually, if a smooth and non-trivial behavior is desired in such a limit, the strong coupling constant should be taken as $g_s\to\bar{g}_s/\sqrt{N_c}$, where $\bar{g}_s$ is kept fixed as 
$N_c\to\infty$. Then, the RG equation for $\bar{\alpha}_s\equiv \bar{g}_s^2/4\pi$ would resemble that in \cref{eq:asympf} with $\beta_0$ defined as   
$\beta_0= \frac{1}{3}\left(11 - \frac{2}{N_c}n_f \right)$ ---otherwise, $\bar{g}_s$ would tend to $0$ inducing a trivial theory\footnote{In addition, this guarantees that the induced QCD 
scale, $\Lambda_{\textrm{QCD}}$, as well as the hadron masses, remain $N_c$-independent. }. 
This means that any interacting process in the large-$N_c$ limit will not survive unless the combinatoric factors of the relevant diagrams are large enough to compensate for the 
$\bar{g}_s/\sqrt{N_c}$ factors. It turns out that only a certain class of diagrams, which can be classified according their topology, survive in this limit (in the purely gluonic theory these 
are the so called planar diagrams). To figure this out, it is convenient to employ the color-line notation introduced by 't Hooft~\cite{'tHooft:1973jz}, 
according to which the quarks propagators can be illustrated as color lines, 
the gluons propagators as color-anticolor lines and a similar representation holds for the vertices, see \cref{fig:cline}.
\begin{figure}
\centering
    \includegraphics[width=0.8\textwidth]{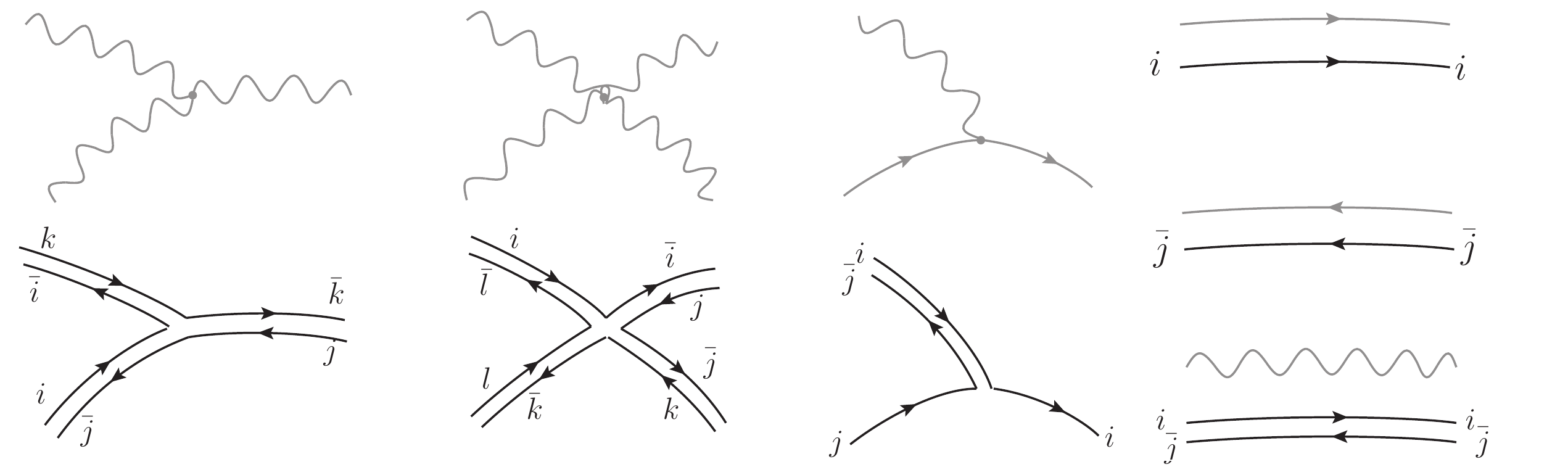}
    \caption{The different QCD vertices and propagators (gray) in the color-lines notation (black). The indices $i,j,k,l$ stand for color indices.\label{fig:cline} }
\end{figure}

As an example, we show in \cref{fig:vpolcl} different contributions to the vacuum polarization appearing in the $\alpha_s$ running together with their $N_c$ counting. 
From those diagrams, only the first and second ones have a combinatoric factor arising from closed color lines large enough to counteract the vertices suppression; 
the third and fourth are suppressed 
with respect to the previous ones by factors of $N_c^{-2}$ and $N_c^{-1}$, respectively. The leading diagrams belong to the so called planar diagrams. In contrast to the 
third one, they can be drawn in such a way that 
color lines do not cross each other and are leading in the large-$N_c$ expansion. Contrary, non-planar diagrams and quark loops are $N_c^{-2}$ and $N_c^{-1}$ suppressed, 
respectively.
\begin{figure}
\centering
    \includegraphics[width=\textwidth]{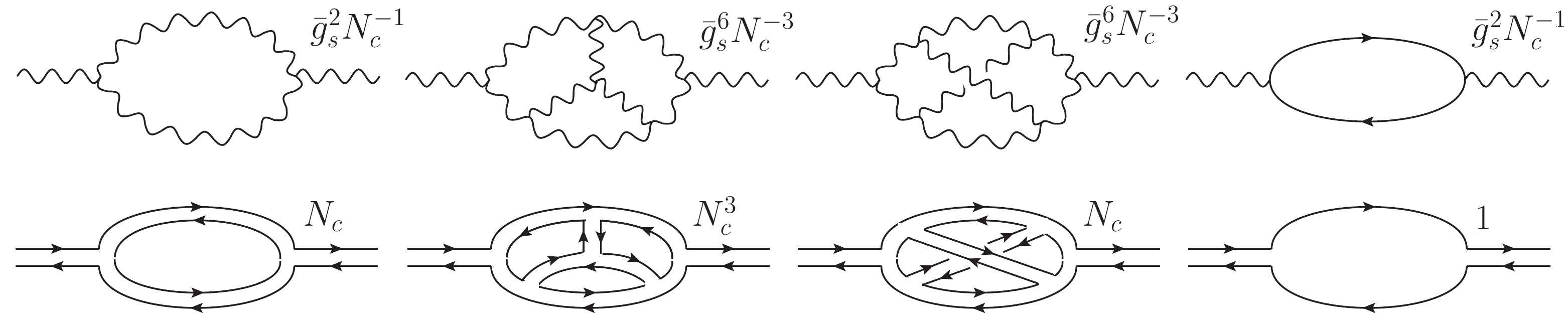}
    \caption{Examples of diagrams contributing to the gluon self-energy. Upper graphs show the vertex suppression, $\bar{g}_s/\sqrt{N_c}$, and the lower ones the combinatoric 
    $SU(N_c)$ enhancement arising from closed color lines $\sim N_c$.\label{fig:vpolcl} }
\end{figure}

Therefore, in order to obtain the gluon self-energy, it would be sufficient, at leading order in the large-$N_c$ expansion, to take the planar diagrams contributions. 
The resummation of all the planar diagrams has only been achieved so far in a $1+1$ space-time dimensions~\cite{'tHooft:1974hx}. Therefore, it is 
difficult to obtain a quantitative answer in large $N_c$. Still, it is possible to obtain a qualitative picture for a variety of QCD phenomena. In this thesis, it is of interest what 
concerns Green's functions involving $\bar{q}\Gamma q$ bilinear currents, where $\Gamma$ is a Dirac bilinear matrix. It turns out that planarity is not enough then. For the case 
of bilinear currents, {\textit{the leading diagrams are the planar diagrams with only a single quark loop which runs at the edge of the diagram}}~\cite{Witten:1979kh}. 
To see this, we refer to \cref{fig:bilin}, where crosses refer to bilinear currents insertions. The first diagram is of order $N_c$; the second, with a gluon at the edge, 
is $N_c^{-2}$ suppressed; the third one, with an internal quark loop, is $N_c^{-1}$ suppressed; the fourth is $N_c^{-1}$ suppressed.
\begin{figure}
\centering
    \includegraphics[width=0.8\textwidth]{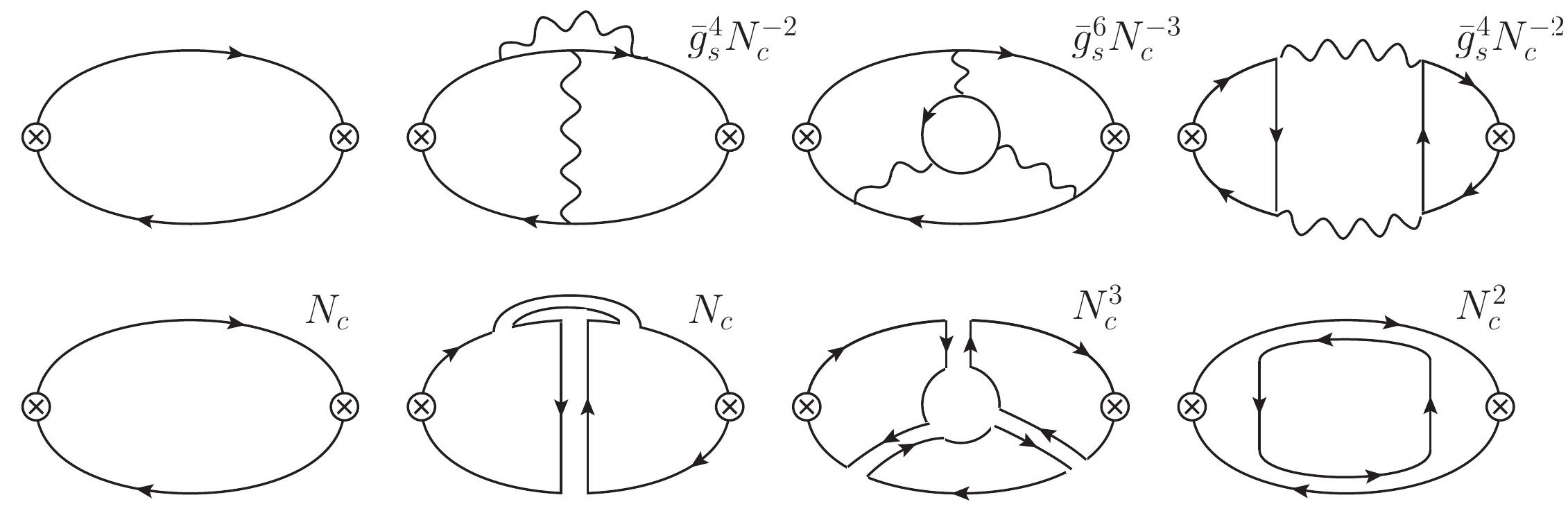}
    \caption{Different contributions to quark bilinear correlation functions, where the insertion is marked by a cross.  Upper graphs indicate the strong-coupling suppression and the 
    lower ones the combinatoric $SU(N_c)$ enhancement.\label{fig:bilin} }
\end{figure}
\\

These observations have far reaching consequences once confinement is assumed: take a typical leading diagram such as that in \cref{fig:singlestate} and cut it through to 
search for possible intermediate states, this is, intermediate quarks and gluons color singlet combinations. First of all, as quark loops are $N_c$ suppressed, any intermediate 
state contains one and only one $q\bar{q}$ pair. 
Second, a closer look to \cref{fig:singlestate} reveals that it is not possible to have two or more singlet configurations, say $q\bar{q}$ and some gluonic state 
---more precisely, these configurations are $N_c$ suppressed.  In conclusion, all the quarks and gluons must bind together to form one particle color-singlet states;
the diagram in \cref{fig:singlestate} represents thereby a perturbative approximation to a single hadron. As a conclusion, bilinear two-point correlation functions ---such as the vacuum 
polarization--- can be expressed in terms of single particle intermediate meson states with the appropriate quantum numbers:
\begin{equation}
\frac{1}{i}\int d^4x e^{i k\cdot x}\bra{0} T\{  J(x) J(0) \} \ket{0} \equiv \langle J(k)J(-k)\rangle = \sum_n \frac{a_n^2}{k^2 - m_n^2 + i\varepsilon},
\end{equation}
where the meson masses, $m_n$, are $N_c$ independent.
Furthermore, it is known that, in the perturbative regime, such function behaves logarithmically, requiring then an infinite number of mesons. In the large-$N_c$ limit, 
correlation functions are given in terms of an infinite sum of narrow-width (stable) meson states. In addition, since the correlation function is of order 
$N_c$, $a_n = \bra{0} J\ket{n} = \sqrt{N_c}$.
\begin{figure}
\centering
    \includegraphics[width=\textwidth]{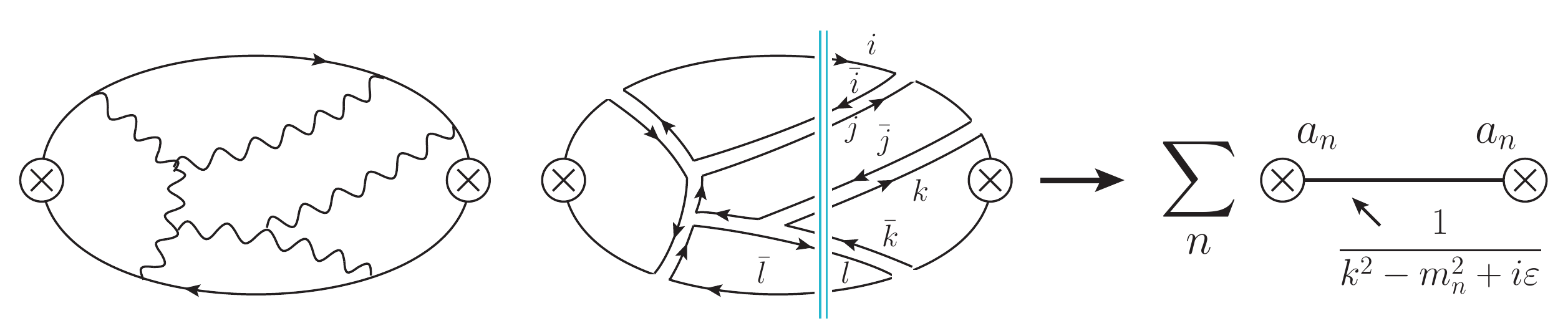}
    \caption{A typical $N_c$-leading contribution to a bilinear two-point function. Multiple singlet color intermediate states cannot appear at the leading order. Two-point functions 
    can be understood then in the large-$N_c$ limit as a sum over single meson states (right).\label{fig:singlestate} }
\end{figure}

The same reasoning can be extrapolated to higher order correlation functions for quark bilinears. As an example, we illustrate this for the three- and four-point 
functions\footnote{Actually, these are of relevance for this thesis, as the pseudoscalar transition form factors can be defined in terms of the $\langle VV\mathcal{P} \rangle$ 
Green's function and the hadronic light-by-light tensor is related to the $\langle VVVV\rangle $ one.} in \cref{fig:npoint}. Again, the only possible intermediate color singlet 
states are single particles ---multiparticle states being $N_c$ suppressed. In addition, the large-$N_c$ counting allows to obtain that the three(four) meson vertex 
is $1/\sqrt{N_c}(1/N_c)$ suppressed, leading to a large-$N_c$ estimate for the meson decay widths.
In general terms, it is found that any Green's function must contain, at the leading order,  single-pole contributions alone ---multiparticle states are suppressed. 
Actually, using crossing symmetry and unitarity arguments, the large-$N_c$ limit implies that, at the leading order, every amplitude can be expressed as if arising from the tree 
level calculation from some local Lagrangian with the following properties: 
%
%
%
\begin{figure}
\centering
    \includegraphics[width=\textwidth]{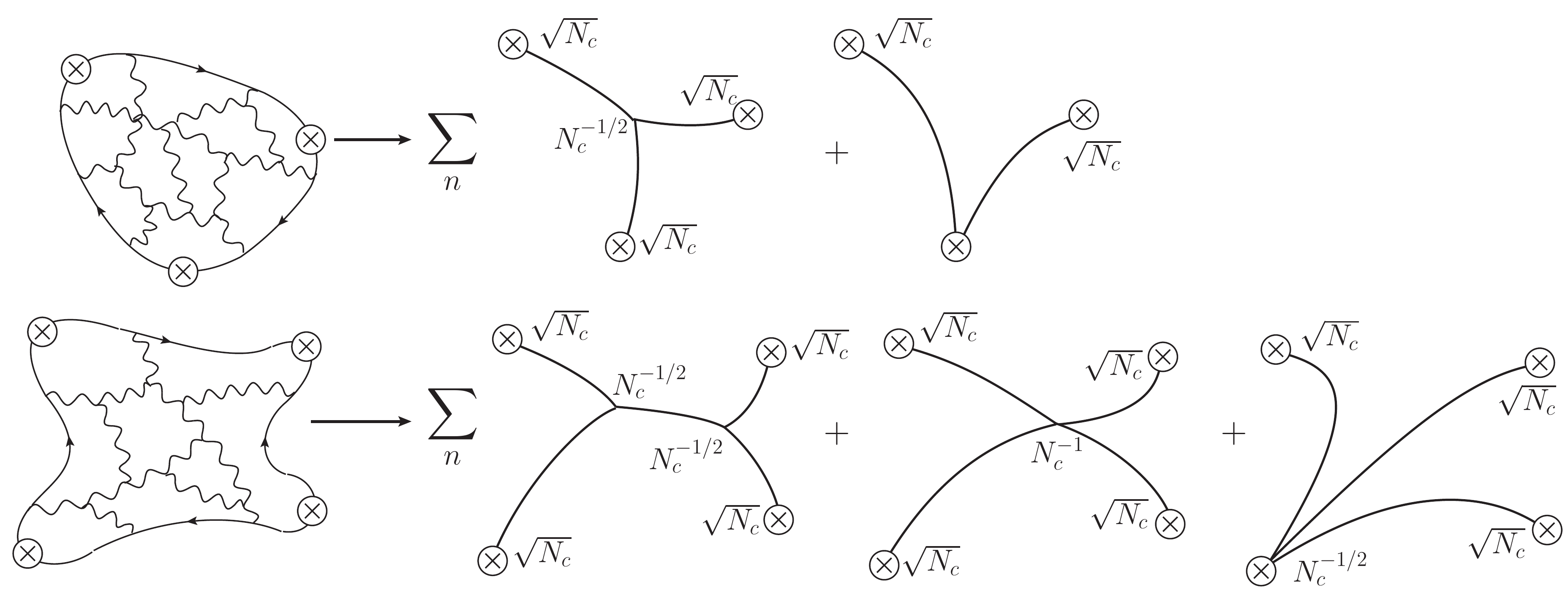}
    \caption{A typical $N_c$-leading contribution to the three- and four- point function (upper and lower row, respectively) and possible meson-exchange decomposition (crossed channels 
    are implied).\label{fig:npoint} }
\end{figure}

\begin{itemize}
\item Green's functions for bilinear quark currents can be expressed as sums over single meson states ---an analogous result holds for purely gluonic currents 
which can be expressed as a sum over purely gluonic bound states (glueballs).
\item  The amplitude for a bilinear current to create $m$ mesons from the vacuum $\bra{0} J \ket{n^m}$ is $\mathcal{O}(N_c^{1-m/2})$ ---similarly, the amplitude for creating $g$ 
                   glueball states $\bra{0} J \ket{n^g} $ is $ \mathcal{O}(N_c^{1-g})$.
\item Vertices involving $m$ mesons are $\mathcal{O}(N_c^{1-m/2})$ ---for $g$ glueball states they are of $\mathcal{O}(N_c^{1-g})$.\
\item Similarly, a meson-glueball vertex ---and thereby meson-glueball mixing--- can be obtained to be $\mathcal{O}(1/\sqrt{N_c})$.
\end{itemize}
The properties above, even qualitative, allow to understand many QCD phenomenological observations supporting the applicability of the large-$N_c$ limit:
\begin{itemize}
\item Most of the observed mesons are (mainly) $q\bar{q}$ states and additional $q\bar{q}$ content seems suppressed.
\item The dominance of narrow resonances over multiparticle continuum.
\item Hadronic decays proceed, dominantly, via resonant states. 
\item It provides a natural explanation (the only one so far)  for Regge phenomenology.
\item It is the only framework justifying the Okubo-Zweig-Iizuka (OZI)\footnote{The OZI rule refers to the suppression of quark-disconnected contributions such as that in the fourth 
      diagram in \cref{fig:bilin}.} rule. 
      It explains for instance why $\phi\to \bar{K}K$ dominates over $\phi\to\rho\pi$ or the approximate nonet symmetry in meson multiplets\footnote{Note that the mass difference 
      from singlet against octet mesons would arise from diagrams such as the last one in \cref{fig:bilin}, which are $N_c$ suppressed.}.
\end{itemize}
Note that in the combined chiral and large-$N_c$ limit not only may one expect the $\eta'$ to be degenerate in mass with the pseudo-Goldstone bosons, but to decouple 
from glueball mixing effects, providing thus an ideal framework to implement the $\eta'$ into the chiral description as previously said.\\

We concluded the previous sections observing that \cpt and pQCD could not provide a complete description of QCD at all scales. The large-$N_c$ limit 
does not provide a quantitative answer either ---as we do not know how to solve it yet--- but it provides a qualitative description. In the region 
of interest between \cpt and pQCD, the relevant physics is provided by the role of intermediate resonances. One could interpolate the QCD Green's functions from the low energies ---calculable within \cpt--- to the high-energies using a rational function incorporating 
the minimum number of resonances required to reproduce the pQCD behavior. This approach has been known as the minimal hadronic 
approximation (MHA)~\cite{Peris:2002wi} and has provided with successful and reasonable descriptions involving different phenomena.

\section{Pad\'e approximants}


The large-$N_c$ limit dictates that, to leading order, the QCD Green's functions are characterized in terms of the different poles arising from intermediate resonance 
exchanges and their residues, motivating the construction of some rational ansatz for them. Specializing to two-point functions, say, 
the hadronic vacuum polarization (HVP)
\begin{equation}
\label{eq:HVPol}
\int d^4x e^{iq\cdot x} \bra{0} T\{ J^{(\mu)}(x)J^{(\nu)}(0) \} \ket{0} \equiv i(q^2g^{\mu\nu} - q^{\mu}q^{\nu})\Pi(q^2),
\end{equation}
the large-$N_c$ limit suggests that, at leading order,
\begin{equation}
\label{eq:mha}
\widehat{\Pi}(q^2) \equiv \Pi(q^2) - \Pi(0) = \frac{A(q^2)}{(q^2-M_{V_1}^2)(q^2-M_{V_2}^2)~...~(q^2-M_{V_n}^2)~...~},
\end{equation}
with $\widehat{\Pi}(q^2)$ the renormalized  HVP, $M_{V_n}$ the $n$-th vector meson resonance mass and $A(q^2)$  
a polynomial\footnote{the $\lim_{q^2\to\infty}\Pi(q^2)\sim \ln(q^2)$ 
behavior requires an inifnite number of resonances and therefore $A(q^2)$ should be an infinite degree polynomial as well.}. Moreover, the residues 
from $\widehat{\Pi}(q^2)$ can be expressed in term of the vector resonances' decay constants $F_{V_n}$ as
\begin{equation}
\label{eq:fv}
\lim_{q^2\to M_{V_n}^2} (q^2 - M_{V_n}^2)\widehat{\Pi}(q^2) \equiv  -F_{V_n}^2, \qquad  \bra{0} J^{\mu} \ket{V_n} \equiv M_{V_n} F_{V_n} \epsilon^{\mu}. 
\end{equation}
Phenomenologically, one may adjust the required number of vector resonances to reproduce the pQCD behavior (MHA) to the physical ($N_c$=3) 
resonance masses\footnote{This is possible with a finite number of resonances if the anomalous dimensions vanish 
so no logarithmic corrections appear or, approximately, if they appear as correction to the leading $Q^2$ behavior~\cite{Peris:2006ds}. 
This is not the case for the HVP; it is the case however for the $\Pi_{LR}$ function~\cite{Masjuan:2007ay} or the TFFs.} and even to determine some 
coefficients in \cref{eq:fv} from the physical vector meson decays.  
Alternatively, as in the real world the vector mesons have a finite width, one may obtain the parameters in \cref{eq:mha} through a data-fitting procedure instead. 
The procedure outlined often provides a reasonable description, which, in some cases, may go beyond the large-$N_c$ expectations. 
Still, when aiming for precision, large $N_c$ is not enough and it would be desirable to be able to implement all the information at hand: the well-known low-energy behavior 
from \cpt~---including multiparticle intermediate states--- and the high-energy behavior from pQCD ---via the operator product 
expansion (OPE). In this section, we introduce Pad\'e approximants (PAs), which can be precisely used for this task. 
Pad\'e theory defines then a rigorous mathematical approach which is applicable, at least, in the space-like region. As an outcome, 
this theory is able to justify the reason why, sometimes, the MHA provides such a good performance beyond large-$N_c$ 
expectations\footnote{This section is based on Ref.~\cite{Queralt:2010sv}. A thorough discussion of PAs can be found 
in Refs.~\cite{Baker,BakerMorris}.}.\\

\subsection{Pad\'e theory essentials}
\label{sec:PadeTheo}

Given a function $f(z)$ of complex variable $z$ with a well defined power expansion around the origin\footnote{The definition is not special for the origin ($z=0$) and generally 
applies to any point $z_0$ in the complex plane as long as the series expansion is well-defined around $z=z_0$.} and a radius of convergence $|z|=R$,
\begin{equation}
\label{eq:sexp}
    f(z) = \sum_{n=0}^{\infty} f_n z^n,
\end{equation}
the Pad\'e approximant~\cite{Baker,BakerMorris} is a rational function\footnote{In the mathematical literature $P^N_M(z)$ is commonly noted as $[N/M]$, $[N|M]$ or $N/M$.}
\begin{equation}
  P^N_M(z) = \frac{Q_N(z)}{R_M(z)} = \frac{ \sum_{n=0}^{N} a_nz^n }{ \sum_{m=0}^{M} b_mz^m }
\end{equation}
with coefficients $a_n, b_m$ defined to satisfy the accuracy-through-order conditions up to order $N+M$
%
\begin{equation}
\label{eq:pasdef}
P^N_M(z) = f_0 + f_1z + ... + f_{M+N}z^{M+N} +  \mathcal{O}(z^{L+M+1}).
\end{equation}
Note that, without loss of generality, one can always choose $b_0=1$. 

If the original function $f(z)$ has a radius of convergence $R\to\infty$, $f(z)$ is said to be an entire function and is given by its power series expansion, \cref{eq:sexp}, 
everywhere in the complex plane. Employing PAs for this kind of functions may accelerate the convergence rate with respect to the series expansion, 
but the gain may not be dramatic. The situation changes for series expansions with a finite radius of convergence $R_0$: the power series \cref{eq:sexp} represents a 
divergent series beyond $R_0$ and convergence deteriorates as one approaches this point. It is in this case where PAs become a powerful tool; they cannot only 
dramatically improve the convergence rate within $|z|<R$ with respect to \cref{eq:sexp}, but may provide convergence in a larger domain $\mathscr{D}\subset\mathds{C}$ 
($\{|z|<R\} \subset \mathscr{D}$), which in some cases could extend (almost) to the whole complex plane ---in such case, PAs would provide in a sense a formal tool to perform an analytic 
continuation of a given series expansion. This is very important, as the functions we want to deal with in QCD, are not analytic in the whole complex plane; 
as we saw, in the large-$N_c$ limit these functions are characterized by an infinite set of resonances or poles, whereas in the real $N_c=3$ world, multiparticle intermediate states imply the existence 
of branch cuts. In that sense, the applicability of \cref{eq:sexp} would be very limited ($R$ would be given by the lowest (multi)particle production point). 
The study of convergence properties for PAs is much more complicated than for the 
cases of power expansions and represents an active field of research in applied mathematics. Nevertheless, there are some classes of functions for which convergence properties are 
very well-known. In the following, we describe the convergence properties for meromorphic and Stieltjes functions, which are representative cases of QCD Green's functions.

\subsubsection{The large-$N_c$ limit of QCD: meromorphic functions}
\label{sec:pasmeromorphic}

A particular class of functions we are interested in are meromorphic functions, this is, functions which are analytic in the whole complex plane except for a set of 
isolated poles and, therefore, represents the case of interest of large-$N_c$ QCD. The convergence properties of PAs to this kind of functions are very well-known 
and can be summarized in terms of Montessus' and Pommerenke's theorems as given in Ref.~\cite{BakerMorris}:\\

\noindent {\textbf{Montessus' theorem}}

Let $f(z)$ be a function which is meromorphic in the disk $|z| \leq R$, with $m$ poles at distinct points $z_1,z_2, ... , z_m$, where $|z_1|\leq|z_2|...\leq|z_m|\leq R$. 
Let the pole at $z_k$ have multiplicity $\mu_k$ and let the total multiplicity $\sum_{k=1}^m \mu_k = M$. Then, 
\begin{equation}
f(z) = \lim_{L\to\infty} P^L_M(z)
\end{equation}
uniformly on any compact subset of 
\begin{equation}
\mathscr{D}_m=\{  z, |z|\leq R, z\neq z_k, k+1,2,...,m  \}.
\end{equation}
When dealing with Green's functions in the large-$N_c$ limit of QCD, this means that a sequence of approximants $P^L_M(z)$ will provide an 
accurate description within a disk $|z|<R$ englobing the first $M$ poles as long as $L\to\infty$. The advantage of the theorem is that it provides uniform convergence, which is a strong property as it implies that no spurious poles or ``defects'' ---see the theorem below--- will appear. 
In particular, the position of the $m$ poles and their residues will be 
correctly determined as $L\to\infty$ (see Refs.~\cite{Peris:2006ds,Masjuan:2007ay}).
The disadvantage is that the theory does not say anything outside $|z|<R$ and the number of poles within must be anticipated ---an 
information which might be unknown. If these requirements were too strong for some specific application, one may resort to Pommerenke's theorem instead.\\

\noindent {\textbf{Pommerenke's theorem}}

Let $f(z)$ be a function which is analytic at the origin and analytic in the entire complex plane except for a countable number of isolated poles and essential singularities. 
Suppose $\varepsilon,\delta>0$ are given. Then, $M_0$ exists such that any $P^L_M$ sequence with $L/M=\lambda$ ($0<\lambda<\infty$) satisfies 
\begin{equation}
\label{eq:pom}
| f(z) - P^{\lambda M}_M | \leq \varepsilon
\end{equation}
for any $M\geq M_0$, on any compact set of the complex plane except for a set  $\mathscr{E}_M$ of measure less than $\delta$. 
Consequently, convergence is found as $M\to\infty$. As an interesting corollary, previous theorem can be generalized to $P^{N+k}_N(z)$ sequences 
with $k\geq-1$ fixed. 

The great advantage of this theorem is threefold: first, the poles do not have to be specified in advance; second, convergence is 
guaranteed for the whole complex plane; third, it includes not only poles, but essential singularities. In contrast, one has to deal with 
the occurrence of artificial poles not present in the original function. Convergence implies that these poles either move away in the complex 
plane, or they pair with a close-by zero, forming what are known as ``defects''\footnote{Defects are regions of the complex plane featuring a 
pole and a close-by zero ---their effect is nevertheless limited to a neighborhood around it and not the whole complex plane.}, 
for which convergence is not uniformly guaranteed but in measure. This means that the region in the complex plane where \cref{eq:pom} is not 
satisfied becomes arbitrarily small. 
See Ref.~\cite{Masjuan:2007ay} for a nice illustration of this feature and the use of Pommerenke's theorem for the 
$\langle VV-AA \rangle$ QCD Green's function.

\subsubsection{Back to the $N_c\!=\!3$ real world: Stieltjes functions}
\label{sec:Stieltjes}

As a consequence of the previous theorems, convergence of PAs to meromorphic functions can be guaranteed, and thereby, the convergence of 
PAs to QCD Green's functions in the large-$N_c$ limit of QCD follows. This allows to reconstruct and to extend the otherwise divergent series defined in \cref{eq:sexp} 
---which may be obtained from \cpt--- up to an arbitrary large domain as long as enough terms in the power-series expansion are known. Of course, this does not 
guarantee an analogous performance in the real world with $N_c=3$. For instance, the hadronic vacuum polarization $\widehat{\Pi}(q^2)$, \cref{eq:HVPol}, does no 
longer consists of an infinite number of resonances; multiparticle channels starting at the $\pi\pi$ threshold manifest themselves instead as a cut along the real axis, allowing to 
express the vacuum polarization through a once-substracted dispersion relation~\cite{Masjuan:2009wy} 
\begin{align}
\label{eq:hvpoldr}
\widehat{\Pi}(q^2) &= q^2\int_{s_{th}}^{\infty} \frac{dt}{t(t-q^2-i\varepsilon)}\frac{1}{\pi}\operatorname{Im}\Pi(t+i\varepsilon) \nonumber \\
              &= z\int_0^1  \frac{du}{1-uz -i\varepsilon}\frac{1}{\pi} \operatorname{Im} \Pi\left(\frac{s_{th}}{u}+i\varepsilon\right),
\end{align}
where $s_{th}=4m_{\pi}^2$ is the lowest threshold for particle production, $z=q^2/s_{th}$, and a change of variables $t=s_{th}u^{-1}$ has 
been performed in the second line of \cref{eq:hvpoldr}. The fact that $ \operatorname{Im} \Pi(q^2)$ is related through the optical theorem 
to the $\sigma(e^+e^-\to\textrm{hadrons})$ cross section, a positive quantity, guarantees that such a function is of the Stieltjes kind.
\\

Stieltjes functions are defined in terms of a Stieltjes integral~\cite{BakerMorris},
\begin{equation}
\label{eq:stieltjes}
    f(z) = \int_0^{\infty} \frac{d\phi(u)}{1+zu}, \qquad  |\operatorname{arg}(z)| < \pi,
\end{equation}
where $\phi(u)$ is a bounded non-decreasing function\footnote{Note that the function $\phi(u)$ is not even required to be continuous. As an example, 
$\phi(u)=\theta(u-u_0)\to d\phi(u) = \delta(u-u_0)du$, which is meromorphic and Stieltjes.} with finite and real-valued moments defining a formal expansion around the 
origin\footnote{In addition, Stieltjes functions can be shown to obey certain determinantal conditions~\cite{BakerMorris,Queralt:2010sv}. See Ref.~\cite{Masjuan:2009wy} 
for an application of them.}
\begin{equation}
\label{eq:stieltseries}
f_j = \int_0^{\infty} u^jd\phi(u), \quad j=0,1,2,... \quad \Rightarrow \quad f(z) = \sum_{j=0}^{\infty} f_j(-z)^{j} .
\end{equation} 
Note that, given a continuous non-zero $d\phi(u)$ function non-vanishing along $0\leq u \leq 1/R$, the Stieltjes function is not well defined in the real $-\infty<z\leq -R$ interval and a 
discontinuity $f(-z-i\varepsilon)\neq f(-z+i\varepsilon)$ appears along this, the reason for which Stieltjes functions are defined in the cut complex plane 
$|\operatorname{arg}(z)| < \pi$. Moreover, the original series expansion, \cref{eq:stieltseries}, is convergent within the $|z|<R$ disk alone; 
for the vacuum polarization, $R=4m_{\pi}^2$ corresponds to the threshold production, and the discontinuity at $4m_{\pi}^2 < z < \infty$ is related to the imaginary part 
or spectral function.

If a given function is of the Stieltjes kind, there is a well-known theorem in the theory of Pad\'e approximants guaranteeing the convergence of the $P^{N+J}_N(z)$ sequence 
in the cut complex plane for $J\geq-1$. In addition, the poles (and zeros) of the approximant are guaranteed to lie along the negative real axis and to have positive residues.

An additional property that Stieltjes functions can be shown to obey is that the diagonal(subdiagonal) $P^N_{N(+1)}(z)$ sequence decreases(increases) 
monotonically as $N$ increases, having a lower(upper) bound. Indeed, if $f(z)$ is a Stieltjes function, 
\begin{equation}
\lim_{N\to\infty} P^{N}_{N+1} \leq f(z) \leq  \lim_{N\to\infty} P^N_N(z), \quad (|\operatorname{arg}(z)| < \pi).
\end{equation}
More generally, any $P^{N+J}_N$ ($J\geq-1$) sequence is monotonically increasing(decreasing) for $J$ odd(even).

The condition that a function is Stieltjes is a very strong one and guarantees the possibility to reconstruct such a function through the use of PAs. Moreover, poles and 
zeros from PAs are guaranteed to pile along the negative real axis, excluding the possibility of defects. This allows to reconstruct certain hadronic functions, like 
the vacuum polarization, in the whole cut complex plane. This reconstruction excludes nevertheless the threshold and resonance region (which is ill-defined as well in the 
original function) and PAs poles cannot be associated therefore to physical resonances but to analytic properties of the underlying function. The PA zeros and poles 
conspire thereby to mimic the effects from the discontinuity at the cut. We illustrate such effect in \cref{fig:stielt} for the Stieltjes function $z^{-1}\ln(1+z)$.
These properties explain therefore the excellent performance of rational approaches beyond the naive large-$N_c$ estimation.
\begin{figure}
 \includegraphics[width=0.325\textwidth]{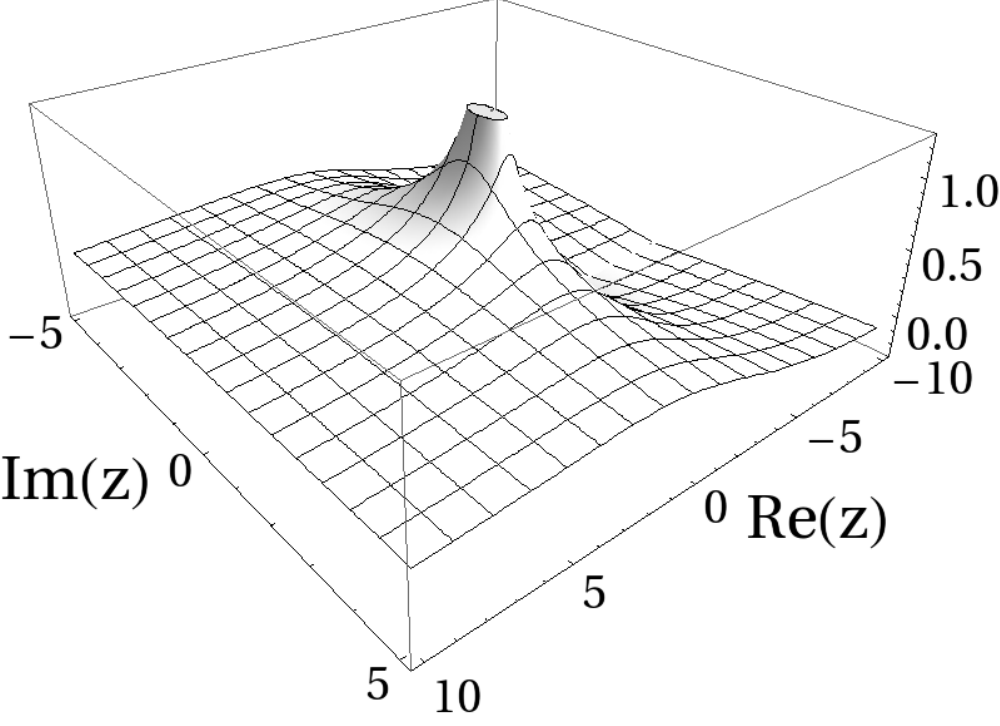}  \includegraphics[width=0.325\textwidth]{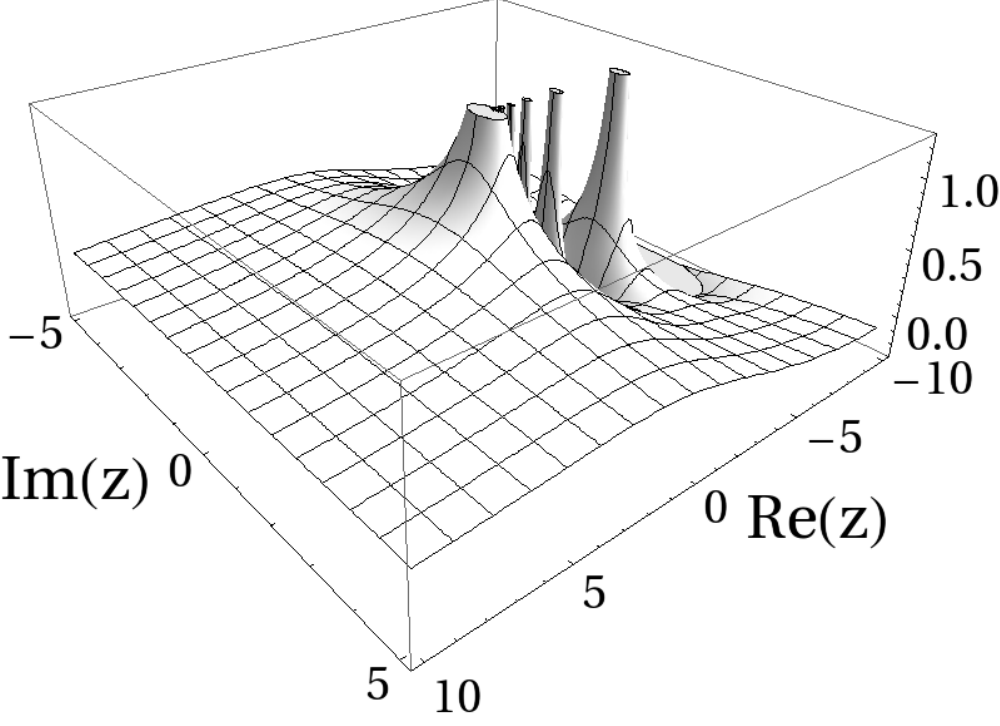}  \includegraphics[width=0.325\textwidth]{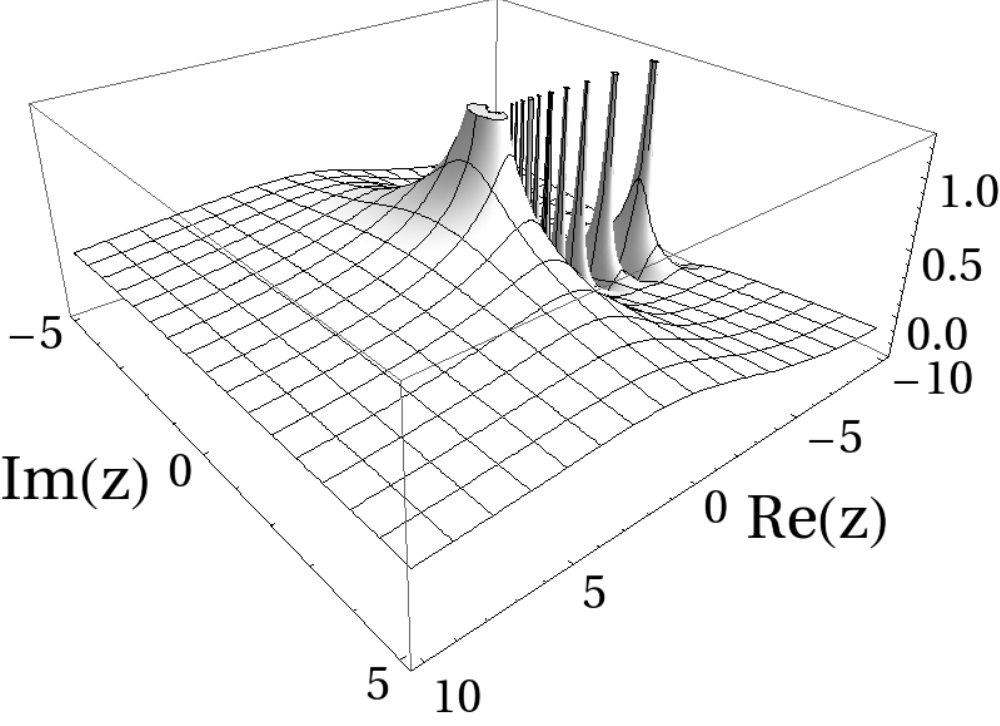}
 \includegraphics[width=0.325\textwidth]{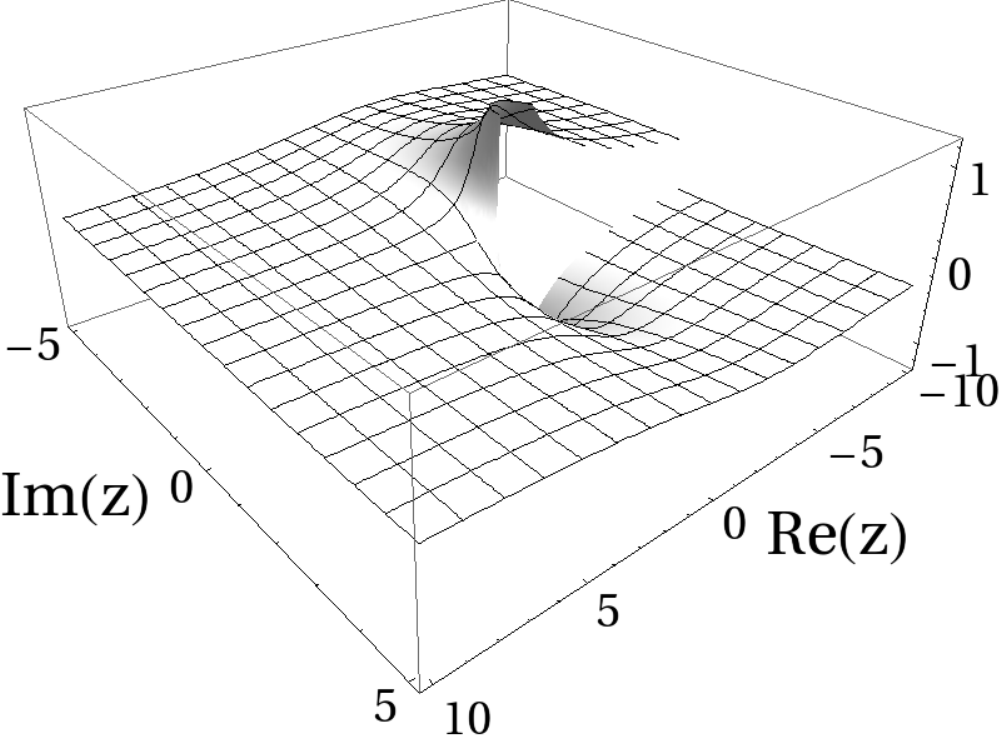}  \includegraphics[width=0.325\textwidth]{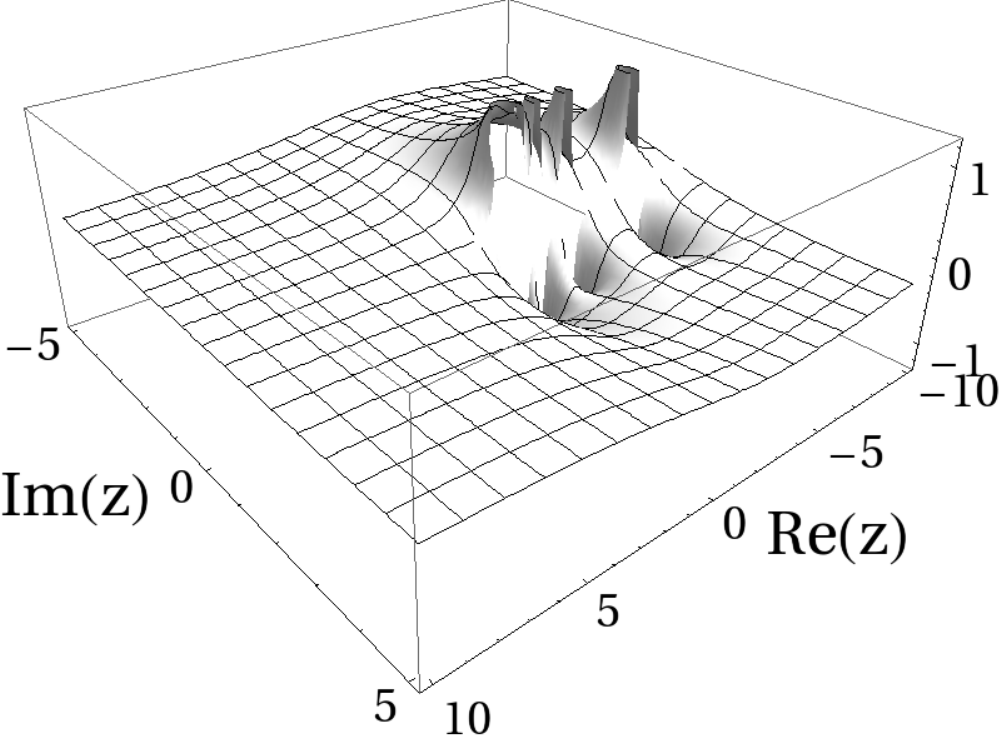}  \includegraphics[width=0.325\textwidth]{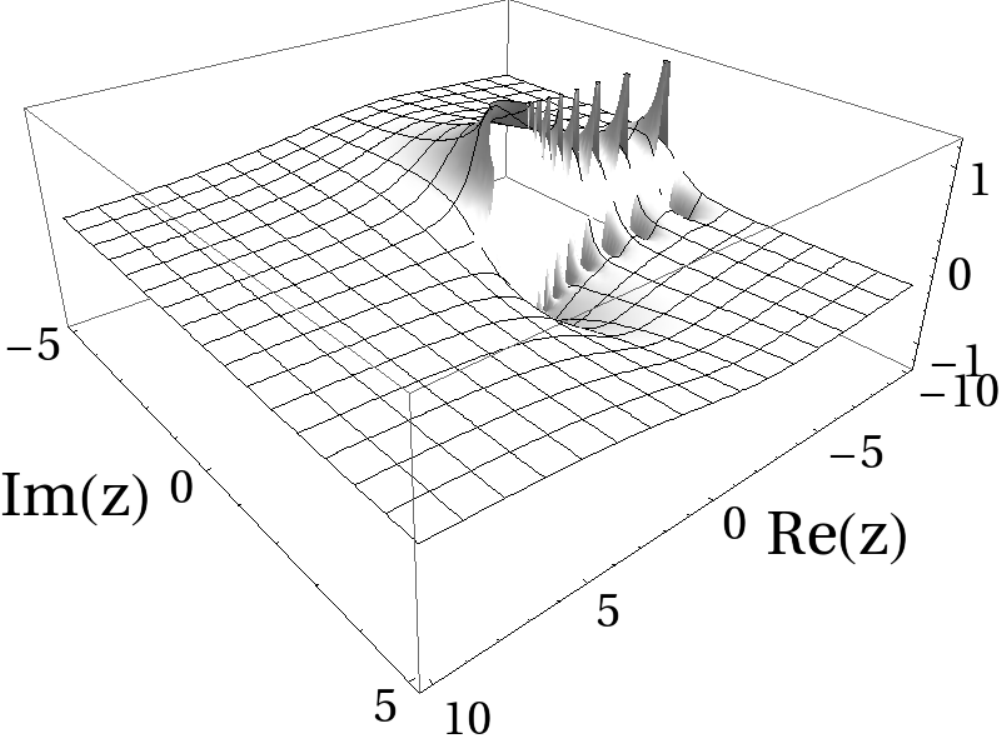}
\caption{The $z^{-1}\ln(1+z)$ function (first column) is compared to the $P^{10}_{11}(z)$ and $P^{30}_{31}(z)$ PAs (second and third column). Upper(lower) row 
  illustrates the real(imaginary) parts. \label{fig:stielt}}
\end{figure}
As a final remark, let us note that a function could be meromorphic and Stieltjes at the same time (i.e., if every pole has a positive-defined residue). In such a case, 
Stieltjes properties would apply as well.

\subsection{Extensions of Pad\'e approximants}

So far, we have only discussed the implementation of PAs based on the low-energy expansion \cref{eq:sexp}. However, in the large-$N_c$ approximation, 
or even in the real $N_c=3$ world, one may wish to include the information about some resonances' position. Additionally, further information away 
from the origin could be available ---the high-energy expansion among others. 
In this section, further extensions of PAs are presented allowing to incorporate this kind of information.

\subsubsection{Pad\'e type and partial Pad\'e approximants}
\label{sec:patype}

As said, from Montessus's and Pommerenke's theorems, it follows that, eventually, the poles and residues of the underlying function are reproduced by the approximant. 
However, it would be interesting to incorporate this information  from the beginning whenever this is known. This possibility is brought by Pad\'e type and partial 
Pad\'e approximants.\\

{\textbf{Partial Pad\'e approximants}}

If the lowest-lying $K$ poles at $z=z_1,z_2,...,z_K$ from the underlying function are known in advance, this information could be incorporated from the beginning using the 
so called Partial Pad\'e approximants defined as 
\begin{equation}
\mathds{P}^N_{M,K}(z) = \frac{Q_N(z)}{R_M(z)T_K(z)},
\end{equation}
where $Q_N(z)$, $R_M(z)$ are degree $N$ and $M$ polynomials and $T_K(z)=(z-z_1)(z-z_2)...(z-z_K)$ is a degree $K$ polynomial defined as to have all the zeros 
exactly at the first $K$-poles location.\\

{\textbf{Pad\'e type approximants}}

\noindent Pad\'e type approximants  is another kind of rational approximant
\begin{equation}
\mathds{T}^N_M(z) = \frac{Q_N(z)}{T_M(z)}
\end{equation}
in which all the poles of the approximant are fixed in advance to the original function lowest-lying poles. This is, $T_M(z)=(z-z_1)(z-z_2)...(z-z_M)$. 
This requires however the knowledge of every pole of the original function if one is aiming to construct an infinite sequence ($N,M\to\infty$).\\

An interesting discussion and illustration of partial Pad\'e and Pad\'e type approximants is illustrated for a physical case, the $\langle VV-AA\rangle$ function, 
in Refs.~\cite{Masjuan:2007ay,Masjuan:2008fr}. Here we only note that these approximants could justify why the MHA has often such a 
good performance---and a slower convergence--- wrt PAs that offer an improvement based on a mathematical framework. 

\subsubsection{N-point Pad\'e approximants}
\label{sec:PAnpoint}

Eventually, one could have analytical information of a particular function, not only at the origin, but at different points, say, $z_0$ and $z_1$ 
\begin{equation}
\label{eq:npoint}
 f(z) = \sum_{n=0}^{\infty} a_n(z-z_0)^n , \qquad  f(z) = \sum_{n=0}^{\infty} b_n(z-z_1)^n, 
\end{equation}
which belongs to what is known as the rational Hermite interpolation problem.
Typical cases is when low-energy, high energy or threshold behavior are known in advance. It is possible then to construct an N-point PA, 
$P^N_M(z)$, in which $J(K)$ terms are fixed from the series expansion around $z_0(z_1)$ from \cref{eq:npoint}, where $J+K=N+M+1$. 
Note that, for $N+M+1$ points, this would correspond to a fitting function interpolating between the given points. 
In general, N-point PAs will produce an improved overall picture with respect to typical (one-point) PAs of the same order, whereas the latter will provide a 
more precise description around their expansion point.

%

\section{The pseudoscalar transition form factors }

The central object of interest in this thesis are the transition form factors (TFFs) describing the interactions of the lowest-lying pseudoscalar mesons 
$(P)$ with two (virtual) photons and as such characterize the internal pseudoscalar structure.
From the $S$-matrix element\footnote{$j^{\mu}_{\textrm{em}} = \frac{2}{3}\bar{u}\gamma^{\mu}u - \frac{1}{3}\bar{d}\gamma^{\mu}d - 
\frac{1}{3}\bar{s}\gamma^{\mu}s  \equiv \mathcal{Q}\bar{q}\gamma^{\mu}q$ defines the electromagnetic current ---sum over quarks and colors is implicit.}
\begin{align}
\label{eq:pggSmat}
   \bra{\gamma^*\gamma^*} S \ket{P} &\equiv  i\mathcal{M}(P\to\gamma^*\gamma^*)(2\pi)^4\delta^{(4)}(q_1+q_2-p)  \\
     &= \frac{(ie)^2}{2!}\int\! d^4x \int d^4y \bra{\gamma^*\gamma^*}  T\left\{  A_{\mu}(x)j^{\mu}_{\textrm{em}}(x), A_{\nu}(y)j^{\nu}_{\textrm{em}}(y) \right\} \ket{P} \nonumber \\
   &= -e^2\!\!\int d^4x \ e^{iq_1\cdot x} \int d^4y \ e^{iq_2\cdot y} \bra{0}  T\left\{  j^{\mu}_{\textrm{em}}(x) , j^{\nu}_{\textrm{em}}(y) \right\} \ket{P} \nonumber  \\
   &= -e^2\!\!\int\! d^4x \ e^{iq_1\cdot x}  \bra{0}  T\left\{  j^{\mu}_{\textrm{em}}(x) , j^{\nu}_{\textrm{em}}(0) \right\} \ket{P} \! (2\pi)^4\delta^{(4)}(q_1\!+\!q_2\!-\!p) \nonumber
\end{align}
where $p,q_1$ and $q_2$ represent the pseudoscalar and photon momenta, 
the relevant amplitude defining the pseudoscalar TFF can be extracted:
\begin{align}
    i\mathcal{M}(P\to\gamma^*\gamma^*) &= -e^2\int d^4x \ e^{iq_1\cdot x}  \bra{0}  T\left\{  j^{\mu}_{\textrm{em}}(x) , j^{\nu}_{\textrm{em}}(0) \right\} \ket{P(p)} \nonumber \\
                          &\equiv ie^2\epsilon^{\mu\nu\rho\sigma}q_{1\rho}q_{2\sigma}F_{P\gamma^*\gamma^*}(q_1^2,q_2^2), \label{eq:tffdef}
\end{align}
which represents a purely hadronic object.
For the case of real photons, the TFFs can be related in the chiral (and, for the $\eta'$, combined large $N_c$) limit to the ABJ anomaly~\cite{Peskin:1995ev}, 
obtaining for $F_{P\gamma^*\gamma^*}(0,0)\equiv F_{P\gamma\gamma}$
\begin{align}
  F_{P\gamma\gamma} = \frac{N_c}{4\pi^2F}\operatorname{tr}(\mathcal{Q}^2\lambda^P) &\Rightarrow \mathcal{M}(P\to\gamma\gamma) = e^2\epsilon^{\mu\nu\rho\sigma}\epsilon_{1\mu}^*\epsilon_{2\mu}^*q_{1\rho}q_{2\sigma}F_{P\gamma\gamma} \nonumber \\
         &\Rightarrow \Gamma(P\to\gamma\gamma) = \frac{\pi\alpha^2 m_P^3}{4}F_{P\gamma\gamma}^2, \label{eq:ABJff}
\end{align}
where $F$ is the decay constant in the chiral limit defined in \cref{eq:Fp} and $\lambda^P=\lambda^{3,8,0}$ for the $\pi$, $\eta_8$ and $\eta_0$, respectively. 
For an elementary particle, the TFF would be constant, whereas for composite particles is expected to exhibit a 
$q^2$-dependency providing valuable information on the pseudoscalar meson structure. 
To study the TFF from first principles in the most general $q^2$ regime poses a formidable task, for which the only firm candidate so far is 
lattice QCD ---there exist some promising results in Refs.~\cite{Feng:2012ck,Cohen:2008ue,Lin:2013im} within a limited energy range. 
Still, there exists some knowledge at some particular energy regimes where different approaches apply.

\subsection{High-energies: perturbative QCD}
\label{sec:tffpqcd}

\begin{figure}
   \includegraphics[width=\textwidth]{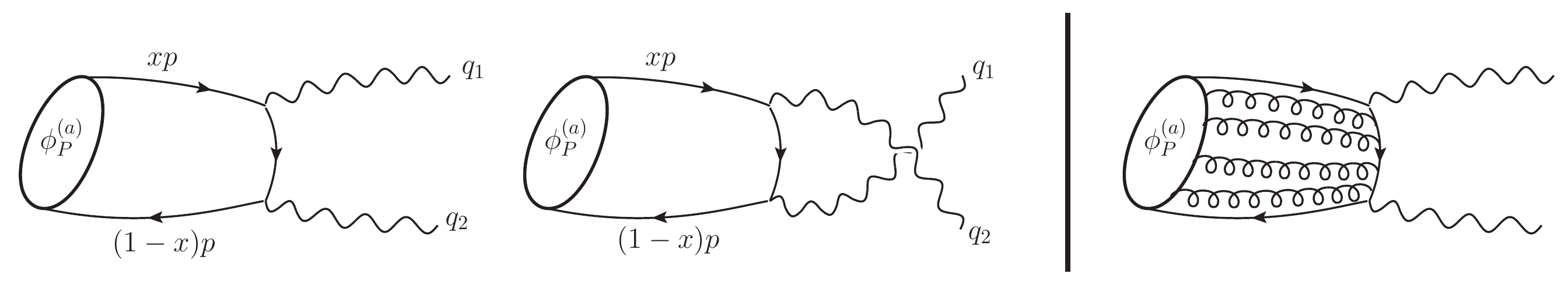}
   \caption{Left: leading order diagrams in pQCD contributing to the hard-scattering amplitude $T_H$. Right: gluon exchanges inducing a gauge link or Wilson line. \label{fig:TFFpQCD}}
\end{figure}
At large space-like energies, the TFF can be calculated as a 
convolution of a perturbatively calculable hard-scattering amplitude $T_H$ and a gauge invariant meson distribution amplitude 
(DA) $\phi_P^{(a)}$ encoding the non-perturbative dynamics of the pseudoscalar bound state~\cite{Lepage:1980fj} 
(summation over flavor $a=3,8,0$ implied; alternatively 
$a=3,q,s$ in the flavor basis, see \cref{chap:mixing}),
%
\begin{equation}
\label{eq:TFFpQCD}
 F_{P\gamma^*\gamma^*}(Q_1^2,Q_2^2) = \operatorname{tr}\left(\mathcal{Q}^2\lambda^{a} \right)F_P^{a} \int_0^1 dx \ T_H(x,Q_{1,2}^2,\mu) \phi_P^{(a)}(x,\mu) ,
\end{equation}
with $\bar{x} = 1-x$. The hard scattering amplitude at LO\footnote{The NLO result was calculated in Refs.~\cite{delAguila:1981nk,Braaten:1982yp}. 
See also Ref.~\cite{Agaev:2010aq}.} (see \cref{fig:TFFpQCD}) reads 
\begin{equation}
\label{eq:TFFTH}
T_H^{\textrm{LO}} = \frac{1}{\bar{x}Q_1^2+xQ_2^2} + (x\to\bar{x}),
\end{equation}
whereas the DA can be defined in terms of the matrix element~\cite{Agaev:2014wna}\footnote{$[z_2,z_1]$ represents a gauge link or Wilson line, see \cref{fig:TFFpQCD}, right.}
\begin{equation}
\label{eq:DAdef}
 \bra{0} \bar{q}(z_2)\gamma^{\mu}\gamma_5[z_2,z_1]\frac{\lambda^{a}}{2}q(z_1) \ket{P(p)} = i p^{\mu}F_P^{a} \int_0^1 dx \ e^{-i z_{21} \cdot p} \phi_P^{(a)}(x,\mu),
\end{equation}
where $z_{21} = \bar{x}z_2 + x z_1$ and obeys $\phi_P^{(a)}(x) = \phi_P^{(a)}(\bar{x})$. 
As a non-perturbative object, its particular shape is unknown from first principles at an arbitrary (renormalization) scale $\mu$. However, its asymptotic behavior 
at large energies is well-known: the DA follows the ERBL evolution~\cite{Efremov:1978rn,Lepage:1980fj} which allows for a convenient decomposition in terms 
of Gegenbauer polynomials 
\begin{equation}
\label{eq:DAgegen}
\phi^{(a)}_P(x,\mu) = 6x(1-x)\left( 1 + \sum_{n=1}^{\infty} c_{2n,P}^{(a)}(\mu)C_{2n}^{3/2}(2x-1) \right),
\end{equation}
with coefficients evolving at LO as\footnote{The LO anomalous dimensions read $\gamma_n = C_F\left( 4\left[\psi(n+2) + \gamma_E\right]  - \left[3 + \frac{2}{(n+1)(n+2)}\right] \right)$, 
$C_F= \frac{N_c^2-1}{2N_c}=\frac{4}{3}$ and $\alpha_s(\mu)$ evolution should be at LO. $\beta_0$ has been defined below \cref{eq:asympf}.}{\textsuperscript{,}}\footnote{An additional 
effect has to be accounted for the singlet component ---a careful description can be found in Ref.~\cite{Agaev:2014wna}. Whereas it has a non-negligible effect, we postpone 
its discussion to \cref{chap:mixing} as it does not change the conclusions outlined below.}
\begin{equation}
c^{(a)}_{n}(\mu) = \left( \frac{\alpha_s(\mu)}{\alpha_s(\mu_0)} \right)^{\gamma_n/\beta_0} c^{(a)}_n(\mu_0),
\end{equation}
As a result, asymptotic freedom implies that at large  
$\mu^2\sim Q_{1}^2+Q_{2}^2$ the DA tends to the asymptotic one, $\phi^{\textrm{as}}(x) = 6x(1-x)$. 
Consequently, the high-energy behavior follows trivially from \cref{eq:TFFpQCD,eq:TFFTH}, implying 
\begin{align}
   \lim_{Q^2\to\infty}   F_{P\gamma^*\gamma^*}(Q^2,0) &= \frac{6F_P^a}{Q^2} \operatorname{tr}(\mathcal{Q}^2\lambda^a), \label{eq:BLlim}\\
   \lim_{Q^2\to\infty}  F_{P\gamma^*\gamma^*}(Q^2,Q^2) &= \frac{2F_P^a}{Q^2} \operatorname{tr}(\mathcal{Q}^2\lambda^a). \label{eq:OPElim}
\end{align}
The first one is known as the Brodsky-Lepage (BL) asymptotic behavior, whereas the second one can be obtained independently from the OPE of two electromagnetic 
currents~\cite{Novikov:1983jt} which are solid pQCD predictions. \\

Even if the DA shape is largely unknown ---the $c_2$ coefficient has been estimated from lattice QCD for the $\pi^0$~\cite{Braun:2006dg,Arthur:2010xf,Braun:2015lfa}--- 
it can be modeled to reproduce the available experimental data for the space-like single-virtual TFF at $Q^2$ large enough (the double-virtual TFF has not been 
measured so far). This has been studied for the $\pi^0$ in light-cone pQCD~\cite{Huang:2013yya}, using light-cone sum rules, both 
for the $\pi^0$~\cite{Stefanis:2012yw,Agaev:2010aq,Agaev:2012tm} and $\eta,\eta'$~\cite{Agaev:2014wna}, or using flat DAs ---which became 
popular after the \babar data release for the $\pi^0$~\cite{Aubert:2009mc}--- among others~\cite{Radyushkin:2009zg,Noguera:2010fe,Noguera:2012aw,Noguera:2011fv}. 
In addition, transverse momentum effects have been studied~\cite{Kroll:2010bf,Radyushkin:2014vla}. 
Alternatively, the TFF has been analyzed using Dyson-Schwinger equations~\cite{Cloet:2013tta}, from Holographic models~\cite{Brodsky:2011xx} and employing 
anomaly sum rules~\cite{Klopot:2012hd,Klopot:2013laa}. 
The agreement among different parameterizations and the conclusions drawn from different authors is not clear at all, except for the solid results \cref{eq:BLlim,eq:OPElim}. 
Particularly, there is no consensus on the range on applicability of pQCD and the onset of the asymptotic behavior.
In addition, the pQCD approach cannot be extended down to $Q^2\to0$ as the theory becomes non-perturbative there. In such limit, an appropriate candidate to describe the TFF is \cpt.

\subsection{Low-energies: \cpt}
\label{sec:tffcpt}

\begin{figure}
\centering
\includegraphics[width=\textwidth]{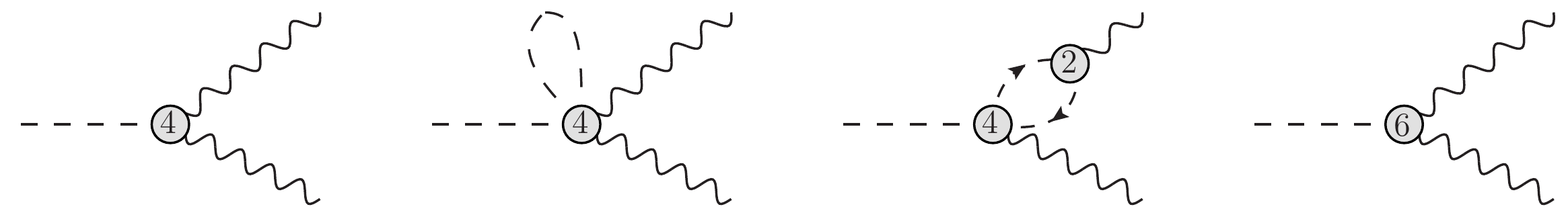}
\caption{The different contributions up to $\mathcal{O}(p^6)$ to the $P\gamma\gamma$ process. First is LO $\mathcal{O}(p^4)$ contributions, while the other are NLO. }
\label{fig:WZW}
\end{figure} 
At low-energies, \cpt can be used to provide the TFF behavior. At leading order $\mathcal{O}(p^4)$ ($\mathcal{O}(1)$ in \lcpt), this is described via the WZW 
Lagrangian \cref{eq:cptwzw}, which exactly reproduces the ABJ result \cref{eq:ABJff}. 
Remarkably, this is a free-parameter prediction once the decay constant $F$ has been fixed from other processes. 
In order to probe the pseudoscalar structure, higher orders bringing mass and $q^2$ (and large-$N_c$) corrections  are required. At NLO in \cpt, $\mathcal{O}(p^6)$, the TFF result 
arises from the diagrams in \cref{fig:WZW} and wave-function renormalization, and can be found in Refs.~\cite{Bijnens:1989jb,Ametller:1991jv,Borasoy:2003yb}. Using the 
$\mathcal{L}_{6,\epsilon}$ Lagrangian from Ref.~\cite{Ebertshauser:2001nj}, the TFF reads ($p_{1,2}^2$ is a time-like quantity)
\begin{align}
F_{P\gamma^*\gamma^*}(p_1^2,p_2^2)\! = & \ \frac{N_c\operatorname{tr}(\mathcal{Q}^2\lambda^P)}{4\pi^2F_P} \bigg(1 - \frac{512\pi^2}{3} \!\! \left[ 2\textrm{L}_8^{6,\epsilon}c_P^8 + 
2\textrm{L}_9^{6,\epsilon}c_P^9 + \textrm{L}_{19}^{6,\epsilon;r}(p_1^2+p_2^2) \right]    \nonumber\\
&   +\frac{1}{96\pi^2F^2}\bigg[ -\left( \ln\left(\tfrac{m_{\pi}^2}{\mu^2}\right) +  \ln\left(\tfrac{m_K^2}{\mu^2}\right) +\frac{2}{3}\right)p_1^2  +(p_1^2-4m_{\pi}^2)\nonumber \\
 &    \times H\!\left(\tfrac{p_1^2}{m_{\pi}^2}\right)\! +\!(p_1^2-4m_{K}^2)H\!\left(\tfrac{p_1^2}{m_{K}^2}\right)\! +\!(p_1^2\to p_2^2)   \bigg]  \bigg). \label{eq:tffcpt}
\end{align}
At $q_1^2=q_2^2=0$, corrections arise from the $\textrm{L}_{8,9}^{6,\epsilon}$ counterterms\footnote{$c_{\pi}^8 = \mathring{M}_{\pi}^2$ 
and $c_{\pi}^9=0$; $c_{\eta_8}^8 = \frac{7\mathring{M}_{\pi}^2-4\mathring{M}_{K}^2}{3}$ and $c_{\eta_8}^9=8(\mathring{M}_{\pi}^2-\mathring{M}_{K}^2)$; in a naive 
$\eta_0$ implementation, $c_{\eta_0}^8 = \frac{2\mathring{M}_{\pi}^2+\mathring{M}_{K}^2}{3}$ and $c_{\eta_0}^9=(\mathring{M}_{\pi}^2+\mathring{M}_{K}^2)$.} 
which are nevertheless not necessary to render the $P\to\gamma\gamma$ amplitude finite since the divergence is reabsorbed in the wave-function renormalization upon 
$F\to F_P$ replacement, a result which holds only at NLO~\cite{Bijnens:2012hf}. 
Furthermore, these corrections vanish in the chiral limit and they are commonly dismissed. 
For finite virtualities, an additional counterterm, $\textrm{L}_{19}^{6,\epsilon;r}$ is required to absorb the 
divergencies\footnote{$\textrm{L}_{19}^{6,\epsilon} \rightarrow \textrm{L}_{19}^{6,\epsilon;r} + \frac{\delta}{8192\pi^4F^2}$ with $\delta=(\frac{2}{\epsilon} 
+ \ln(4\pi\mu^2) + \gamma_E +1)$.}, incorporating a $p_{1,2}^2$-dependency together with the $H(s)$ loop function (see Eq.~(3.10) in Ref.~\cite{Donoghue})
\begin{equation}
\label{eq:loopfunc}
H(s) = 
   \begin{cases}
       2+ \beta(s) \ln\left(\frac{\beta(s)-1}{1+\beta(s)}\right) ,&  s\leq0   \\
       2+  |\beta(s)|\left( 2\tan^{-1}(|\beta(s)|) - \pi  \right) ,&  0<s<4 \\
      2 + i\pi\beta(s) + \beta \ln\left(\frac{1-\beta(s)}{1+\beta(s)}\right),&  s\geq4 \\
   \end{cases} \quad 
\end{equation}
with $\beta(s)=\sqrt{1-4s^{-1}}$. A naive extrapolation to incorporate the $\eta_0$ singlet state would yield an analogous result to that in \cref{eq:tffcpt} with 
an extra $1/2$ factor in the second line. However, the inclusion of the singlet state requires invoking large $N_c$. It turns out that, in the \lcpt counting, loops 
and the $\textrm{L}_9^{6,\epsilon}$ contributions are $N_c$ suppressed. Consequently, the chiral logarithms and loop function should be absent together with 
$\textrm{L}_9^{6,\epsilon}$. Moreover, an additional purely singlet OZI-violating term $\Lambda_3$~\cite{Leutwyler:1997yr} 
appears\footnote{Which amounts to replace $\frac{3\operatorname{tr}(Q^2\lambda^0)}{4\pi^2F_0} (1 - [...] ) \rightarrow 
\frac{3\operatorname{tr}(Q^2\lambda^0)}{4\pi^2F_0} (1+\Lambda_3 - [...] )$ in \cref{eq:tffcpt}.}, 
which in contrast to $\textrm{L}_{8}^{6,\epsilon}$ cannot be avoided in order to cancel the $F_0$ QCD scale dependency.
Describing the physical $\eta$ and $\eta'$ TFFs requires though to introduce the mixing, which we discuss in \cref{chap:mixing}.
It is well known that the TFF $p_{1,2}^2$ dependency in \cref{eq:tffcpt} is fully dominated by $\textrm{L}_{19}^{6,\epsilon}$ instead of the (a priori large) chiral 
logarithms~\cite{Bijnens:2012hf,Kampf:2012pa} ---a sign that such a process is dominated from vector resonance effects, with the consequent breakdown of the 
chiral expansion at energies close to the resonance. Given the lowest-lying $\rho$ and $\omega$ resonances, one cannot expect the chiral theory to work beyond 
$0.6$~GeV$^2$ ---even if including an infinite number of terms--- and this cannot be matched to pQCD to provide a full-energy range description.

\subsection{Alternatives approaches and Pad\'e theory}

As previously stated, the presence of resonances limits the applicability of the chiral effective field theory which begs for the presence of 
additional degrees of freedom. One possibility is to parametrize these contributions into the chiral theory in terms of pseudoscalar mesons 
rescattering effects which are experimentally known~\cite{Borasoy:2003yb}. Actually, this can be generalized advocating for a fully a dispersive 
framework~\cite{Gorchtein:2011vf,Hoferichter:2014vra,Hanhart:2013vba,Xiao:2015uva} incorporating different time-like information. 
Note however that such approaches have in practice either a limited range of applicability or require some modeling assumptions.

Alternatively, the situation can be analyzed within the large-$N_c$ limit of QCD in which the resonances are far more important 
than those effects which may be accounted for in \cpt or pQCD. 
From this point of view, one could describe the TFFs through modeling the infinite tower of vector resonances~\cite{RuizArriola:2006jge}. 
Alternatively, it has been customary to employ the MHA to saturate the TFF with a minimal finite amount of well-known resonances~\cite{Knecht:2001xc,Husek:2015wta}. 
Furthermore, there have been attempts to incorporate the resonances (within large $N_c$) explicitly into \cpt in what is known as resonance chiral perturbation 
theory~\cite{Roig:2014uja,Czyz:2012nq}.

From an orthogonal point of view, PAs can be used to directly address the problem posed at the end of \cref{sec:tffpqcd,sec:tffcpt}, 
this is, to provide an interpolation between \cpt and pQCD (at least in the full space-like region) without the necessity of invoking 
large $N_c$ ---which is ultimately an approximation and requires some modeling. Recall that PAs do not only apply in the large-$N_c$ limit 
of meromorphic functions, but offer an opportunity to go beyond this and to apply them to the real world, as it was shown for the case of 
Stieltjes functions. In this way, PAs allow to improve upon ideas as old as the MHA or the Brodsky-Lepage (BL) interpolation 
formula~\cite{Brodsky:1981rp}. Moreover, having a limited amount of information, they provide improved convergence properties with respect to 
typical resonant approaches used nowadays. For the case of the TFF, the analytic properties of the function 
are much more intricate than for two-point Green's functions, and therefore we cannot anticipate convergence ---note however that the salient features 
such as the $\pi\pi$ elastic rescattering and different resonances are of the Stieltjes kind. We can however check this a 
posteriori and estimate a systematic error from the convergence pattern, which we anticipate to be excellent, which provides an advantage with 
respect to previous methods. From this point of view, all the required information is encapsulated in the TFF series expansion 
\begin{equation}
\label{eq:tffpadeseries}
F_{P\gamma^*\gamma}(Q^2) = F_{P\gamma\gamma}(0,0)\left(1 - b_P \frac{Q^2}{m_P^2} + c_P\frac{Q^4}{m_P^4} - d_P\frac{Q^6}{m_P^6} + ... \right),
\end{equation}
which can be determined from data as it is explained in the next chapter. 
\\

\chapter{Data analysis with Pad\'e approximants}
\label{chap:data}
\minitoc

\section{Introduction}

For the phenomenological applications of pseudoscalar transition form factors (TFFs) covered in this thesis, 
we find that a very accurate description of these TFFs at very low energies ---where no available experimental data exists--- is required. For this reason, and regarding 
our approach based on Pad\'e approximants (PAs) to reconstruct the TFFs, it is extremely important to our work to know the series expansion for the TFF at zero energies. 
For the moment, we will restrict ourselves to the simpler single virtual case 
\begin{equation}
\label{eq:series}
F_{P\gamma^*\gamma}(Q^2) \equiv F_{P\gamma^*\gamma^*}(Q^2,0) = F_{P\gamma\gamma}\left(1 - b_P \frac{Q^2}{m_P^2} + c_P\frac{Q^4}{m_P^4} - d_P\frac{Q^6}{m_P^6} + ... \right) ,
\end{equation}
where $b_P$, $c_P$ and $d_P$ are referred to as slope, curvature and third derivative, respectively. 
The value for $F_{P\gamma\gamma}\equiv F_{P\gamma^*\gamma}(0)$ is well known for every pseudoscalar, as it is related to the 
Adler~\cite{Adler:1969gk}-Bell-Jackiw~\cite{Bell:1969ts} anomaly and 
can be theoretically related in \cpt to the meson decay constants for the $\pi^0, \eta$ and $\eta'$ 
(the mixing parameters are required for the last two though, see \cref{chap:mixing}). Furthermore, they can be experimentally extracted from the measured 
$P\rightarrow\gamma\gamma$ two-photon decays\cite{Larin:2010kq,Babusci:2012ik,Acciarri:1997yx,Agashe:2014kda}. 
In contrast, the additional low-energy parameters (LEPs) $b_P, c_P, ... $ cannot be obtained from first principles in QCD or  predicted 
from \cpt, as their values are given in terms of unknown low-energy constants. 
Moreover, they are not directly related to any experimental quantity.
Consequently, these parameters have always been obtained after modelization. For instance, with quark-loop models~\cite{Ametller:1991jv}, Brodsky-Lepage interpolation 
formula~\cite{Brodsky:1981rp,Ametller:1991jv}, resonance models~\cite{Czyz:2012nq} or \cpt supplied with vector meson dominance (VMD) ideas~\cite{Bijnens:1988kx,Bijnens:1989jb}. \\


A possible venue to address this problem would be to use low-energy experimental data so that \cpt or the series expansion, \cref{eq:series}, apply. 
Then, the above parameters could be extracted from a fitting procedure in a model-independent way. 
However, these data at very low energies are, in general, not available, rather scarce, or not precise, and one relies then on fits to models from high-energy data to extract 
these parameters. Such procedure is model-dependent and implicitly includes 
a systematic error which has never been considered. 
Actually, depending on the fitted data set, inconsistencies seem to appear in some cases, for 
instance, when comparing space-like and time-like data-based extractions for the slope parameter $b_{\eta}$.\\

In this chapter, we show how PAs can be used as a data-fitting tool to extract valuable information of the underlying function ---the single-virtual 
TFF--- including, among others, the desired LEPs in \cref{eq:series}.
 We illustrate that current inconsistencies cannot only be understood, but actually solved within a Pad\'e framework.
The results from this chapter represents the starting point of the following ones, as it provides the basic inputs for reconstructing the TFFs which are used in our 
calculations, as well as for extracting the $\eta-\eta'$ mixing parameters.
We proceed as follows: in \cref{sec:PAbasics}, we outline the procedure to obtain the LEPs from a fitting procedure. The corresponding systematic error is estimated  
in \cref{sec:pasys} through the use of different well-motivated models. Then, we apply our approach to the real case for the $\eta$ and $\eta'$ 
mesons using space-like data in \cref{sec:SL}. In \cref{sec:TL}, we argue, in view of the recent Dalitz decay measurements, why PAs could be applied to 
the low-energy time-like region as well, reevaluating our LEPs extraction. We give our conclusions and main results in \cref{sec:concl}.

\section{Pad\'e approximants as a fitting tool}
\label{sec:PAbasics}

Traditionally, the lack of low-energy data for the TFF has implied that the LEPs have been determined from phenomenological fits to 
high-energy data. There, the vector meson dominance (VMD) fitting function
\begin{equation}
\label{eq:VMDfit}
   F^{VMD}_{P\gamma^*\gamma}(Q^2) = F_{P\gamma\gamma}\frac{\Lambda^2}{\Lambda^2+Q^2}
\end{equation}
has been employed~\cite{Behrend:1990sr,Gronberg:1997fj}, which then ---upon expansion--- allowed to extract a determination for the slope $b_P$ parameter, 
which for this model is given by 
$b_P=m_P^2/\Lambda^2$. Additional LEPs were not discussed in this context though as they are all fixed in the ansatz above (i.e. $c_P=b_P^2$). 
Moreover, given the quality and precision of previous data, this discussion was irrelevant then, a situation which has changed with the recent release 
of new and more precise data in a wider energy regime, which makes timely a study of this kind. The possible deficiencies and model dependencies from this 
approach can be easily understood from Pad\'e theory, 
where the old VMD determinations can be understood as the simplest step in a systematic and convergent expansion~\cite{Masjuan:2007ay,Masjuan:2008fv}. 
As such, this implies that previous fitting approaches ---implying large systematic uncertainties as we illustrate below--- can be systematically improved, which makes 
possible not only a more accurate determination for $b_P$, but a meaningful extraction for additional parameters such as $c_P$ and $d_P$ in a model-independent way after 
performing the expansion of \cref{eq:series} for the fitted approximants.\\

Certainly, previous assertion relies on the assumption that the underlying function is such that some convergence to a given PA sequence exists, so our fits and the LEPs 
extracted from them will converge to the real ones.
Having incomplete analytical information about the TFFs, this cannot be guaranteed beforehand. 
Note however that the prominent features around $Q^2=0$ and the space-like region seem dominated by the role of the lowest-lying resonances ---of almost meromorphic nature--- 
and the $\pi\pi$ rescattering effects, essentially accounting for the $\rho$ width ---basically of Stieltjes nature. The existing convergence theorems, see \cref{sec:PadeTheo}, 
would justify then an excellent performance, at least, in the space-like region, of main interest for our applications.
%
%
Finally, even in the case where convergence to the underlying function cannot be guaranteed ---nor disproved---, the PAs practitioner 
can still judge on the convergence of a given sequence {\textit{a posteriori}} after the fitting procedure. We illustrate this in the following with the help of three different 
well-motivated models, where the different scenarios described above apply. This exercise will provide not only helpful to describe and get familiar with the procedure we use to 
extract the LEPs, but to assess a systematic error that we will employ when determining the LEPs from real data.
\\


The last point to discuss is the kind of sequences that should be used then for the fitting procedure. 
A glance at time-like data reveals that the first resonance effects are dominating the low-energy time- and space-like description,
meaning that including a single pole 
is enough to achieve a precise description and motivates the use of the $P^N_{1}$ sequence. It is important to note however that such description 
violates unitarity at high energies as it diverges as $(Q^2)^{N-1}$. This motivates the use of a second sequence, $P^N_{N+1}$, which can be thought of as a two-point 
PA (see \cref{sec:PAnpoint}) and incorporates the appropriate 
high-energy BL behavior, see \cref{eq:BLlim}, $F_{P\gamma^*\gamma}(Q^2) \sim Q^{-2}$. Given the uncertainty about convergence, cross-checking the results 
from both sequences will reassure the consistency of the method.
In what follows, we restrict our attention to these two sequences. Alternative choices exist, as for instance, Pad\'e- or partial Pad\'e-type 
approximants, see \cref{sec:patype}. Their arbitrariness in choosing a pole, and their slower convergence as compared to the previous ones 
make them less attractive, and we do not further consider their study.

\section{Estimation of a systematic error}
\label{sec:pasys}

For testing the convergence of our chosen $P^N_1$ and $P^N_{N+1}$ sequences, we propose the use of three different motivated theoretical models out of 
the vast literature, which will illustrate the performance of our approach in different representative situations. These are, the large-$N_c$ Regge model from 
Refs.~\cite{RuizArriola:2006jge,Arriola:2010aq}, the logarithmic model in Ref.~\cite{Masjuan:2012wy} ---which finds inspiration in flat distribution amplitudes~\cite{Radyushkin:2009zg} 
and quark models~\cite{Ametller:1983ec}--- and the holographic model proposed in Refs.~\cite{Brodsky:2011yv,Brodsky:2011xx}. For studying 
the convergence pattern, we generate a set of pseudo-data points in a similar manner to how real experimental data are 
distributed. As this is to represent the ideal case where the function is known up to arbitrary precision, we don't ascribe any error to the data for our fitting procedure, 
and therefore it does not make sense to give the $\chi^2$ from the fits. Finally, we perform the expansion in \cref{eq:series} to extract the LEPs from our fits and compare. 

\subsection{Large-$N_c$ Regge model}
\label{sec:reggemodel}

The large-$N_c$ model from Refs.~\cite{RuizArriola:2006jge,Arriola:2010aq} consists of an infinite sum of vector resonances, which sum can be expressed in terms of the polygamma function $\psi^{(n)}=\frac{d^{n+1}}{z^{n+1}}\ln \Gamma(z)$ with $\Gamma(z)$ the Gamma function,
\begin{equation}
   F_{P\gamma^*\gamma}(Q^2) = \frac{a F_{P\gamma\gamma}}{Q^2\psi^{(1)}\left(\frac{M^2}{a}\right)} \left[  \psi^{(0)}\left( \frac{M^2+Q^2}{a} \right)  -   \psi^{(0)}\left( \frac{M^2}{a} \right) \right].
\end{equation}
The parameters above have been slightly renamed for convenience with respect to those appearing in Refs.~\cite{RuizArriola:2006jge,Arriola:2010aq}. To reproduce the physical case, 
we choose the experimental $F_{P\gamma\gamma}\equiv F_{P\gamma\gamma}(0,0)$ together with $a=1.3~\textrm{GeV}^2$ and $M^2=\lambda\times0.64~\textrm{GeV}^2$~\cite{Masjuan:2012wy,Escribano:2013kba} 
where $\lambda=1,0.95,1.05$ for the $\pi^0, \eta, \eta'$, respectively\footnote{The parameter $a$ is taken from the analysis of different Regge trajectories in Ref.~\cite{Masjuan:2012gc}. For the $\pi^0$, $M$ is roughly the $\rho$ and $\omega$ meson masses. For the $\eta$ and $\eta'$ there is an interplay of $\rho,\omega$ and $\phi$ resonances~\cite{Landsberg:1986fd} which effectively translates in the $\lambda$ parameter.}. In what follows, we focus in the $\eta$ case, though very similar results are obtained for the $\eta'$ 
as it shares a similar TFF and available data sets. Actually, the results for the $\pi^0$ are similar too, see Ref.~\cite{Masjuan:2012wy}. For this model convergence is expected as it is a 
meromorphic function, which in addition represents the interesting case in which the large-$N_c$ limit applies. \\

Adopting the points defined in Ref.~\cite{Escribano:2013kba} ---10 points in the region $0.6<Q^2<2.2~\textrm{GeV}^2$, 15 points in the region $2.7<Q^2<7.6~\textrm{GeV}^2$ 
and 10 points in the region $8.9<Q^2<34~\textrm{GeV}^2$--- which resembles the experimental situation, we obtain the results in \cref{tab:RggSL}.
\begin{table}[t]
\centering
\tiny
\begin{tabular}{cccccccccccc} \toprule
      & $P^0_1$ & $P^1_1$ & $P^2_1$ & $P^3_1$ & $P^4_1$ & $P^5_1$ & $P^6_1$ & $P^7_1$ & $P^1_2$ & $P^2_3$ & Exact \\\cmidrule(r){1-2}\cmidrule(lr){3-9}\cmidrule(lr){10-11}\cmidrule(l){12-12}  
   $F_{P\gamma\gamma}$ & $0.268$ & $0.273$ & $0.274$ & $0.275$ & $0.275$ & $0.275$ & $0.275$ & $0.275$ & $0.275$ & $0.275$ & $0.275$ \\
          $b_{P}$       & $0.332$ & $0.373$ & $0.394$ & $0.404$ & $0.411$ & $0.416$ & $0.419$ & $0.421$ & $0.413$ & $0.425$ & $0.426$ \\
          $c_{P}$       &  --- & $0.143$ & $0.163$ & $0.173$ & $0.182$ & $0.188$ & $0.192$ & $0.195$ & $0.185$ & $0.201$ & $0.204$ \\
          $d_{P}$       & --- & --- & $0.067$ & $0.074$ & $0.081$ & $0.085$ & $0.088$ & $0.091$ & $0.083$ & $0.097$ & $0.100$ \\\bottomrule
\end{tabular}
\caption{LEPs determination from the space-like pseudo-data set for the large-$N_c$ Regge model. $F_{P\gamma\gamma}$ is expressed in GeV$^{-1}$; additional quantities are dimensionless.}
\label{tab:RggSL}
\end{table}
We find the expected convergence pattern we anticipated (note that the curvature and third derivative are not extracted up to $P^1_1$ and $P^2_1$, respectively). 
Moreover, we find an hierarchy: there is a faster convergence for $F_{P\gamma\gamma}\equiv F_{P\gamma^*\gamma}(0)$, then for $b_P$, and 
so on. An important observation at this point is that no matter whether strong correlations and, possibly, a tiny $\chi^2$ value in the real case appear, the highest the 
element within a sequence, the better the extraction for the LEPs becomes ---a feature common to all the models and characteristic of PAs. Therefore, we should aim for the 
largest possible element in our sequence when fitting real data for extracting our desired parameters. In addition, we find that the $P^N_{N+1}$ sequence has the better 
performance---note though that this sequence increases its number of parameters in units of two, so the $P^1_2$ should be compared with the $P^2_1$ and so on. 
This can be understood from the fact that, even if at these energies it is the influence of the first pole that dominates, there are additional higher resonances. Not less, 
this sequence implements as well  the appropriate high-energy behavior, relevant for the data range we are using.
In \cref{sec:TL}, we will employ also some very low-energy time-like data points in addition to the space-like ones. Given their small $q^2$ values, we expect that they 
significantly improve the accuracy from our determination, which demands a new systematic error evaluation. For this, we add to our previous pseudo-data set 8 points in the 
$(0.045)^2<q^2<(0.100)^2~\textrm{GeV}^2$ region, 15 points in the $(0.115)< q^2<(0.200)^2~\textrm{GeV}^2$ region and  31 points in the $(0.230)^2<q^2<(0.470)~\textrm{GeV}^2$ 
region in order to reproduce the experimental situation~\cite{Escribano:2015nra}. Our results are displayed in \cref{tab:RggTL}. We find very similar conclusions together 
with an improved accuracy ---to be expected from the increased amount of low-energy data points.
\begin{table}[t]
\centering
\tiny
\begin{tabular}{cccccccccccc} \toprule
      & $P^0_1$ & $P^1_1$ & $P^2_1$ & $P^3_1$ & $P^4_1$ & $P^5_1$ & $P^6_1$ & $P^7_1$ & $P^1_2$ & $P^2_3$ & Exact \\\cmidrule(r){1-2}\cmidrule(lr){3-9}\cmidrule(lr){10-11}\cmidrule(l){12-12}  
   $F_{P\gamma}$ & $0.279$ & $0.276$ & $0.275$ & $0.275$ & $0.275$ & $0.275$ & $0.275$ & $0.275$ & $0.275$ & $0.275$ & $0.275$ \\
          $b_{P}$       & $0.415$ & $0.433$ & $0.437$ & $0.437$ & $0.436$ & $0.435$ & $0.435$ & $0.434$ & $0.435$ & $0.433$ & $0.426$ \\
          $c_{P}$       &  --- & $0.196$ & $0.204$ & $0.207$ & $0.209$ & $0.210$ & $0.210$ & $0.210$ & $0.210$ & $0.210$ & $0.204$ \\
          $d_{P}$       & --- & --- & $0.095$ & $0.098$ & $0.101$ & $0.102$ & $0.102$ & $0.103$ & $0.102$ & $0.104$ & $0.100$ \\\bottomrule
\end{tabular}
\caption{LEPs determination from the space- and time-like pseudo-data set for the large-$N_c$ Regge model. $F_{P\gamma\gamma}$ is expressed in GeV$^{-1}$; additional quantities are dimensionless.}
\label{tab:RggTL}
\end{table}

\subsection{Logarithmic model}
\label{sec:QM}

This model finds inspiration in quark models~\cite{Ametller:1983ec} or flat distribution amplitudes~\cite{Radyushkin:2009zg}, which have been proposed ever since the puzzling 
\babar data for the $\pi^0$ TFF~\cite{Aubert:2009mc} were released. The model includes a logarithmic enhancement with respect to the BL asymptotic behavior, 
\begin{equation}
 F_{P\gamma^*\gamma^*}(Q^2) = \frac{F_{P\gamma\gamma}M^2}{Q^2} \ln\left( 1+\frac{Q^2}{M^2} \right),
\end{equation}
with $M^2=0.6~\textrm{GeV}^2$~\cite{Radyushkin:2009zg} to reproduce \babar data~\cite{Aubert:2009mc} and $F_{P\gamma\gamma}$ to reproduce the physical value. 
This second model is known to belong to the class of 
Stieltjes functions, which guarantees the performance of the method and allows to test the effects of perturbative logarithms as well, representing therefore an interesting 
case of study. Taking the pseudo-data points discussed above, we find the results in ~\cref{tab:QMSL}.
\begin{table}[t]
\centering
\tiny
\begin{tabular}{cccccccccccc} \toprule
      & $P^0_1$ & $P^1_1$ & $P^2_1$ & $P^3_1$ & $P^4_1$ & $P^5_1$ & $P^6_1$ & $P^7_1$ & $P^1_2$ & $P^2_3$ & Exact \\\cmidrule(r){1-2}\cmidrule(lr){3-9}\cmidrule(lr){10-11}\cmidrule(l){12-12}  
   $F_{P\gamma}$ & $0.268$ & $0.273$ & $0.274$ & $0.275$ & $0.275$ & $0.275$ & $0.275$ & $0.275$ & $0.275$ & $0.275$ & $0.275$ \\
          $b_{P}$       & $0.159$ & $0.196$ & $0.217$ & $0.227$ & $0.233$ & $0.238$ & $0.241$ & $0.243$ & $0.233$ & $0.247$ & $0.250$ \\
          $c_{P}$       &  --- & $0.040$ & $0.052$ & $0.058$ & $0.063$ & $0.068$ & $0.071$ & $0.073$ & $0.064$ & $0.078$ & $0.083$ \\
          $d_{P}$       & --- & --- & $0.012$ & $0.015$ & $0.017$ & $0.020$ & $0.021$ & $0.023$ & $0.018$ & $0.026$ & $0.031$ \\\bottomrule
\end{tabular}
\caption{LEPs determination from the space-like pseudo-data set for the logarithmic model. $F_{P\gamma\gamma}$ is expressed in GeV$^{-1}$; additional quantities are dimensionless.}
\label{tab:QMSL}
\end{table}
Again, we can reach to very similar conclusions as those in the Regge model. Moreover, we find that in this case the $P^N_{N+1}$ sequence has even better performance 
with respect to the $P^N_1$ than in the previous case. This can be understood from the convergence theorems for Stieltjes functions existing for the  $P^N_{N+1}$ sequence
(see \cref{sec:Stieltjes}) and from the much more involved analytic structure which a cut implies with respect to a single pole.
Once more, we reanalyze the systematic error for the case where we include the time-like data points on top and display the results in \cref{tab:QMTL}
\begin{table}[t]
\centering
\tiny
\begin{tabular}{cccccccccccc} \toprule
      & $P^0_1$ & $P^1_1$ & $P^2_1$ & $P^3_1$ & $P^4_1$ & $P^5_1$ & $P^6_1$ & $P^7_1$ & $P^1_2$ & $P^2_3$ & Exact \\\cmidrule(r){1-2}\cmidrule(lr){3-9}\cmidrule(lr){10-11}\cmidrule(l){12-12}  
   $F_{P\gamma}$ & $0.281$ & $0.278$ & $0.276$ & $0.276$ & $0.275$ & $0.275$ & $0.275$ & $0.275$ & $0.281$ & $0.275$ & $0.275$ \\
          $b_{P}$       & $0.199$ & $0.230$ & $0.245$ & $0.251$ & $0.253$ & $0.253$ & $0.253$ & $0.253$ & $0.253$ & $0.251$ & $0.250$ \\
          $c_{P}$       &  --- & $0.057$ & $0.068$ & $0.075$ & $0.079$ & $0.081$ & $0.082$ & $0.083$ & $0.081$ & $0.084$ & $0.083$ \\
          $d_{P}$       & --- & --- & $0.019$ & $0.022$ & $0.025$ & $0.027$ & $0.028$ & $0.029$ & $0.027$ & $0.031$ & $0.031$ \\\bottomrule
\end{tabular}
\caption{LEPs determination from the space- and time-like pseudo-data set for the logarithmic model. $F_{P\gamma\gamma}$ is expressed in GeV$^{-1}$; additional quantities are dimensionless.}
\label{tab:QMTL}
\end{table}

\subsection{Holographic model}
\label{sec:hol}

Finally, we take a model based on light-front holographic QCD from Ref.~\cite{Brodsky:2011xx} and we restrict ourselves, for simplicity, to the simplest leading twist 
result (though similar patterns are found for the other models in~\cite{Brodsky:2011xx})  
\begin{equation}
   F_{P\gamma^*\gamma^*}(Q^2) = \frac{P_{q\bar{q}}}{\pi^2F_{\pi}} \int_0^1 \frac{dx}{(1+x)^2} x^{Q^2 P_{q\bar{q}} /(8\pi^2F_{\pi}^2)}
\end{equation} 
with $P_{q\overline{q}}=1/2$ in order to reproduce the anomaly result. This model represents a particularly interesting case as, in contrast to previous models,  no 
convergence-theorem is known for it, which represents a similar situation to that in the real world. Taking the previous pseudo-data points, we obtain the results in 
\cref{tab:HSL}.
\begin{table}[t]
\centering
\tiny
\begin{tabular}{cccccccccccc} \toprule
      & $P^0_1$ & $P^1_1$ & $P^2_1$ & $P^3_1$ & $P^4_1$ & $P^5_1$ & $P^6_1$ & $P^7_1$ & $P^1_2$ & $P^2_3$ & Exact \\\cmidrule(r){1-2}\cmidrule(lr){3-9}\cmidrule(lr){10-11}\cmidrule(l){12-12}  
   $F_{P\gamma}$ & $0.279$ & $0.277$ & $0.276$ & $0.276$ & $0.275$ & $0.275$ & $0.275$ & $0.275$ & $0.275$ & $0.275$ & $0.275$ \\
          $b_{P}$       & $0.376$ & $0.357$ & $0.342$ & $0.334$ & $0.327$ & $0.322$ & $0.319$ & $0.316$ & $0.307$ & $0.311$ & $0.311$ \\
          $c_{P}$       &  --- & $0.126$ & $0.114$ & $0.107$ & $0.101$ & $0.096$ & $0.092$ & $0.090$ & $0.078$ & $0.083$ & $0.083$ \\
          $d_{P}$       & --- & --- & $0.038$ & $0.034$ & $0.031$ & $0.028$ & $0.026$ & $0.025$ & $0.018$ & $0.017$ & $0.021$ \\\bottomrule
\end{tabular}
\caption{LEPs determination from the space-like pseudo-data set for the holographic model. $F_{P\gamma\gamma}$ is expressed in GeV$^{-1}$; additional quantities are dimensionless.}
\label{tab:HSL}
\end{table}
Remarkably, we can reach similar conclusions to those in previous sections even if no convergence theorem could be provided in this case. 
Again, we use in addition the combined time- and space-like pseudo-data points obtaining the results in \cref{tab:HTL}
\begin{table}[t]
\centering
\tiny
\begin{tabular}{cccccccccccc} \toprule
      & $P^0_1$ & $P^1_1$ & $P^2_1$ & $P^3_1$ & $P^4_1$ & $P^5_1$ & $P^6_1$ & $P^7_1$ & $P^1_2$ & $P^2_3$ & Exact \\\cmidrule(r){1-2}\cmidrule(lr){3-9}\cmidrule(lr){10-11}\cmidrule(l){12-12}  
   $F_{P\gamma}$ & $0.271$ & $0.273$ & $0.274$ & $0.274$ & $0.275$ & $0.275$ & $0.275$ & $0.275$ & $0.275$ & $0.275$ & $0.275$ \\
          $b_{P}$       & $0.334$ & $0.323$ & $0.316$ & $0.313$ & $0.310$ & $0.310$ & $0.310$ & $0.310$ & $0.311$ & $0.311$ & $0.311$ \\
          $c_{P}$       &  --- & $0.102$ & $0.095$ & $0.091$ & $0.087$ & $0.085$ & $0.084$ & $0.083$ & $0.082$ & $0.083$ & $0.083$ \\
          $d_{P}$       & --- & --- & $0.028$ & $0.026$ & $0.024$ & $0.023$ & $0.022$ & $0.022$ & $0.020$ & $0.020$ & $0.021$ \\\bottomrule
\end{tabular}
\caption{LEPs determination from the space- and time-like pseudo-data set for the holographic model. $F_{P\gamma\gamma}$ is expressed in GeV$^{-1}$; additional quantities are dimensionless.}
\label{tab:HTL}
\end{table} 

\subsection{Final results}

To summarize, we find after comparing to different well-motivated models that, both, $P^N_1$ and $P^N_{N+1}$ sequences have a great performance for extracting the LEPs 
through a fitting procedure from experimental data. We emphasize that this may be the case even when convergence is not guaranteed, see \cref{sec:hol}.
As an important result, we show that in order to have the most accurate prediction, we should reach the 
highest element in each sequence regardless of correlations or $\chi^2_{\nu}\ll1$ values. In this respect, we remark that we do not aim for the best fitting-function rather than for the 
best LEPs extraction. In addition, we find that it is the $P^N_{N+1}$ sequence which has the better performance, though it increases its number of parameters 
in units of two, making the fitting procedure more complicated than for the $P^N_1$ sequence. 
In order to estimate the systematic errors, we adopt a conservative approach and take those arising from the quark model, which shows the slowest convergence pattern, 
obtaining the results in \cref{tab:syst}
\begin{table}
\centering
\scriptsize
\begin{tabular}{ccccccccccc} \toprule
      & $P^0_1$ & $P^1_1$ & $P^2_1$ & $P^3_1$ & $P^4_1$ & $P^5_1$ & $P^6_1$ & $P^7_1$ & $P^1_2$ & $P^2_3$ \\\cmidrule(r){1-2}\cmidrule(lr){3-9}\cmidrule(l){10-11}  
          $F_{P\gamma}$ & $6/0$ & $2/0$ & $1/0$ & $0.5/0$ & $0.1/0$ & $0/0$ & $0/0$ & $0/0$ & $0.1/0$ & $0/0$   \\
          $b_{P}$       & $40/20$ & $20/10$ & $15/5$ & $10/5$ & $5/1$ & $5/1$ & $5/1$ & $5/0.5$ & $5/1$ & $1/0$   \\
          $c_{P}$       &  --- & $50/30$ & $40/20$ & $30/10$ & $25/5$ & $20/1$ & $15/1$ & $10/1$ & $25/3$ & $5/0$   \\
          $d_{P}$       & --- & --- & $60/40$ & $50/30$ & $45/20$ & $40/15$ & $30/10$ & $30/5$ & $45/15$ & $15/2$   \\\bottomrule
\end{tabular}
\caption{Our final systematic errors in $\%$ for the SL/(SL+TL) data sets \label{tab:syserror}}
\label{tab:syst}
\end{table} 
---such table represents the main result from this section. At this point, we find that it is possible to have a meaningful extraction for the slope and curvature parameters 
in both data sets, whereas an accurate extraction for the third derivative is only possible and considered in our combined space- and time-like data set study. 
In addition, an analogous procedure shows that experimental determinations for $b_P$ based on time-like data alone should be ascribed an additional $5\%$ systematic error. 

\section{Space-like data: $\eta$ and $\eta'$ LEPs}
\label{sec:SL}

Having demonstrated the excellent performance of Pad\'e approximants as fitting functions to extract the LEPs, and having estimated a systematic error for the procedure, we 
proceed to their extraction in the real case. This was done for the $\pi^0$ in Ref.~\cite{Masjuan:2012wy} and 
we extend this approach to the $\eta$ and $\eta'$ below~\cite{Escribano:2013kba}. 
For that, we start using all the available data in the space-like region in which convergence is expected. This comprises the measurements from 
CELLO~\cite{Behrend:1990sr}, CLEO~\cite{Gronberg:1997fj} and \babar~\cite{BABAR:2011ad} for the $\eta$ and $\eta'$, and the additional L3 
Collaboration~\cite{Acciarri:1997yx} data-set at low-energies for the $\eta'$. In addition, we use the measured two-photon decay 
widths $\Gamma_{\eta\rightarrow\gamma\gamma}=0.516(18)~\textrm{keV}$~\cite{Agashe:2014kda} 
and $\Gamma_{\eta'\rightarrow\gamma\gamma}=4.35(14)~\textrm{keV}$~\cite{Agashe:2014kda}, dominated from the recent KLOE-2~\cite{Babusci:2012ik} 
and L3~\cite{Acciarri:1997yx} measurements, respectively.
For the fitting procedure, we take the function $Q^2F_{\eta^{(\prime)}\gamma^*\gamma}(Q^2)$ rather than $F_{\eta^{(\prime)}\gamma^*\gamma}(Q^2)$, since this is the standard 
way in which experimental data has been published, with the exception of L3 and CELLO\footnote{The CELLO Collaboration does not report a systematic 
error for each bin of data. While for the $\eta'$ case such error is $16\%$ of the total number of events (which we translate into $32\%$ for each bin), for the 
$\eta$ case only $12\%$ for the two-photon channel is reported. Accounting for all the different systematic sources we could find in the publication, we ascribe a 
$12\%$ of systematic error for the hadronic $\eta$ decay which leads to a $6\%$ error for the global number of events (implying $12\%$ systematic error for each bin).} 
collaborations. For these, we transform their results into $Q^2F_{\eta^{(\prime)}\gamma^*\gamma}(Q^2)$. Moreover, we relate the two-photon decay widths to 
$F_{\eta^{(\prime)}\gamma^*\gamma}(0)$ using the relation
\begin{equation}
\label{eq:dwtoff}
|F_{P\gamma\gamma}|^2 = \frac{64\pi}{(4\pi\alpha)^2}\frac{\Gamma_{P\rightarrow\gamma\gamma}}{m_P^3},
\end{equation}
obtaining $F_{\eta\gamma\gamma}=0.2738(47)~\textrm{GeV}^{-1}$ and $F_{\eta'\gamma\gamma}=0.3437(55)~\textrm{GeV}^{-1}$.
\\

For the fitting procedure, we employ the $P^N_1$ and $P^N_{N+1}$ sequences motivated in the previous sections, which 
translate into the $P^N_1$ and $P^N_N$ sequences for the $Q^2F_{\eta^{(')}\gamma^*\gamma}(Q^2)$ published data. 
Then, we must reach the highest possible element within a sequence as to maximally reduce the systematic uncertainty for the LEPs extraction 
as shown in the previous section. However, when using real data, it is not possible to go all the way up to an arbitrary 
large $N$ element. At some point, some of these parameters from which our PAs are built become statistically compatible with zero, meaning that its 
extraction is meaningless. We must stop at this point and take this result as our better extraction, and ascribe a systematic error as estimated from our results in 
 the previous section. \\

In order to show the performance of our method, we employ a bottom-up approach. We start fitting the $Q^2F_{\eta^{(')}\gamma^*\gamma}(Q^2)$ space-like 
data without any information at $Q^2=0$. This 
means in particular that the mathematical limit $\lim_{Q^2\to0}Q^2F_{\eta^{(\prime)}\gamma^*\gamma}(Q^2)=0$ is not imposed but extracted from data. In a second step, 
we impose such limit making use of PAs whose numerator starts at order $Q^2$ (i.e. there is no constant term). This study allows then to extract the TFFs at zero, and 
therefore predict the two-photon partial decay widths in addition to the slope and curvature parameters. Finally, as a last step, we incorporate the measured two-photon 
partial widths in our set of data, to be fitted together with the space-like data points. This approach will show the robustness of our results. \\

Starting then without constraining the $\lim_{Q^2\to0}Q^2F_{\eta^{(\prime)}\gamma^*\gamma}(Q^2)=0$ limit, we find that our fits ``see the zero'' for the $\eta$ and 
$\eta'$ cases within two and one standard deviations for the $\eta$ and $\eta'$, respectively. Particularly, we find $P^1_1(0)=0.059(29)$ and  
$P^3_1(0)=-0.02(3)$ for the $\eta$ and $\eta'$, respectively. Once this is seen to be zero, the next coefficient in its series expansion is associated with the TFF normalization. 
We find $F_{\eta'\gamma\gamma}(0)=0.38(6)~\textrm{GeV}^{-1}$, which translates into $\Gamma_{\eta'\rightarrow\gamma\gamma}=5.3(1.7)~\textrm{keV}$. This illustrates the 
potential of space-like data, which are ranging from $0.6$ to $35~\textrm{GeV}^2$ in the $\eta$ case and from $0.06$ to $35~\textrm{GeV}^2$ for the $\eta'$, to predict LEPs, which 
are our main aim for further applications in this work~\cite{Escribano:2013kba}.\\

Next, we make use of $\lim_{Q^2\to0}Q^2F_{\eta^{(\prime)}\gamma^*\gamma}(Q^2)=0$, meaning that the PAs numerator starts already at order $Q^2$. This simple 
constraint allows for an improved LEPs determination, shown in \cref{tab:fitnogg}. 
\begin{table}[t]
\scriptsize
\centering
\begin{tabular}{@{\hspace{0em}}c@{\hspace{0.999em}}c@{\hspace{0.999em}}c@{\hspace{0.999em}}c@{\hspace{0.6em}}c@{\hspace{0.3em}}c@{\hspace{0.999em}}c@{\hspace{0.999em}}c@{\hspace{0.999em}}c@{\hspace{0.6em}}c@{\hspace{0.3em}}c@{\hspace{0em}}}\toprule
 & \multicolumn{5}{c}{$\eta$} & \multicolumn{5}{c}{$\eta'$}\\ \cmidrule(r){2-6} \cmidrule(){7-11}
& $N$ & $b_{\eta}$  & $c_{\eta}$ & $F_{\eta\gamma\gamma},$ {\tiny{GeV$^{-1}$}}& $\chi^2_{\nu}$ & $N$ & $b_{\eta'}$ & $c_{\eta'}$ & $F_{\eta\gamma\gamma},$ {\tiny{GeV$^{-1}$}}& $\chi^2_{\nu}$ \\\midrule
$P^N_1(Q^2)$ & $2$ & $0.45(13)$ & $0.20(12)$ & $0.235(53)$ & $0.79$ & $5$ & $1.25(16)$ & $1.57(42)$ & $0.339(17)$ & $0.70$ \\
$P^N_N(Q^2)$ & $1$ & $0.36(6)$  & $0.13(4)$  & $0.201(28)$ & $0.78$ & $1$ & $1.19(6)$  & $1.42(15)$ & $0.332(15)$ & $0.68$ \\\bottomrule
\end{tabular}
\caption{LEPs for the $\eta$ and $\eta^\prime$ TFFs obtained from our fits \emph{without including} 
information on $\Gamma_{P\rightarrow\gamma\gamma}$. The first column indicates the type of sequence used for the fit and $N$ is the highest order 
achieved. 
We also present the quality of the fits in terms of $\chi^2_{\nu}$. Errors are only statistical and symmetrized.\label{tab:fitnogg}}
\end{table}
%
%
%
%
%
In this case, we reach up to the second and fifth elements of 
the $P^L_1$ sequence for the $\eta$ and $\eta'$, respectively, which TFFs are shown in \cref{fig:etaetap}. 
\begin{figure}[t]
\centering
\includegraphics[width=0.49\textwidth]{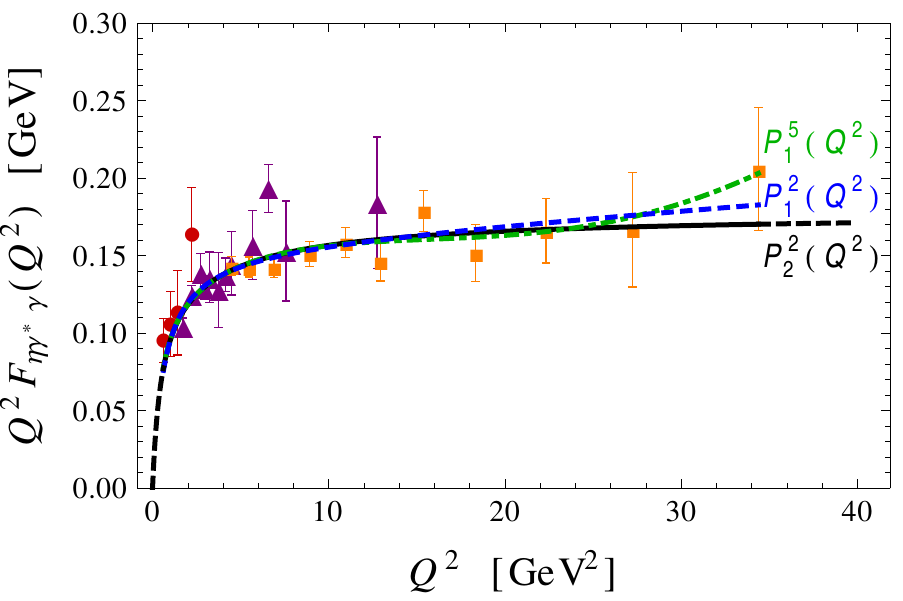}
\includegraphics[width=0.49\textwidth]{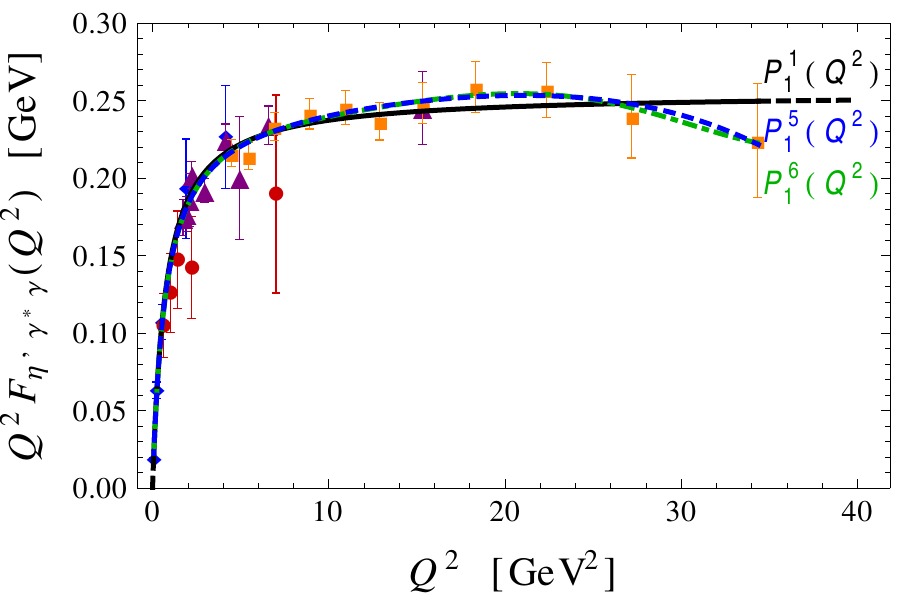}
\caption{$\eta$ (left panel) and $\eta^\prime$ (right panel) TFFs best fits. Blue-dashed lines show our best $P^L_{1}(Q^2)$ {\textit{without including}} the 
$\Gamma_{P\rightarrow\gamma\gamma}$ information in our fits; green-dot-dashed lines show our best $P^L_{1}(Q^2)$ when {\textit{including}} 
the $\Gamma_{P\rightarrow\gamma\gamma}$ information in our fits;
black-solid lines show our best $P^N_{N}(Q^2)$ in the latter case, which extrapolation down to $Q^2=0$ and $Q^2\to\infty$ 
is shown as a black-dashed line. Experimental data points are from CELLO (red circles) \cite{Behrend:1990sr}, CLEO (purple 
triangles) \cite{Gronberg:1997fj}, L3 (blue diamonds) \cite{Acciarri:1997yx}, and \babar (orange squares)~\cite{BABAR:2011ad} collaborations. \label{fig:etaetap}}
\end{figure}
Unfortunately, it is not possible to go beyond the first element for the $P^N_N$ sequence in both cases: 
for higher elements, the fit places poles in the space-like region, mimicking statistical fluctuations in the 
data ---such results should not be considered and the sequence should be truncated at this point. 
Remarkably, in this approach we obtain $\Gamma_{\eta\rightarrow\gamma\gamma}=0.38(17)~\textrm{keV}$ and 
$\Gamma_{\eta'\rightarrow\gamma\gamma}=4.22(42)~\textrm{keV}$, at $0.8$ and $0.3$ standard deviations from their physical values. 
\\

Finally, we include in our fits the two-photon decay widths through \cref{eq:dwtoff}. In this case, we reach, for the $P^L_1$ sequence, up to the fifth and sixth element 
for the $\eta$ and $\eta'$, respectively. On the other hand, {\textit{including}} $\Gamma_{P\rightarrow\gamma\gamma}$ allows to reach up to the second element in the $P^N_N$ 
sequence for the $\eta$, whereas this is not possible for the $\eta'$. The obtained TFFs are shown in \cref{fig:etaetap}. The LEPs obtained for these 
cases are shown in \cref{tab:fitgg}. 
\begin{table}[t]
\centering
\footnotesize
\begin{tabular}{ccccccccc}\toprule
 & \multicolumn{4}{c}{$\eta$} & \multicolumn{4}{c}{$\eta'$}\\ \cmidrule(lr){2-5} \cmidrule(l){6-9}
& $N$ & $b_{\eta}$  & $c_{\eta}$ & $\chi^2_{\nu}$ & $N$ & $b_{\eta'}$ & $c_{\eta'}$ & $\chi^2_{\nu}$ \\\midrule
$P^N_1(Q^2)$ & $5$ & $0.58(6)$  & $0.34(8)$  & $0.80$ & $6$ & $1.30(15)$ & $1.72(47)$ & $0.70$ \\
$P^N_N(Q^2)$ & $2$ & $0.66(10)$ & $0.47(15)$ & $0.77$ & $1$ & $1.23(3)$  & $1.52(7)$  & $0.67$ \\
    Final    &     & $0.60(6)$  & $0.37(10)$ &        &     & $1.30(15)$ & $1.72(47)$ & \\\bottomrule
\end{tabular}
\caption{LEPs for the $\eta$ and $\eta^\prime$ TFFs obtained from our fits when \emph{including} 
information on $\Gamma_{P\rightarrow\gamma\gamma}$. The first column indicates the type of sequence used for the fit and $N$ is the highest order 
achieved. The last row shows our final result for each LEP ---find details in the text. We also present the quality of the fits in terms of 
$\chi^2_{\nu}$. Errors are only statistical and symmetrized.\label{tab:fitgg}}
\end{table}
The similarity for these results and those found previously {\textit{without including}} the two-photon decay widths, \cref{tab:fitgg}, are quite reassuring. 
On top, we show our convergence results  for the slope $b_P$ and curvature $c_P$ parameters within the $P^L_1$ sequence in \cref{fig:slope,fig:curv}, 
\begin{figure}[t]
\small
\centering
\includegraphics[width=0.49\textwidth]{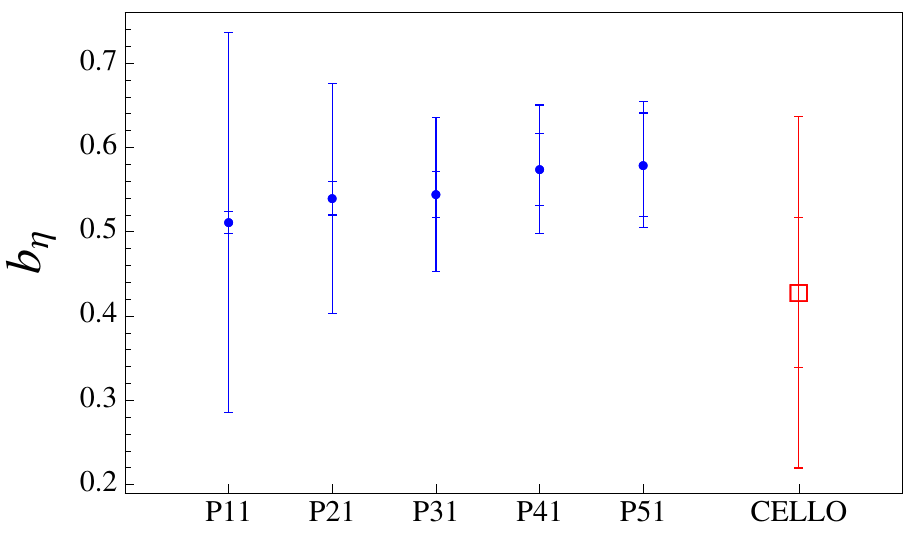}
\includegraphics[width=0.49\textwidth]{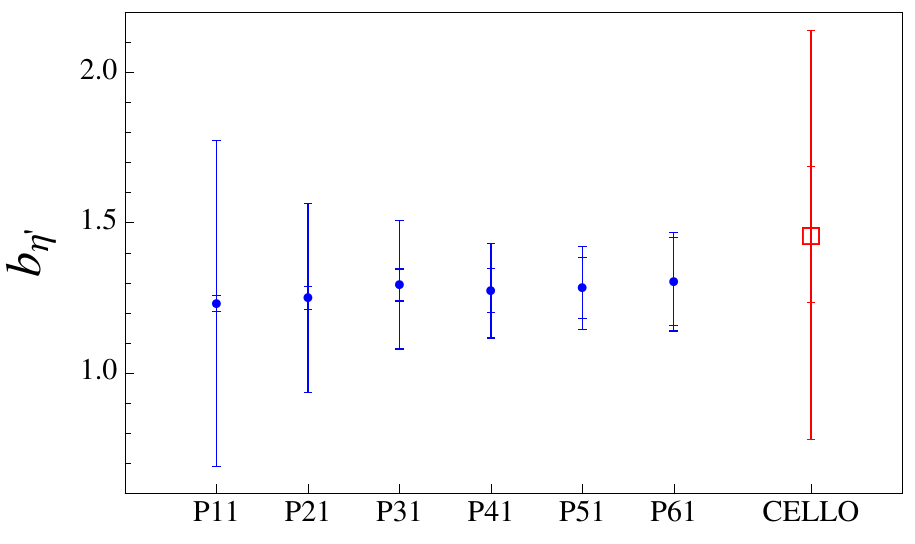}
\caption{Slope predictions for the $\eta$ (left panel) and $\eta^\prime$ (right panel) TFFs using the $P^L_1(Q^2)$ sequence (blue circles). 
The inner error bars correspond to the statistical error of the different fits. The outer error bars are the combination of statistical and systematic 
errors determined as explained in the main text. The CELLO determination is also shown for comparison (empty-red squares). \label{fig:slope}}
\end{figure}
\begin{figure}[t]
\centering
\includegraphics[width=0.49\textwidth]{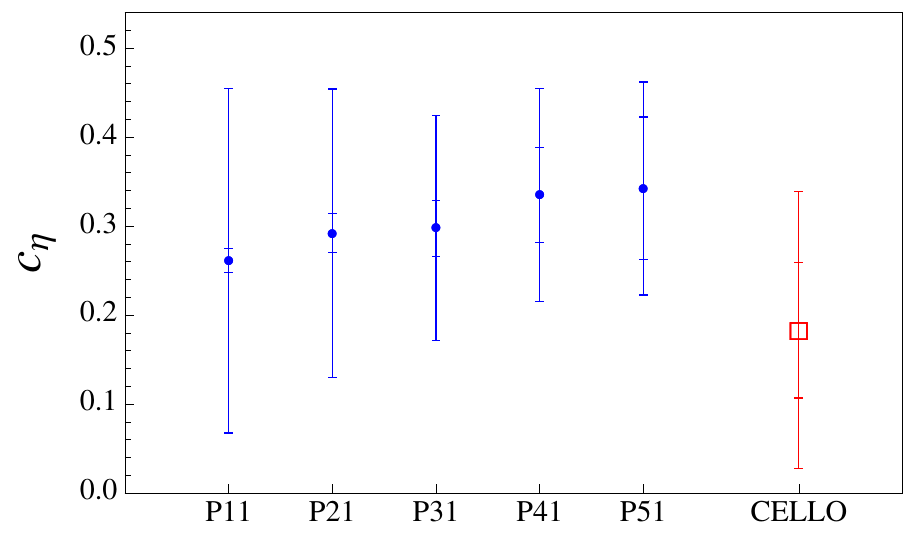}
\includegraphics[width=0.49\textwidth]{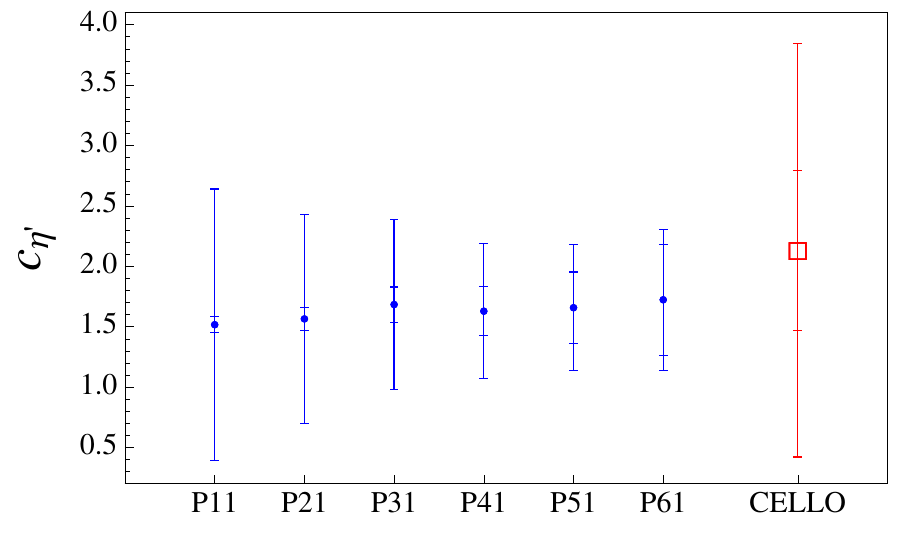}
\caption{Curvature predictions for the $\eta$ (left panel) and $\eta^\prime$ (right panel) TFFs using the $P^L_1(Q^2)$ sequence (blue circles). 
The inner error bars correspond to the statistical error of the different fits. The outer error bars are the combination of statistical and systematic errors 
determined as explained in the main text. The CELLO determination is also shown for comparison (empty-red squares). \label{fig:curv}}
\end{figure}
where the systematic errors are taken from \cref{tab:syserror}.
The observed pattern shows an excellent convergence. In these plots, we show in addition the results from CELLO for $b_{\eta(\eta')}$ obtained from a VMD 
model fit~\cite{Behrend:1990sr}. To perform an appropriate comparison, we add to their determinations an additional $40\%$ error corresponding to the $P^1_1$ 
element as determined in \cref{tab:syst}.
\\

In addition, we comment on the fitted poles obtained from the $P^N_1$ sequence, which we show in \cref{fig:PL1poles}, and range from $\sqrt{s_p}=(0.71-0.77)~\textrm{GeV}$ and $\sqrt{s_p}=(0.83-0.86)~\textrm{GeV}$ for the $\eta$ and $\eta'$ respectively. 
We note that such pole does not correspond to a particular physical resonance. It corresponds instead to an effective parameter which effectively accounts for the presence of 
different resonances, threshold effects and analytic structure of the whole function in general.   
\begin{figure}
\centering
  \includegraphics[width=0.49\textwidth]{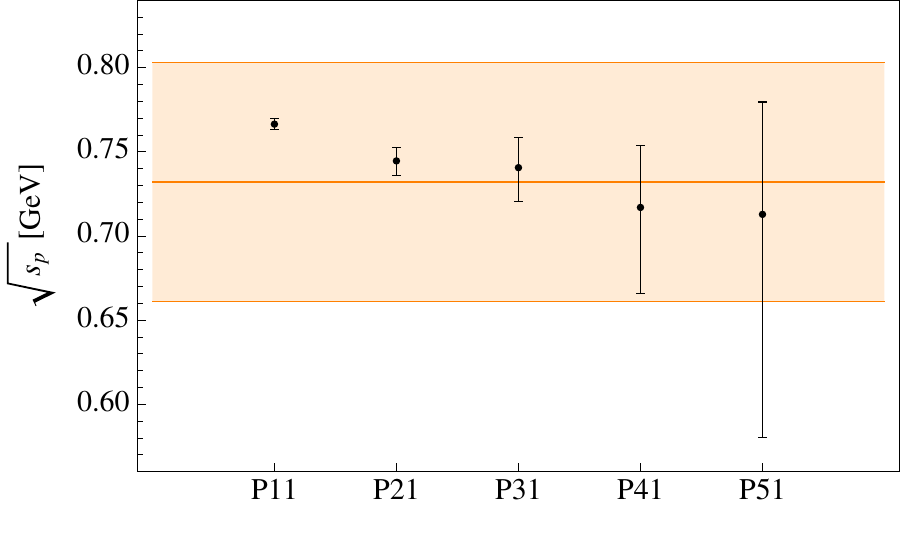}
  \includegraphics[width=0.49\textwidth]{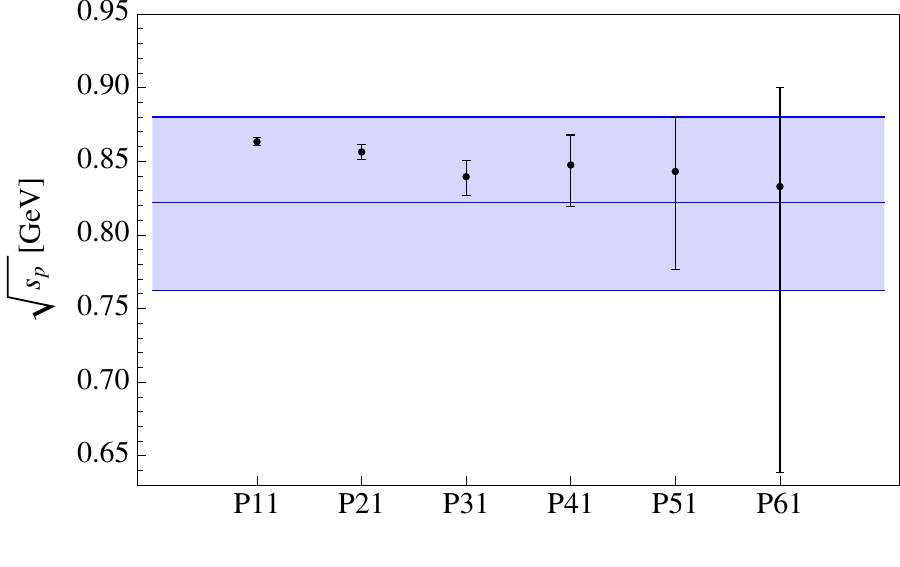}
\caption{Pole-position predictions for the $\eta$ (left panel) and $\eta^\prime$ (right panel) TFFs using the $P^L_1(Q^2)$ sequence. For comparison, we 
also display (orange and blue bands) the range $m_{\textrm{eff}}\pm\Gamma_{\textrm{eff}}/2$ corresponding to the effective VMD meson resonance evaluated using the 
half-width rule (see main text for details).\label{fig:PL1poles}}
\end{figure}
For comparison, we show as orange and blue bands what would correspond to the effective VMD meson resonance, 
$m_{\textrm{eff}}$~\cite{Landsberg:1986fd} using $m_{\rho}=0.775~\textrm{GeV}$, $\Gamma_{\rho}=0.148~\textrm{GeV}$, 
$m_{\omega}=0.783~\textrm{GeV}$, $\Gamma_{\omega}=0.008~\textrm{GeV}$, 
$m_{\phi}=1.019~\textrm{GeV}$ and $\Gamma_{\phi}=0.004~\textrm{GeV}$ and a mixing angle in the flavor basis $\phi=39^{\circ}$ (see \cref{chap:mixing}). Alternatively, see
 Ref.~\cite{Feldmann:1998vh} or the updated values of Ref.~\cite{Escribano:2005qq} in~\cite{Escribano:2013kba}). The bands represent the range of such mass implied by 
the half-width rule~\cite{Masjuan:2012gc,Masjuan:2012sk,Masjuan:2013xta}, i.e., $m_{\textrm{eff}}\pm\Gamma_{\textrm{eff}}/2$, which offers a nice estimation of large-$N_c$ 
corrections to typical 
resonance approaches. We obtain $m_{\textrm{eff}}=0.732(71)~\textrm{GeV}$ and $m_{\textrm{eff}}=0.822(58)~\textrm{GeV}$ for the $\eta$ and $\eta'$. As 
already indicated in Refs.~\cite{Masjuan:2012wy,Masjuan:2008fv,Masjuan:2007ay,Masjuan:2008fr}, fitting space-like data in resonant models does not produce 
an accurate determination for resonance parameters. We do not recommend then this method for such determination. For an alternative model-independent 
method, we refer to the interested readers to Refs.~\cite{Masjuan:2013jha,Masjuan:2014psa}.\\

Finally, it is possible from the $P^N_N$ sequences to extrapolate beyond the available data up to arbitrary large $Q^2$ values (dashed lines in \cref{fig:etaetap}), which 
allows for extracting the asymptotic behavior. For the $\eta$, we reach up to the second element, while for the $\eta'$ we reach only up to the first element due to the 
appearance, once more, of space-like poles mimicking statistical noise in the data. We obtain~\cite{Escribano:2013kba}
\begin{align}
\lim_{Q^2\to\infty}Q^2F_{\eta\gamma^*\gamma}(Q^2)& =  0.160(24)~\textrm{GeV}, \label{eq:etaSLinf} \\
\lim_{Q^2\to\infty}Q^2F_{\eta'\gamma^*\gamma}(Q^2)& =  0.255(4)~\textrm{GeV}.  \label{eq:etapSLinf}
\end{align}
We emphasize here that previous errors ---which are statistical alone--- could be deceptive. While the results for the $\eta$, arising from a higher element, suggest a larger error than for the 
$\eta'$ counterpart, the last has an intrinsic larger systematic error and it would be desirable as well for the $\eta'$ to reach a higher element, an achievement which is 
not possible with the available data so far. For completeness, if we would have used the $P^1_1$ element for extracting the asymptotic behavior of the $\eta$, we would have obtained 
$0.160(3)~\textrm{GeV}$ instead. The similarity with \cref{eq:etaSLinf} is reassuring.
\\

Finally, combining in weighted average our results in \cref{tab:fitgg} from the different sequences when {\textit{including}} the information on 
$\Gamma_{P\rightarrow\gamma\gamma}$, we obtain~\cite{Escribano:2013kba}
\begin{align}
  b_{\eta}& = 0.60(6)_{\textrm{stat}}(3)_{\textrm{syst}}&    c_{\eta}& = 0.37(10)_{\textrm{stat}}(7)_{\textrm{syst}}, \label{eq:etaSLleps}\\
  b_{\eta'}& = 1.30(15)_{\textrm{stat}}(7)_{\textrm{syst}}&  c_{\eta'}& = 1.72(47)_{\textrm{stat}}(34)_{\textrm{syst}}, \label{eq:etapSLleps}
\end{align}
where the second error is systematic, of the order of $5\%$ and $20\%$ for $b_P$ and $c_P$, respectively. 
When the spread of the central values for the weighted averaged 
result is larger than the error after averaging, we enlarge this error to cover the spread\footnote{We thank C.F.~Redmer for discussions on the average procedure.}.
For the $\eta'$ case, we could only reach the first element within the $P^N_N$ sequence. Since the first error of each sequence has a large systematic uncertainty, this 
should not be used, and consequently, we do not include it in our averaged result.
\\

For the $\eta$, the slope of the TFF obtained in \cref{eq:etaSLleps} can be compared with
$b_{\eta}=0.428(89)$ from CELLO \cite{Behrend:1990sr} and $b_{\eta}=0.501(38)$ from CLEO \cite{Gronberg:1997fj}.
The TFF was also measured in the time-like region with the results
$b_{\eta}=0.57(12)$ from Lepton-G \cite{Dzhelyadin:1980kh}, 
 $b_{\eta}=0.585(51)$ from NA60 \cite{Arnaldi:2009aa}, 
$b_{\eta}=0.58(11)$ from A2 \cite{Berghauser:2011zz}, and 
$b_{\eta}=0.68(26)$ from WASA \cite{Hodana:2012rc}. Recently, the A2 Collaboration reported $b_{\eta}=0.59(5)$~\cite{Aguar-Bartolome:2013vpw}, 
the most precise experimental extraction up to date.
Note the tendency among space- and time-like determinations, the former always smaller than the latter. 
This can be understood having a look at \cref{fig:slope,fig:curv}, which shows the poor result which is obtained from VMD-like fits ($P^1_1$) to space-like data. 
Our rigorous mathematical and systematical approach improves on this issue. 
For the $\eta^\prime$, the slope in \cref{eq:etapSLleps} can be compared with
$b_{\eta^\prime}=1.46(23)$ from CELLO \cite{Behrend:1990sr},
$b_{\eta^\prime}=1.24(8)$ from CLEO \cite{Gronberg:1997fj},
$b_{\eta^\prime}=1.7(4)$ from the time-like analysis by the Lepton-G Collaboration 
(cited in Ref.~\cite{Landsberg:1986fd}) and $b_{\eta'}=1.58(35)$ from BES-III~\cite{Ablikim:2015wnx}.
One should notice that all the previous collaborations used a VMD model fit to extract the slopes.
In order to be consistent when comparing with our results, a systematic error of about $40\%$ should be added to the experimental determinations based on space-like 
data (see \cref{tab:syst}) and a smaller one of about $5\%$ on the ones based on time-like data.
We present all these results in \cref{fig:slopecomp}, 
\begin{figure}[t]
\centering
 \includegraphics[width=0.49\textwidth]{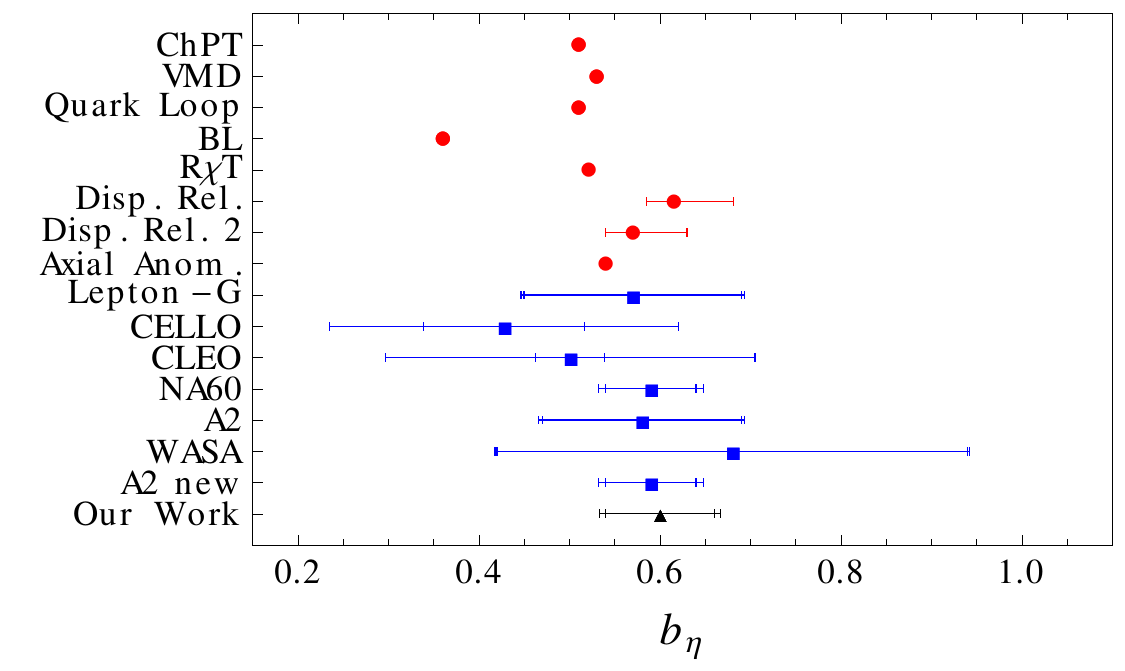}
 \includegraphics[width=0.49\textwidth]{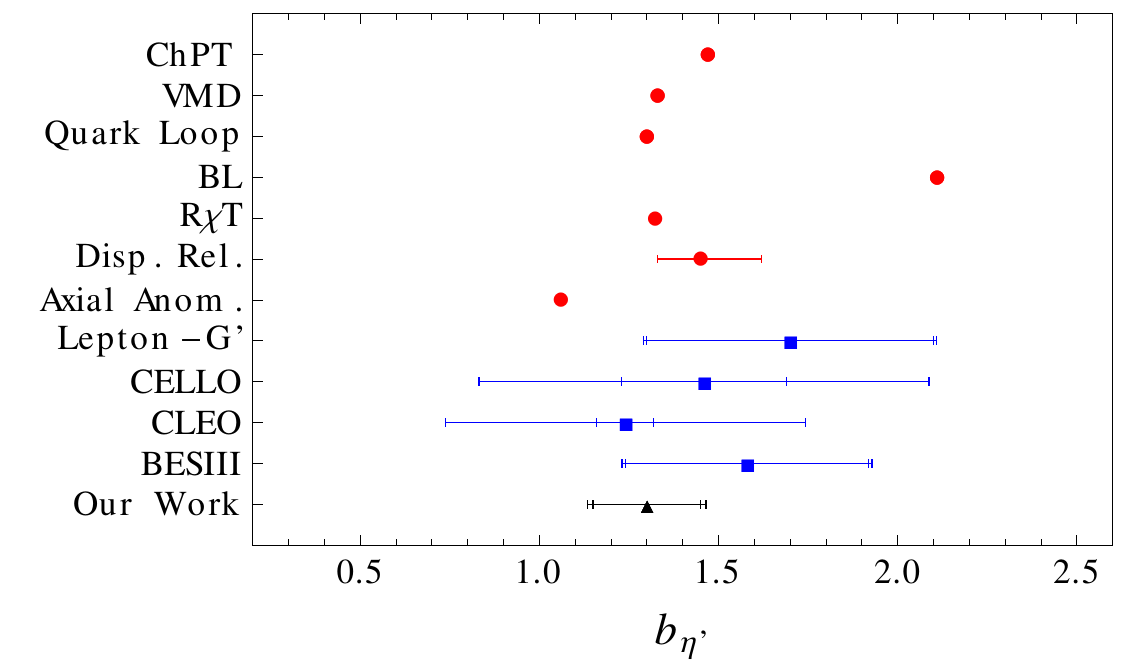}
\caption{Slope determinations for  $\eta$ (left panel) and $\eta^\prime$ (right panel) TFFs from different theoretical (red circles) and 
experimental (blue squares) references discussed in the text. Inner error is the statistical one and larger error is the combination of 
statistical and systematic errors. ChPT~\cite{Bijnens:1988kx,Bijnens:1989jb}, VMD, Quark Loop, BL~\cite{Ametller:1991jv}, R$\chi$T~\cite{Czyz:2012nq}, 
Disp.Rel~\cite{Hanhart:2013vba}, Disp.Rel 2~\cite{Kubis:2015sga,Xiao:2015uva}, Axial Anom.~\cite{Klopot:2013laa}, Lepton-G~\cite{Dzhelyadin:1980kh}, Lepton-G'~\cite{Landsberg:1986fd}, 
CELLO~\cite{Behrend:1990sr}, CLEO~\cite{Gronberg:1997fj}, NA60~\cite{Arnaldi:2009aa}, A2~\cite{Berghauser:2011zz}, WASA~\cite{Hodana:2012rc}, A2 new~\cite{Aguar-Bartolome:2013vpw}, 
BESIII~\cite{Ablikim:2015wnx}, Our Work~\cite{Escribano:2013kba}.
\label{fig:slopecomp}}
\end{figure}
where the smaller error is the statistical and the larger the quadratic combination of both statistical and systematic.
For completeness, we include different existing theoretical results, 
$b_{\eta}=0.51$ and  $b_{\eta^\prime}=1.47$ from \cpt~\cite{Bijnens:1988kx,Bijnens:1989jb}
for $\sin\theta_P=-1/3$ \cite{Ametller:1991jv},
being $\theta_P$ the $\eta-\eta^\prime$ mixing angle in the octet-singlet basis defined at lowest order;
$b_{\eta}=0.53$ and $b_{\eta^\prime}=1.33$, from vector meson dominance (VMD)~\cite{Ametller:1991jv};
$b_{\eta}=0.51$ and $b_{\eta^\prime}=1.30$, from constituent-quark loops; 
$b_{\eta}=0.36$ and $b_{\eta^\prime}=2.11$, from the Brodsky-Lepage interpolation formula \cite{Brodsky:1981rp}; 
$b_{\eta}=0.521(2)$ and $b_{\eta^\prime}=1.323(4)$, from resonance chiral theory \cite{Czyz:2012nq};
and recently, while our work in Ref.~\cite{Escribano:2013kba} was in progress,  $b_{\eta}=0.61^{+0.07}_{-0.03}$ and $b_{\eta^\prime}=1.45^{+0.17}_{-0.12}$
from a dispersive analysis~\cite{Hanhart:2013vba}\footnote{The dispersive results~\cite{Hanhart:2013vba} neglected the $a_2$ tensor meson 
contribution~\cite{Kubis:2015sga}. After accounting for this, they obtain $b_{\eta}=0.57(^{+6}_{-3})$~\cite{Kubis:2015sga,Xiao:2015uva}.}.\\

Eventually, we want to comment on the effective single-pole mass determination $\Lambda_P$ which \cref{eq:etaSLleps} 
implies for the $P^1_1$ reconstruction. 
Using $b_P=m_P^2/\Lambda_{P}^2$ and the values in \cref{eq:etaSLleps} ,
we obtain $\Lambda_{\eta}=0.706$ GeV and $\Lambda_{\eta^\prime}=0.833$ GeV. 
These values together with $\Lambda_{\pi}=0.750$ GeV obtained in Ref.~\cite{Masjuan:2012wy}
lead to $\Lambda_{\eta}<\Lambda_{\pi}<\Lambda_{\eta^\prime}$, in agreement with constituent-quark loops and VMD model approaches \cite{Ametller:1991jv}.\\

Notoriously, our results for the LEPs would not be affected to the quoted precision if the additional high-energy data points measured by \babar Collaboration 
at $q^2=112~\textrm{GeV}^2$~\cite{Aubert:2006cy} are included through the duality assumption that 
$\lim_{q^2\to\infty}F_{P\gamma^*\gamma}(q^2) = \lim_{Q^2\to\infty}F_{P\gamma^*\gamma}(Q^2)$ extends to large but finite energies. 
One would expect similarly that this is the case for the space-like \babar data in the $(4-35)~\textrm{GeV}^2$ range~\cite{BABAR:2011ad}.
However, this is not the case: the high-energy data are relevant in order to reach higher PA sequences leading to more constrained values of the LEPs.
In the case at hand, only the \babar Collaboration provides precise measurements in the region between $5$ and $35$~GeV$^2$.
For instance, the value of the $\eta$ slope parameter shown in \cref{eq:etaSLleps}, $b_{\eta} = 0.60(6)(3)$,
turns out to be $b_{\eta} = 0.65(9)(7)$ when the \babar data are not included in the fits.
In view of this behavior and having in mind the $\pi^0$ TFF controversy after the measurements of the \babar~\cite{Aubert:2009mc} and Belle~\cite{Uehara:2012ag} 
collaborations, a second experimental analysis by the Belle Collaboration covering this high-energy region would be very welcome. Remarkably, we will find in \cref{sec:TL} 
that even when including very low-energy time-like data, the \babar data points are still of relevance.
\\

For convenience, we also provide our parametrization of the highest $P^L_1$ fits, which can be used to predict the TFF low-energy behavior. Defining the $P^L_1(Q^2)$ 
for $F_{\eta^{(\prime)}\gamma^*\gamma}(Q^2)$ as
\begin{equation}
\label{eq:etaTFF}
P^L_1(Q^2) = \frac{t_0+t_1Q^2 + ... t_L(Q^2)^L}{1+r_1Q^2},
\end{equation}
the corresponding coefficients are given in \cref{tab:SLcoeff}
%
\begin{table}[t]
\scriptsize
\centering
\begin{tabular}{ccccccccc}\toprule
 &  $t_0$  & $t_1$ & $t_2$ & $t_3$ & $t_4$ & $t_5$ & $r_1$\\\midrule
$\eta$ & $0.274$ & $0.011$ & $-0.789\times10^{-3}$ & $0.229\times10^{-4}$ & $-0.169\times10^{-6}$ & --- & $1.968$\\
$\eta'$ & $0.343$ & $0.007$ & $-0.986\times10^{-3}$ & $0.744\times10^{-4}$ & $-0.252\times10^{-5}$ & $0.290\times10^{-7}$ & $1.442$ \\\bottomrule
\end{tabular}
\caption{Fitted coefficients for our best $P^L_1(Q^2)$ for the $\eta$ and $\eta'$ TFFs in units of GeV$^{-2i}$ for $t(r)_i$ and GeV$^{-1}$ for $t_0$.\label{tab:SLcoeff}}
\end{table}

\section{Time-like data: $\eta$ and $\eta'$ LEPs}
\label{sec:TL}

Our space-like data-based description above~\cite{Escribano:2013kba} provides an accurate description for the TFF in the low-energy range, which 
is the reason why we could obtain such an accurate extraction for the LEPs. Of course, there is no special analytic property at $Q^2=0$ which prevents us to 
make a prediction for low time-like energies. It remains the question then on what low means here. It is well known that at larger time-like energies
the appearance  of production thresholds, starting with $\pi^+\pi^-$, imply the appearance of additional singularities and cuts. 
The analytic structure of PAs in turn is given by a set of isolated poles, which would in principle forbid its use above threshold production and  
would question then the applicability of our approach to the $\eta^{(\prime)}\to \bar{\ell}\ell \gamma$ Dalitz decays above threshold.
\\

Very recently, the A2 Collaboration at MAMI~\cite{Aguar-Bartolome:2013vpw} reported a new measurement of the $\eta\rightarrow e^+e^-\gamma$ 
Dalitz decay with the best statistical precision up to date, which allowed them to extract the (normalized) $\eta$ TFF, $\tilde{F}_{\eta\gamma^*\gamma}(q^2)$, 
in the low-energy time-like region, $q^2\in(4m_{e}^2,m_{\eta}^2)$. In their study, they performed a comparison with different theoretical models, obtaining the results in \cref{fig:A2}.
\begin{figure}[t]
\centering
\includegraphics[width=\textwidth]{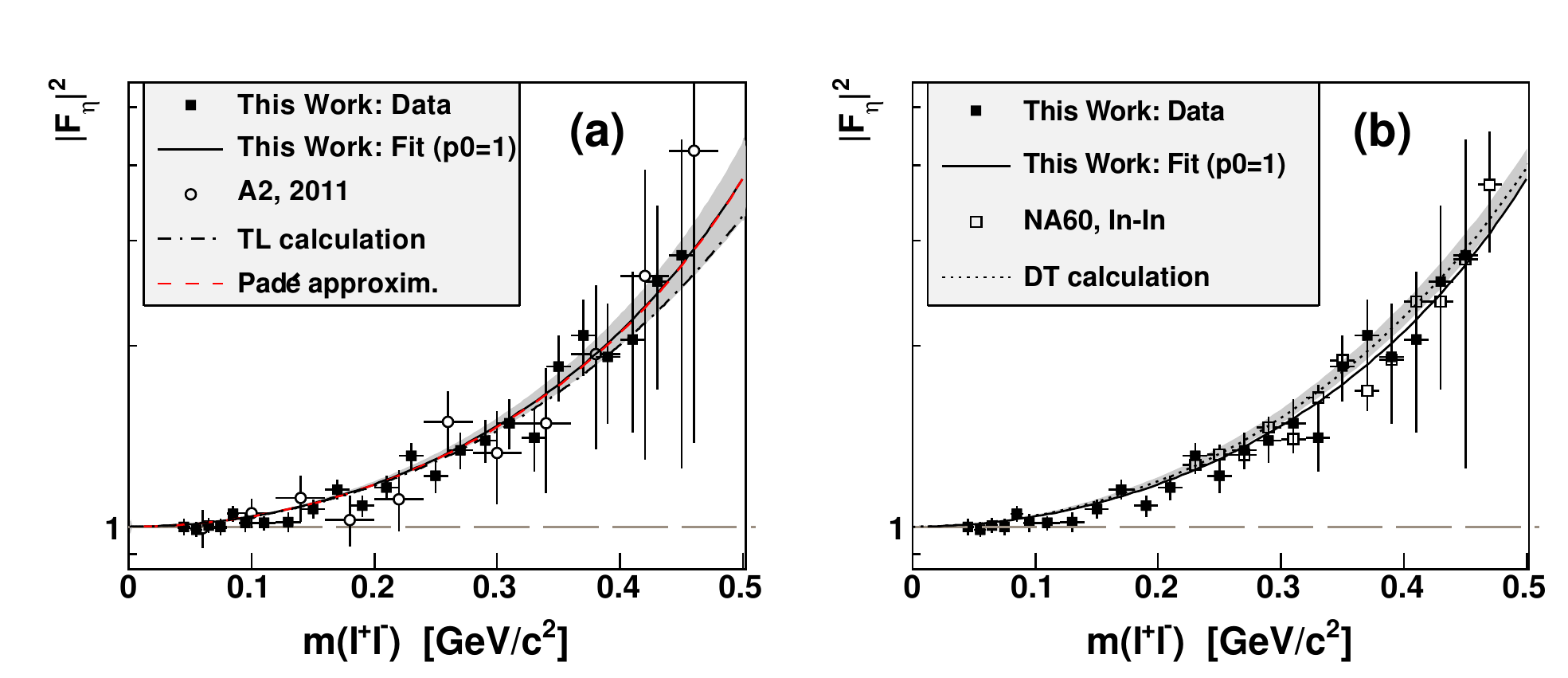}
\caption{The normalized $\eta$ TFF results obtained from A2 Collaboration at MAMI. Plot taken from Fig.~10~in~\cite{Aguar-Bartolome:2013vpw}. Their results displayed as 
solid squares~\cite{Aguar-Bartolome:2013vpw} are compared to NA60~\cite{Arnaldi:2009aa} (open squares in (b)) and former A2 results~\cite{Berghauser:2011zz} 
(open circles in (a)). The data and their fit is compared to different theoretical calculations: TL~\cite{Terschlussen}, dispersive theory (DT)~\cite{Hanhart:2013vba} 
and our results~\cite{Escribano:2013kba} from \cref{eq:etaTFF} (red line in (a) with gray error band).\label{fig:A2}}
\end{figure}
The agreement with our parameterization, \cref{eq:etaTFF}, is excellent (we note that this would not have been the case for the simplest $P^0_1$ element). 
Moreover, we can see that our parameterization is superior compared with the different theoretical 
models considered in \cite{Aguar-Bartolome:2013vpw}, though the precision from data does not allow to discard any of them. 
\\

Furthermore, new time-like data are also available from BESIII. They have been able to measure, for the first time, the $\eta'\rightarrow e^+e^-\gamma$ Dalitz 
decay\footnote{The $\eta'\rightarrow\mu^+\mu^-\gamma$ was measured before~\cite{Dzhelyadin:1979za,Dzhelyadin:1980ki} though with less 
precision, and in the higher range $q^2\in(4m_{\mu}^2,m_{\eta'}^2)$.}, allowing them to extract the normalized $\eta'$ TFF in the $q^2\in(4m_{e}^2,m_{\eta'}^2)$ 
region~\cite{Ablikim:2015wnx}. Since their last bin is at $0.75$~GeV and our approximant pole, \cref{eq:etaTFF}, lies at $0.83$~GeV, we can extrapolate up to 
their last point, obtaining again an excellent agreement ---though the current precision is not comparable to that in the $\eta$ Dalitz decay--- see \cref{fig:besiii}.\\


The excellent agreement displayed above challenged our understanding of PAs and the underlying reason behind these 
findings~\cite{Escribano:2015nra,Escribano:2015vjz,Escribano:2015yup}. Since PAs are analytic functions in the whole complex plane except at their poles location,
they cannot reproduce the analytical structure which a branch cut requires. As an example, our construction above would not 
allow to open the second Riemann sheet and, consequently, it cannot be used to determine resonance parameters. The latter would be possible if constructing the 
approximant above the threshold~\cite{Masjuan:2013jha,Masjuan:2014psa}, which however would forbid the LEPs determination. For the particular case 
when the original function to be approximated is Stieltjes with a finite radius $R$ of convergence around the origin, 
it is a well-known result in the theory of Pad\'e approximants that the sequence $P^{N+ J}_N (z)$ (with $J \geq -1$)
converges to the original function
as $N \to \infty$ on any compact set in the complex plane, excluding the cut at $R \leq z < \infty $, see \cref{sec:PadeTheo} 
---where the poles of the approximant locate to emulate the cut effects, cf. \cref{fig:stielt}.
In other words, even though the $\pi\pi$ unitary cut driving the decay is of Stieltjes nature, there is \emph{a priori} no reason
why the PA should work above the branch cut.
The surprising situation is, however, that at least the $P^L_1(s)$ sequence does seem to work well above the cut (cf. \cref{fig:A2,fig:besiii}) for the two observables.
One could speculate about the good agreement found above.

\begin{figure}[t]
\centering
  \includegraphics[width=0.6\textwidth]{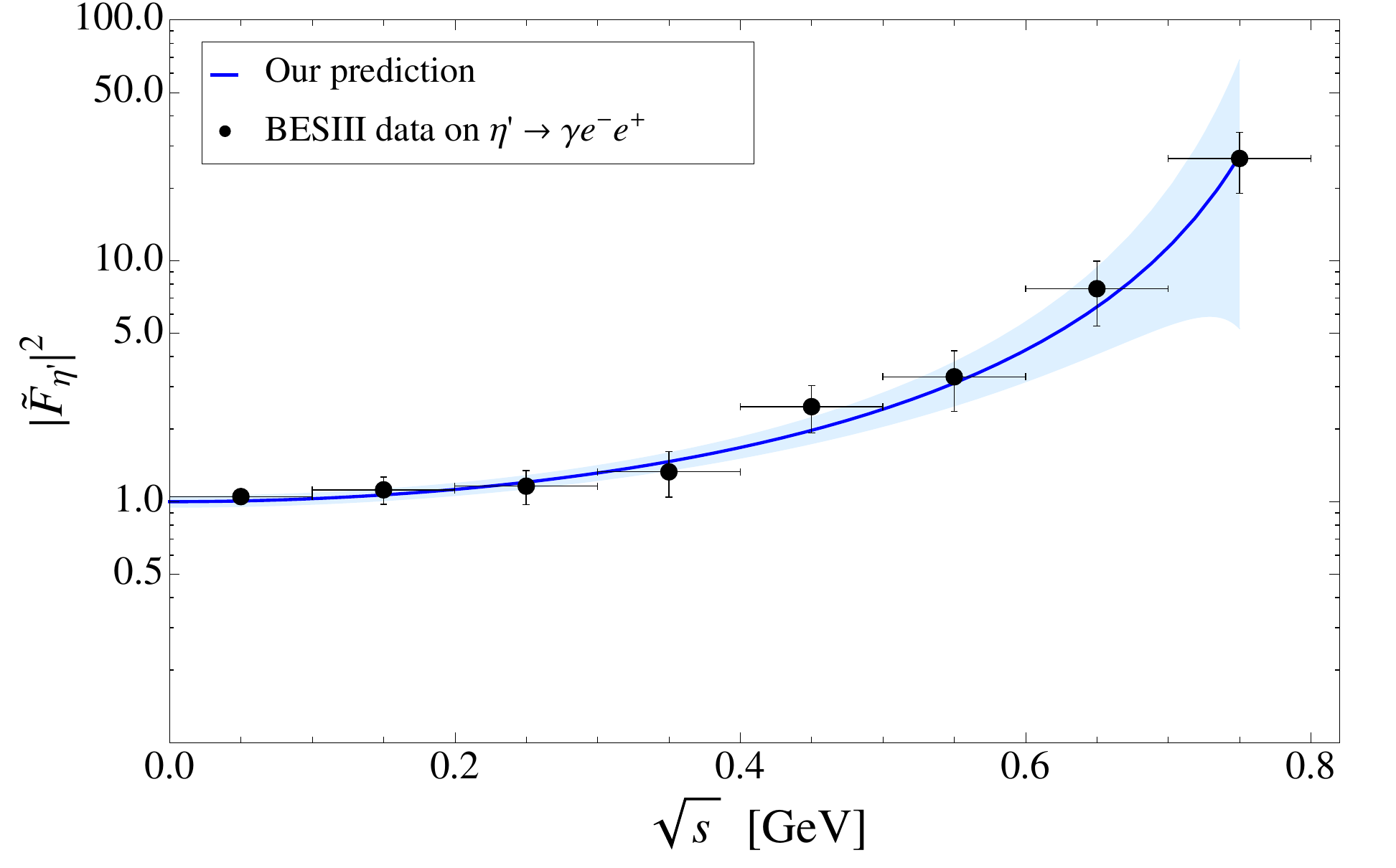}
  \caption{Our space-like $P^N_1$ prediction (blue line), \cref{eq:etaTFF}, for the $\eta'$ TFF including statistical errors (blue-band) compared with the recent BESIII results~\cite{Ablikim:2015wnx}.}
\label{fig:besiii}
\end{figure}

To qualitatively understand the situation, as a first approximation, it would be fair to say that the TFF is a meromorphic function ---as it would 
in the large-$N_c$ limit of QCD. In such scenario, PAs are an excellent approximation tool~\cite{Masjuan:2007ay}. Particularly, if the 
TFF contains a single and isolated pole, the $P^L_1(s)$ sequence reproduces the pole of the TFF with infinite precision.
As soon as the width is again switched on, the $\pi \pi$ threshold opens a branch cut responsible for that width.
Then, at first, no mathematical theorem will guarantee convergence on this scenario.
On the contrary, if the convergence theorem is to be satisfied, one would expect the single pole of the $P^L_1(s)$
to be located closer and closer to the threshold point as soon as $L \to \infty$, since this is the first singular point the PA is going to find.
However, the behavior of this $\pi\pi$ branch cut at threshold is well known  
as it comes from the $\pi\pi$ P-wave, implying the imaginary part expansion at threshold 
to behave as $(s-4m_{\pi}^2)^{3/2}$ ---such behavior can be easily obtained from \cref{eq:tffcpt,eq:loopfunc} and gives an estimate for the discontinuity size.  
Beyond, the well-studied $\pi\pi$ P-wave rescattering will be responsible to modulate such discontinuity, which is related to the well-studied $\pi\pi$ vector 
form factor. It is the smoothness of such discontinuity that explains the excellent performance found above.   
More precisely, taking the definition of a $P^L_1(s)$ given by
\begin{equation}
\label{PAeq}
P_1^L(s)=\sum_{k=0}^{L-1}a_k(s)^k+\frac{a_L(s)^L}{1-\frac{a_{L+1}(s)}{a_L(s)}} ,
\end{equation}
we would expect the PA pole to effectively account for the TFF pole, whereas the polynomial part would accurately reproduce the induced 
$\pi\pi$ P-wave effects  subthreshold. The latter would guarantee a reasonable approximation above threshold as long as the discontinuity is mild, this is, as this 
does not become resonant. This happens basically at a distance of the pole given by the half-width rule~\cite{Masjuan:2012sk},  which
can provide a simple estimate of the PAs applicability range.
In a realistic situation with multiple cuts, the picture will develop new features, but the final result would be similar.
The PA pole becomes an effective pole resulting from the combination of the absolute values of the different resonances entering the process,
closer to the one with larger coupling in the particular reaction and with shifts produced by their respective widths.
\\

For a quantitative discussion, we focus on the particular case of the $\eta$ TFF. To illustrate our statements, we choose the dispersive approach from 
Ref.~\cite{Hanhart:2013vba}, which has the appropriate $\pi\pi$ branch cut implementation along with $\pi\pi$ rescattering effects\footnote{Our study 
requires an unsubstracted version of \cite{Hanhart:2013vba}; though this may deteriorate the accuracy to which the data is reproduced, it does not 
affect our discussion.}. We generate then a space-like data set analog to that in \cref{sec:pasys}  with such model and perform a fit 
using the $P^N_1(Q^2)$ sequence of approximants. The results are shown in \cref{fig:DispAn} left and display a perfect agreement below threshold with 
respect to the dispersive model and a smooth offset above. Both the dispersive model and the PA extrapolations to the time-like region can be 
compared to real experimental data for that channel. Interestingly enough, the observed offset is below the experimental resolution as can be 
inferred from \cref{fig:DispAn} left, supporting our previous comments and justifying the observed performance of PAs. In addition, the relative 
difference of the fitted approximants with respect to the dispersive model is plotted in \cref{fig:DispAn} right. The latter suggests that a precision around 
$5\%$ and $10\%$ could be achieved from our results at energies above threshold and close to the $\eta$ mass, respectively; below this precision, 
it seems unlikely that experiments could spot deviations from our approach predictions.
\begin{figure}[t]
\centering
\includegraphics[width=0.49\textwidth]{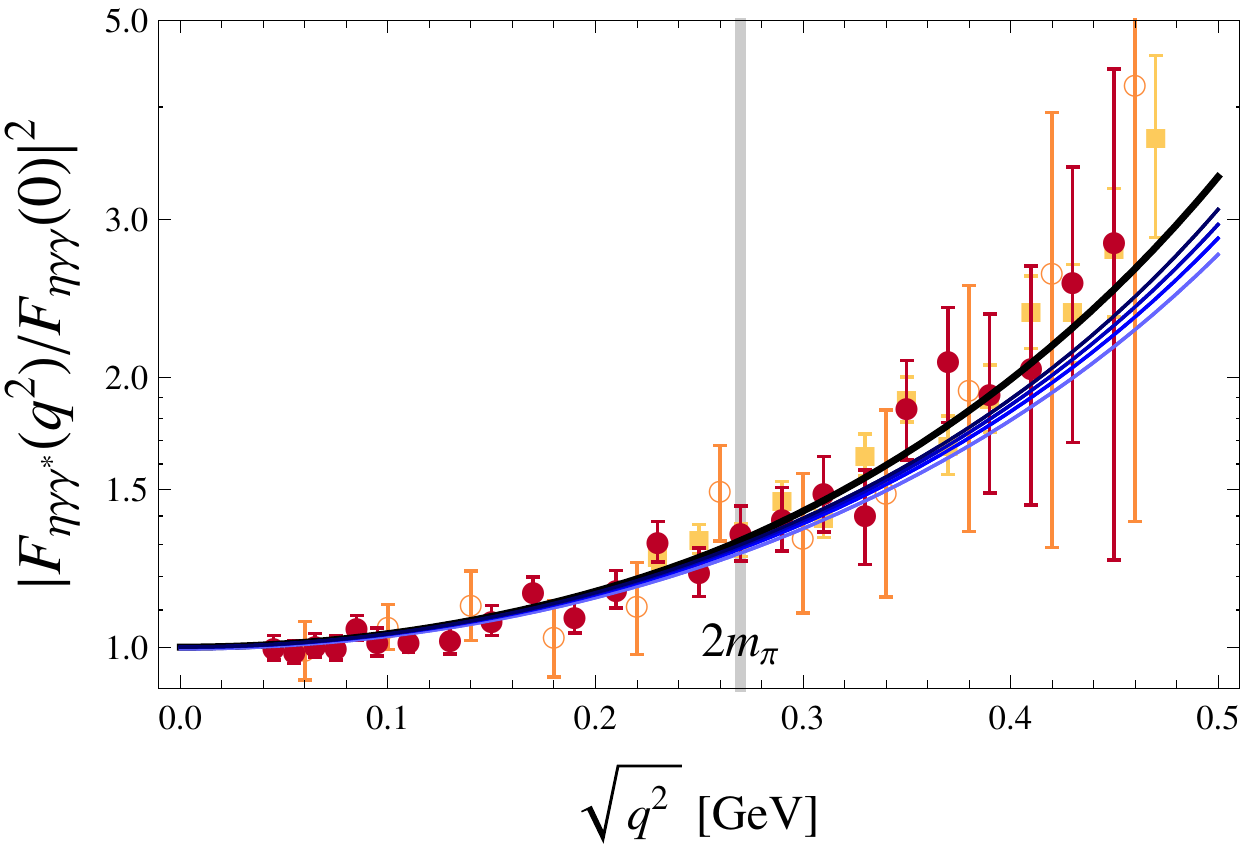}
\includegraphics[width=0.49\textwidth]{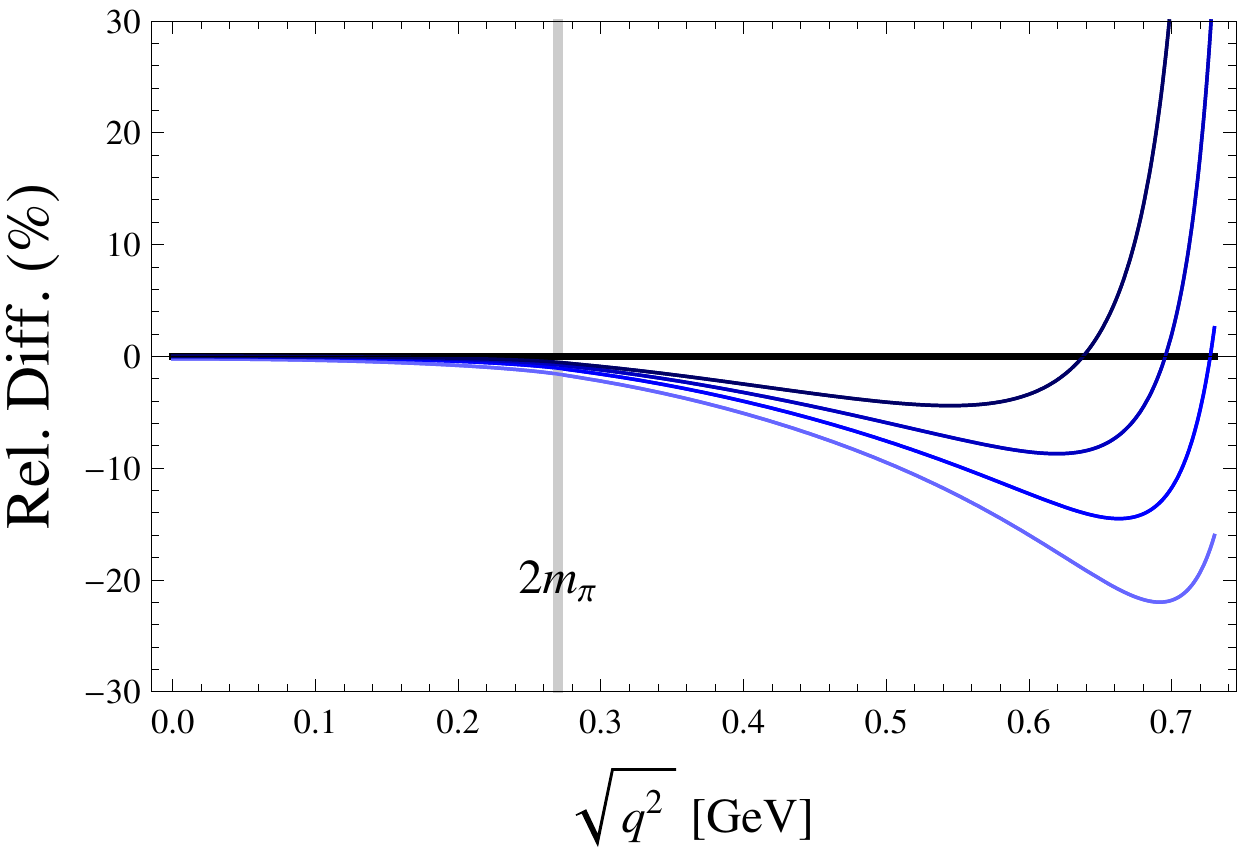}
\caption{Dispersive model for the $\eta$ TFF in the time-like region (thick-black line) compared to the fit to that model in the space-like region with $P^N_1(q^2)$ approximants, which is extrapolated to the time-like region (light to dark blue lines for $N=0,3,7,14$, respectively). The gray band represents the $\pi\pi$ threshold. The data corresponds to A2 2011 (empty orange circles) \cite{Berghauser:2011zz}, NA60 (yellow squares) \cite{Arnaldi:2009aa} and A2 2013 (red circles) \cite{Aguar-Bartolome:2013vpw} and are included to provide a context for the differencies among the model and PAs.}
\label{fig:DispAn}
\end{figure}
Still, PAs cannot differentiate among the different weights of the different contributions appearing in the TFF.
However, being fitted to experimental data, all the possible pieces are included ---as they are in the data.
An interesting exercise would be to compare our predictions below threshold against dispersive approaches, 
where each contribution must be explicitly included. 
Incorporating every single contribution represents though a formidable task, for which only those expected to play the main role are included. 
In this respect, our approach would help on identifying if relevant pieces should be included, as well as potential model dependencies in such formalisms. 
More comments later in this section.
\\

The discussion above already excludes the generalization of our results to any arbitrary Stieltjes function since one can immediately conclude that
the clue feature of the function that would allow the PA to provide a good performance above the branch cut is its behavior around the threshold point. 
As an example, for a scalar resonance the effects would be larger. Particularly, the imaginary part behavior at threshold starts at order 
$(s-4m_{\pi}^2)^{1/2}$. This, together with the broadness of scalar resonances~\cite{Agashe:2014kda}, would anticipate an early and large disagreement above 
threshold between data and PAs.\\

In the light of the excellent prediction that PAs provide for the available time-like data and the discussion above, we proceed to include 
the time-like data in our study~\cite{Escribano:2015nra,Escribano:2015yup}\footnote{The size of current errors for the TFFs in the time-like region played a  
relevant role in the previous discussion. If in the near future more precise data with discriminating power enough to discern branch cut effects become available, 
it may be necessary to carefully reconsider which data points could be used.}. 
We take, on top of the previous space-like data set, the current available experimental results for the $\eta$ and $\eta'$ Dalitz-decays. 
For the first, this includes the $\eta\rightarrow\gamma e^+e^-$ results from A2 Collaboration in 2011~\cite{Berghauser:2011zz}, together with the more 
recent ones~\cite{Aguar-Bartolome:2013vpw}, as well as the NA60 Collaboration results~\cite{Arnaldi:2009aa} obtained from the $\eta\rightarrow\gamma\mu^+\mu^-$ 
Dalitz decay\footnote{More recently, NA60 presented an improved preliminary result, 
$\Lambda^{-2}=(1.951\pm0.059_{\textrm{stat}}\pm0.042)_{\textrm{syst}}~\textrm{GeV}^{-2}$~\cite{Uras:2012qk}, but the corresponding data are not yet published.}.
These collaborations include as well their fitted VMD $\Lambda$ parameter (cf. ~\cref{eq:VMDfit}) which includes both, statistic and systematic errors. 
Unfortunately, such systematic error is not included in the data. In order to obtain the combined statistical and systematic published error, one can define a new 
source of error defined in the following way: $\Delta_{\textrm{final}} = \sqrt{\Delta_{\textrm{stat}}^2 +(\epsilon|F(Q_i)|^2)^2 }$ for each $Q_i^2$ datum, with $\epsilon$ some 
percentage. The specific value for $\epsilon$ is chosen as to reproduce their combined statistical and systematical error\footnote{We thank Marc Unverzagt for 
discussions on this subject.}. We find that for the different collaborations, $\Lambda^{-2}=(1.92\pm35_{\textrm{stat}}\pm13_{\textrm{syst}})$~\cite{Berghauser:2011zz}, 
$\Lambda^{-2}=(1.95\pm17_{\textrm{stat}}\pm5_{\textrm{syst}})$~\cite{Arnaldi:2009aa} and $\Lambda^{-2}=(1.95\pm15_{\textrm{stat}}\pm10_{\textrm{syst}})$~\cite{Aguar-Bartolome:2013vpw}, 
require $\epsilon=6.8\%, \ 1.9\%$ and $4.8\%$, respectively. For the $\eta'$, the time-like data comprise only the BESIII 
results~\cite{Ablikim:2015wnx}\footnote{As said, previous results from Lepton-G from $\eta'\rightarrow\mu^+\mu^-\gamma$ have rather large errors and are 
not available in their publication~\cite{Dzhelyadin:1979za,Dzhelyadin:1980ki}.}. Fortunately, this time they provided a systematic error for the data points.
For the fitting procedure, we employ the $\chi^2$ function
\begin{align}
\chi^2 = & \   \sum_{\textrm{SL}} \left(\frac{Q^2P^{N-1}_M(Q^2) - Q^2F_{P\gamma^*\gamma}^{\textrm{exp}}(Q^2)}{\sigma_{\textrm{exp}}}\right)^2 + \nonumber\\
   & \ \sum_{\textrm{TL}} \left(\frac{\tilde{P}^{N-1}_M(Q^2) - \tilde{F}_{P\gamma^*\gamma}^{\textrm{exp}}(Q^2)}{\sigma_{\textrm{exp}}}\right)^2 
  + \left( \frac{P^{N-1}_M(0) - F_{P\gamma^*\gamma}(0)}{\sigma_\textrm{exp}} \right)^2,
\end{align}
where $P^N_M(Q^2)$ is the PA to fit $Q^2F_{P\gamma^*\gamma}^{\textrm{exp}}(Q^2)$ and $\tilde{f}(Q^2)$ means that $\tilde{f}(0) = 1$.\\

We next report on our results. 
We start by fitting with a $P^L_1(Q^2)$ sequence. We reach up to $L=7$ both for $\eta$ and $\eta'$, which is shown in \cref{SLTLfit} as a green-dashed line. 
The smaller plot in \cref{SLTLfit} is a zoom into the time-like region. The obtained LEPs are collected in \cref{tab:psresults} 
and shown in \cref{fig:slopeSLTL,fig:slopeSLTLetap} together with our previous results in \cref{fig:slope,fig:curv} when only space-like data were included in our 
fits. The stability observed for the LEPs with the $P^L_1(Q^2)$ sequence is remarkable, and the impact of the 
inclusion of time-like data is clear since it not only allows us to reach higher precision on each PA but also enlarges our PA sequence 
by two and one elements for the $\eta$ and $\eta'$, respectively. The stability of the result is also reached earlier, the systematic error is reduced and 
our method allows to extract, for the first time, the LEPs from a combined fit to all the available 
data\footnote{The only exception is the Novosibirsk data~\cite{Achasov:2000zd,Akhmetshin:2004gw,Achasov:2006dv,Achasov:2007kw,Achasov:2013eli}
 in the resonant region around $0.700-1.400$~GeV.}. In order
\begin{figure}[t]
\centering
\includegraphics[width=0.49\textwidth]{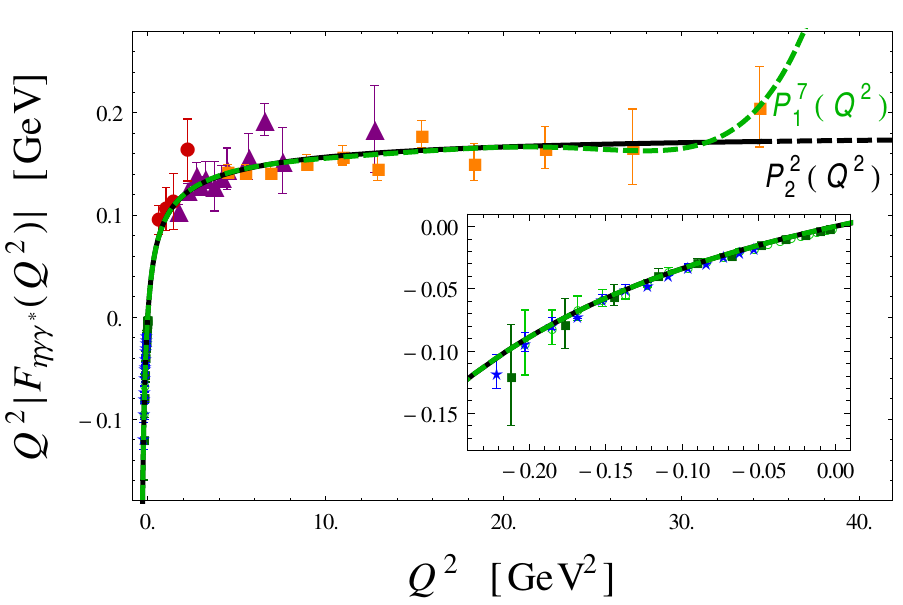}
\includegraphics[width=0.49\textwidth]{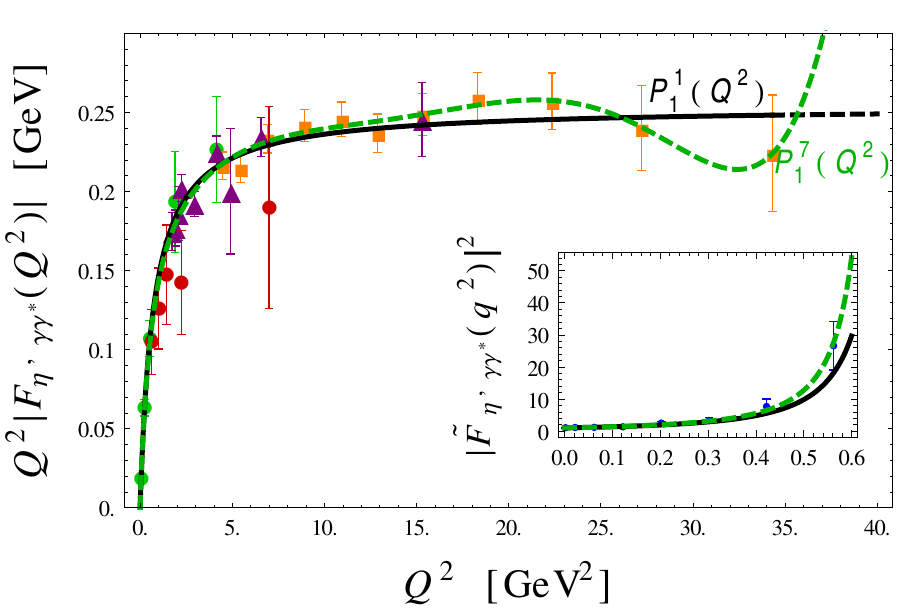}
\caption{$\eta$ and $\eta'$ TFF best fits. Green-dashed line shows our best $P^L_{1}(Q^2)$ fit and black line our best $P^N_N(Q^2)$ fit. Experimental data points in the 
space-like region are from CELLO (red circles) \cite{Behrend:1990sr}, CLEO (purple triangles) \cite{Gronberg:1997fj}, L3 (green points) \cite{Acciarri:1997yx}, 
and \babar \ (orange squares) \cite{BABAR:2011ad} collaborations. Experimental data points in the time-like region are from NA60 (blue stars)  \cite{Arnaldi:2009aa}, 
A2 2011 (dark-green squares) \cite{Berghauser:2011zz}, A2 2013 (empty-green circles) \cite{Aguar-Bartolome:2013vpw}, and BESIII (blue points) \cite{Ablikim:2015wnx}. 
The inner plot shows a zoom into the time-like region.}
\label{SLTLfit}
\end{figure}
\begin{table}[t]
\scriptsize
\centering
\begin{tabular}{@{\hspace{0em}}c@{\hspace{0.999em}}c@{\hspace{0.999em}}c@{\hspace{1em}}c@{\hspace{0.999em}}c@{\hspace{1em}}c@{\hspace{0.999em}}c@{\hspace{1em}}c@{\hspace{0.999em}}c@{\hspace{1em}}c@{\hspace{0.999em}}c@{\hspace{0em}}c}\toprule
 & \multicolumn{5}{c}{$\eta$} & \multicolumn{5}{c}{$\eta'$}\\ \cmidrule(r){2-6} \cmidrule(){7-11}
& $N$ & $b_{\eta}$ & $c_{\eta}$ & $d_{\eta}$ & $\chi^2_{\nu}$ & $N$ & $b_{\eta'}$ & $c_{\eta'}$ & $d_{\eta'}$ & $\chi^2_{\nu}$ \\\midrule
$P^N_1(Q^2)$ & $7$ & $0.575(16)$ & $0.338(22)$ & $0.198(21)$ & $0.6$  & $7$ & $1.31(4)$ & $1.74(9)$ & $2.30(19)$ & $0.7$ \\
$P^N_N(Q^2)$ & $2$ & $0.576(15)$ & $0.340(20)$ & $0.201(19)$ & $0.6$  & $1$ & $1.25(3)$ & $1.56(6)$ & $1.94(12)$ & $0.7$ \\
Final & $ $ & $0.576(11)$ & $0.339(15)$ & $0.200(14)$ & $ $ & $ $ & $1.31(4)$ & $1.74(9)$ & $2.30(19)$ & $ $ \\\bottomrule
\end{tabular}
\caption{Low-energy parameters for the $\eta$ and $\eta'$ TFFs obtained from the PA fits to experimental data. The first column indicates the type of sequence 
used for the fit and $N$ is its highest order. The last row shows the weighted average result for each LEP. We also present the quality of the 
fits in terms of $\chi^2_{\nu}$. Errors are only statistical and symmetrical.\label{tab:psresults}}
\end{table}
\begin{figure}[t]
\centering
\includegraphics[width=0.47\textwidth]{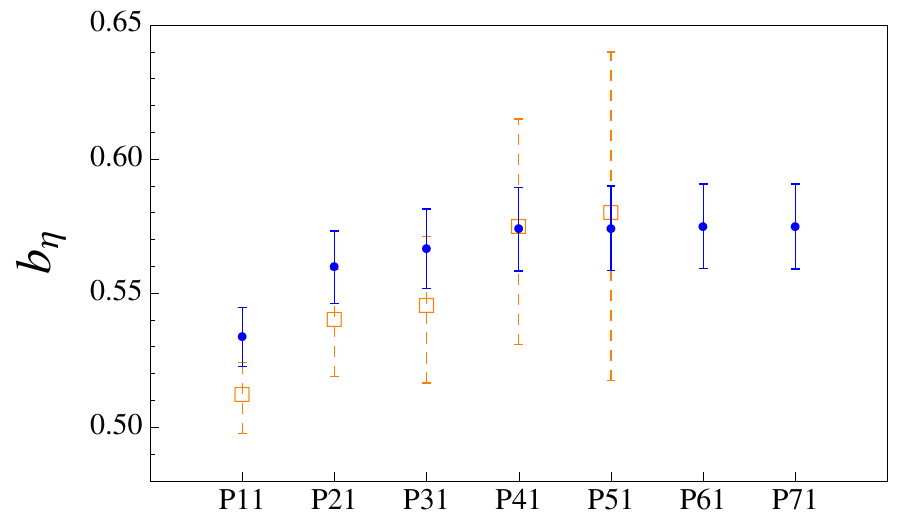}
\hspace{0.04\textwidth}
\begin{minipage}[b]{0.47\textwidth}
\caption{ Slope (top), curvature (bottom-left), and third derivative (bottom-right) predictions for the $\eta$ TFF using the $P^L_1(Q^2)$ (blue points). 
Previous space-like data results, \cref{fig:slope}, are also shown (empty-orange squares). Only statistical errors are shown. \label{fig:slopeSLTL}}
\end{minipage}
\bigskip
\includegraphics[width=0.47\textwidth]{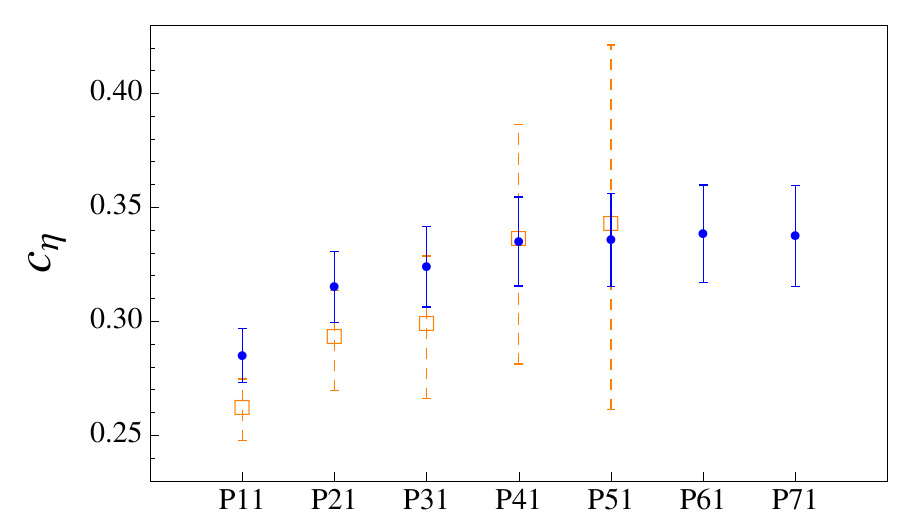}
\hspace{0.04\textwidth}
\includegraphics[width=0.47\textwidth]{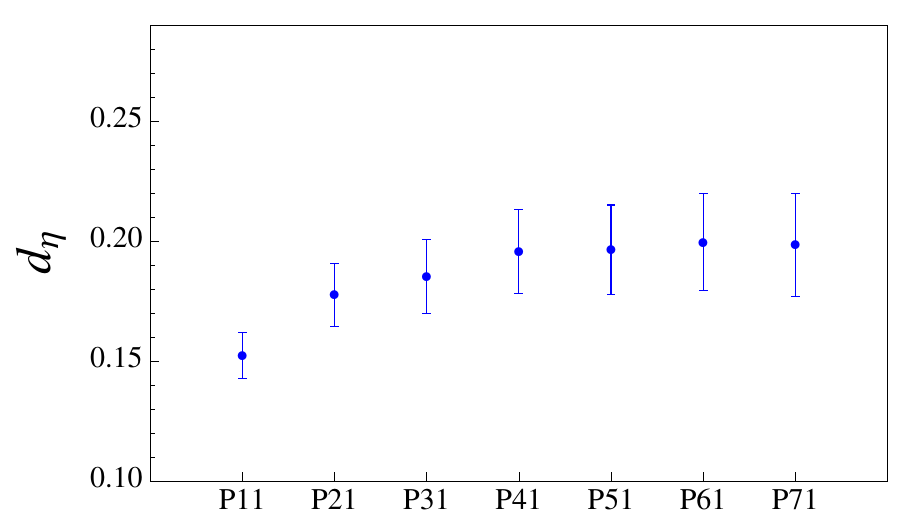}
\end{figure}
\begin{figure}[t]
\centering
\includegraphics[width=0.47\textwidth]{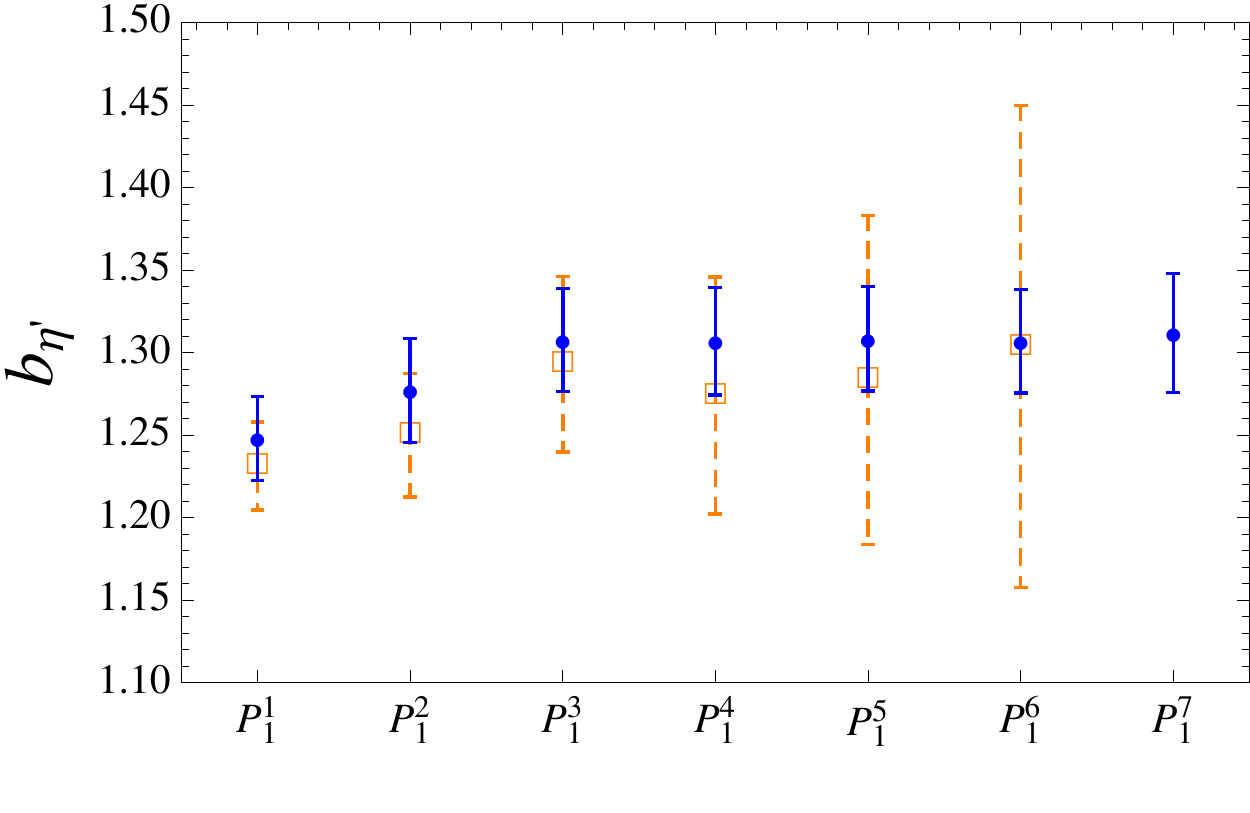}
\hspace{0.04\textwidth}
\begin{minipage}[b]{0.47\textwidth}
\caption{ Slope (top), curvature (bottom-left), and third derivative (bottom-right) predictions for the $\eta'$ TFF using the $P^L_1(Q^2)$ (blue points). 
Previous space-like data results, \cref{fig:slope}, are also shown (empty-orange squares). Only statistical errors are shown. \label{fig:slopeSLTLetap}}
\end{minipage}
\bigskip
\includegraphics[width=0.47\textwidth]{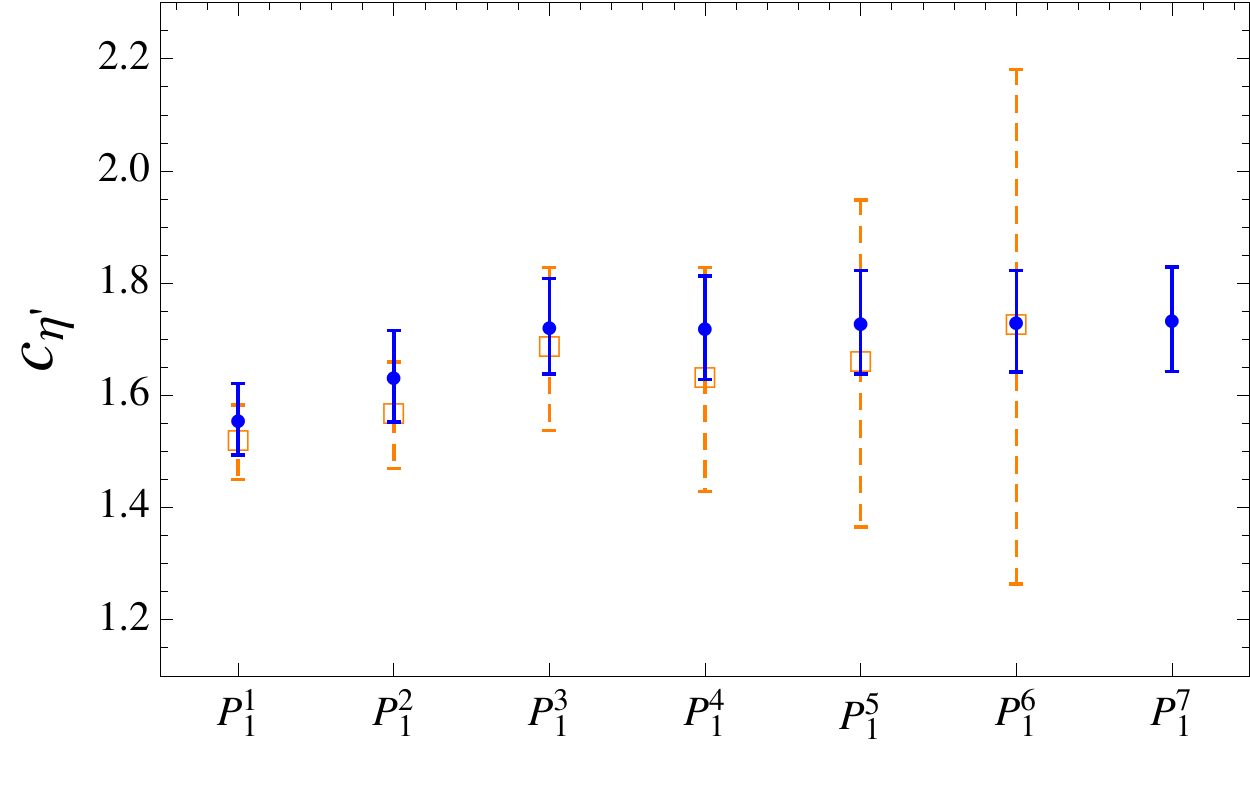}
\hspace{0.04\textwidth}
\includegraphics[width=0.47\textwidth]{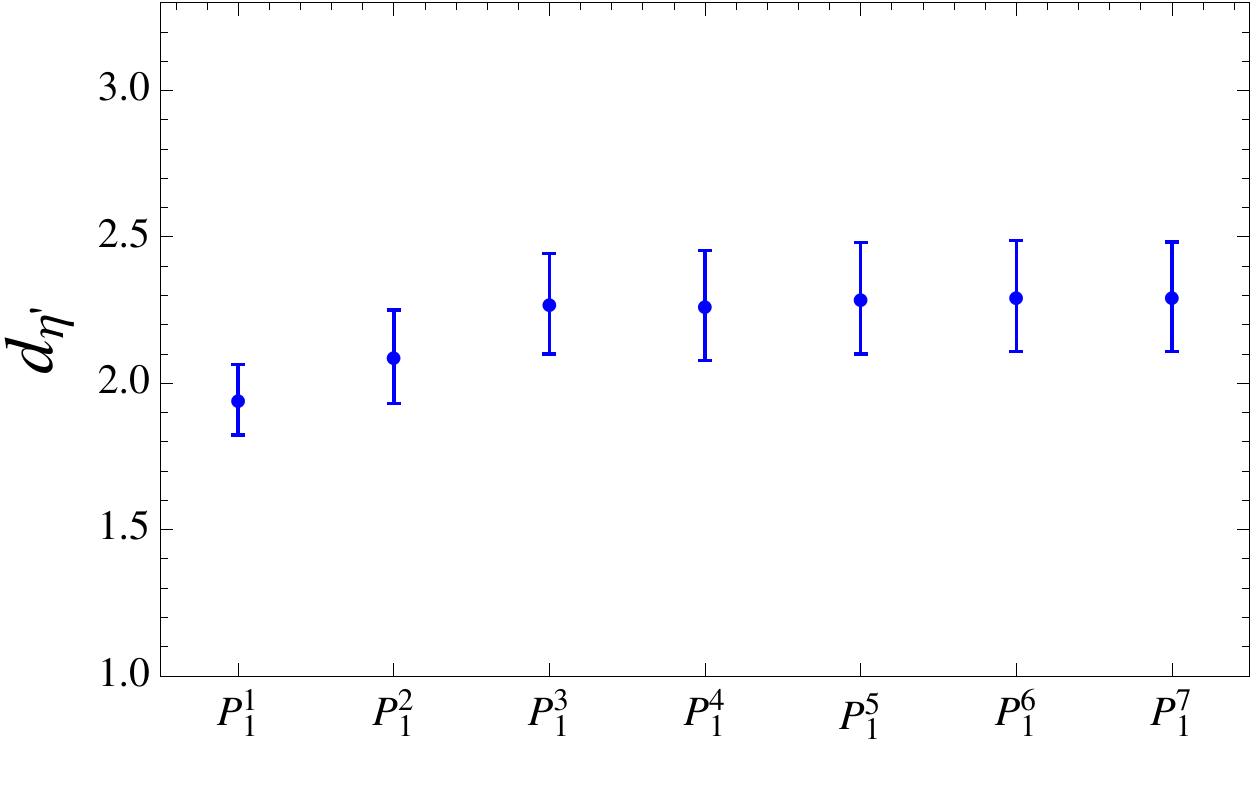}
\end{figure}
\hspace{-2pt}to reproduce the asymptotic  behavior of the TFF, we have also considered the $P^N_{N}(Q^2)$ sequence (second row in \cref{tab:psresults}). 
The results obtained are in very nice agreement with our previous determinations. The best fit is shown as black-solid line in \cref{SLTLfit}. 
We reach $N=2(1)$ for the $\eta(\eta')$. Since these approximants contain the correct high-energy behavior built-in, 
they can be extrapolated up to infinity (black-dashed line in \cref{SLTLfit}) and then predict the leading $1/Q^2$ coefficient~\cite{Escribano:2015nra,Escribano:2015yup}
\begin{align}
\label{eq:etainfSLTL}
\lim_{Q^2\to\infty }Q^2F_{\eta\gamma^*\gamma}(Q^2) &=0.177^{+0.020}_{-0.009}~\textrm{GeV},\\
\label{eq:etapinfSLTL}
\lim_{Q^2\to\infty }Q^2F_{\eta'\gamma^*\gamma}(Q^2) &=0.254(4)~\textrm{GeV}.
\end{align}
Even though the prediction for the $\eta$ is larger ---but compatible within errors--- than our previous result from the space-like data, \cref{eq:etaSLinf}, it is still 
far below the \babar \  time-like measurement  
at $q^2=112$~GeV$^2$, $F_{\eta\gamma^*\gamma}(112)=0.229(30)(8)$~GeV~\cite{Aubert:2006cy}. The result for the $\eta'$ is on the other hand similar to the 
previous space-like determination \cref{eq:etapSLinf}. See more discussions on \babar time-like measurements below.\\

Our combined weighted average results from Table \ref{tab:psresults}, taking into account both types of PA sequences, give~\cite{Escribano:2015nra,Escribano:2015yup}
\begin{align}
\label{etaetapvaluesSLTLslope}
 b_{\eta} &= 0.576(11)_{\rm stat}(4)_{\rm sys}  &   b_{\eta'} &= 1.31(4)_{\rm stat}(1)_{\rm sys} \\
\label{etaetapvaluesSLTLcurv}
 c_{\eta} &= 0.339(15)_{\rm stat}(5)_{\rm sys}  &   c_{\eta'} &= 1.74(9)_{\rm stat}(3)_{\rm sys} \\
\label{etaetapvaluesSLTLthird}
 d_{\eta} &= 0.200(14)_{\rm stat}(18)_{\rm sys} &   d_{\eta'} &= 2.30(20)_{\rm stat}(21)_{\rm sys}
\end{align}
where the first error is statistic and the second systematic, see \cref{tab:syst}. 
These results can be compared to our previous results from space-like data, \cref{eq:etaSLleps,eq:etapSLleps}, which shows the great improvement not only on the statistical 
error, but on the systematic one as well, both by an order of magnitude. Our results, \cref{etaetapvaluesSLTLslope,etaetapvaluesSLTLcurv,etaetapvaluesSLTLthird}, 
represent the most precise determination to date for the LEPs. 
As a further check, for the $\eta'$, we have checked the relevance of including the last data points in the time-like region. 
We have found that omitting them yields very similar results. Therefore, we believe this justifies their inclusion in our fitting procedure. 
For comparison, we update in \cref{fig:slopecompSLTL} our previous \cref{fig:slopecomp} to include this additional determination .
\begin{figure}[t]
\centering
 \includegraphics[width=0.49\textwidth]{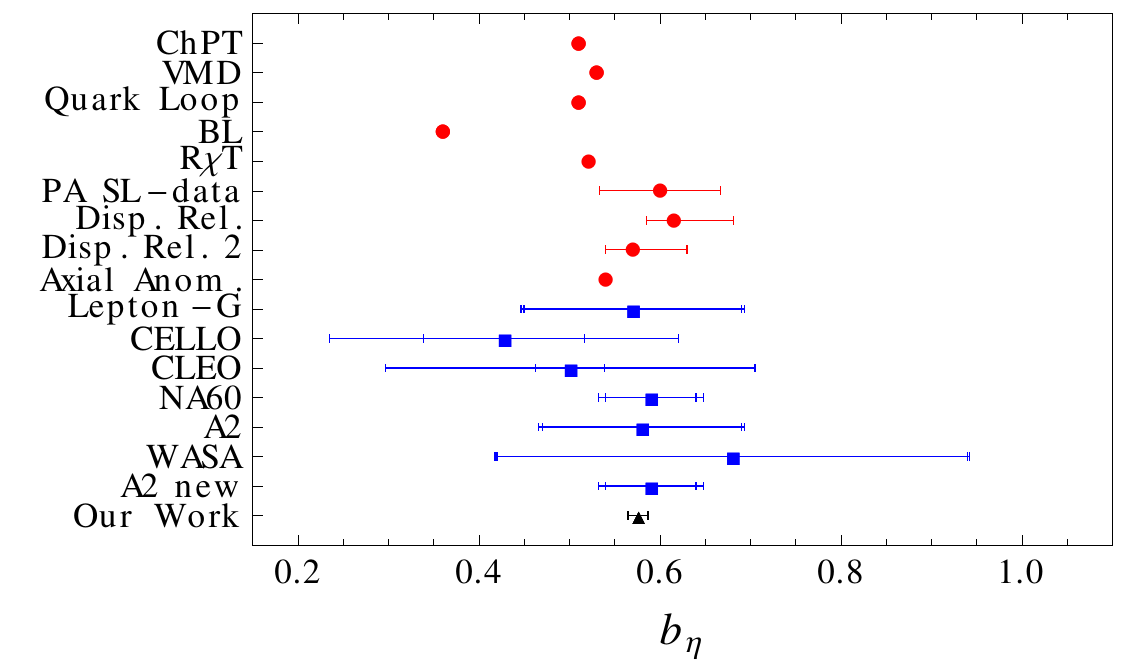}
 \includegraphics[width=0.49\textwidth]{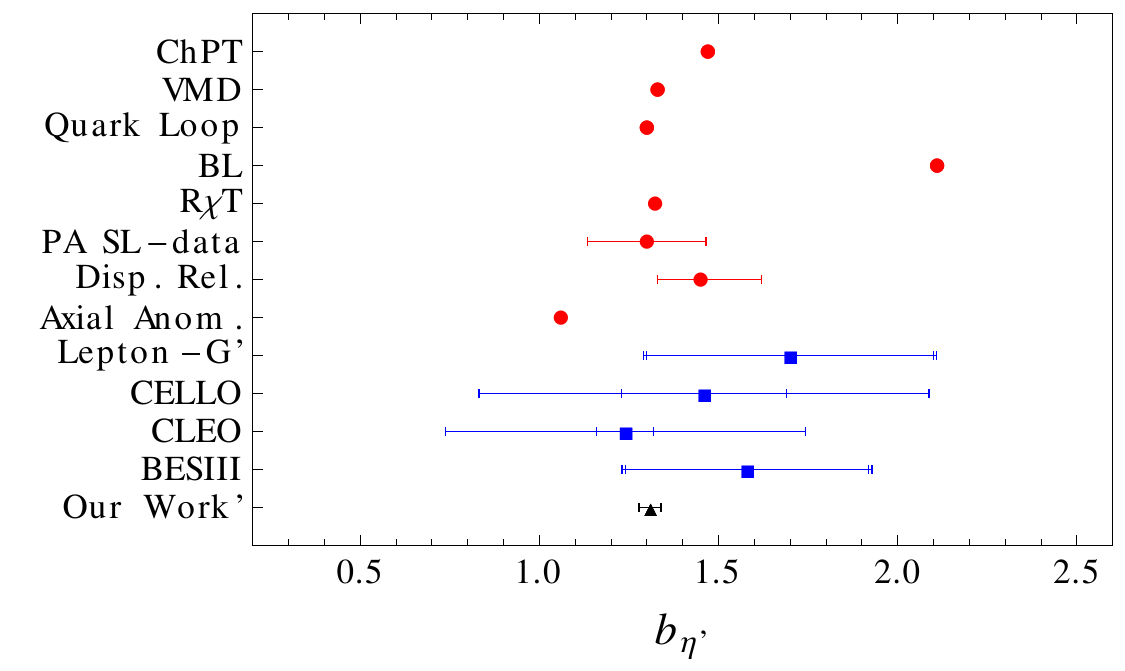}
\caption{Slope determinations for the $\eta$ from different theoretical (red circles) and 
experimental (blue squares) references discussed in the text. Inner error is the statistical one and larger error is the combination of 
statistical and systematic errors. 
ChPT~\cite{Bijnens:1988kx,Bijnens:1989jb}, VMD, Quark Loop, BL~\cite{Ametller:1991jv}, R$\chi$T~\cite{Czyz:2012nq}, Disp. Rel.~\cite{Hanhart:2013vba}, 
Disp. Rel. 2~\cite{Kubis:2015sga,Xiao:2015uva}, Axial Anom.~\cite{Klopot:2013laa}, Lepton-G~\cite{Dzhelyadin:1980kh}, Lepton-G'~\cite{Landsberg:1986fd}, 
CELLO~\cite{Behrend:1990sr}, CLEO~\cite{Gronberg:1997fj}, NA60~\cite{Arnaldi:2009aa}, A2~\cite{Berghauser:2011zz}, WASA~\cite{Hodana:2012rc}, A2 new~\cite{Aguar-Bartolome:2013vpw}, 
BESIII~\cite{Ablikim:2015wnx}, PA SL-data~\cite{Escribano:2013kba}, Our Work~\cite{Escribano:2015nra}, Our Work'~\cite{Escribano:2015yup}.
\label{fig:slopecompSLTL}}
\end{figure}
Note as said, that previous dispersive results~\cite{Hanhart:2013vba} stands at one standard deviation from ours, both for $\eta$ and $\eta'$. 
The reason being the omission of the $a_2$ tensor meson contribution, which was observed in~\cite{Kubis:2015sga} and recently included 
in their later analysis for the $\eta$~\cite{Xiao:2015uva}, bringing their result closer to our value and confirming thereby our determination 
---the $\eta'$ modified result has not been reported--- which could have been predicted from our $b_P$ determination and
shows the potential of our method to estimate unaccounted effects in dispersive approaches. In addition, 
this result could be used as an input to perform further subtractions in their method.
%
\\

After showing the excellent precision achieved in our study, we would like to comment on the role of data in our results. 
The models studied in \cref{sec:PAbasics} suggest that, due to the large amount of low-energy data, the presence of new data 
will not improve on the systematic errors achieved so far (except for the $d_{\eta(\eta')}$ parameter if higher elements are reached, see \cref{tab:syst}). 
However, since the current limitation, except for $d_{\eta(\eta')}$, 
is the statistical one, new precise data will be very welcome. In principle, one may think that it is the low-energy data 
which may be preferred. We notice however, that in order to reach large PA sequences ---which allow for more accurate extractions--- 
the high-energy data, which from $5$ to $35~\textrm{GeV}^2$ is dominated by \babar, is also very important. To show the role 
of each collaboration, we report for the $\eta$ case 
(similar results are obtained for the $\eta'$) on the different results for the slope and asymptotic values arising from each one in \cref{tab:coll}.
We find that a fit exclusively to \babar data yields similar results both for the slope and asymptotic values than other 
space-like configurations. This contrast for instance for the asymptotic value obtained when only CELLO or time-like data is used. 
The role of \babar data is then twofold, allowing to reach larger approximants, such as $P^2_2(Q^2)$ and determining basically 
the asymptotic value. In view of the $\pi^0$ puzzle between \babar~\cite{Aubert:2009mc} and Belle~\cite{Uehara:2012ag} results, a second 
experimental measurement covering the high-energy region would be very welcome here.  
In the future, the Belle II Collaboration may be able to provide such measurements.\\ 

\begin{table}[t]
\scriptsize{
\centering
\begin{tabular}{ccccccc}\toprule
& Data range & \multicolumn{2}{c}{$P^L_1(Q^2)$} & \multicolumn{3}{c}{$P^N_N(Q^2)$} \\\cmidrule(r){3-4}\cmidrule(l){5-7}
& (GeV$^2$) & $L$ & $b_{\eta}$ & $N$ & $b_{\eta}$ &  $\eta_{\infty}$ \\\midrule
CELLO~\cite{Behrend:1990sr} & 0.62--2.23  &$2$ & $0.48(20)$ & $1$ & $0.427(66)$ & $0.193(30)$ \\
CLEO~\cite{Gronberg:1997fj} & 1.73--12.74&$3$ & $0.73(12)$ & $1$ & $0.522(19)$ & $0.157(5)$ \\
\babar~\cite{BABAR:2011ad} & 4.47--34.38&$4$ & $0.53(9)$ & $1$ & $0.509(14)$ & $0.162(3)$ \\
CELLO,CLEO &0.62--12.74 & $3$ & $0.65(9)$ & $2$ & $0.704(87)$ & $0.25(10)$ \\
SL &0.62--34.38 & $5$ & $0.58(6)$ & $2$ & $0.66(10)$ & $0.161(24)$ \\ \midrule
A2-11,A2-13~\cite{Berghauser:2011zz,Aguar-Bartolome:2013vpw}  & -0.212 -- -0.002& $2$ & $0.475(76)$ & $1$ & $0.551(40)$ & $0.149(11)$ \\
NA60~\cite{Arnaldi:2009aa} & -0.221 -- -0.053 & $3$ & $0.640(77)$ & $1$ & $0.582(19)$ & $0.141(5)$ \\
TL  & -0.221 -- -0.002& $3$ & $0.565(87)$ & $1$ & $0.576(17)$ & $0.143(5)$ \\ \midrule
CELLO,TL & -0.221 -- 2.23 &$5$ & $0.531(39)$ & $2$ & $0.533(30)$ & $0.203(58)$ \\
CELLO,CLEO,TL &  -0.221 -- 12.74 & $6$ & $0.567(22) $ & $1$ & $0.550(13)$ & $0.152(3)$ \\
A2-11,A2-13,SL &   -0.212 -- 34.38 & $7$ & $0.561(35)$ & $2$ & $0.569(28)$ & $0.178(16)$ \\ \midrule
\textbf{TL,SL} & \textbf{ -0.221 -- 34.38}  & $\mathbf{7}$ & $\mathbf{0.575(16)}$ & $\mathbf{2}$ & $\mathbf{0.576(15)}$ & $\mathbf{0.177(15)}$ \\\bottomrule
\end{tabular}
}
\caption{Role of the different sets of experimental data in determining slope and asymptotic values ($\eta_{\infty}$) of the $\eta$ TFF. 
SL refers the the space-like data set, i.e., data from CELLO,CLEO,\babar\cite{Behrend:1990sr,Gronberg:1997fj,BABAR:2011ad} 
collaborations, and TL refers to the time-like data set, i.e., data from NA60+A2-11+A2-13~\cite{Arnaldi:2009aa,Berghauser:2011zz,Aguar-Bartolome:2013vpw} 
collaborations. Bold numbers are our final result. No systematic errors included.}
\label{tab:coll}
\end{table}

To complete our previous discussion, we comment as well on the role of $\Gamma_{\eta\to\gamma\gamma}$ in our extractions given the current 
discrepancy among $e^+e^-$ collider results and Primakoff measurements for this quantity. 
We find that our previous results are rather stable though mildly depend on this input. For instance, if we would 
have used the value measured through the Primakoff mechanism omitted in the PDG average~\cite{Agashe:2014kda} 
(i.e., $\Gamma_{\eta\gamma\gamma}^{\textrm{Primakoff}}=0.476(62))$~keV~\cite{Agashe:2014kda}), we would find $b_{\eta}=0.570(13)$, which represents 
half a standard deviation with respect to our result, \cref{etaetapvaluesSLTLslope}. Even though this does not represent a puzzle as everything 
agrees within uncertainties, it may suggest to look again for a Primakoff measurement\footnote{This kind of measurement is part of the experimental programme of $GlueX$ 
Collaboration at CLAS in Jefferson Lab~\cite{Dudek:2012vr}.}, specially given 
that both, $\Gamma_{\eta\gamma\gamma}$ and $b_{\eta}$,  play a central role in our following calculations: $\eta-\eta'$ mixing, $P\rightarrow\bar{\ell}\ell$ decays and $(g-2)$.
\\

Finally, we comment on the result from the \babar Collaboration at very large time-like energies~\cite{Aubert:2006cy}. As already mentioned before, \babar measured 
the process $e^+e^-\rightarrow\gamma^*\rightarrow\eta^{(\prime)}\gamma$ at the center of mass energies 
$\sqrt{s}=10.58$~GeV. Its relation to the TFF~\cite{Rosner:2009bp}, 
\begin{equation}
\label{eq:cccontinuum}
  \sigma(e^+e^-\to P\gamma) = \frac{2\pi^2\alpha^3}{3} \left(1-\frac{m_P^2}{s}\right)^3 \left\vert F_{P\gamma^*\gamma}(s)\right\vert^2  ,
\end{equation}
where $s$ the center of mass energy squared, allowed them to extract a measurement for the TFF absolute value in the time-like region for  
$q^2=112~\textrm{GeV}^2$, obtaining $q^2F_{\eta\gamma^*\gamma}(q^2)=0.229(31)$~GeV and $q^2F_{\eta'\gamma^*\gamma}(q^2)=0.251(21)$~GeV, where 
statistical and systematic errors have been added in quadrature. 
Taking into account the kinematical factor $(1-m^2_P/s)^3$ (see Ref.~\cite{Rosner:2009bp}) that was missing in the \babar expression, and 
assuming that duality $F_{P\gamma^*\gamma}(Q^2)=F_{P\gamma^*\gamma}(q^2)$~\cite{Aubert:2006cy} holds at large but finite energies, implies
\begin{equation}
\label{eq:babar112}
\begin{split}
 |Q^2F_{\eta\gamma^*\gamma}(Q^2)|_{Q^2=112~\textrm{GeV}^2} &= 0.231(31)~\textrm{GeV}, \\
 |Q^2F_{\eta'\gamma^*\gamma}(Q^2)|_{Q^2=112~\textrm{GeV}^2} &= 0.254(21)~\textrm{GeV}.
\end{split}
\end{equation}
This suggests to include these data points in our fitting procedure, assuming that at this high-momentum transfer, the duality between 
space- and time-like region holds, and no extra error should be included. For the $\eta'$, given our results in \cref{eq:etapSLinf,eq:etapinfSLTL}, in excellent 
agreement with \babar results, it is clear that this won't change much. For the $\eta$ case, its inclusion will mainly modify the asymptotic prediction from $P^2_2$ 
increasing its value up to $\lim_{Q^2\to\infty}Q^2F_{\eta\gamma^*\gamma}(Q^2)=0.247$~GeV, higher than the \babar result and with a good $\chi^2_{\nu}<1$. 
Curiously enough, the fit function at $Q^2=112~\textrm{GeV}^2$ is $Q^2F_{\eta\gamma^*\gamma}(Q^2)=0.219$~GeV, below \cref{eq:babar112}. Even worse is 
the prediction (assuming duality) for the time-like counterpart at $q^2=112~\textrm{GeV}^2$, $q^2F_{\eta\gamma^*\gamma}(q^2)=0.307$~GeV.
One may speculate in light of these results on the validity of duality assumptions and whether the asymptotic regime is reached or not. Actually, 
a recent analysis of the $\eta$ and $\eta'$ TFFs based on perturbative corrections~\cite{Agaev:2014wna} concludes that the difference between the time- 
and space-like form factors at $Q^2=112~\textrm{GeV}^2$ can be of the order $(5-13)\%$ for different pseudoscalar distribution amplitudes, and can be 
enhanced by Sudakov-type corrections. 
It may be surprising to find such a large error on duality assumptions at these energies. Notice however that, even at these high-energies, the TFFs 
are sensitive to soft scales for $x\simeq0(1)$, see \cref{sec:tffpqcd}. These corrections become relevant if the pseudoscalar 
DAs are relatively broad, which seems the case for the $\pi^0$ and $\eta$ cases, which TFFs, definitely not VMD-like, seems to require a broad DA~\cite{Agaev:2014wna}. 
Similar results are found from CLEO results~\cite{Pedlar:2009aa}, which measured cross sections at $q^2=14~\textrm{GeV}^2$ ---assuming continuum contribution and duality--- 
lead to 
\begin{equation}
\label{eq:cleo14}
\begin{split}
 |Q^2F_{\eta\gamma^*\gamma}(Q^2)|_{Q^2=14~\textrm{GeV}^2} &= 0.203(41)~\textrm{GeV}, \\
 |Q^2F_{\eta'\gamma^*\gamma}(Q^2)|_{Q^2=14~\textrm{GeV}^2} &= 0.249(29)~\textrm{GeV},
\end{split}
\end{equation}
even though with potentially larger corrections being at lower energies. On the other hand, the $\eta'$ 
seems not that affected, which may suggest a much narrower DA less sensitive to the end-point behavior. This would be reasonable given its heavier singlet 
nature, introducing an explicit scale that would drive the DA away from a flat shape. Still, to draw firmer conclusions, further and more precise experimental 
results are required. There is at the moment an ongoing analysis at BES III to measure such processes at $q^2=18.5$~GeV$^2$~\cite{Marc:private}. 

Alternatively, we can use our TFF description to extract the cross section which duality arguments would imply for these processes when using \cref{eq:cccontinuum}.
This contribution is of relevance when estimating background contribution to $\psi(nS)\rightarrow \gamma \eta^{(\prime)}$ decays. We obtain at the center of mass energies 
of the different resonances, the cross sections quoted in \cref{tab:ccprod}, where, for completeness, we include the $\pi^0$ results obtained from the work 
in Ref.~\cite{Masjuan:2012wy}.
\begin{table}[t]
\centering
\footnotesize
   \begin{tabular}{cccc} \toprule
                  & \multicolumn{3}{c}{$\sigma(e^+e^-\to P\gamma)$ (fb)} \\ 
     $P$     & $J/\psi$   &   $\psi(2S)$   &   $\psi(3770)$ \\ \midrule
   $\pi^0$   &   $324(16)$  &  $192(14)$   &  $180(14)$   \\
   $\eta$   &   $237(24)$  &  $130(16)$   &  $120(14)$   \\
   $\eta'$   &   $456(34)$  &  $264(27)$  &   $245(28)$ \\  \bottomrule
   \end{tabular}
\caption{The continuum cross sections for $\sigma(e^+e^-\to P\gamma)$ processes in fb at the center of mass energies of different charmonium resonances. \label{tab:ccprod}}
\end{table}
This represents an improvement with respect to Ref.~\cite{Rosner:2009bp} as the latter assumes the asymptotic behavior to extrapolate down to the charmonium energies. 
Still, we note that these predictions are only valid in the case that duality holds (strictly as $Q^2\to\infty$) and would require a more refined analysis 
in line of \cite{Agaev:2014wna} in order to estimate for these corrections.

\section{Conclusions}
\label{sec:concl}

In this chapter we have described how PAs can be used as fitting functions in order to extract relevant information from the pseudoscalar TFFs, 
namely the LEPs and the asymptotic behavior. 
We have demonstrated this using three different models for the TFF, illustrating the PAs performance in cases where convergence theorems exist or not, that
has allowed on top to estimate a systematic error, an unique property of our approach. 
The proposed method has been applied then to the real $\eta$ and $\eta'$ cases, obtaining an excellent performance in the space-like region.
Moreover, we have discussed that our previous description can be extrapolated for these TFFs into the low-energy time-like region up to an excellent accuracy, 
allowing for the first combined description as well as an improved LEPs determination.
All in all, our method has allowed a systematic and model-independent robust extraction for the central 
quantities that we need for later reconstructing the (single-virtual) pseudoscalar TFFs. 
\begin{table}[t]
\centering
\footnotesize
\begin{tabular}{cccccc}\toprule
 & $F_{P\gamma\gamma}$ & $b_P$  & $c_P$  & $d_P$ & $P_{\infty}$\\ 
 & $(\textrm{GeV}^{-1})$ &   &    &   & (GeV)\\\midrule
$\pi^0$~\cite{Masjuan:2012wy} & $0.2725(29)$ & $0.0324(12)(19)$  & $0.00106(9)(25)$   & ---  & $2F_{\pi}$\\ 
$\eta$~\cite{Escribano:2015nra} &   $0.2738(47)$   & $0.576(11)(4)$ & $0.339(15)(5)$  &  $0.200(14)(18)$  & $0.177(15)$\\ 
$\eta'$~\cite{Escribano:2015yup} & $0.3437(55)$ & $1.31(3)(1)$  & $1.74(9)(2)$   & $2.30(20)(21)$  & $0.254(4)$\\\midrule
$\eta^{SL}$~\cite{Escribano:2013kba} &   $0.2738(47)$   & $0.60(6)(3)$ & $0.37(10)(7)$  &  ---  & $0.160(24)$\\ 
$\eta'^{SL}$~\cite{Escribano:2013kba} & $0.3437(55)$ & $1.30(15)(7)$  & $1.72(47)(34)$   & ---  & $0.255(4)$\\\bottomrule
\end{tabular}
\caption{The main results from our work in this chapter. The numbers come from the combined space- and time-like data, \cref{sec:SL,sec:TL}. 
         We include the $\pi^0$ results from Ref.~\cite{Masjuan:2012wy} and the TFFs at zero energies implied by experiments. In addition, we quote what 
         would be obtained from space-like data alone, which is labelled as $P^{SL}$. \label{tab:chap1mainres}}
\end{table}
Moreover, we were able to explain the existing discrepancies among space- and time-like data analysis from different collaborations 
on the basis of a systematic error. 
Our main results are the low-energy parameters for the TFF expansion 
\begin{equation}
F_{P\gamma^*\gamma}(Q^2) = F_{P\gamma\gamma}\left(1 - b_P \frac{Q^2}{m_P^2} + c_P\frac{Q^4}{m_P^4} - d_P\frac{Q^6}{m_P^6} + ... \right),
\end{equation}
as well as the asymptotic behavior, $P_{\infty}\equiv\lim_{Q^2\to\infty}Q^2F_{P\gamma^*\gamma}(Q^2)$.
We recapitulate them together with the $\pi^0$ results from space-like data, which were not analyzed here, but in Ref.~\cite{Masjuan:2012wy},
in \cref{tab:chap1mainres}. We expect to reanalyze the $\pi^0$ TFF as well in the near future once the new data from BESIII~\cite{Adlarson:2014hka} in the 
low-energy space-like $(0.3\leq Q^2\leq 10)~\textrm{GeV}^2$ range and time-like data from NA62~\cite{Hoecker:2016lxt} and A2~\cite{Marc:private} 
collaborations from the $\pi^0\to\gamma e^+e^-$ decay become available. Moreover, there are prospects to measure the $\pi^0$ TFF at even lower space-like 
energies at KLOE-2~\cite{Babusci:2011bg} and $GlueX$~\cite{Gan:2015nyc} collaborations. This would allow for a statistical and systematic improvement for the $\pi^0$ LEPs. 
Additional data for the $\eta$ and $\eta'$ mesons is expected too in a similar range. Although this would not improve much the systematic error, an improvement on the 
statistical one ---the dominant at the moment--- is to be expected.
%
%
For completeness, we also show the $\eta$ and $\eta'$ results using space-like data alone, labelled as $\eta^{(\prime)SL}$, in order to compare the 
effects of including the time-like data. We remark that the value shown for the TFF at zero energies, $F_{P\gamma\gamma}$ in \cref{tab:chap1mainres}, 
is the experimental one obtained from the $\Gamma_{P\gamma\gamma}$ decay widths from PDG~\cite{Agashe:2014kda}. Actually, this result has changed 
for the $\pi^0$ with respect to Ref.~\cite{Masjuan:2012wy}, where the $\Gamma_{\pi^0\gamma\gamma}^{\textrm{PrimEx}}$~\cite{Larin:2010kq} 
value was used. We include however the subsequent PDG combination~\cite{Agashe:2014kda}  
including, among others, the value from Ref.~\cite{Larin:2010kq}. In addition, the asymptotic behavior was not extracted there but included, since its theoretical 
prediction, $\pi_{\infty}=2F_{\pi}$, is a clean one as compared to the $\eta$ and $\eta'$, where the mixing and effects related to their singlet component obscure their calculation.
This represents the first step in order to reconstruct our PAs describing the pseudoscalar TFFs in next chapters.

\chapter{Canterbury Approximants}
\label{chap:CA}
\minitoc

\section{Introduction}

So far, we have carefully described how to reconstruct the single-virtual transition form factor (TFF) from the theory of Pad\'e approximants (PAs). 
However, for almost every practical application in this thesis, see \cref{chap:PLL,chap:gm2}, it is the double-virtual TFF that is required. 
From the very basic principle of Bose symmetry, we know that $F_{P\gamma^*\gamma^*}(Q_1^2,Q_2^2)=F_{P\gamma^*\gamma^*}(Q_2^2,Q_1^2)$. Such symmetry 
principle certainly simplifies the most general form that the double-virtual TFF could have, but it is not constrictive enough as to fully predict the double-virtual TFF from its 
single-virtual version alone. We illustrate this assertion using two simple ans\"atze. A simple extension of the single-virtual TFF, which respects Bose symmetry, 
is the factorization approach 
\begin{equation}
\label{eq:fact}
F_{P\gamma^*\gamma^*}^{\textrm{fact}}(Q_1^2,Q_2^2) = \frac{F_{P\gamma^*\gamma^*}(Q_1^2,0)\times F_{P\gamma^*\gamma^*}(Q_2^2,0)}{F_{P\gamma\gamma}}.
\end{equation}
This construction was
proposed back in the 60's based on vector meson dominance ideas~\cite{Landsberg:1986fd,Sakurai:1960ju,Sakurai} ---and recently reconsidered in~\cite{Xiao:2015uva}. 
There, the form factor was given through vector resonance exchanges 
as depicted in \cref{fig:factres} left, which implicitly uses factorization. Note however that in a large-$N_c$ framework 
additional diagrams exist ---see \cref{fig:factres} right or Ref.~\cite{Roig:2014uja}--- which break factorization. Still, from the study in Ref.~\cite{Bijnens:2012hf}, it seems that the leading 
logarithms in \cpt support the 
factorization approach at low energies, corrections appearing one loop higher than expected ---and even two loops higher in the chiral limit.
However, \cref{eq:fact} cannot reproduce at the same time the high-energy single- and double-virtual behavior which is implied from pQCD, see \cref{sec:tffpqcd}.
Namely, if the single-virtual TFF falls as $Q^{-2}$ ---as the BL, \cref{eq:BLlim}, implies--- the double-virtual factorized version, \cref{eq:fact}, necessarily falls as $Q^{-4}$, 
in conflict with the OPE which predicts $Q^{-2}$, \cref{eq:OPElim}.
This implies that, even if factorization would be appropriate at low-energies, it must fail at energies large enough.
\begin{figure}[htbp]
\centering
   \includegraphics[width=0.8\textwidth]{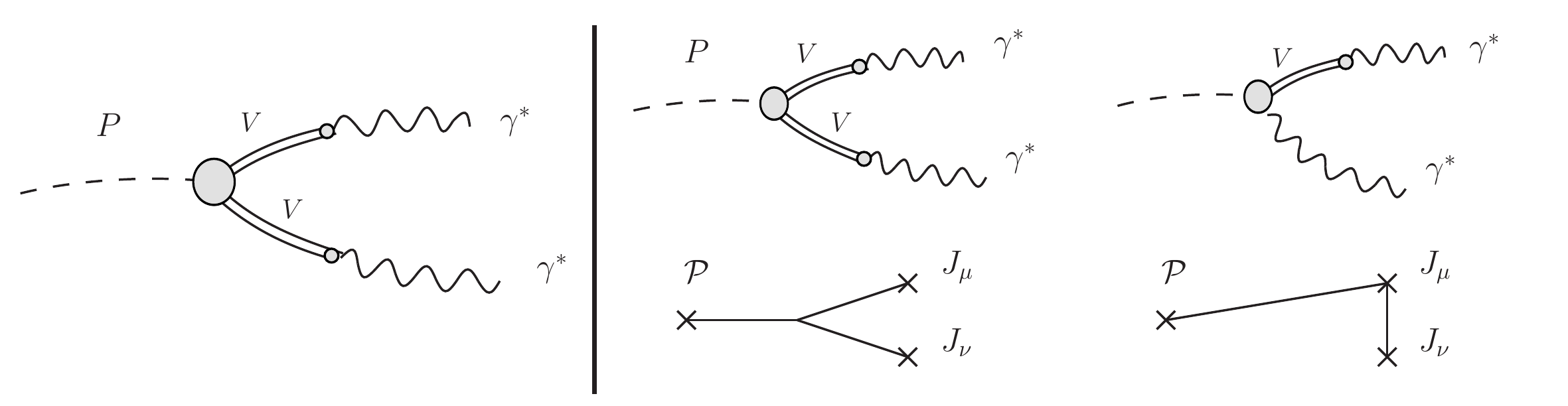}
\caption{Left: standard vector meson dominance conception; factorization is implied. Right: resonant approach to the TFF; factorization is not implied. The graphics on top arise from 
the large-$N_c$ pseudoscalar pole contribution to the (large-$N_c$) Green's functions sketched below (cf. \cref{fig:npoint}). \label{fig:factres}}
\end{figure}

An alternative idea, which would keep Bose symmetry without spoiling the high energy behavior, would be to extend the TFF as  
$F_{P\gamma^*\gamma^*}(Q_1^2,Q_2^2) = F_{P\gamma^*\gamma^*}(Q_1^2+Q_2^2,0)$.
However, this would imply that, if the high-energy behavior for the single-virtual TFF is given as $\lambda Q^{-2}$, its double-virtual counterpart would read 
$(\lambda/2) Q^{-2}$, whereas pQCD requires $(\lambda/3) Q^{-2}$ instead, see \cref{eq:OPElim,eq:BLlim}. These examples illustrate that the TFF double-virtual extension cannot be 
trivially reconstructed from the single-virtual one, but will require a dedicated effort. 
From a Pad\'e theory point of view, this amounts to the observation that, given the most general double-virtual TFF series expansion,
%
\begin{equation}
\label{eq:dvseries}
F_{P\gamma^*\gamma^*}(Q_1^2,Q_2^2) = F_{P\gamma\gamma}\left( 1 - b_{P}\frac{Q_1^2+Q_2^2}{m_P^2} + c_{P}\frac{Q_1^4+Q_2^4}{m_P^4} + a_{P;1,1}\frac{Q_1^2Q_2^2}{m_P^4} +... \right), 
\end{equation}
Bose symmetry only dictates that $a_{P;i,j}=a_{P;j,i}$, but does not enforce additional relations among the single-virtual parameters, $b_P,c_P,...$, and the double-virtual ones, 
$a_{P;i,j}$, which therefore must be provided as an additional input.
%
In this chapter, we explore how to consistently generalize in the spirit of Pad\'e theory our previous approach, which would provide then a model-independent 
framework to reconstruct the most general double-virtual TFF from the parameters in \cref{eq:dvseries}. Our method is described in \cref{sec:ca}, while its 
performance and properties are explored along \cref{sec:examples} using practical examples. 
Once more, experimental data, when available, would provide then the external required input to reconstruct the TFF. We investigate this possibility, in analogy to 
\cref{chap:data}, in \cref{sec:cadata}. Finally, we summarize the main results in \cref{ca:concl}.

\section{Canterbury approximants}
\label{sec:ca}

To extend the PAs to the bivariate case, we follow the approach from the Canterbury Group, started by Chisholm in Refs.~\cite{Chisholm:1973,Chisholm:1974} 
and giving birth to what is known as Canterbury approximants (CAs)~\cite{BakerMorris,Jones:1976}. 
This approach requires symmetrizing some equations, which is ideal in our case of study given the symmetry of our function. In this section, we 
review the basics of the method when applied to symmetric functions.
Let's define a function $f(x,y)=f(y,x)$ analytic in a certain domain around $x=y=0$, which series expansion reads
\begin{equation}
\label{eq:fseries}
f(x,y) = \sum_{\alpha,\beta}c_{\alpha,\beta}x^{\alpha}y^{\beta}, \quad (c_{\alpha,\beta} = c_{\beta,\alpha}).
\end{equation}
The Canterbury approximant is constructed from the rational function 
\begin{equation}
\label{eq:ca}
C^N_M(x,y) =  \frac{P_N(x,y)}{Q_M(x,y)} =  \frac{\sum_{i,j=0}^{N}  a_{i,j}x^iy^j}{\sum_{k,l=0}^{M}  b_{k,l}x^ky^l} \qquad (b_{1,1}=1). 
\end{equation}
Note that the rational function is constructed as to have the maximum power in each variable rather than a total maximum power in $x^iy^j$ with $i+j\leq N(M)$,
essential for the construction~\cite{Chisholm:1973}.
Next, we need to set the defining equations for the bivariate approximant in analogy to \cref{eq:pasdef}. A natural extension from the univariate case would be
%
\begin{equation}
\label{eq:chish1}
\sum_{i,j}^{M}b_{i,j}x^{i}y^{j} \sum_{\alpha,\beta}^{\infty}c_{\alpha,\beta}x^{\alpha}y^{\beta}=   \sum_{k,l}^{N}a_{k,l}x^{k}y^{l} + \mathcal{O}\left(x^{\gamma}y^{n+m+1-\gamma}\right),
\end{equation}
with $\gamma\in(0,n+m+1)$. 
Such set of equations define (Bose symmetry is implied)
\begin{gather}
\qquad\qquad \sum_{i=0}^{N+M}  = N+M+1 \qquad\qquad\qquad \ \ (c_{i,0} \ \textrm{terms}) \\
\sum_{(i\geq j)=1}^{i+j=N+M}   = \left\{ \begin{array}{ll} \frac{(N+M)^2}{4}, & N+M\in\textrm{even} \\ \frac{(N+M)^2-1}{4},  & N+M\in\textrm{odd} \end{array} \right.  \qquad (c_{i,j} \ \textrm{terms})
\end{gather}
constraints for the single and double-virtual parameters, respectively, the first of which are reminiscent from the univariate case. 
To obtain the number of equations for the double-virtual terms, note that each order $\mathcal{O}(L)\equiv \mathcal{O}(x^{L-i}y^i)$ involves, 
after using Bose symmetry, $L/2((L-1)/2)$ coefficients  for $L\in \textrm{even(odd)}$, implying $\sum_{i=1}^{L/2}i + \sum_{i=1}^{L/2-1}i = L^2/4$ terms for 
$L\in \textrm{even}$ and  $2\sum_{i=1}^{(L-1)/2}i = (L^2-1)/4$ terms for $L\in \textrm{odd}$.
In turn, \cref{eq:ca} involves 
\begin{gather}
\qquad\qquad\qquad \sum_{i=0}^N  +  \sum_{j=1}^M = N+M+1  \ \quad\qquad\qquad (a_{i,0}, b_{i,0} \ \textrm{terms}) \\
\sum_{(i\geq j)=1}^{N} + \sum_{(i\geq j)=1}^{M}   = \frac{1}{2}N(N+1) + \frac{1}{2}M(M+1) \qquad (a_{i,j}, b_{i,j} \ \textrm{terms}) \label{eq:cadvterms}
\end{gather}
terms for the single-virtual and double-virtual parameters, respectively ---to obtain the number of double-virtual terms, note that 
$\sum_{(i\geq j)=1}^{L} = \sum_{i=1}^L i = L(L+1)/2$. Expressing \cref{eq:cadvterms} as 
$(N+M)^2/4 + (N-M)^2/4+(N+M)/2$, it becomes clear that additional constraints beyond \cref{eq:chish1} are required to fix the double-virtual terms 
as Chisholm noted~\cite{Chisholm:1973}.

In the following, we illustrate how to find the defining set of equations for CAs as in Refs.~\cite{Jones:1976,Morris:1974}. For this, take a $C^N_M(x,y)$ 
approximant for which $N\geq M$ (an identical procedure applies for $M\geq N$). Its $P_N(x,y)$ numerator polynomial involves $N(N+1)/2$ double-virtual 
terms, which are classified according their total order in \cref{tab:coef}. All the terms $\sim x^Ny^{M\leq N}$ are present and need to be included therefore 
in the defining equations. However, we find that for a given order $\mathcal{O}(L\leq2N)$ not all the terms need to be filled in \cref{tab:coef};
the additional terms up to $\mathcal{O}(N+M+1)$ represent $M(M+1)/2$ terms 
which can be exactly matched from the $Q_M(x,y)$ polynomial double-virtual parameters, fixing every coefficient in \cref{eq:ca}. 
These represent the defining equations for CAs, which can be summarized as 
\begin{gather}
\sum_{i,j}^{M}b_{i,j}x^{i}y^{j} \sum_{\alpha,\beta}^{\infty}c_{\alpha,\beta}x^{\alpha}y^{\beta} - \sum_{k,l}^{N}a_{k,l}x^{k}y^{l} = \sum_{\gamma,\delta}^{\infty}d_{\gamma,\delta}x^{\gamma}y^{\delta}, \label{eq:cas} \\
  {\begin{array}{rll}
     d_{\gamma,\delta} = & 0  &  \ 0\leq \gamma\!+\!\delta \leq M\!+\!N \\  
     d_{\gamma,\delta} = & 0  &  \ 0\leq \gamma \leq \textrm{max}(M,N), \\  
                                      &     &  \ 0\leq  \delta \leq \textrm{max}(M,N) \\  
     d_{\gamma,\delta} = & 0 &  \  1 \leq \gamma \leq \textrm{min}(M,N), \\
                                      &     &  \ \delta=M\!+\!N\!+\!1\!-\!\gamma. 
  \end{array}} \label{eq:cadeq}
\end{gather}

\begin{table}[t]
\centering  
\begin{tabular}{cccccccc}\toprule
         & $\mathcal{O}(2)$ & $\mathcal{O}(3)$ & $\mathcal{O}(4)$ & $\mathcal{O}(5)$ &   $\mathcal{O}(6)$  &   $\mathcal{O}(7)$  &      $\mathcal{O}(8)$     \\ \midrule
 $C^0_M$ &        $-$       &                  &                  &                  &                     &                     &                          \\ 
 $C^1_M$ &    $_{11}$       &        $-$       &        $-,-$     &                 &                      &                     &                          \\ 
 $C^2_M$ &    $_{11}$       &    $_{21}$       &    $_{22},-$     &     $-,-$        &                     &                     &                           \\ 
 $C^3_M$ &    $_{11}$       &    $_{21}$       & $_{22},_{31}$    &    $_{32},-$     &     $_{33},-,-$     &     $-,-,-$         &        $-,-,-,-$             \\ 
 $C^4_M$ &    $_{11}$       &    $_{21}$       & $_{22},_{31}$    & $_{32},_{41}$    &  $_{33},_{42},-$    &        $_{43},-,-$  &             $_{44},-,-,-$            \\ 
 $C^5_M$ &    $_{11}$       &    $_{21}$       & $_{22},_{31}$    & $_{32},_{41}$    & $_{33},_{42},_{51}$ &    $_{43},_{52},-$  &           $_{44},_{53},-,-$          \\ 
 $C^6_M$ &    $_{11}$       &    $_{21}$       & $_{22},_{31}$    & $_{32},_{41}$    & $_{33},_{42},_{51}$ & $_{43},_{52},_{61}$ &        $_{44},_{53},_{62},-$        \\ 
 $C^7_M$ &    $_{11}$       &    $_{21}$       & $_{22},_{31}$    & $_{32},_{41}$    & $_{33},_{42},_{51}$ & $_{43},_{52},_{61}$ &    $_{44},_{53},_{62},_{71}$      \\ \bottomrule
\end{tabular}
\caption{Coefficients $b_{i,j}=b_{j,i}\equiv_{i,j}$ appearing in the degree $N$  polynomial $P_N(x,y)$ from $C^N_M(x,y)$. The order $\mathcal{O}$ stands for $i+j$.}
\label{tab:coef}
\end{table}
The defining equations, \cref{eq:cas,eq:cadeq}, represent the most important definition in this chapter as it is the basis to reconstruct the bivariate approximants. 
The definition above corresponding to the Canterbury group fulfills several properties~\cite{Chisholm:1973,Morris:1974,Jones:1976}:
\begin{itemize}
\item If either $x$ or $y$ is taken to vanish, CAs reduce to PAs.
\item If the original function is symmetric, this is $f(x,y)=f(y,x)$, the resulting CAs preserve this symmetry as well.
\item If the original function can be written $f(x,y)=g(x)h(y)$, the resulting CAs factorize in terms of the PAs for $g(x)$ and $h(y)$.
\item The $C^N_M(x,y)$ approximant for $1/f(x,y)$ is identical to $1/\tilde{C}^M_N(x,y)$, being $\tilde{C}^M_N(x,y)$ the approximant for $f(x,y)$. 
\item The diagonal approximants are invariant under the group of homographic transformations, this is, if $C^N_N(x,y)$ is the approximant for $f(\frac{A x}{1-B x},\frac{A y}{1-C y})$, this 
         is identical to  $\tilde{C}^N_N(\frac{A x}{1-B x},\frac{A y}{1-C y})$, where $\tilde{C}^N_N(x,y)$ is the approximant to $f(x,y)$ ---a well known property of diagonal PAs.
\end{itemize}
These properties are of relevance for us. In particular, 
reduction to PAs allows us to connect to our previous work; the second condition guarantees Bose symmetry; the third one is interesting regarding factorization discussions, 
whereas the last properties are reassuring in the sense that they extend important and well known properties of PAs to the bivariate case. 
In addition, Montessus theorem (cf. \cref{sec:pasmeromorphic}) as well as convergence to Stieltjes functions have been proved for CAs as 
well~\cite{BakerMorris,Cuyt:1990,Alabiso:1974vk}. Note that the former guarantees convergence of CAs for the pseudoscalar TFFs  in the large-$N_c$ limit of QCD.
As a final comment, there exist additional extensions of PAs to the multivariate case.
Their relevance can be understood for example if considering non-symmetric functions, which substantially complicates the procedure outlined above 
(for more details see Ref.~\cite{Guillaume:1990} and references therein). Note however that alternative approaches may not respect several of the properties quoted above.

\section{Practical examples}
\label{sec:examples}

In this section, we illustrate the performance and operation of CAs for the particular cases of two functions already discussed in \cref{chap:data} in their univariate case 
(i.e., one of their variables is taken to be zero) in the context of PAs, where excellent results were obtained\footnote{As an additional 
source for practical applications and discussions, the reader is referred to a similar study of the Euler's Beta function in~\cite{Morris:1974}.}.
These are the Regge and logarithmic models discussed in \cref{chap:data}. 

The first one reads in its bivariate (double-virtual) form~\cite{RuizArriola:2006jge}
\begin{equation}
\label{eq:ReggeDV}
F^{\textrm{Regge}}_{P\gamma^*\gamma^*}(Q_1^2,Q_2^2) = \frac{aF_{P\gamma\gamma}}{Q_1^2-Q_2^2}
                                         \frac{\left[ \psi^{(0)}\left(\frac{M^2+Q_1^2}{a}\right) -\psi^{(0)}\left(\frac{M^2+Q_2^2}{a}\right) \right]}{\psi^{(1)}\left(\frac{M^2}{a}\right)},
\end{equation}
and we take $M=0.8$~GeV and $a=1.3~\textrm{GeV}^2$, see \cref{sec:reggemodel}. We note that, whereas QCD evolution is necessary to restore the BL asymptotic behavior for 
one large virtuality~\cite{RuizArriola:2006jge}, the asymptotic behavior for two equal and large virtualities is already built-in in the model. To see this, take
\begin{equation}
\label{eq:ReggeDVeq}
\lim_{Q_2^2\to Q_1^2\equiv Q^2}F^{\textrm{Regge}}_{P\gamma^*\gamma^*}(Q_1^2,Q_2^2) = \frac{F_{P\gamma\gamma}}{\psi^{(1)}\left(\frac{M^2}{a}\right)} \psi^{(1)}\left(\frac{M^2+Q^2}{a}\right),
\end{equation}
which asymptotic behavior \cref{eq:ReggeDVeq} reads
\begin{equation}
\label{eq:reggeas}
\lim_{Q^2\to\infty}F^{\textrm{Regge}}_{P\gamma^*\gamma^*}(Q^2,Q^2) = \frac{aF_{P\gamma\gamma}}{\psi^{(1)}\left(\frac{M^2}{a}\right) } Q^{-2} + \mathcal{O}(Q^{-4}).
\end{equation}

The second (logarithmic) model is generalized to the bivariate (double-virtual) version as
\begin{equation}
\label{eq:appell}
F^{\textrm{log}}_{P\gamma^*\gamma^*}(Q_1^2,Q_2^2) = \frac{F_{P\gamma\gamma}M^2}{Q_1^2-Q_2^2}\ln\left( \frac{1+Q_1^2/M^2}{1+Q_2^2/M^2} \right),
\end{equation}
with $M^2=0.6~\textrm{GeV}^2$, see \cref{sec:QM}. We note that this function arises as a natural extension of flat distribution amplitudes, in the line of~\cite{Radyushkin:2009zg,Noguera:2010fe}, 
to the double-virtual case. To see this, consider the representation
\begin{equation}
\label{eq:logflat}
F^{\textrm{log}}_{P\gamma^*\gamma^*}(Q_1^2,Q_2^2) = F_{P\gamma\gamma}M^2 \int_0^1 dx \frac{1}{xQ_1^2 + (1-x)Q_2^2 + M^2},
\end{equation}
which essentially corresponds to a flat DA $\phi_P(x)\equiv1$ in \cref{eq:TFFpQCD,eq:TFFTH}. In addition, \cref{eq:appell} corresponds, up to normalization, to a particular 
case of the Appell hypergeometric function $F_1(1,1,1,2; -Q_1^2/M^2 , -Q_2^2/M^2)$. This function has a singularity at 
$Q_1^2=Q_2^2=-M^2$ and branch cut discontinuities for $Q_{1(2)}^2<-M^2$, disappearing whenever both virtualities meet such condition at the same time. A nice feature from 
this model is again obtained in the limit
\begin{equation}
\label{eq:appelleq}
\lim_{Q_2^2\to Q_1^2\equiv Q^2}F^{\textrm{log}}_{P\gamma^*\gamma^*}(Q_1^2,Q_2^2) = F_{P\gamma\gamma} \frac{M^2}{M^2+Q^2},
\end{equation}
which fulfills the  appropriate asymptotic behavior, even if the BL limit was not reproduced. A final interesting property, is that \cref{eq:logflat} can be re-expressed as 
\begin{equation}
\label{eq:hamburger}
 \frac{F_{P\gamma\gamma}M^2 }{\frac{1}{2}(Q_1^2+Q_2^2)+M^2} \int_{-\frac{1}{2}}^{+\frac{1}{2}}  \frac{du}{uz+1}; \quad z=\frac{Q_1^2-Q_2^2}{\frac{1}{2}(Q_1^2+Q_2^2)+M^2},
\end{equation} 
which represents an extended Stieltjes function ---see section 5.6 from Ref.~\cite{BakerMorris}

\subsection{Branch cuts: Stieltjes functions}
\label{sec:stieltjes}

The Stieltjes theorem for PAs proved to be a powerful tool in physical 
applications~\cite{Masjuan:2009wy,Aubin:2012me}. It provides convergence for the whole complex plane ---except for the cut, where the 
original function itself is ill-defined--- as well as bounds ($P^N_{N+1}(x)\leq f(x)<P^N_N(x)$) for the (Stieltjes) function to be approximated, \cref{sec:Stieltjes}. 
In this subsection, we illustrate its performance for the bivariate case through the use of the logarithmic model in \cref{eq:appell}, which corresponds to a generalized 
Stieltjes function, for which convergence is guaranteed~\cite{Alabiso:1974vk,BakerMorris}. 

As a first analysis, we check the convergence for the diagonal $C^N_{N}(Q_1^2,Q_2^2)$ and subdiagonal $C^N_{N+1}(Q_1^2 , Q_2^2)$ sequences. 
The lowest order elements read
\begin{align}
   C^0_1(Q_1^2,Q_2^2) = &\frac{F_{P\gamma\gamma}}{1+\frac{Q_1^2+Q_2^2}{2M^2} + \frac{Q_1^2Q_2^2}{6M^4}}, \\
   C^1_1(Q_1^2,Q_2^2) = &\frac{F_{P\gamma\gamma}( 1+\frac{Q_1^2+Q_2^2}{6M^2} + \frac{Q_1^2Q_2^2}{18M^4} )}{1+\frac{2(Q_1^2+Q_2^2)}{3M^2} + \frac{7Q_1^2Q_2^2}{18M^4}}, \\
   C^1_2(Q_1^2,Q_2^2) = &\frac{F_{P\gamma\gamma}( 1+\frac{Q_1^2+Q_2^2}{2M^2} + \frac{4Q_1^2Q_2^2}{5M^4} )}{1+\frac{Q_1^2+Q_2^2}{M^2} + \frac{14Q_1^2Q_2^2}{15M^4} +  \frac{Q_1^4+Q_2^4}{6M^4} +\frac{2Q_1^2Q_2^2(Q_1^2+Q_2^2)}{15M^6} + \frac{Q_1^4Q_2^4}{90M^8}}.  
\end{align}
The performance for these sequences is excellent up to large $Q^2$ values as it is illustrated in \cref{fig:conappell}, where the relative deviation, defined as 
$C^N_M(Q_1^2,Q_2^2)/F^{\textrm{log}}_{P\gamma^*\gamma^*}(Q_1^2,Q_2^2)-1$, is shown for two selected cases. There, we observe ---as anticipated--- that 
the diagonal and subdiagonal sequences approach the original function from above and below, respectively. Recall in this respect 
that the $C^N_{N}(Q_1^2,Q_2^2)$ sequence behaves 
as a constant for large $Q^2$ values, the $C^N_{N+1}(Q_1^2,Q_2^2)$ falls as $Q^{-2}(Q^{-4})$ for one (two) large virtualities, and the original function, as 
$\ln(Q^2)Q^{-2}$ and $Q^{-2}$, respectively, for one and two large virtualities.

\begin{figure}[t]
\centering
  \includegraphics[width=0.4\textwidth]{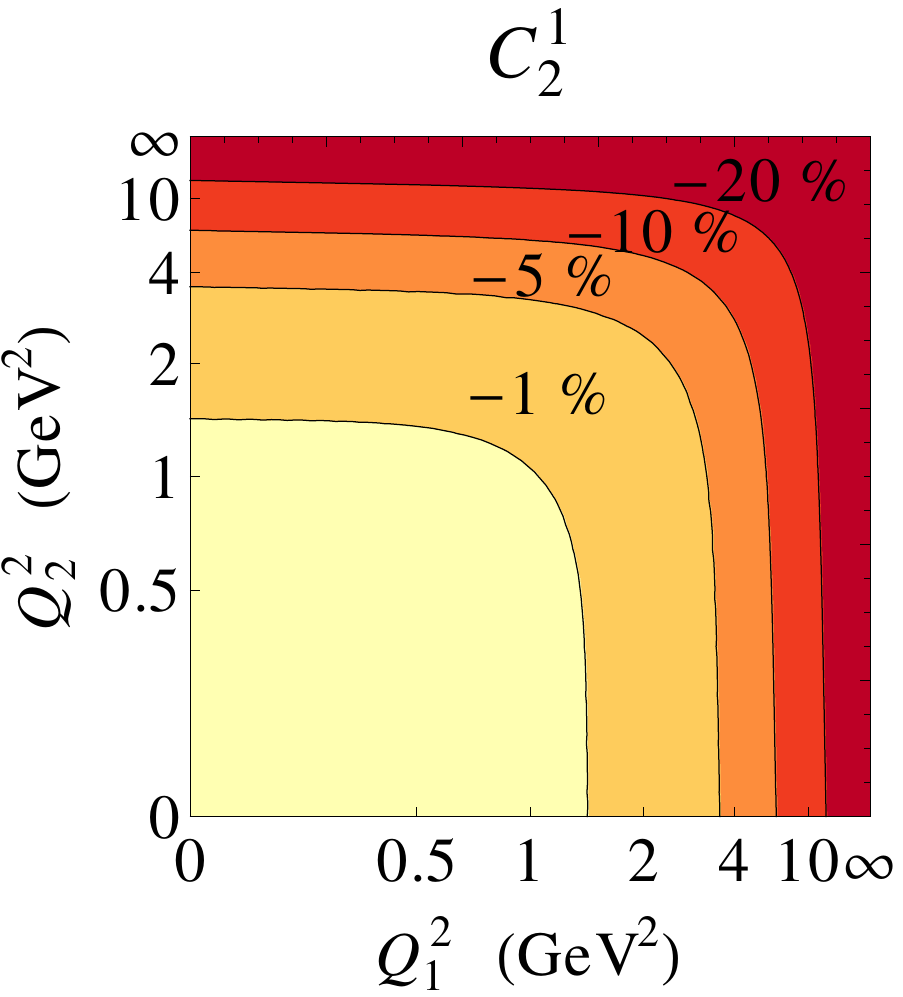}
    \includegraphics[width=0.4\textwidth]{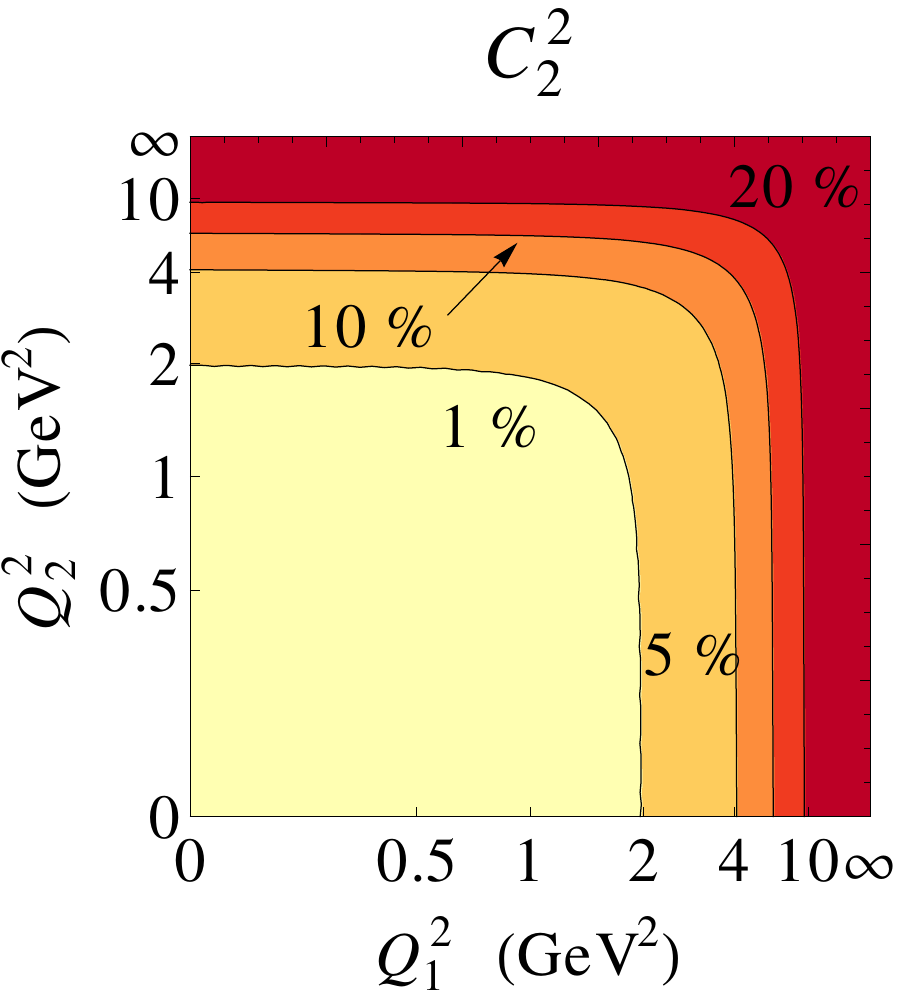}
  \caption{Convergence of the $C^N_{N+1}(Q_1^2,Q_2^2)$ and $C^N_{N}(Q_1^2,Q_2^2)$ sequences to the logarithmic model. We show the $C^1_2$ (left) and $C^2_2$ 
                 (right) elements, respectively.  The first, second, third, and fourth contours, from light to dark red, stand for the relative $\mp1,\mp5,\mp10$ and $\mp20\%$ 
                 deviations. Both axis have been scaled as $Q^2/(1+Q^2)$. See discussion in the  text. \label{fig:conappell}}
\end{figure}

An interesting implication from Stieltjes theorem is that the poles and zeros from the approximant must be located along the branch cut discontinuity, 
where the function itself is ill-defined. We check as a second step this property, and illustrate the poles and zeros for some elements of the diagonal and 
subdiagonal sequences in \cref{fig:cuts}.
\begin{figure}
\centering
  \includegraphics[width=0.4\textwidth]{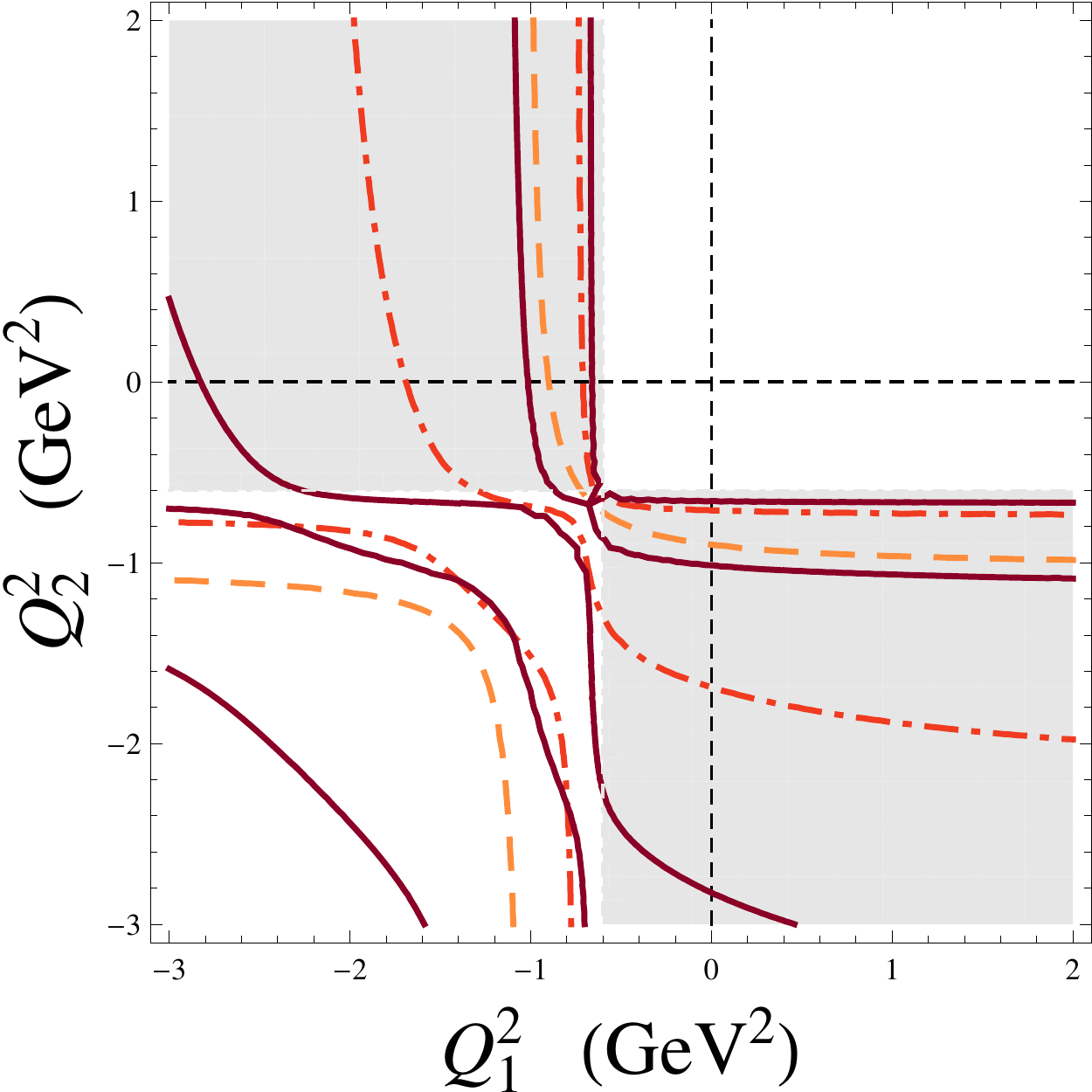}
    \includegraphics[width=0.4\textwidth]{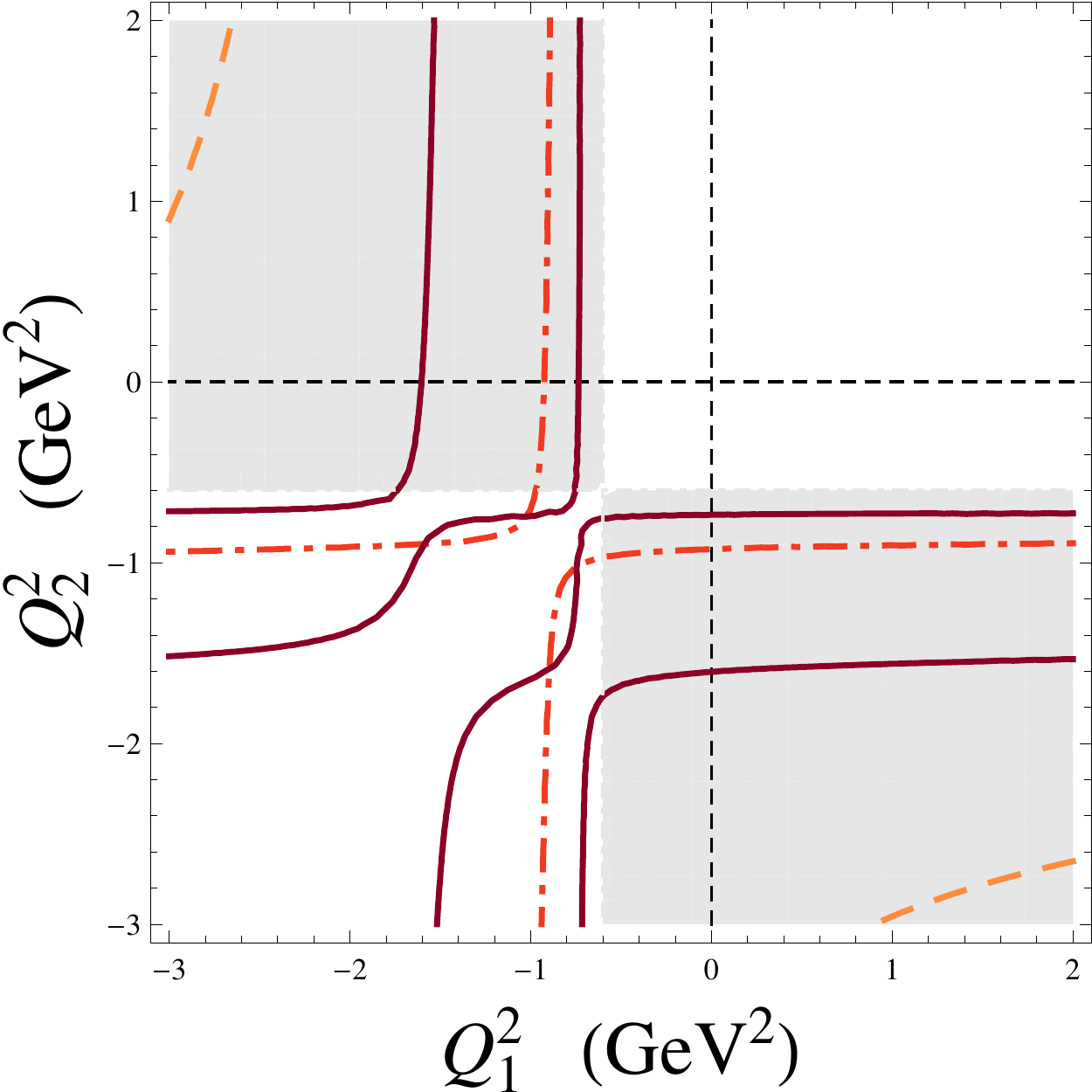}
  \caption{The poles (left) and zeros (right) for the $C^1_1, C^2_2$ and $C^3_3$ elements as dashed, dash-dotted and full lines, respectively. The gray-shaded areas represent the regions for which a branch cut exists.\label{fig:cuts}}
\end{figure}
There is no pole or zero in the space-like region and, in addition, these approach to the branch cut locations, as expected from the univariate case. 
There is an interesting remark though. As observed, there exist poles and zeros in the time-like region where no cut exists (light shaded time-like region in \cref{fig:cuts}). 
Still, these poles and zeros are spurious in the sense that they approach the gray-shaded regions in \cref{fig:cuts} ---where these should be located---  
as the order of the approximant increases, but indicate a slower convergence within this region. It would be interesting in this respect
to find whether it is possible to accelerate such convergence. 
We note in this respect that the logarithmic model in \cref{eq:appell} enjoys an additional symmetry, 
\begin{equation}
F_{P\gamma^*\gamma^*}^{\textrm{log}}(-Q_1^2-2M^2,-Q_2^2-2M^2) = -F_{P\gamma^*\gamma^*}^{\textrm{log}}(Q_1^2,Q_2^2),
\end{equation}
which actually relates the two space- and time-like light-shaded regions in \cref{fig:cuts}. It is intuitive that, constraining such symmetry into the approximant, 
the excellent convergence which is obtained for the space-like region will be translated into the time-like one. We find that such symmetry can only be 
implemented ---at least for the lowest approximants--- for the subdiagonal sequence, which lowest elements read
\begin{align}
C^0_1(Q_1^2,Q_2^2)  = & \frac{F_{P\gamma\gamma}}{1+\frac{(Q_1^2+Q_2^2)}{2M^2}},\label{eq:c01sym} \\
C^1_2(Q_1^2,Q_2^2)  =   & \frac{F_{P\gamma\gamma}\left(1+ \frac{(Q_1^2+Q_2^2)}{2M^2}\right) }{ 1+\frac{(Q_1^2+Q_2^2)}{M^2} + \frac{(Q_1^4+Q_2^4)}{6M^4} +\frac{2Q_1^2Q_2^2}{3M^4} },\label{eq:c12sym} \\
C^2_3(Q_1^2,Q_2^2) = & \frac{F_{P\gamma\gamma}\left(1+\frac{(Q_1^2+Q_2^2)}{M^2} + \frac{11(Q_1^4+Q_2^4)}{60M^4} + \frac{19Q_1^2Q_2^2}{30M^4}\right)}{1+\frac{3(Q_1^2+Q_2^2)}{2M^2} +\frac{3(Q_1^4+Q_2^4)}{5M^4} +\frac{(Q_1^6+Q_2^6)}{20M^6}+\frac{9Q_1^2Q_2^2}{5M^4}+\frac{9Q_1^2Q_2^2(Q_1^2+Q_2^2)}{20M^6}}. \label{eq:c23sym}
\end{align}
It is amusing to check that, in addition, for $Q_1^2=Q_2^2$ the equal-virtual behavior \cref{eq:appelleq} is exactly reproduced in \cref{eq:c01sym,eq:c12sym,eq:c23sym} 
even if this was not imposed.
Incidentally,  we find that the polynomials in our approximants, \cref{eq:c01sym,eq:c12sym,eq:c23sym}, can be constructed as $\sum_{i+j=0}^N c_{i,j}Q^{2i\phantom{j}}_{1}\!\! Q_2^{2j}$, 
missing the elements $Q^{2i\phantom{j}}_{1}\!\! Q^{2j}_2$ with $i+j>N$. We remark that this is a particular feature for this model, which cannot be generalized to other 
functions~\cite{Alabiso:1974vk}.

To end our discussion, we show the poles and zeros of \cref{eq:c01sym,eq:c12sym,eq:c23sym} in \cref{fig:symcuts}.
\begin{figure}[t]
\centering
  \includegraphics[width=0.4\textwidth]{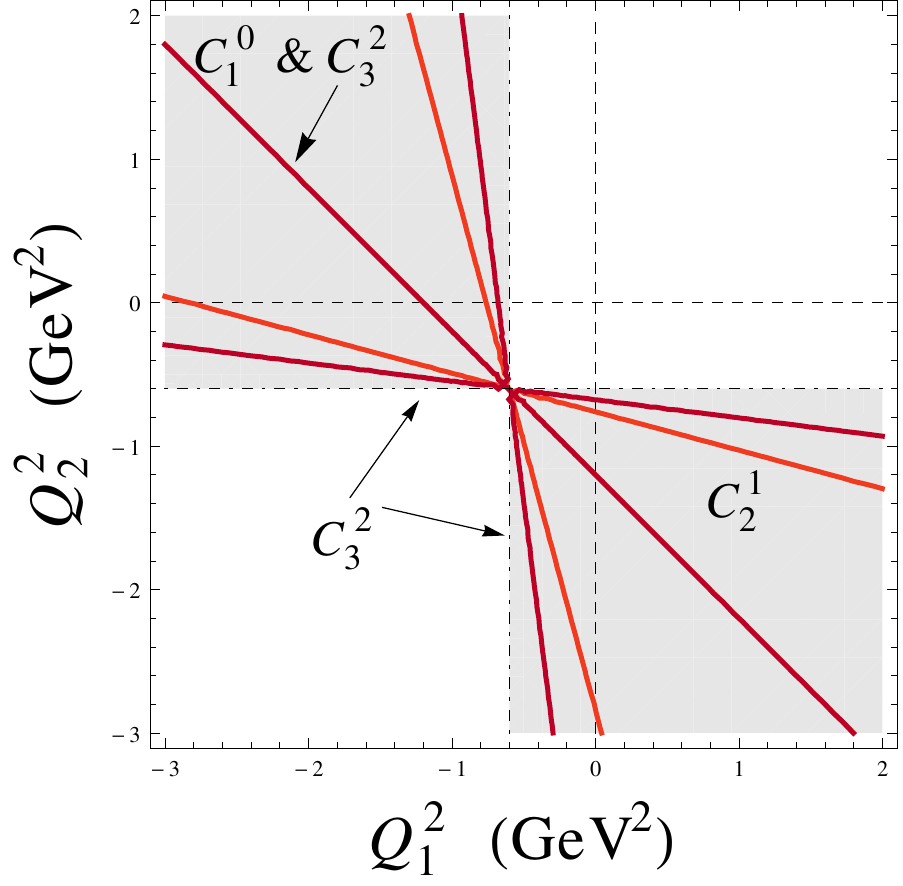}
    \includegraphics[width=0.4\textwidth]{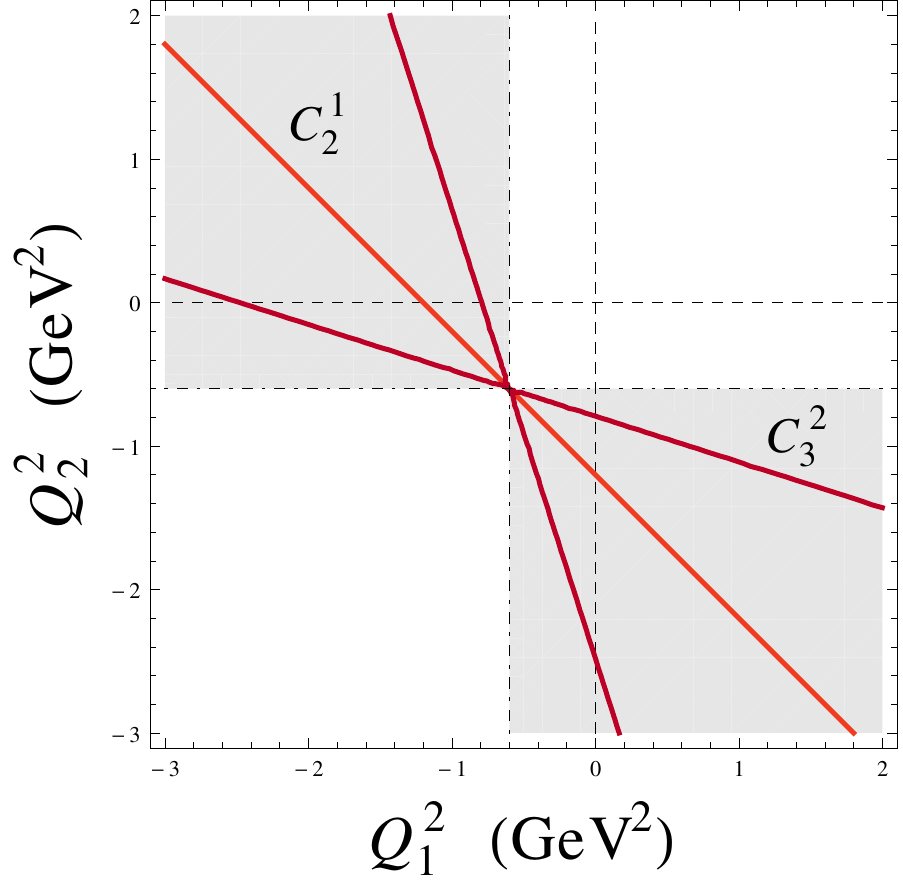}
  \caption{From lighter to darker full red lines, the poles (left) and zeros (right) for the $C^0_1, C^1_2$ and $C^2_3$ elements once the symmetry of the original function has 
                been constrained. The dotted-dashed lines represent the original logarithmic function branch cuts. We note that the pole for the $C^0_1$ approximant overlaps with one 
                pole of the $C^2_3$ approximant. They gray-shaded areas represent the regions for which a cut is opened. \label{fig:symcuts}}
\end{figure}
In contrast to \cref{fig:cuts}, there are no poles or zeros in the region $(x,y)<-(M^2,M^2)$ (time-like light shaded region in \cref{fig:symcuts}), which can now be described 
---as anticipated--- to the same precision as the space-like one. 
As a conclusion, whenever a symmetry principle exists, its inclusion improves convergence. We shall not forget that such symmetry necessarily implies a connection among the 
single- and double-virtual parameters in \cref{eq:dvseries}. An interesting discussion along these lines is found in Ref.~\cite{Cuyt:2006}.

\subsection{The large-$N_c$ limit and meromorphic functions theorems: Montessus and Pommerenke}

In this subsection, we employ the Regge model in \cref{eq:ReggeDV} to discuss additional convergence theorems which apply to the large-$N_c$ limit of QCD, in which the 
Green's functions become meromorphic. These are the Montessus' theorem and Pommerenke's theorem. 
As a brief summary from \cref{sec:pasmeromorphic}, we recall that, for the special case of meromorphic functions, Montessus theorem implies convergence within a disk containing 
$M$ poles for the $P^N_M(x)$ sequence, whereas Pommerenke's theorem implies convergence in the whole complex plane for the $P^{N+M}_N(x)$ sequence. We shall not 
forget that Montesus theorem has been obtained already for the multivariate case~\cite{BakerMorris,Cuyt:1990}. 
In addition, we recall that, if a meromorphic function have only positive residues (the same applies if all are negative), this is of the Stieltjes kind. As such condition is fulfilled for the  
Regge model, Stieltjes theorem applies here as well.

To discuss Montessus theorem, we reconstruct the $C^N_1$ sequence for the Regge model, which for the first elements read
\begin{align}
\frac{C^0_1(Q_1^2,Q_2^2)}{F_{P\gamma\gamma}}& = \frac{1}{ 1-\frac{(Q_1^2+Q_2^2)\psi^{(2)}}{2a\psi^{(1)}}  - \frac{Q_1^2Q_2^2}{2a^2}( \frac{\psi^{(3)}}{3\psi^{(1)}} - (\frac{\psi^{(2)}}{\psi^{(1)}})^2 )},\\
\frac{C^1_1(Q_1^2,Q_2^2)}{F_{P\gamma\gamma}}& = \frac{1- \frac{ Q_1^2+Q_2^2 }{ a }(\frac{ \psi^{(3)} }{ 3\psi^{(2)} } - \frac{1}{2}\frac{ \psi^{(2)} }{ \psi^{(1)} } ) + \frac{Q_1^2Q_2^2}{3a^2}( \frac{2}{3}(\frac{\psi^{(3)}}{\psi^{(2)}})^2 -\frac{\psi^{(3)}}{2\psi^{(1)}} -\frac{\psi^{(4)}}{4\psi^{(2)}}  )}{
 1-\frac{(Q_1^2+Q_2^2)\psi^{(3)}}{3a\psi^{(2)}}  - \frac{Q_1^2Q_2^2}{3a^2}( \frac{\psi^{(4)}}{4\psi^{(2)}} - \frac{2}{3}(\frac{\psi^{(3)}}{\psi^{(2)}})^2 )},
\end{align}
where $\psi^{(n)}\equiv\psi^{(n)}(M^2/a)$. The performance, as expected, resembles that of the univariate case. As an example, we show how the poles of the 
$C^N_1(Q_1^2,Q_2^2)$ sequence approach those of the 
original function at $Q_1^2(Q_2^2)=-M^2$ in \cref{fig:cn1} left. As we move either further from the first pole, or far into the space-like region, convergence deteriorates and is 
eventually lost as we move away 
from the convergence disk. This is in accordance to Montessus theorem, and can be easily understood for this particular case from the power-like behavior of the 
approximant, which rapidly diverges as $N$ is increased, in contrast to the original function. To enlarge such convergence disk beyond the second pole from the model, we 
need to go to the $C^N_2$ sequence. The poles from such approximant are illustrated in \cref{fig:cn1} (right panel), where it can be observed the hierarchical convergence for the 
poles, which approach faster to those closer to the expansion point. This is to be expected, as the imprint from the poles far from the origin should be small.

%
%
%
%
\begin{figure}[t]
\centering
  \includegraphics[width=0.44\textwidth]{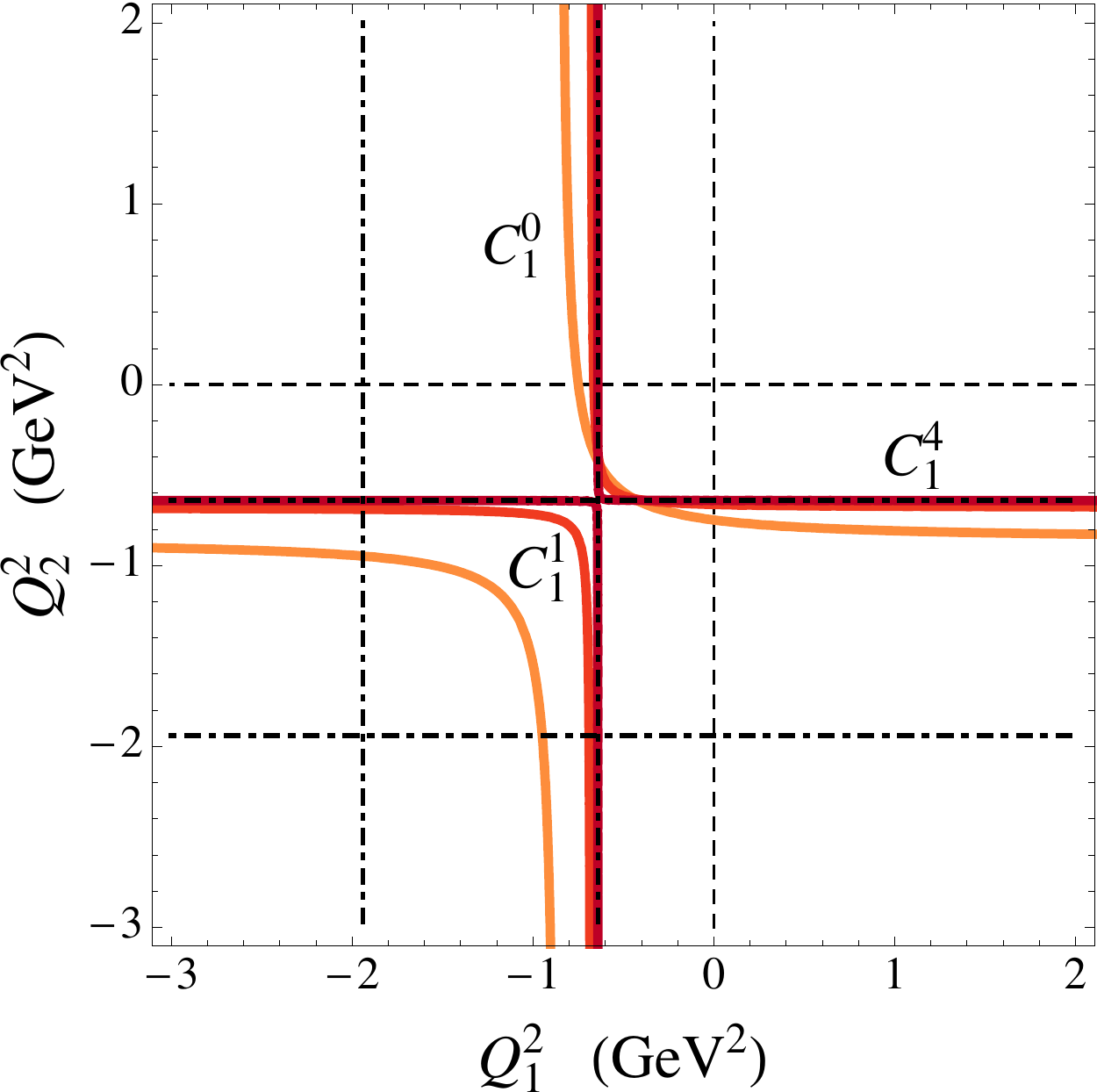}
    \includegraphics[width=0.44\textwidth]{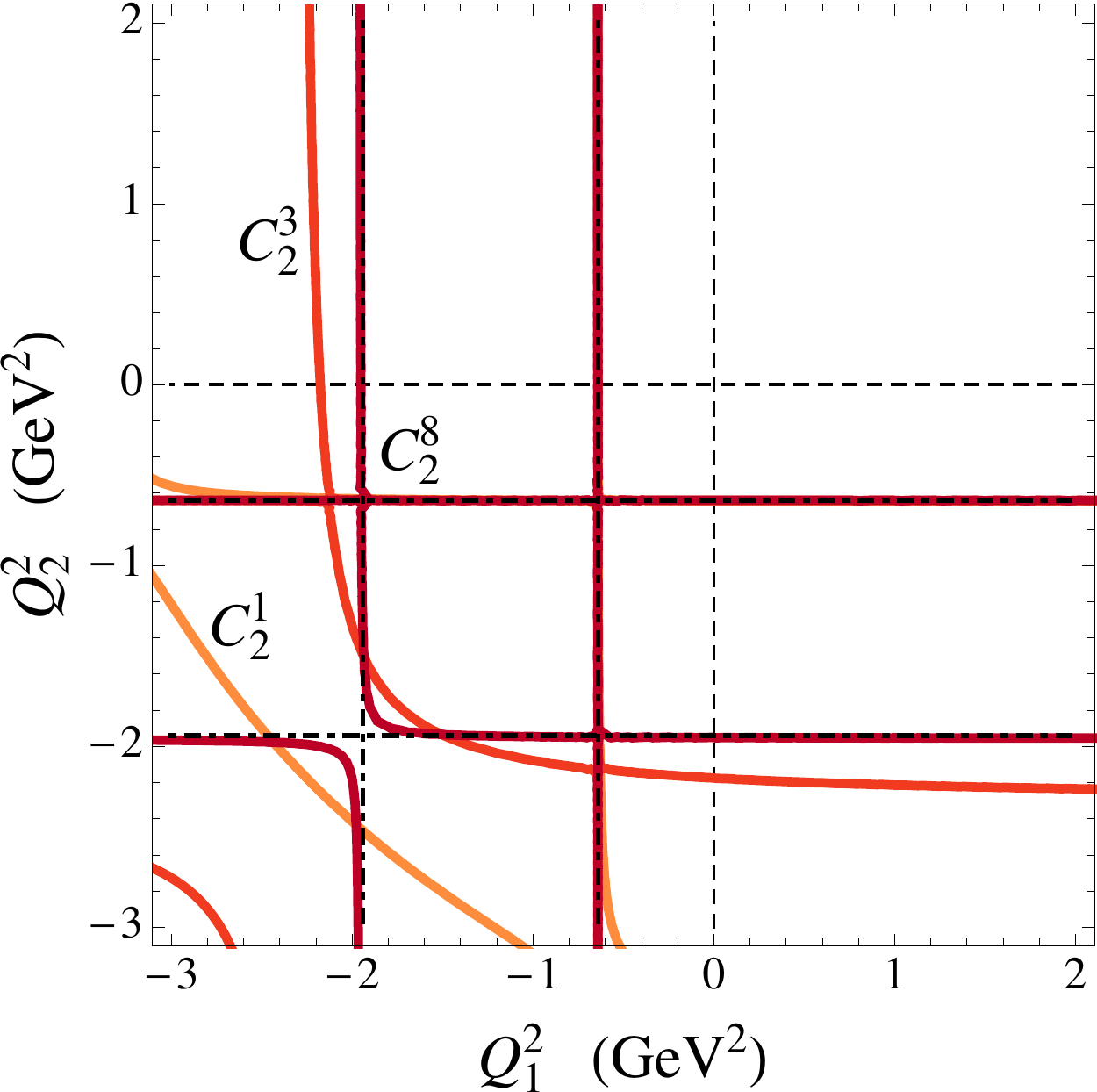}
    \caption{The poles from the $C^N_1(Q_2^2,Q_2^2)$ (left) and $C^N_2(Q_2^2,Q_2^2)$ (right) sequences for the $C_1^0, C_1^1$, $C_1^4$ and $C^1_2$, $C^3_2$, $C^8_2$ elements, 
    respectively, (light to dark red lines). The original first and second poles are displayed as dashed-dotted lines. \label{fig:cn1}}
\end{figure}
%
Eventually, our goal is to reproduce the function in the whole complex plane or, at least, in the whole space-like region. To this aim, and dealing with 
meromorphic functions, we can appeal to Pommerenke's theorem and check if this seems to extend to the bivariate case too. As an example, we use the subdiagonal 
$C^{N-1}_N(Q_1^2,Q_2^2)$ sequence, for which the theorem applies. We show the relative error, defined as in the previous subsection in \cref{fig:cnn1regge}, 
obtaining excellent results and suggesting that Pommerenke's applies to the bivariate case too. Moreover, there we find that the 
original function is always approached from below in this sequence. The opposite would have been found for the diagonal sequence. 
This was to be anticipated as this function is not only meromorphic but Stieltjes, which places stronger contraints.\\
\begin{figure}[t]
\centering
  \includegraphics[width=0.325\textwidth]{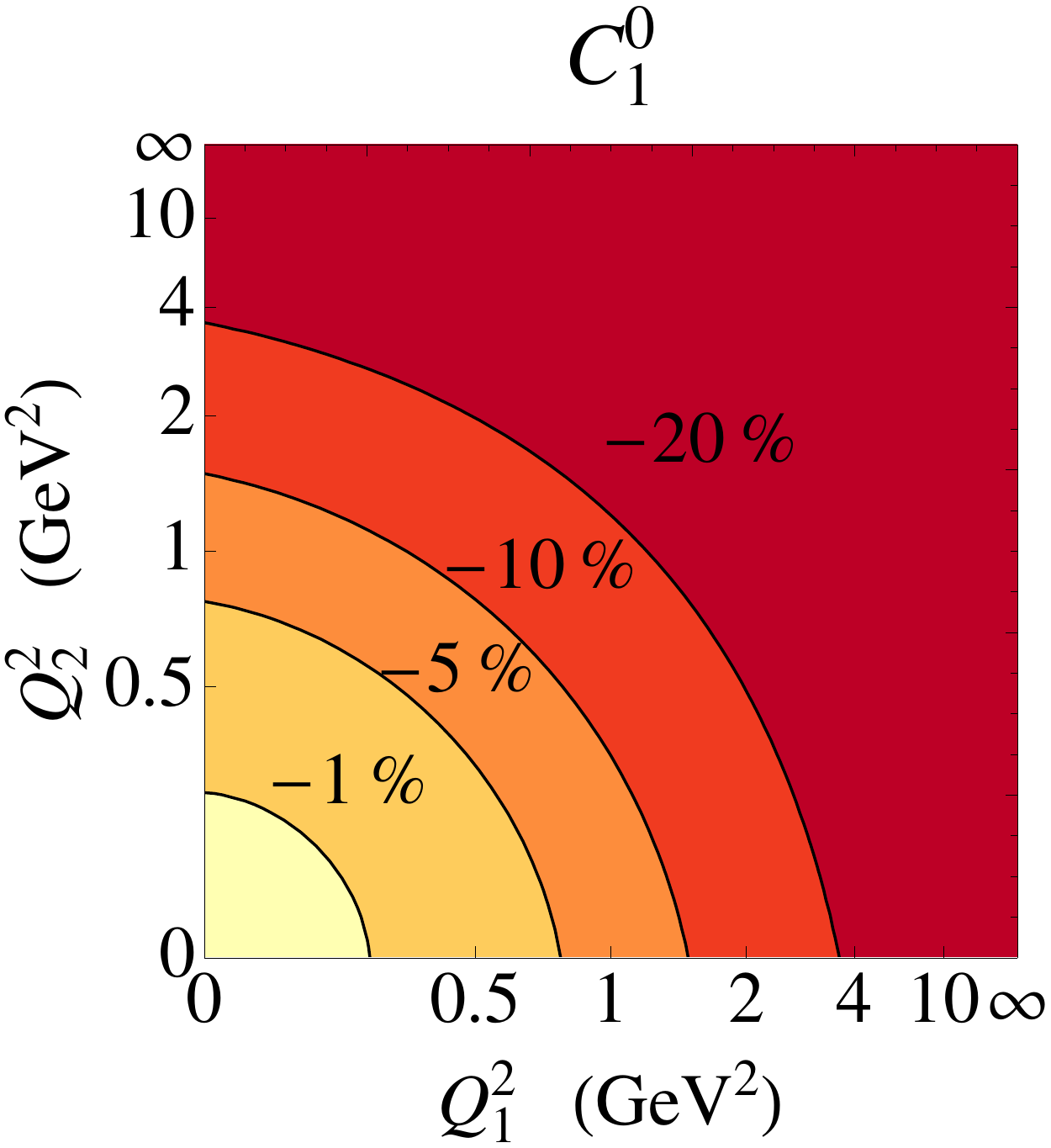}
  \includegraphics[width=0.325\textwidth]{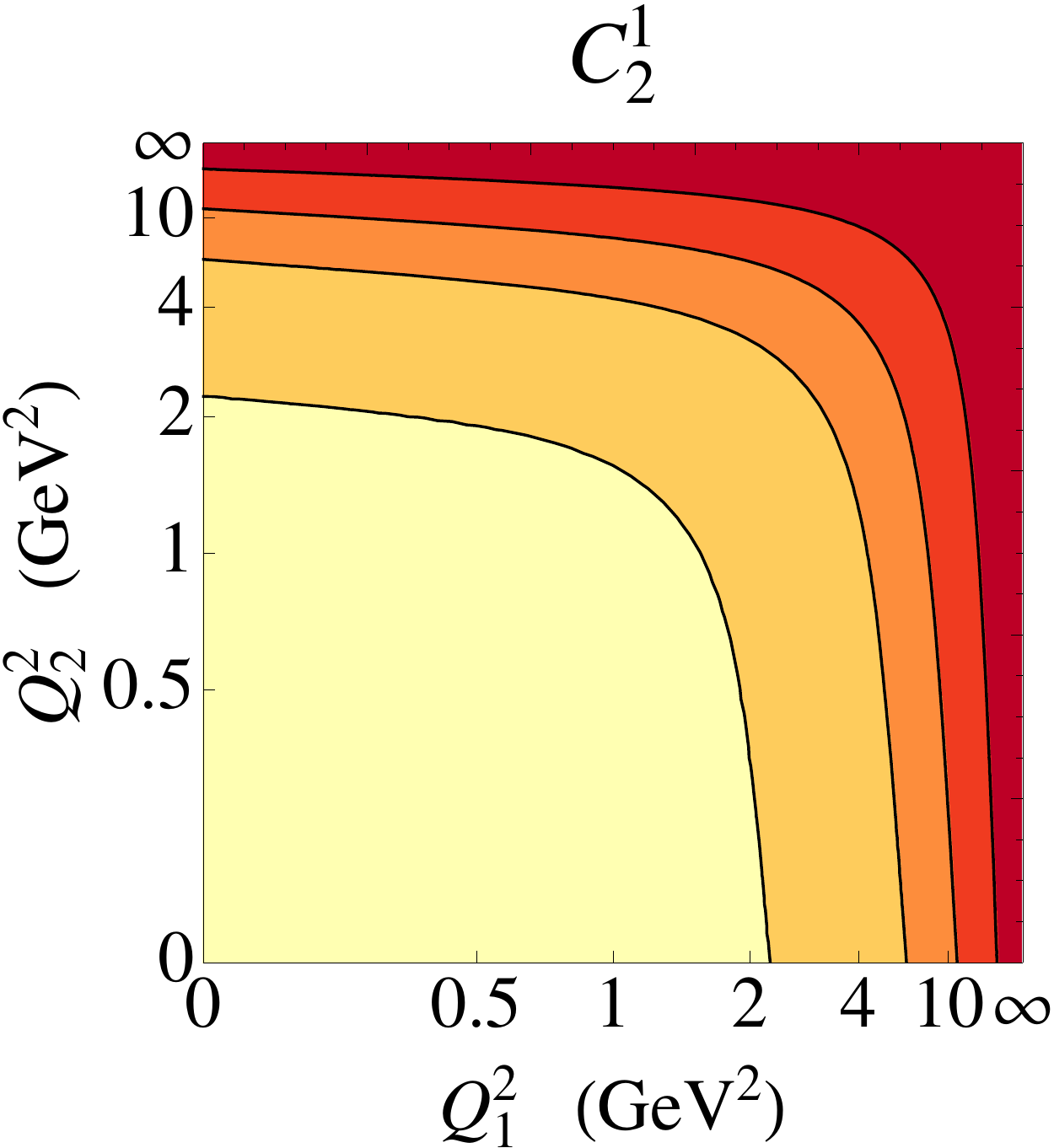}
  \includegraphics[width=0.325\textwidth]{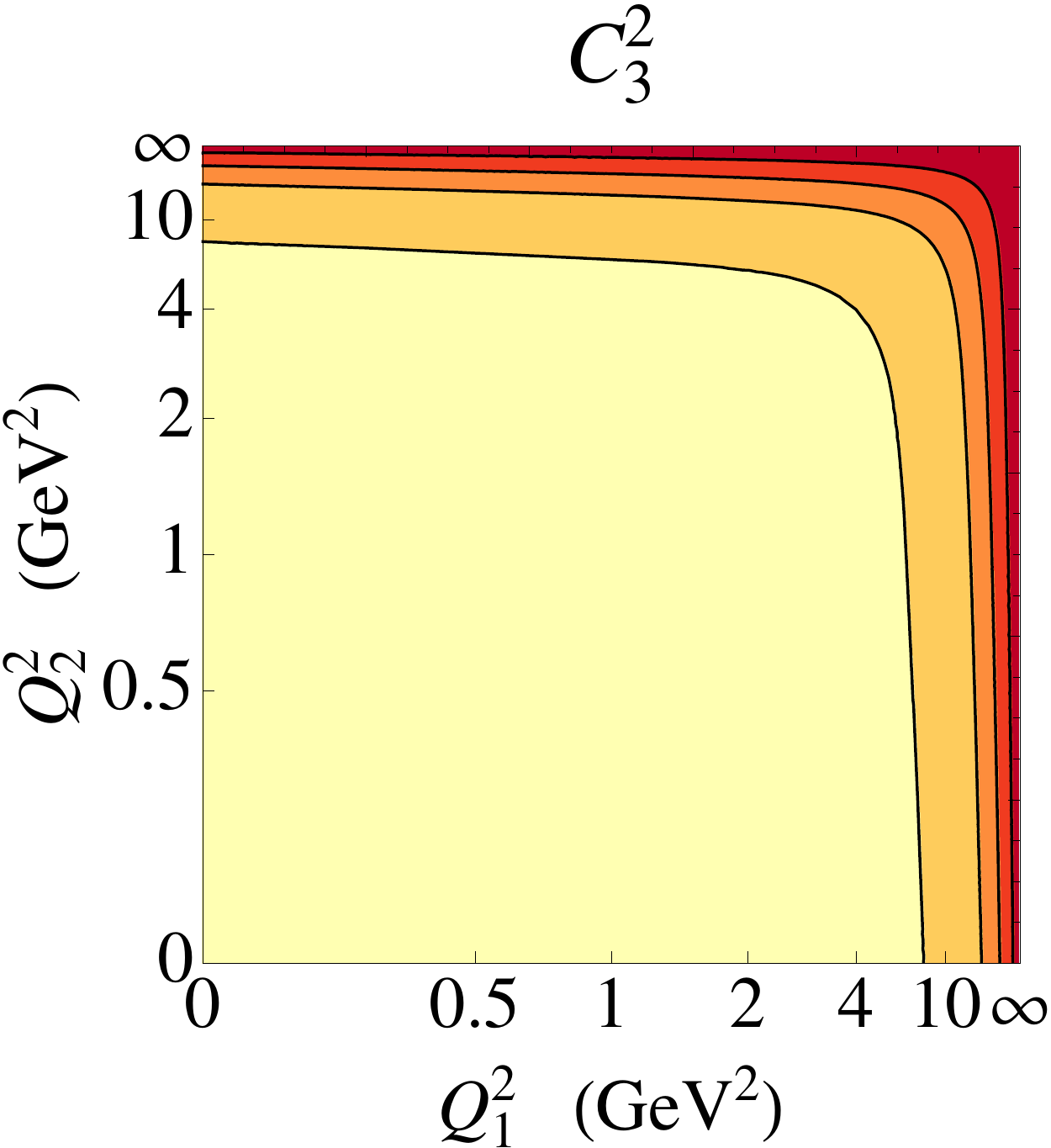}
  \caption{Convergence of the $C^{N-1}_{N}(Q_1^2,Q_2^2)$ sequence to the Regge model for different elements. The first, second, third, and fourth contours, from light to dark red, 
  stand for the relative $-1,-5,-10$ and $-20\%$ deviations. Both axis have been scaled as $Q^2/(1+Q^2)$. \label{fig:cnn1regge}}
\end{figure}
Given the observed ability of the approximants to reproduce the original pole, it is natural to ask ourselves whether its residue is approached at a similar 
convergence rate. Actually, this quantity is of physical relevance too. As an example, in our Regge model for the TFF, this would represent some vector meson form factor, 
say, the $\omega\pi^0\gamma^*$ TFF ---of course, in the real world with finite-width resonances, this identification is misleading, and would only hold, approximately, for 
extremely narrow resonances.
To this end, we take the residue from our $C^N_1(Q_1^2,Q_2^2)$, which is illustrated in \cref{fig:res}.
%
\begin{figure}[t]
\centering
  \includegraphics[width=0.6\textwidth]{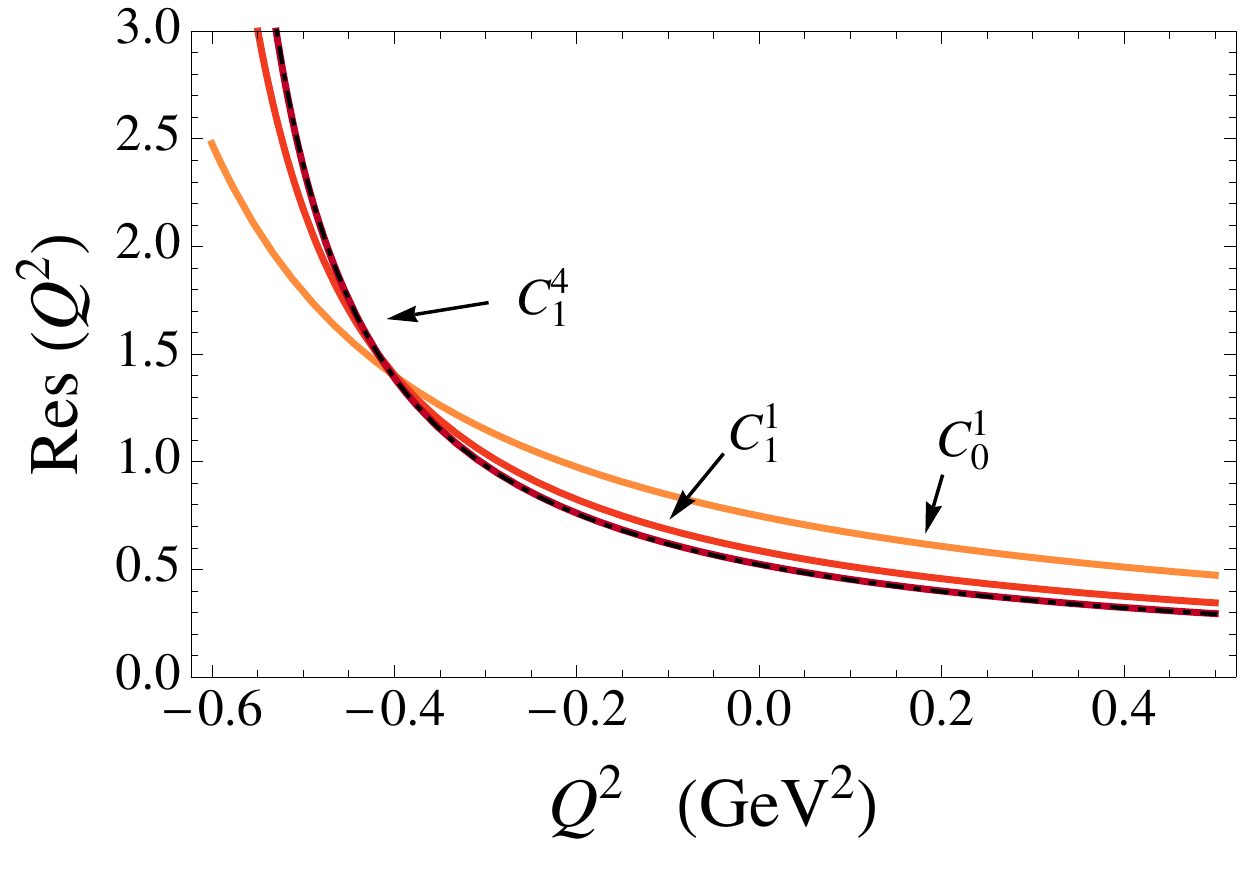}
  \caption{From lighter to darker full-red lines, the residue associated to the $C^0_1,C_1^1,C^4_1$ approximants whenever some virtuality hits a pole. The original residue, 
  overlapping with the $C^4_1$ element, is plotted as dotted-dashed black line.\label{fig:res}}
\end{figure}
%
We find an excellent convergence too, even if the accuracy is smaller than that found for the pole position. If we would repeat the same exercise for the $C^N_2$ sequence, we 
would find an excellent convergence for extracting the first pole ---see \cref{fig:cn1}--- and its residue. For the second pole, as illustrated in \cref{fig:cn1}, the convergence 
is slower and an even slower convergence rate is found for its residue. We conclude that, as in the univariate case of PAs, 
Canterbury approximants provide an excellent description for meromorphic functions in the space-like region, they are able to predict the poles position and, eventually, describe 
their residues as well, this is, they provide a complete description of the original function.

\subsection{Resonant approaches: Pad\'e Type extensions}
\label{sec:ptype}

From the previous discussion, it seems that if the poles would have been known {\textit{a priori}}, these could have been used from the very 
beginning, bringing additional parameters to our approach. This is interesting, as in the real situation we often know several poles
from our function\footnote{It must be noted that, in the real world, many of these poles may have a significant width. Including them as real 
zero-width poles implies then an additional error.}, but not its series expansion. In this section, we study the implications from this approach, 
in which the poles of the approximant are given in advance, and are in correspondence with the lowest-lying poles from the original function. 
This is known in the univariate case as Pad\'e-Type approximants, see \cref{sec:patype}, and have been implicitly  used in the past 
years in resonant approaches. 
For reconstructing these approximants, we build in our case the denominator from our Canterbury-Type approximant, ${C_T}^N_M$, as
\begin{equation}
\prod_{n=0}^{M-1}(Q_1^2+M^2+na)(Q_2^2+M^2+na),
\end{equation}
whereas the remaining parameters from the $P_N(Q_1^2,Q_2^2)$ polynomial, \cref{eq:ca}, are fixed from the series expansion.
The obtained results for the first approximants are shown in \cref{fig:ct}. Comparing with \cref{fig:cnn1regge}, it is easy to see that the achieved convergence rate 
is not as satisfactory as in the previous case, and the resulting systematic error from the approach is larger.
\begin{figure}[t]
\centering
  \includegraphics[width=0.325\textwidth]{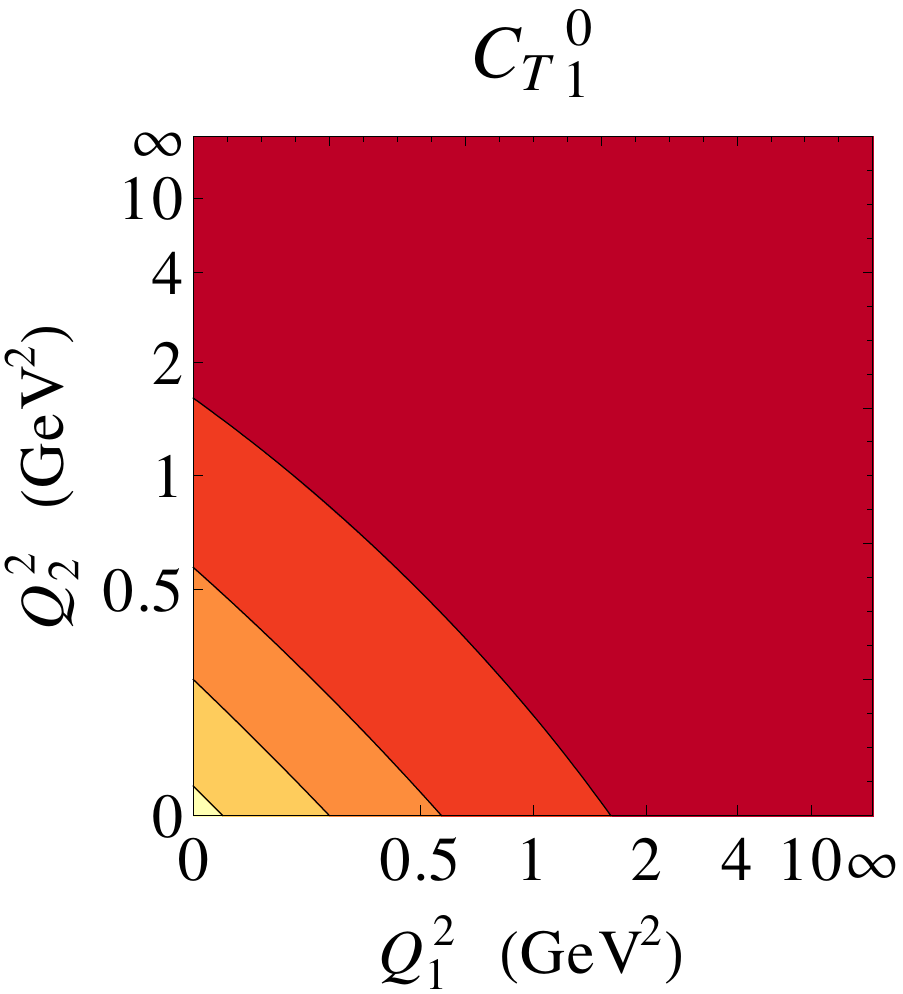}
  \includegraphics[width=0.325\textwidth]{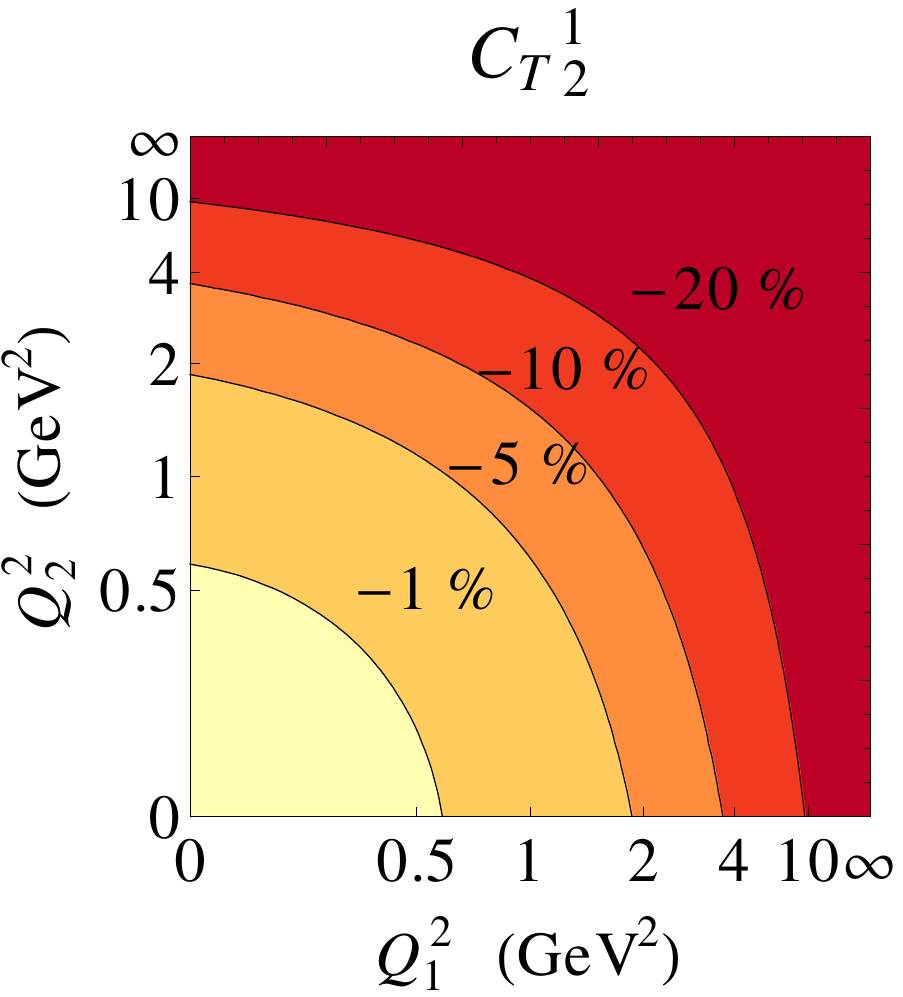}
  \includegraphics[width=0.325\textwidth]{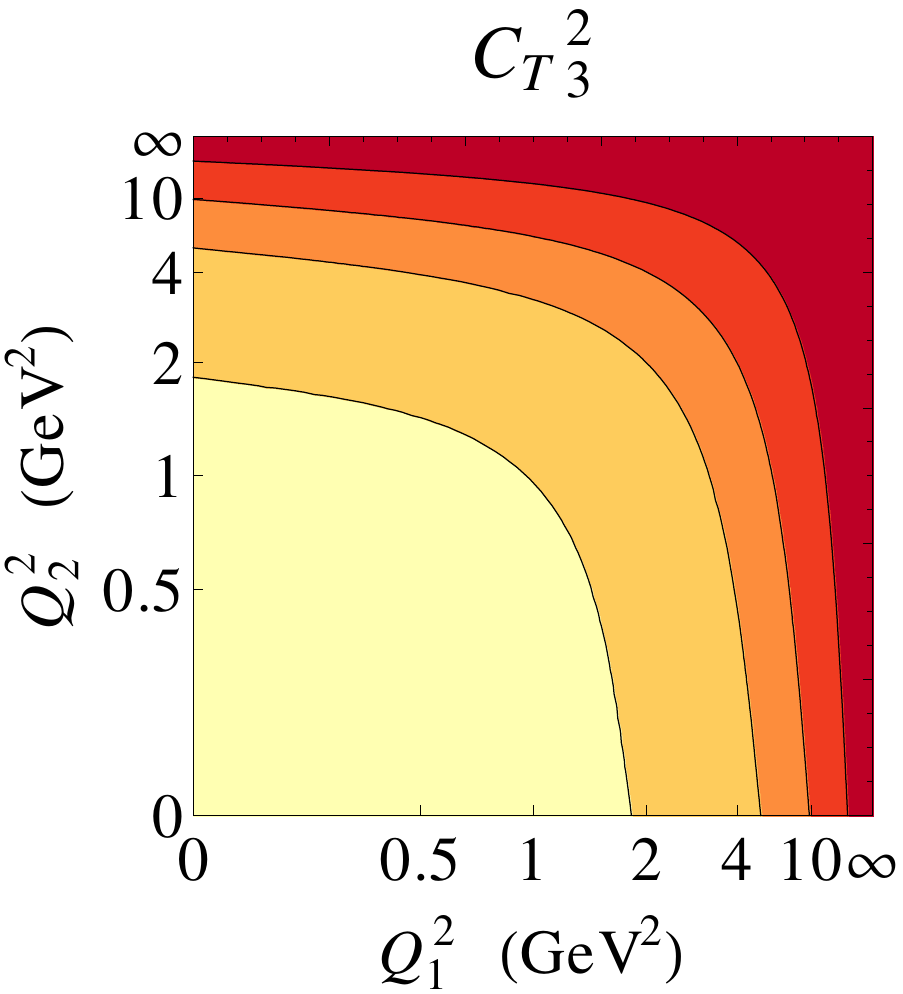}
  \caption{Convergence of the ${C_T}^N_{N+1}(Q_1^2,Q_2^2)$ sequence to the Regge model for different elements. The first, second, third, and fourth contours, from light to dark red, 
  stand for the relative $-1,-5,-10$ and $-20\%$ deviations. Both axis have been scaled as $Q^2/(1+Q^2)$.\label{fig:ct}}
\end{figure}
This was easy to anticipate, as the position of the poles ---specially those far from the expansion point--- did not exactly correspond to the original pole location in our previous examples. 
This kind of approach may better reproduce the resonant region which is close to the fixed poles, 
but this comes at cost of the space-like region which we are interested in. 
We conclude then that fixing the poles in advance is not 
the best strategy to find a fast convergence, and we warn against its generalized use in phenomenological applications. An intermediate choice which may be competitive is that 
of using Partial-Pad\'e approximants, \cref{sec:patype}, in which only a finite number of poles is fixed in advance, the others being constrained from the series expansion as usual.

\subsection{High energy limit: two-point approximants}
\label{sec:caope}

In the previous subsections, we found that the convergence from our approximants deteriorated at very large $Q^2$ values. This was easy to anticipate, as our 
models (\cref{eq:ReggeDV,eq:appell}) approached $0$ for $Q_{1}^2=Q_{2}^2\equiv Q^2\rightarrow\infty$ as $Q^{-2}$ (cf. \cref{eq:reggeas,eq:appelleq}), 
whereas none of the approximants constructed above implemented such behavior. 
In this subsection, we discuss how such behavior ---which could have been anticipated from the OPE expansion--- can be implemented into our approximant. 
To this object, we review the concept of two-point PAs, see \cref{sec:PAnpoint}, applied to CAs, which in our case allows to describe both, the low- and the high-energy 
expansions, providing then a tool to unify our knowledge from \cpt and pQCD. 

Our two expansions of interest for the Regge and logarithmic models are
\begin{align}
 F_{\pi^0\gamma^*\gamma^*}^{\textrm{Model}}(Q_1^2,Q_2^2)& = \sum_{n,m=0}^{\infty} c_{n,m}Q_1^{2n}Q_2^{2m}     \qquad (Q_{1,2}^2\rightarrow0), \label{eq:modellecs} \\
  F_{\pi^0\gamma^*\gamma^*}^{\textrm{Model}}(Q^2,Q^2)& = \sum_{n=0}^{\infty} c^{\textrm{OPE}}_{n}Q^{-2n}  \qquad (Q_1^2=Q_2^2\rightarrow\infty). \label{eq:modelope}
\end{align}
The first one represents the expansion at the origin of energies used in previous sections, whereas the second one represents the 
OPE expansion for equal large virtualities $Q_1^2=Q_2^2\equiv Q^2\rightarrow\infty$. For illustrating the construction of two-point CAs, we make use 
of the diagonal and subdiagonal sequences, which high-energy behavior expansion reads (see \cref{eq:ca})
\begin{align}
C^N_N(Q^2,Q^2) &= \frac{a_{N,N}}{b_{N,N}} + \frac{2a_{N,N-1}b_{N,N}-2b_{N,N-1}a_{N,N}}{b_{N,N}^2}Q^{-2} + ... \ ,  \label{eq:HEexpD}\\
C^{N}_{N+1}(Q^2,Q^2) &= \frac{a_{N,N}}{b_{N+1,N+1}}Q^{-4} + ... \qquad\qquad\qquad(b_{N+1,N+1}\neq 0), \label{eq:HEexpSD1} \\
C^{N}_{N+1}(Q^2,Q^2) &= \frac{a_{N,N}}{2b_{N+1,N}}Q^{-2} +  ... \ \qquad\qquad\qquad (b_{N+1,N+1}=0).  \label{eq:HEexpSD2} 
\end{align}
For both of our models $c^{\textrm{OPE}}_0=0$, the first non-vanishing term in the high-energy expansion \cref{eq:modelope} 
being $c^{\textrm{OPE}}_1$. This implies $a_{N,N}=0$ and $b_{N+1,N+1}=0$ for the diagonal and subdiagonal sequences, respectively (cf.  
\cref{eq:HEexpD,eq:HEexpSD1,eq:HEexpSD2}). If additional terms from the high-energy expansion are to be included in our two-point CA, say 
$c^{\textrm{OPE}}_1$, additional constraints are present. The resulting equations are taken instead those arising from the higher order terms in the low-energy expansion.
However, in contrast to PAs, for the bivariate case there are many different terms $c_{N,M}Q_1^{2N}Q_2^{2M}$ of the same order $L=N\!+\!M$. 
For our models, we find that the best convergence is achieved when the most asymmetric terms are replaced for the high-energy ones, this is, 
the terms $c_{L-1,1}, c_{L-2,2}, ...$ are replaced by $c^{\textrm{OPE}}_0, c^{\textrm{OPE}}_1, ... $ .

\begin{figure}
\centering
  \includegraphics[width=0.325\textwidth]{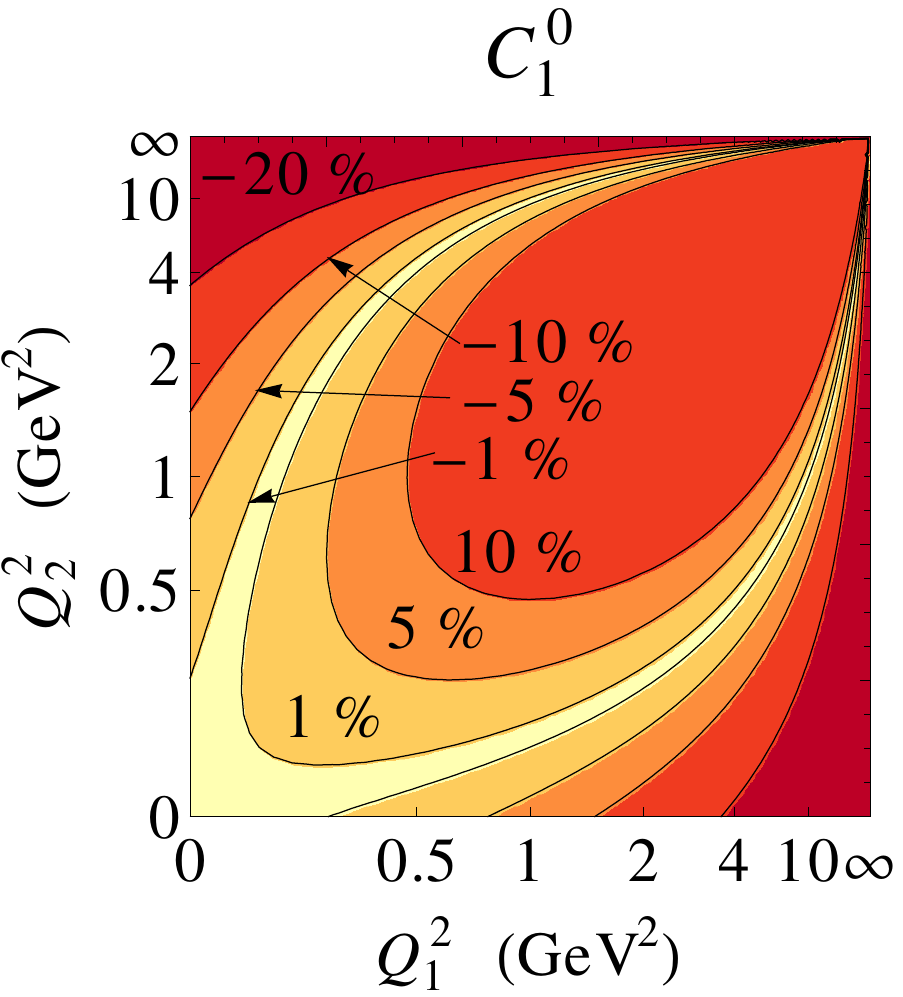}
  \includegraphics[width=0.325\textwidth]{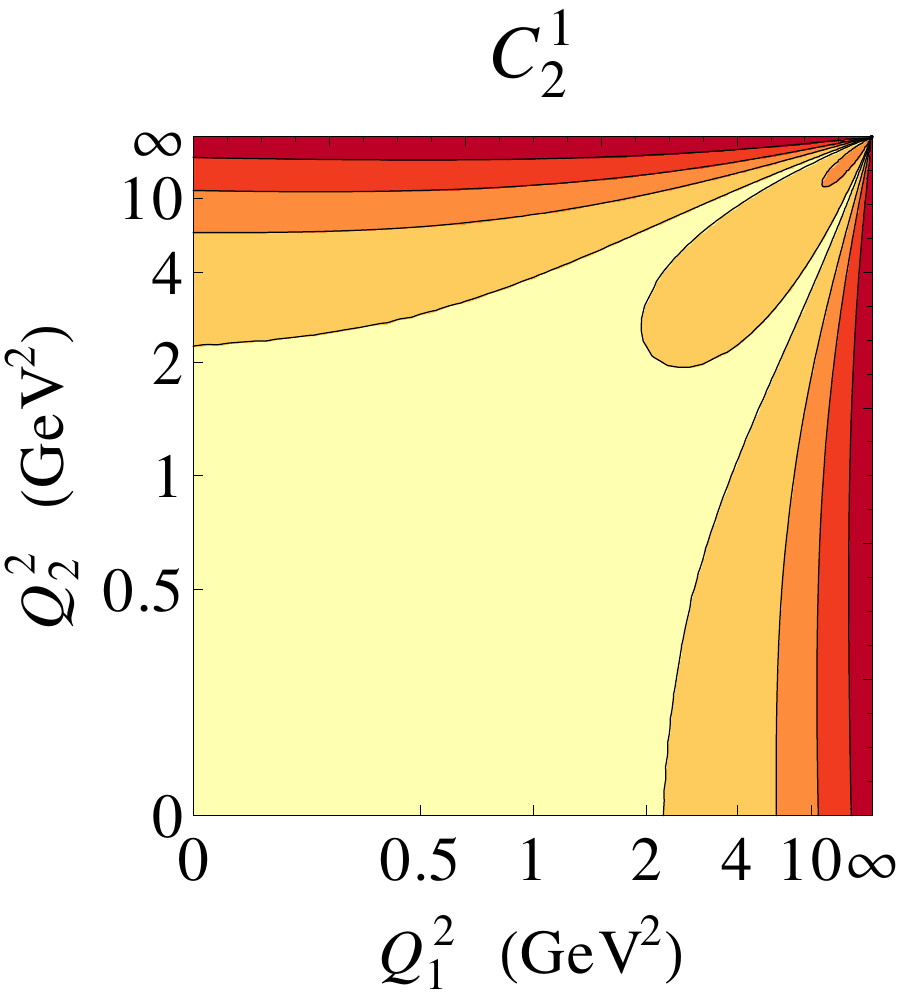}
  \includegraphics[width=0.325\textwidth]{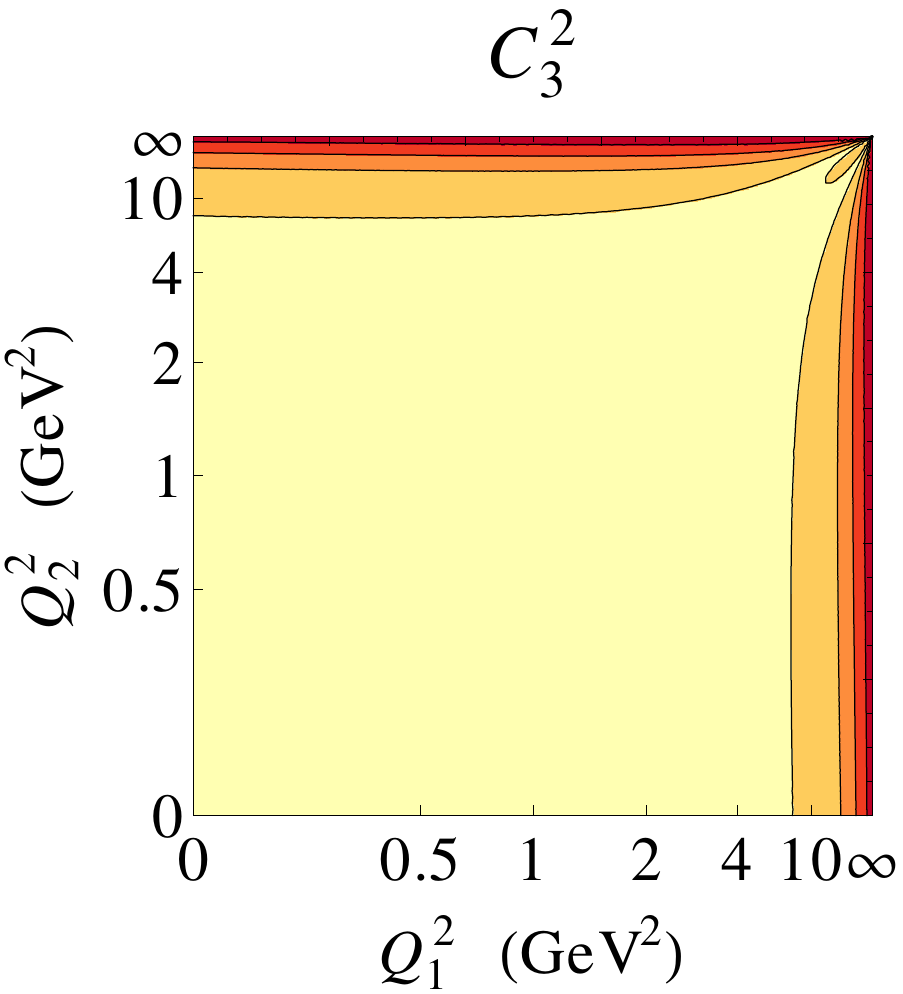}
  \caption{Convergence of the $C^N_{N+1}(Q_1^2,Q_2^2)$ sequence  with the appropriate high-energy behavior to the Regge model for different 
  elements. The first, second, third, and fourth outer(inner) contours, from light to dark red, stand for the relative $\mp1,\mp5,\mp10$ and $\mp20\%$ deviations.  
   Both axis have been scaled as $Q^2/(1+Q^2)$. \label{fig:cnn1ope}}
\end{figure}
As an illustration, we show the result from matching $c^{\textrm{OPE}}_0 = 0$ alone. The resulting equation replaces   
the $c_{2N,1}$ ($c_{2N+1,1}$) matching condition for the diagonal (subdiagonal) sequence, respectively. The results obtained for the Regge 
model are illustrated in \cref{fig:cnn1ope}, and show the expected improved convergence along the $Q_1^2=Q_2^2$ region. A similar improvement 
is achieved for the logarithmic model as well. Actually, we find that in this case the equal-virtual behavior \cref{eq:appelleq} is exactly satisfied, 
reproducing then all the terms in \cref{eq:modelope} and reaching an infinite precision along $Q_1^2=Q_2^2$. 
For the Regge model this is no longer possible, as its equal-virtual behavior \cref{eq:ReggeDVeq} is not represented by a rational function, requiring then 
an infinite sequence to reproduce it. Still, additional terms in the high-energy expansion may be predicted even if these were not matched. As an example, 
we show the prediction for the $c^{\textrm{OPE}}_1$ term in \cref{eq:modelope} in \cref{tab:ope}.
\begin{table}
\centering
\begin{tabular}{cccccccc} \toprule
 & $C^1_1$ & $C^2_2$ & $C^3_3$ & $C^0_1$ & $C^1_2$ & $C^2_3$ & Exact \\\cmidrule(r){1-4}\cmidrule(lr){5-7}\cmidrule(l){8-8}
$c^{\textrm{OPE}}_1$  & $0.172$ & $0.234$ & $0.246$ & $0.374$ & $0.276$ & $0.264$ & $0.257$ \\\bottomrule
\end{tabular}
\caption{The prediction for the leading $c^{\textrm{OPE}}_1$ term in the high-energy expansion for the diagonal and subdiagonal sequences compared to the exact result for the 
Regge model. \label{tab:ope}  }
\end{table}
As a conclusion, we find that CAs are able as well to use the information at zero and infinity, providing a reliable description of the underlying function in the whole-energy range.

\section{Canterbury approximants as a fitting tool}
\label{sec:cadata}

Our knowledge about the double-virtual TFF is rather scarce. Theoretically, the situation resembles that of the single-virtual TFF. At high-energies, pQCD can be used 
to predict the leading $Q_1^2=Q_2^2\equiv Q^2$ behavior in similarity to the BL limit, see \cref{eq:OPElim}. 
At low energies, \cpt can be used to obtain the TFF series expansion at zero virtualities, leading to a clear prediction for $F_{P\gamma\gamma}$. A higher order calculation 
could be performed to obtain the single-virtual leading $Q_{1(2)}^2$ behavior, however, some unknown low-energy constants were required to regularize the theory, thus 
losing predictive power. The situation does not ameliorate for the double-virtual expansion, where an even higher order calculation is required  to obtain the coefficients for 
$Q_{1(2)}^4$ and $Q_1^2Q_2^2$ with the consequent proliferation of additional unknown low-energy constants. 

The experimental situation for the double-virtual case is even more complicated. Whereas for the 
single-virtual case the theoretical ignorance was alleviated with a rich experimental knowledge of the TFF in a wide kinematical regime, there is at the moment not a single 
measurement for the double-virtual TFF. As a result, it is difficult to assess the different theoretical ideas. 
This situation is related to the particular kinematics of the processes in which the double-virtual TFF can be experimentally accessed. 

In the space-like region, such measurement can be accessed at $e^+e^-$ colliders in the $e^+e^-\rightarrow e^+e^-\gamma^*\gamma^*\rightarrow e^+e^-P$ reaction.
Such cross section is two-fold suppressed. On the one hand, the photon emission from the $e^{\pm}$ is suppressed for large photon virtualities. On the other hand, 
the TFF $F_{P\gamma^*\gamma^*}(Q_1^2,Q_2^2)$ receives an additional $Q_2^2$ suppression with respect to $F_{P\gamma^*\gamma^*}(Q_1^2,0)$. Therefore, to 
obtain a significant amount of events, it is necessary to look at low virtualitites, this is, at small $e^{\pm}$ scattering angles. However, this kinematic regime is experimentally 
extremely challenging due to the detector geometry and Bhabha scattering background. Remarkably, there is an ongoing effort at BES III to measure this process at low 
energies~\cite{Adlarson:2014hka}, which will provide valuable information.

In the time-like region, the double-virtual TFF can be accessed at energies below the pseudoscalar mass in the double Dalitz decay process 
$P\rightarrow\gamma^*\gamma^*\rightarrow\ell^+\ell^-\ell'^+\ell'^-$. However, its large suppression due to the additional electromagnetic couplings with 
respect to the two photons and Dalitz decays, makes such process very challenging. In addition, even though its BR would provide valuable information, 
it is the differential decay width which gives direct access to the TFF, which measurement requires even higher statistics. 
Moreover, the presence of the photon propagators greatly enhances low energies relative to the high energies, hiding the double-virtual effects, of order
$(\mathcal{O}(q_1^2q_2^2))$, as compared to the single-virtual ones, of order $\mathcal{O}(q_{1(2)}^2)$, encoded in  the slope parameter $b_P$ 
and playing the main role in this decay\footnote{This handicap would be alleviated using the $\ell=\mu$ channel for the $\eta$ and $\eta'$, which is insensitive to the 
very low-energy dynamics given the $\mu$ mass~\cite{Escribano:2015vjz}.}.

\subsection{Extracting the low-energy parameters from data}

It is evident that a first measurement on the double-virtual TFF is required to improve our current knowledge, but is equally important to perform 
an appropriate and reliable theoretical analysis from these data. In this section we discuss, in analogy to \cref{chap:data}, how CAs provide an 
excellent tool to perform such analysis and extract the relevant low- (and high-) energy parameters in \cref{eq:dvseries} in a systematic and model-independent fashion, and 
assess on the precision which would be achieved.

For this purpose, we speculate about a possible measurement for the double-virtual TFF corresponding to 36 points in the 
$[(0,5)\times(0,5)]~\textrm{GeV}^2$ region\footnote{We take a square grid with $1$~GeV$^2$ spacing starting at $(0,0)$~GeV$^2$ and ending at $(5,5)$~GeV$^2$.} and investigate 
what could be obtained for the double-virtual parameters from a fitting procedure similar to that in \cref{chap:data} for the single-virtual case. We emphasize that $11$ of 
the $36$ data points, corresponding to the single-virtual TFF, are already available at even finer gridding, and will be improved in the future thanks to 
BESIII~\cite{Adlarson:2014hka}, NA62~\cite{Hoecker:2016lxt}, A2~\cite{Marc:private}, KLOE-2~\cite{Babusci:2011bg} and $GlueX$~\cite{Gan:2015nyc} collaborations. 
The purely double-virtual data-points are then reduced to 25. 
Moreover, only 15 of them are truly independent data points which need to be measured, as half of the square grid can be obtained by reflection from Bose symmetry. 

To show the performance of the method, we employ the different sequences which have been revised in this chapter, $C^N_1(Q_1^2,Q_2^2), C^N_N(Q_1^2,Q_2^2)$ and 
$C^N_{N+1}(Q_1^2,Q_2^2)$. We quote the extracted values for the different parameters of the series expansion, \cref{eq:dvseries}, in \cref{tab:fitlog,tab:fitregge} for the logarithmic 
and Regge model, respectively.
\begin{table}[t]
\scriptsize
\centering
\begin{tabular}{ccccccccccc} \toprule
  & $C^0_1$  & $C^1_1$  & $C^2_1$  & $C^3_1$  &  $C^1_2$  & $C^2_3$  & $C^1_1$  & $C^2_2$  & $C^3_3$ & Exact \\ \cmidrule(r){1-2} \cmidrule(lr){3-5}\cmidrule(lr){6-7}\cmidrule(lr){8-10}\cmidrule(l){11-11}
$F_{P\gamma\gamma}$  & $0.270$  & $0.275$    & $0.275$   & $0.275$    & $0.275$  & $0.275$ &  $0.275$   & $0.275$  & $0.275$ & $0.275$ \\
           $b_P$                   & $0.614$   & $0.743$   &  $0.787$  &  $0.806$    & $0.805$  & $0.828$  &  $0.743$  & $0.821$  & $0.832$ & $0.833$ \\
           $c_P$                   & $0.377$  & $0.609$   &  $0.727$   & $0.792$     & $0.793$  & $0.893$  &  $0.609$  & $0.856$  & $0.914$ & $0.926$ \\
       $a_{P;1,1}$              & $0.612$   &  $0.798$  &  $0.861$  & $0.887$     & $0.881$  & $0.918$  &  $0.797$  & $0.906$  & $0.925$ & $0.926$ \\
       $a_{P;2,1}$              & $0.520$  & $0.827$   & $0.964$  & $1.0319$     & $1.019$  & $1.124$  &  $0.827$   & $1.087$  & $1.150$ & $1.157$ \\\bottomrule
\end{tabular}
\caption{Convergence for the logarithmic model parameters. The parameters are defined according to \cref{eq:dvseries} with $m_P=1$~GeV. \label{tab:fitlog}}
\end{table}
\begin{table}[t]
\scriptsize
\centering
\begin{tabular}{ccccccccccc} \toprule
  & $C^0_1$  & $C^1_1$  & $C^2_1$  & $C^3_1$  &  $C^1_2$  & $C^2_3$  & $C^1_1$  & $C^2_2$  & $C^3_3$ & Exact \\ \cmidrule(r){1-2} \cmidrule(lr){3-5}\cmidrule(lr){6-7}\cmidrule(lr){8-10}\cmidrule(l){11-11}
           $F_{P\gamma\gamma}$  & $0.273$  & $0.275$   & $0.275$   & $0.275$        & $0.275$  & $0.275$        &  $0.275$   & $0.275$  & $0.275$     & $0.275$ \\
           $b_P$                & $1.085$   & $1.246$  &  $1.295$  &  $1.315$       & $1.316$  & $1.334$         &  $1.246$  & $1.330$  & $1.335$      & $1.336$ \\
           $c_P$                & $1.186$  & $1.632$   &  $1.813$  & $1.900$        & $1.911$  & $2.005$         &  $1.632$  & $1.978$  & $2.009$      & $2.016$ \\
       $a_{P;1,1}$              & $1.824$   &  $2.038$ &  $2.050$  & $2.042$        & $2.627$  & $2.017$         &  $2.034$  & $2.020$  & $2.017$      & $2.016$ \\
       $a_{P;2,1}$              & $2.690$  & $3.239$   & $3.267$   & $3.238$        & $3.177$  & $3.124$        &  $3.239$   & $3.141$  & $3.122$    & $3.119$ \\\bottomrule
\end{tabular}
\caption{Convergence for the Regge model parameters.  The parameters are defined according to \cref{eq:dvseries} with $m_P=1$~GeV. \label{tab:fitregge}}
\end{table}
The agreement and convergence obtained is excellent, meaning that we have the chance to have a decent extraction once the first measurement for the 
double-virtual TFF is performed. 
Naturally, the systematic accuracy that may be achieved depends on whether the quantity of data points is larger or smaller than that used here, but equally important is the 
measured energy range. On the one hand, if we would have enlarged the interval beyond $5~\textrm{GeV}^2$, but keeping the same number of data points, the quality of 
the extraction would have deteriorated. On the other hand, 
taking a smaller interval ---while keeping the number of data points--- would improve the result and convergence of the sequence. 
Once more, we emphasize that the systematic error from the first element, the $C^0_1(Q_1^2,Q_2^2)$, is not negligible, 
which stress the necessity of using larger approximants. 
In particular, this means that, even if we employ the $C^0_1(Q_2^2,Q_2^2)$ approximant to describe the TFF in some calculation, we should not take the parameters which 
are obtained from a direct fit to this last, but those obtained for the highest approximants. 
This is a well-known feature in PAs and its oversight would result in a large systematic error.

\subsection{Implementing and extracting the high-energy behavior}

Given the large amount of unknowns in our approximants, it may be useful, specially regarding the real case in which data contain non-negligible statistical errors, 
to reduce the quantity of free parameters. One possibility is to implement the high-energy double-virtual behavior, 
which is dictated by pQCD as explained in \cref{sec:caope}. In this way, we do not only get rid of one parameter, but we can extract the high-energy expansion as 
well, see \cref{sec:caope}.
We find that  this approach results in an improved extraction of the low-energy parameters as compared to \cref{tab:fitlog,tab:fitregge}, with the exception 
of the $C^0_1(Q_2^2,Q_2^2)$ approximant, which often involves a poor description. In addition, we extract the $c^{\textrm{OPE}}_1$ parameter from the high-energy 
expansion \cref{eq:modelope}, which result is shown in \cref{tab:cafitope} for the logarithmic and Regge models for different approximants.
\begin{table}[t]
\footnotesize
\centering
\begin{tabular}{cccccccc}\toprule
$c^{\textrm{OPE}}_1$  & $C^0_1$ & $C^1_2$ & $C^2_3$ & $C^1_1$ & $C^2_2$ & $C^3_3$ & Exact \\\cmidrule(r){0-0}\cmidrule(lr){2-4}\cmidrule(lr){5-7}\cmidrule(l){8-8}
Log        &  $0.191$  &  $0.174$  &  $0.171$      &  $0.137$  &  $0.152$  &  $0.148$  &  $0.165$ \\
Regge   &  $0.103$  &  $0.078$  &  $0.074$      &  $0.043$  &  $0.061$  &  $0.063$  &  $0.071$ \\\bottomrule
\end{tabular}
\caption{The $c^{\textrm{OPE}}_1$ coefficient in \cref{eq:modelope} extracted from different approximants for each model. The last column represents the exact value. \label{tab:cafitope}}
\end{table}
In particular, we find that the diagonal (subdiagonal) sequence seems to provide a lower (upper) bound for this value ---in accordance with \cref{sec:caope}---, offering a 
powerful method to obtain an estimate for the systematic error. \\

The method presented here provides a powerful mathematical approach not only to reconstruct or extract the TFF, but for the experimentalists to analyze their data without 
any theoretical prejudice and to estimate reliable systematic errors in an easy way. This is actually not only of relevance for the double-virtual measurement projected at BESIII, 
but for those collaborations measuring the single-virtual TFF. 
In this sense, we have to recall that these experiments always involve a deeply virtual photon together with a quasi-real one; the virtuality from the latter 
is certainly small but does not need to vanish. 
As an example, for the Belle $\pi^0$ measurement~\cite{Uehara:2012ag} this is mainly less than $0.01~\textrm{GeV}^2$, whereas for 
\babar it may be as large as $0.6~\textrm{GeV}^2$~\cite{Aubert:2009mc,BABAR:2011ad}. 
To assess the corrections from the quasi-real photon effects, the experimental community requires then some model parametrizing the double-virtual TFF. 
The chosen parametrization is not unique, for instance, Belle 
uses a factorized approach, whereas \babar takes a $1/(Q_1^2+Q_2^2)$ parametrization.
Our method would be of help for these experiments in order to improve in precision and systematics. 
In addition, this may allow to extract some information about the double-virtual TFF. 
Finally, there are ongoing lattice studies for the $\pi^0$ TFF~\cite{antoine:private}; such approaches do requrie as well some function to fit their results. Our apporach would provide then a valuable tool for them as well.

\section{Conclusions}
\label{ca:concl}


In this chapter, we have introduced a generalization of PAs to the bivariate case. This generalization extends the previous ideas on Pad\'e theory for the single-virtual
to the most-general double-virtual TFF. For the case of symmetric functions, as the TFF, the use of Canterbury approximants is natural and straightforward, it guarantees 
the convergence to meromorphic functions (representing the large-$N_c$ limit of QCD), respects factorization without imposing it (which may approximately holds 
at low-energies for the TFF), reproduces well-known properties from PAs and provides convergence to Stieltjes functions.

In addition, the performance of the approach has been illustrated through the use of two different models previously employed in the univariate case. 
We have found that the intuition from PAs when dealing with poles and cuts can be extrapolated to this case. Moreover, in similarity to PAs, the poles may be given in 
advance, though this implies again larger systematic errors. Once more, the method allows to implement not only the low-, but 
the high-energy information as well. As a final remark, we have shown that the underlying symmetries of the original function may help to improve on convergence. 
Regretfully, there is no clear symmetry or relation among the low-energy expansion parameters for the TFF beyond that imposed from Bose symmetry, though a deeper 
study along this line would be of interest.

In analogy to PAs, our method allows then to extract the (theoretically unknown) low- and high-energy parameters entering the TFF from experimental data through a fitting procedure 
in a systematic and model-independent fashion. 
The ongoing experimental effort at BESIII to perform such a measurement would provide then the last required piece of information to reconstruct the double-virtual TFF. 
As an outcome, our method may be of interest for the experimental community (which often has to deal with the double-virtual TFF even if measuring the single-virtual one) 
and for the lattice community.

This chapter closes the theoretical framework which has been developed for describing the pseudoscalar TFFs. With all the required ingredients at hand, we 
proceed to discuss in the next chapters different applications in which these  TFFs represent the main input in the calculation.

\chapter{$\eta-\eta'$ mixing}
\label{chap:mixing}
\minitoc

\section{Introduction}

The $\eta-\eta'$ mixing has been a subject of deep investigation since the advent of the quark model. Early attempts to describe the $\eta-\eta'$ 
structure through the use of $SU(3)_F$ symmetry and Gell-Mann-Okubo (GMO) mass formulas appeared in Refs.~\cite{Isgur:1976qg,Fritzsch:1976qc}, which 
obtained a mixing angle  $\theta_P\approx-10^{\circ}$. Later on, as \cpt was established as the low-energy effective field theory of QCD and calculations 
at NLO became available, it was realized that corrections to the GMO mass formula shifted the mixing angle to $\theta_P \approx -20^{\circ}$, which was in better agreement with experimental results~\cite{Donoghue:1986wv,Gilman:1987ax}. However, in the years to come, different phenomenological analysis 
appeared, questioning such result and suggesting values from $\theta_P = -20^{\circ}$ to $\theta_P = -10^{\circ}$, depending on the observables taken into 
account and on the models assumptions~\cite{Bramon:1989kk,Ball:1995zv,Bramon:1997mf,Venugopal:1998fq,Bramon:1997va}. 
This situation was understood after the development of large-$N_c$ \cpt (\lcpt), which provides a framework to bring the $\eta'$ meson 
into \cpt. It was clear after the publication of~\cite{Schechter:1992iz,Leutwyler:1997yr}, and subsequent works~\cite{Kaiser:1998ds,Kaiser:2000gs,HerreraSiklody:1997kd}, 
that the $\eta-\eta'$ mixing requires two angles to parametrize their decay constants as a consequence  of $SU(3)_F$ breaking. 
This feature has been incorporated in subsequent phenomenological 
analysis~\cite{Feldmann:1998vh,Feldmann:1998sh,Feldmann:1999uf,Benayoun:1999au,Bramon:2000fr,Escribano:2005qq,Escribano:2007cd} 
resulting in different values depending on the modeling procedure.\\

In the following, we take our previous results from \cref{chap:data} in order to provide a new alternative  determination for the $\eta-\eta'$ mixing parameters. 
As an advantage, this approach is free of the simplifying assumptions required in previous approaches. 
In \cref{sec:1angle}, we provide a brief reminder of the mixing at LO in \lcpt, whereas the necessity of a two-angle description at NLO is 
discussed in \cref{sec:2angle}, where we introduce the octet-singlet and quark-flavor basis.
Our novel approach for determining the mixing parameters is discussed in \cref{sec:MixRes}. As an innovation, we sequentially include the effects of 
OZI-violating parameters and, in general, the full NLO corrections in a comprehensive way. Applications concerning the mixing 
are discussed in \cref{sec:appl}. Finally, we present our conclusions in \cref{sec:mixingconcl}.

\section{One-angle approximation}
\label{sec:1angle}

From the \lcpt Lagrangian $\mathcal{L}^{(0)}$ \cref{eq:lcptLO}, we extract the LO result 
for the kinetic and mass terms for the (bare) $\eta_8$ and $\eta_{0}$ fields, $\eta_B\equiv(\eta_8,\eta_0)^T$~\cite{Escribano:2010wt,Bickert:2015cia},
%
%
%
\begin{gather}
\label{eq:kinmass}
\mathcal{L}^{(0)} = \frac{1}{2}\partial_{\mu}\eta_B^T \mathcal{K} \partial^{\mu}\eta_B  -  \frac{1}{2}\eta_B^T \mathcal{M}^2 \eta_B, \\
\label{eq:LOmix}
\mathcal{K}=\mathds{1}_{2\times2}, \qquad 
 \mathcal{M}^2 = 
\begin{pmatrix}
 M_8^2    & M_{80}^2 \\
 M_{80}^2 & M_0^2 + M_{\tau}^2 
\end{pmatrix}, 
\end{gather}
which entries can be expressed in terms of the LO $\pi$ and $K$ masses, \cref{eq:cptmass,eq:toppmass}, as
\begin{align}
M_8^2 = \frac{2B_0}{3}(\hat{m} + 2m_s) = \frac{1}{3}(4\mathring{M}_K^2 - \mathring{M}_{\pi}^2), \label{eq:m8}\\ 
M_0^2 = \frac{2B_0}{3}(2\hat{m} + m_s) =  \frac{1}{3}(2\mathring{M}_K^2 + \mathring{M}_{\pi}^2), \label{eq:m0} \\ 
M_{80}^2 = \frac{2\sqrt{2}}{3}(\hat{m}-m_s)  =  -\frac{2\sqrt{2}}{3}(\mathring{M}_K^2 - \mathring{M}_{\pi}^2), \label{eq:m80} 
\end{align}
%
%
%
and $M_{\tau}^2 = \frac{6\tau}{F^2}$. 
It is clear then from $\hat{m}\neq m_s$ ---equivalently, $\mathring{M}_K^2\neq\mathring{M}_{\pi}^2$---, that the $\eta_8$ and $\eta_{0}$ fields will mix among each other 
into the physical $\eta$ and $\eta'$. At this order, \cref{eq:kinmass} can be diagonalized through the rotation matrix 
\begin{equation}
\label{eq:rotmat}
R(\theta_P) = 
\begin{pmatrix}
\cos\theta_P & -\sin\theta_P \\
\sin\theta_P & \cos\theta_P
\end{pmatrix},
\end{equation}
allowing to express the physical $\eta_{P} = (\eta,\eta')^T$ fields in terms of the bare ones in \cref{eq:kinmass} 
as $\eta_P = R(\theta_P) \eta_B$, where~\cite{Escribano:2010wt,Bickert:2015cia}
\begin{equation}
\sin(2\theta_P) = \frac{2 M_{80}^2}{M_{\eta'}^2 - M_{\eta}^2},
\end{equation}
and $M_{\eta}^2, M_{\eta'}^2$ are the eigenvalues solving the previous system, this is, the prediction for the physical masses. 
The mixing introduced above when diagonalizing the mass term $\mathcal{M}$ is referred to as the state-mixing and involves a single angle $\theta_P$,  
not only at this order, but at any order. At LO in \lcpt, one obtains the result $\theta_P=-19.6^{\circ}$~\cite{Bickert:2015cia}. 
However, non-negligible corrections are found at higher orders in the systematic \lcpt expansion~\cite{Guo:2015xva,Bickert:2015cia} shifting this value 
towards $\theta_P\approx-10^{\circ}$~\cite{Bickert:2015cia}.\\

Of special interest  for our later discussions are the pseudoscalar decay constants. These are defined in terms of the QCD axial current as
\begin{equation}
\label{eq:defaxialSO}
   \bra{0} J_{5\mu}^a \ket{P}=ip_{\mu}F_P^a \qquad J_{5\mu}^a = \overline{q}\gamma_{\mu}\gamma_5 \frac{\lambda^a}{2} q  \qquad \operatorname{Tr}(\lambda^a\lambda^b)=2\delta^{ab},
\end{equation}
where $\lambda^a$ is a Gell-Mann matrix in flavor space and $\lambda^0 = \sqrt{2/3} \ \mathds{1}_{3\times3}$. We remark that our normalization for the axial current yields 
$F_{\pi}=92.21(14)$~MeV \cite{Agashe:2014kda}. At LO in \lcpt, one finds $F_{\pi} = F_K = F$. 
For the $\eta$ and $\eta'$, due to the mixing, the decay constants are conveniently expressed, 
following Ref.~\cite{Leutwyler:1997yr} at LO as
\begin{equation}
\label{eq:FLO}
 \! F_P^{80}  \! \equiv  \!
\begin{pmatrix}
F_{\eta}^8 & F_{\eta}^{0} \\
F_{\eta'}^8 & F_{\eta'}^{0}
\end{pmatrix}
 \! = \! F  \!
\begin{pmatrix}
\cos\theta_P & -\sin\theta_P \\
\sin\theta_P & \cos\theta_P
\end{pmatrix}
 \! =  \!
R(\theta_P)  \!
\begin{pmatrix}
F & 0 \\
0 & F 
\end{pmatrix}
 \! \equiv  \! 
R(\theta_P)\hat{F}, \! \!
\end{equation}
where $\hat{F}=\textrm{diag}(F_8,F_0)$ and $F_8=F_0=F$ at LO. Consequently, the $\eta$ and $\eta'$ couple both, to the octet and singlet axial 
currents. It follows then that, at LO, their couplings to these currents $(F_P^{80})$ can be expressed in terms of the octet and singlet 
$\hat{F}$ decay constants using the same rotation matrix we used for the state mixing, this is, $F_P^{80}=R(\theta_P) \hat{F}$,  cf.~\cref{eq:rotmat} and comments below.
In the jargon of $\eta-\eta'$ mixing, the decay constants follow the state mixing.
This situation is particular to the LO case. As we illustrate below, at higher orders, $SU(3)_F$ breaking effects destroy this simple picture, requiring a two-angle 
description to express the decay constants.

\section{Two-angle mixing schemes}
\label{sec:2angle}

\subsection{Octet-singlet scheme}
\label{sec:singletoctet}

When moving on to NLO, the $\eta-\eta'$ mixing becomes more involved as now the kinetic matrix $\mathcal{K}$ in \cref{eq:kinmass} becomes non-diagonal 
too~\cite{Feldmann:1999uf,Escribano:2005qq,Kaiser:1998ds,Leutwyler:1997yr}, a fact which was pointed out for the first time in~\cite{Leutwyler:1997yr,Kaiser:1998ds}. 
Actually, $\mathcal{K}$ and $\mathcal{M}^2$ cannot be simultaneously diagonalized within a single rotation. 
The diagonalization is performed then, perturbatively, in two sequential steps~\cite{Escribano:2010wt,Bickert:2015cia}. 
First, a field redefinition for the bare fields $\eta_{B} = {Z^{1/2}}^T \hat{\eta}$ allows to diagonalize the kinetic term $\mathcal{K}$. 
Then, the resulting mass matrix, ${Z^{1/2}} \mathcal{M} \  {Z^{1/2}}^T$, is diagonalized through a rotation $\eta_P = R(\theta_P)\hat{\eta}$;
the required angle in this rotation defines the state-mixing angle in analogy to \cref{eq:rotmat}. 
Note however that the overall transformation $\eta_P = R(\theta_P)({Z^{1/2}}^T)^{-1}\eta_B$ includes the non-diagonal $Z^{1/2}$ matrix. 
For these reasons, the pseudoscalar decay constants cannot be expressed in a simple form analog to \cref{eq:FLO} as four parameters 
are now required. Instead, they are defined as
\begin{equation}
\label{eq:FP80}
F_P^{80}\equiv
\begin{pmatrix}
F_{\eta}^{8}&F_{\eta}^{0}\\
F_{\eta^{\prime}}^{8}&F_{\eta^{\prime}}^{0}
\end{pmatrix}
\equiv
\begin{pmatrix}
F_{8}\cos\theta_{8}&-F_{0}\sin\theta_{0}\\
F_{8}\sin\theta_{8}&F_{0}\cos\theta_{0}
\end{pmatrix}
\neq R(\theta_P)
\begin{pmatrix}
F_{8} & 0 \\
0 &F_{0} 
\end{pmatrix}
.
\end{equation}
We emphasize again that the state-mixing involves a single mixing angle, $\theta_P$, at any order. 
It is the decay constants $F_P^a$ description that requires two-angles or, alternatively, four independent quantities.
\lcpt provides then the appropriate framework to relate these decay constants to other quantities in the mesonic sector of QCD. 
Among others,  the mixing-angle and additional decay constants $F_{\pi}$ and $F_{K}$.
Particularly, at NLO, the following relations hold~\cite{Feldmann:1999uf,Escribano:2010wt}\footnote{To obtain these relations, the relevant LECs 
defining these quantities have been traded for $F_{\pi}$ and $F_K$. Moreover, multiplicative factors such as $(F_K/F_{\pi}-1)$ have been neglected as they can be 
understood as NNLO effects.}
%
\begin{gather}
  F_8^2 = \frac{4F_K^2 - F_{\pi}^2}{3},  \qquad F_{0}^2 = \frac{2F_K^2 + F_{\pi}^2}{3} + F_{\pi}^2\Lambda_1, \label{eq:F8F0} \\
  F_8 F_{0} \sin(\theta_8 - \theta_{0}) = -\frac{2\sqrt{2}}{3}\left( F_K^2 - F_{\pi}^2 \right) \label{eq:sin80},
\end{gather}
%
%
\begin{equation}
\label{eq:angles}
\theta_8 + \theta_{0} = 2\theta_P, \qquad  \theta_8 - \theta_{0} = -\frac{4\sqrt{2}}{3}\left( \frac{F_K}{F_{\pi}} -1 \right),
\end{equation}  
%
with $\Lambda_1$ an OZI-violating parameter. \cref{eq:FP80} defines the so-called octet-singlet mixing scheme and relations~\labelcref{eq:F8F0,eq:sin80,eq:angles} hold up to NNLO corrections in the combined \lcpt expansion.
Given that $F_K/F_{\pi}=1.198(5)$~\cite{Agashe:2014kda}, it follows from \cref{eq:sin80,eq:angles} that $SU(3)_F$ breaking implies $\theta_8 \neq \theta_{0}$.
It was the neglected $SU(3)_F$ breaking encoded in the GMO formula and $F_8/F_{\pi}$ ---not included up to~\cite{Donoghue:1986wv,Gilman:1987ax}--- that lead to bad results in the 
earlier years~\cite{Isgur:1976qg,Fritzsch:1976qc}. The same effect, this time encoded in $(\theta_8 - \theta_0)\neq0$, lead to different extractions for the decay constants 
from different observables~\cite{Bramon:1989kk,Ball:1995zv,Bramon:1997mf,Venugopal:1998fq,Bramon:1997va}, which often require the decay constants rather than the state-mixing.\\

At this point, there is a further property which must be discussed. Given the anomalous dimension of the singlet axial current, the singlet decay constants 
defined via $\bra{0} J_{5\mu}^{0} \ket{P} = ip_{\mu}F_P^{0}$ will inherit the scale-dependency which is dictated from QCD~\cite{Leutwyler:1997yr,Kaiser:1998ds,Kaiser:2000gs}
\begin{equation}
\label{eq:F0run}
\mu \frac{dF_{0}}{d\mu}  = \gamma_A(\mu) F_{0} = - \frac{3 C_2(r) N_F \alpha_s^2}{8\pi^2} F_{0}   + \mathcal{O}(\alpha_s^3)  = -N_F \! \left( \frac{\alpha_s(\mu)}{\pi} \right)^2 \! F_{0}.
\end{equation}
Here, $\mu$ is the renormalization scale, $\gamma_A(\mu)$ the axial current anomalous dimension~\cite{Espriu:1982bw} given in terms of the 
group invariant $C_2(r)$ ---for the fundamental representation $C_2(r)= (N_c^2-1)/(2N_c)$--- 
and $N_F$ is the number of active flavors $(u,d,s,...)$ at that scale. The solution to this equation is given, at $\mathcal{O}(\alpha_s)$ 
as~\cite{Espriu:1982bw,Leutwyler:1997yr,Kaiser:1998ds}
\begin{equation}
\label{eq:F0runII}
F_{0}(\mu) =  F_{0}(\mu_0)  \left(1+\frac{2N_F}{\beta_0}  \left(  \frac{\alpha_s(\mu)}{\pi} - \frac{\alpha_s(\mu_0)}{\pi} \right)  \right)  \equiv   F_{0}(\mu_0)(1+ \delta_{\textrm{RG}}(\mu)),
\end{equation}
where $\mu_0$ is some reference scale and we have used the LO result for the $\alpha_s$ running, involving at this order the beta function coefficient  
$\beta_0 = 11N_c/3 - 2N_F /3$. Of course, physical observables are scale independent, and $F_{0}(\mu)$-dependent terms will be accompanied by additional 
terms in such a way that the scale-dependency is cancelled. In the \lcpt Lagrangian, this is easy to see, as these ($\Lambda_i$ OZI-violating) terms are explicitly included 
in order to make the (bare) Lagrangian scale-independent. As an example, the WZW part requires an additional term~\cite{Leutwyler:1997yr}\footnote{Note our 
$\epsilon^{0123}=1$ convention and the replacement with respect to~\cite{Leutwyler:1997yr} $\psi\rightarrow(\sqrt{6}/F)\eta_0$.}
\begin{equation}
\label{eq:wzw2}
 \mathcal{L}^{(2)}_{\textrm{WZW}} \supset \frac{N_c\alpha\Lambda_3}{6\sqrt{6}\pi F}\epsilon^{\mu\nu\rho\sigma}F_{\mu\nu}F_{\rho\sigma}\eta_0,
\end{equation}
where $\Lambda_3$ is a scale-dependent OZI-violating parameter with running $\Lambda_3(\mu)=\Lambda_3(\mu_0)(1+ \delta_{\textrm{RG}}(\mu))$ analogous 
to that in \cref{eq:F0runII} and renders the two-photon decays in \cref{eq:Feta0,eq:Fetap0} scale-independent. Alternatively, heavy processes involving 
$\eta(\eta')$ in final states are often expressed in terms of $\sum_i C_i(\mu)\bra{0}\mathcal{O}_i\ket{P}$ matrix elements, where $\mathcal{O}_i$ is a local 
operator ---for instance, $\mathcal{O}_j = \overline{q}\gamma^{\mu}\gamma_5q$--- and $C_i(\mu)$ is the so-called Wilson coefficient, which accounts for the operator 
evolution from the heavy $(\mu=M_H)$ to the low $(\mu=\mu_0)$ scale. The latter should match that of $F_0(\mu_0)$, implying that any shift $\mu_0\to\mu_0'$ would not 
alter the result. This is the case for the TFF asymptotic behavior discussed in \cref{sec:MixRes}.

\subsection{Quark-flavor basis}
\label{sec:qfbasis}

The features outlined above make the description of any physical process involving the singlet sector much involved. 
For this reason, later on, the quark-flavor mixing scheme was proposed in Ref.~\cite{Feldmann:1998vh}. 
This scheme was motivated by the fact that vector and tensor singlet mesons ---where the axial anomaly plays no role--- can be pretty well described in terms of light and strange 
quark singlet components. Actually, we show below that such assumption agrees with NLO \lcpt provided that OZI-violating effects 
are obviated. In such approximation, the physical states and 
decay constants follow the same mixing and can therefore be described in terms of one angle alone, which greatly simplifies our description. Defining the 
light and strange axial currents $J_{5\mu}^{q,s} = \overline{q}\gamma_{\mu}\gamma_5 \frac{\lambda^{q,s}}{2} q$, with $\lambda^q = \textrm{diag}(1,1,0)$ and 
$\lambda^s = \textrm{diag}(0,0,\sqrt{2})$, the pseudoscalar decay constants $\bra{0}J_{5\mu}^{q,s}\ket{P(p)} \equiv i p_{\mu}F_P^{q,s}$ read
%
%
\begin{equation}
\label{eq:Fqs}
(F_P^{qs})\equiv
\begin{pmatrix}
F_{\eta}^{q}&F_{\eta}^{s}\\
F_{\eta^{\prime}}^{q}&F_{\eta^{\prime}}^{s}
\end{pmatrix}
\equiv
\begin{pmatrix}
F_{q}\cos\phi_{q}&-F_{s}\sin\phi_{s}\\
F_{q}\sin\phi_{q}&F_{s}\cos\phi_{s}
\end{pmatrix}.
\end{equation}
\\
Relating the decay constants in both basis is rather simple as it only amounts to a rotation of our fundamental QCD currents. From the above definition, it is easy to check that the 
octet-singlet and quark-flavor basis are related via rotation matrix
\begin{equation}
\begin{pmatrix}
 J_{5\mu}^8 \\
 J_{5\mu}^{0} 
\end{pmatrix}
= \frac{1}{\sqrt{3}}
\begin{pmatrix}
1 & -\sqrt{2} \\
\sqrt{2} & 1
\end{pmatrix}
\begin{pmatrix}
 J_{5\mu}^q \\
 J_{5\mu}^s 
\end{pmatrix}
\ \ \Rightarrow  \ \
 (J_{5\mu}^{80})_{\alpha} = U(\theta_{ideal})_{\alpha a} (J_{5\mu}^{qs})_a.
\end{equation}
Here, the equation on the left-hand side has been expressed in matricial form in the right one with obvious identifications. The indices $\alpha$ and $a$ denote octet-singlet and 
flavor indices, respectively (summation assumed if repeated indices). Then, the decay constants in \cref{eq:FP80,eq:Fqs} can be related as 
\begin{equation}
\label{eq:80qs}
(F_P^{qs})_{Pa} = (F_P^{80})_{P\alpha}U(\theta_{ideal})_{\alpha a},
\end{equation}
where the index $P=\{\eta,\eta'\}$ and, again, summation over repeated indices is assumed. Relation~\eqref{eq:80qs} will be our dictionary when relating results in different basis. 
In this way, we can translate~\cref{eq:F8F0,eq:sin80} to their analogues in the quark-flavor basis obtaining~\cite{Feldmann:1999uf,Escribano:2005qq}
\begin{gather}
  F_q^2 = F_{\pi}^2 + \frac{2}{3}F_{\pi}^2\Lambda_1,  \qquad F_s^2 = 2F_K^2 - F_{\pi}^2 + \frac{1}{3}F_{\pi}^2\Lambda_1, \label{eq:FqFs} \\
  F_q F_s \sin(\phi_q - \phi_s) = \frac{\sqrt{2}}{3}F_{\pi}^2\Lambda_1. \label{eq:sinqs}
\end{gather}
It is clear, as anticipated, that neglecting the OZI-violating $\Lambda_i$ parameters implies $\phi_q=\phi_s\equiv\phi$, achieving a simpler 
one-angle description for the decay constants, $F_P^{qs}=R(\phi)\textrm{diag}(F_q,F_s)$. Indeed, there is a strong phenomenological success 
supporting this idea~\cite{Feldmann:1999uf,Escribano:2005qq}. Under the assumption $\phi_q=\phi_s$ ---that is commonly known as the 
FKS scheme~\cite{Feldmann:1998vh,Feldmann:1998sh,Feldmann:1999uf}---, this basis has become a standard choice given its simplicity and the predictive 
power with respect to the octet-singlet one. This assumption is specially useful for studying the TFFs~\cite{Agaev:2014wna} within pQCD.\\

An alternative approach to understand this situation follows from the pQCD picture in Ref.~\cite{Feldmann:1998sh} when considering the Fock state description of the 
$\eta$ and $\eta'$. Given that $(m_u \simeq m_d) \ll m_s$, it seems reasonable that $\eta$ and $\eta'$ may be described in terms of light and strange quarks degrees 
of freedom
\begin{equation}
\label{eq:qfWF}
\ket{\eta_q} = \Psi_q\frac{1}{\sqrt{2}}\ket{u\overline{u} +d\overline{d}} + ..., \quad \ket{\eta_s} = \Psi_s\ket{s\overline{s}} + ...,
\end{equation}
%
where the ellipses stand for additional Fock states including gluons and sea quarks, and $\Psi_q$ and $\Psi_s$ stand for the wave-functions, 
which are in general different from each other, i.e., $\Psi_q\neq\Psi_s$. Finally, $F_P^{q,s}$ is related to the $\Psi_{q,s}$ wave function  normalization, cf. 
\cref{eq:DAdef}. Assuming further that 
\begin{equation}
\label{eq:qfFS}
\ket{\eta} = \cos\phi \ket{\eta_q} - \sin\phi \ket{\eta_s} , \quad \ket{\eta'} = \sin\phi \ket{\eta_q} + \cos\phi \ket{\eta_s} ,
\end{equation}
%
implies that, when rotating back to the octet-singlet basis, an analogous $\eta-\eta'$ description along the lines of \cref{eq:qfFS},
\begin{equation}
\label{eq:soFS}
\ket{\eta} = \cos\theta_P \ket{\eta_8} - \sin\theta_P \ket{\eta_0} , \quad \ket{\eta'} = \sin\theta_P \ket{\eta_8} + \cos\theta_P \ket{\eta_0} ,
\end{equation}
would require defining the corresponding Fock states as 
\begin{align}
\label{eq:soWF}
\ket{\eta_8} &= \frac{\Psi_q + 2\Psi_s}{3} \frac{\ket{u\overline{u} + d\overline{d} - 2s\overline{s}} }{\sqrt{6}}  +  \frac{\sqrt{2}(\Psi_q - \Psi_s)}{3} \frac{\ket{u\overline{u} + d\overline{d} +s\overline{s}} }{\sqrt{3}} , \\
\ket{\eta_0} &= \frac{\sqrt{2}(\Psi_q - \Psi_s)}{3} \frac{\ket{u\overline{u} + d\overline{d} - 2s\overline{s}} }{\sqrt{6}}  + \frac{2\Psi_q + \Psi_s}{3}  \frac{\ket{u\overline{u} + d\overline{d} +s\overline{s}} }{\sqrt{3}} ,
\end{align}
so what has been defined as the octet(singlet) $\ket{\eta_{8(0)}}$ component is an admixture of the octet and singlet Fock states unless $SU(3)_F$-symmetry represents a 
good approximation and $\Psi_q=\Psi_s$ holds. This represents a result analogous to that in \cref{eq:sin80}. Conversely, in such $SU(3)_F$-symmetric case, where 
$\theta_8 = \theta_{0}=\theta_P$, we could start with an analogous single-octet description. Rotating back to the flavor basis, we would find an analogous result to that in 
\cref{eq:soWF}, namely, that the light(strange) quark state is an admixture of light and strange quark Fock states unless $\Psi_8=\Psi_0$. 
In this language, this is easy to see, as $\ket{q\overline{q}}$-like states get mixed via the QCD anomaly, an OZI-violating effect analogous to the result in Eq.~\eqref{eq:sinqs}. 
\\

To summarize, the quark-flavor basis provides a simpler choice ---in terms of a single angle--- whenever the precision we aim for does not require to include 
OZI-violating effects in our framework and has become the most popular choice in phenomenological 
analyses~\cite{Feldmann:1998vh,Feldmann:1998sh,Feldmann:1999uf,Bramon:2000fr,Escribano:2005qq,Escribano:2007cd}. 
In the case where the required precision may become sensitive to OZI-violating effects, both basis involve the use of two-angles 
---alternatively, four independent decay constants--- and the octet-singlet basis may become simpler for incorporating such effects.

\section{Determining the $\eta-\eta'$ mixing from the TFFs}
\label{sec:MixRes}

The different analyses used in the literature to extract the mixing parameters defined in the previous section ---$F_8,F_0,\theta_8,\theta_0$ in the 
octet-singlet basis or, alternatively, $F_q,F_s,\phi_q,\phi_s$ in the quark-flavor basis---  find often non-compatible values among their extractions. 
As an illustration, we refer to the approaches from Refs.~\cite{Leutwyler:1997yr,Feldmann:1998vh,Benayoun:1999au,Escribano:2005qq} which are depicted in \cref{Fig:MixRes}. 
It would be desirable then to have an alternative approach which is defined in terms of \lcpt quantities alone ---the decay constants--- and has control over the OZI-violating 
parameters. This requires avoiding, for instance, models for the $VP\gamma$ transitions ---more comments on them in \cref{sec:gvpg}--- which are widely used to 
extract the mixing parameters, or, eventually, the popular $J/\Psi\to\gamma\eta(\eta')$ decays ---further comments on this point in \cref{sec:jpsirad}.
We suggest that this is possible using the available information on the $\eta$ and $\eta'$ TFFs from \cref{chap:data}. Moreover, it is possible to account for the OZI-violating 
parameters, whose impact we discuss below. Actually our approach does not only allow to extract the above-mentioned mixing parameters 
but the additional OZI-violating parameter $\Lambda_3$, cf. \cref{eq:wzw2}.
\\

The starting point in our approach is the remarkable observation that, not only the low-energy behavior for the $\eta$ and $\eta'$ TFFs 
---related to their two photon decays---, but their high-energy behavior $\lim_{Q^2\to\infty}F_{P\gamma^*\gamma}(Q^2)$ dictated by pQCD \cref{eq:BLlim} is given, essentially, in terms of the desired mixing parameters.
Particularly, at NLO, the two-photon decays can be calculated from \lcpt, 
obtaining~\cite{Feldmann:1999uf,Kaiser:2000gs,Escribano:2015yup} 
\begin{align}
 F_{\eta\gamma\gamma} \equiv  F_{\eta\gamma\gamma}(0) = & \ \frac{1}{4\pi^2}\frac{\hat{c}_8(1+K_2^8) F_{\eta'}^{0} - \hat{c}_{0}(1+K_2^0+\Lambda_3) F_{\eta'}^8}{F_{\eta'}^{0} F_{\eta}^8 - F_{\eta'}^8 F_{\eta}^{0}}, \label{eq:Feta0} \\
 F_{\eta'\gamma\gamma} \equiv  F_{\eta'\gamma\gamma}(0) =  & \ \frac{1}{4\pi^2}\frac{-\hat{c}_8(1+K_2^8) F_{\eta}^{0} + \hat{c}_{0}(1+K_2^0+\Lambda_3) F_{\eta}^8}{F_{\eta'}^{0} F_{\eta}^8 - F_{\eta'}^8 F_{\eta}^{0}}, \label{eq:Fetap0}
\end{align}
where $\hat{c}_8 = 1/\sqrt{3}$ and $\hat{c}_{0} = 2\sqrt{2}/\sqrt{3}$ are charge factors. Besides, $K_2^8\equiv K_2\frac{7\mathring{M}_{\pi}^2-4\mathring{M}_{K}^2}{3}$ 
and $K_2^0\equiv K_2\frac{2\mathring{M}_{\pi}^2+\mathring{M}_{K}^2}{3}$ are related to the LEC $K_2$ in the \lcpt Lagrangian~\cite{Kaiser:2000gs}\footnote{The $K_2$ LEC represents the \lcpt version for the $SU(3)_F$ \cpt $L_8^{6\epsilon}$ LEC, see Ref.~\cite{Kaiser:2000gs}. Particularly, it compares to ~\cref{eq:tffcpt} via $K_2\to-(1024\pi^2/3)L_8^{6\epsilon}$.}. The latter appear as well in the $\pi^0$ TFF via
\begin{align}
F_{\pi\gamma\gamma} \equiv  F_{\pi\gamma\gamma}(0) = & \ \frac{1+K_2 \mathring{M}_{\pi}^2}{4\pi^2F_{\pi}}.
\end{align}
From the experimental $\pi^0\to\gamma\gamma$ result~\cite{Agashe:2014kda}, we obtain $K_2=-0.45(58)$, which is small and compatible with zero and has been often 
neglected in previous analyses. 

It must be emphasized that, in \cref{eq:Feta0,eq:Fetap0},
the $\Lambda_3$ OZI-violating parameter from \cref{eq:wzw2} must be included to render the result scale-independent. To see this, note that both $F_P^0$ and 
$\Lambda_3$, unlike $F_P^8$ and $K_2$, scale as $(1+\delta_{\textrm{RG}}(\mu))$. This produces overall factors in the numerator and denominator canceling the 
scale-dependency. 
To obtain the expression for the high-energy behavior, we have first to take into account the running of the axial 
current, \cref{eq:F0run}, which implies an additional running effect on top of that of the Gegenbauer coefficients, \cref{eq:DAgegen}. 
From Eq.~\eqref{eq:F0runII}, and taking as the reference scale for the $(\eta)\eta'\rightarrow\gamma\gamma$ decays $\mu_0=1$~GeV, we obtain for 
$F_P^{0}$ at $Q^2\rightarrow\infty$ the relation 
\begin{equation}
\label{eq:runinf}
F_P^{0}(\infty) = F_P^{0}\left( 1 - \frac{2N_F}{\beta_0}\frac{\alpha_s}\pi{} \right) = F_P^{0}(1+\delta_{\textrm{RG}}(\infty)) \equiv F_P^{0}(1+\delta),
\end{equation}
where $\alpha_s$ is to be evaluated at $1$~GeV and $F_P^{0}$ is the decay constant appearing in the $\eta(\eta')\rightarrow\gamma\gamma$ decays, to be taken at $\mu_0=1$~GeV. 
Taking into account corrections from higher orders by using the $\alpha_s$-running to four-loops accuracy~\cite{Kniehl:2006bg} as well as considering threshold effects, we obtain that 
$\delta=-0.17$.
The high-energy behavior ---assuming that asymptotic behavior is reached--- then reads~\cite{Agaev:2014wna} 
\begin{align}
 \eta_{\infty} \equiv \lim_{Q^2\to\infty} Q^2F_{\eta\gamma^*\gamma}(Q^2) =  & \ 2(\hat{c}_8 F_{\eta}^8 + \hat{c}_{0}(1+\delta) F_{\eta}^{0}), \label{eq:Infeta} \\
  \eta'_{\infty} \equiv \lim_{Q^2\to\infty} Q^2F_{\eta'\gamma^*\gamma}(Q^2) = & \ 2(\hat{c}_8 F_{\eta'}^8 + \hat{c}_{0}(1+\delta) F_{\eta'}^{0}).\label{eq:Infetap}
\end{align}
The resulting effect is by no means negligible and, to our best knowledge, was implemented for the first time in Ref.~\cite{Agaev:2014wna}.\\

We have at this stage a set of four equations at our disposal (\cref{eq:Feta0,eq:Fetap0,eq:Infeta,eq:Infetap}) to extract the 
four mixing parameters we are interested in. 
It seems then a straightforward task to determine the mixing parameters ---at least, if we neglect the {\textit{a priori}} small parameter $\Lambda_3$ 
and either neglect or take $K_2$ from the $\pi^0\to\gamma\gamma$ decay. 
However, there is a subtle connection among the different equations which avoids for such an easy solution. As noted for the first time in our work in 
Refs.~\cite{Escribano:2013kba,Escribano:2015nra}, the system 
of equations is degenerate. To see this, we can obtain an expression for $F_{\eta}^8$ and $F_{\eta'}^8$ from \cref{eq:Infeta,eq:Infetap}. Then, 
substituting in \cref{eq:Feta0,eq:Fetap0}, we can linearize the system, which may be expressed in matrix form as
%
%
\begin{equation}
\label{eq:degmat}
A \left(F_{\eta}^8 , F_{\eta'}^8 , F_{\eta}^{0} , F_{\eta'}^{0} \right)^T  =  
\left( \eta_{\infty} ,  \eta'_{\infty} , 0 , 0 \right)^T, \\
\end{equation}
where the $A$ matrix is defined as 
\begin{equation}
\label{eq:degmat}
A=
\begin{pmatrix}
2\hat{c}_8  &  0 &  2\hat{c}_{0}(1+\delta)  & 0 \\
0 & 2\hat{c}_8  &  0 &  2\hat{c}_{0}(1+\delta)  \\
0  &  \tilde{c}_{0}  & -\frac{2\pi^2}{\hat{c}_8}\eta'_{\infty}F_{\eta\gamma\gamma} & \frac{2\pi^2}{\hat{c}_8}\eta_{\infty}F_{\eta\gamma\gamma} - \tilde{c}_8  \\
-\tilde{c}_{0}  & 0 & \tilde{c}_8 - \frac{2\pi^2}{\hat{c}_8}\eta'_{\infty}F_{\eta'\gamma\gamma}  & \frac{2\pi^2}{\hat{c}_8}\eta_{\infty}F_{\eta'\gamma\gamma} &  
\end{pmatrix}, 
\end{equation}
\\
where $\tilde{c}_8 = \hat{c}_8(1+K_2^8)$ and $\tilde{c}_0 = \hat{c}_0(1+K_2^0+\Lambda_3)$. Then, the degeneracy is inferred from the determinant, which is proportional to
\begin{equation}
\label{eq:degm}
\left(\hat{c}_8^2(1+K_2^8) + \hat{c}_{0}^2(1+\delta)(1+K_2^0+\Lambda_3)\right) - 2\pi^2\left(F_{\eta\gamma\gamma}\eta_{\infty} + F_{\eta'\gamma\gamma}\eta'_{\infty}\right).
\end{equation}
It may look that \cref{eq:degm} is in general non-vanishing.
However, it turns out that
\begin{multline}
F_{\eta\gamma\gamma}\eta_{\infty} + F_{\eta'\gamma\gamma}\eta'_{\infty} = \frac{ \hat{c}_8^2(1+K_2^8) + \hat{c}_{0}^2(1+\delta)(1+K_2^0+\Lambda_3)}{2\pi^2} \\ 
= \frac{3}{2\pi^2}\left( 1 + \frac{1}{9}\left[  K_2^8 + 8\left(\delta + (K_2^0+\Lambda_3)(1+\delta) \right)  \right]   \right), \label{eq:deg}
\end{multline}
yields a vanishing value for \cref{eq:degm}, where in the last term we have replaced the charge factors $\hat{c}_i$. As an alternative approach, we can find that there is a null 
space for the system in Eq.~\eqref{eq:degmat},
\begin{equation}
\left( \hat{c}_{0} F_{\eta'\gamma\gamma}(1+\delta) \ , \  - \hat{c}_{0}(1+\delta)F_{\eta\gamma\gamma} \ ,  \ -\hat{c}_8F_{\eta'\gamma\gamma} \ ,  \ \hat{c}_8F_{\eta\gamma\gamma}  \right)^T.
\end{equation}
All in all, we have to deal with a degenerate system, which may look like a dead-end for our approach. However, contrary to the expectations, it turns out that one can 
take advantage of \cref{eq:deg} to solve all these problems. Curiously enough, the OZI-violating $\Lambda_i$ parameters play a central role in this discussion. 
In order to illustrate their impact and conceptual relevance, we first set $K_2=0$ and sequentially include these parameters one by one. 
First, we set $\Lambda_1=\Lambda_3=0$ and discuss the results. Second, we let $\Lambda_3\neq0$ but, still, $\Lambda_1=0$. Third, we let $\Lambda_1,\Lambda_3\neq0$ 
and obtain them through a fitting procedure. Finally, we include the parameter $K_2$, which completes the full list of NLO LECs which are relevant to our study. The latter is the main result from this chapter and represents, to our best knowledge, the first result fully consistent with \lcpt at NLO. Finally, we discuss our findings and compare to previous phenomenological approaches.

\subsection{The $\eta-\eta'$ mixing: $K_2=\Lambda_1 \! = \! \Lambda_3 \! =\! 0$}
\label{sec:MixI}

The simplest choice one can take to solve for the mixing parameters, see Ref.~\cite{Escribano:2015nra}, is to set all the OZI-violating $\Lambda_i$ parameters present in 
our equations to $0$, this is $\Lambda_1=\Lambda_3=0$ (as well as $K_2=0$). This choice implies, via \cref{eq:sinqs}, that $\phi_q=\phi_s\equiv \phi$. This does not only break the degeneracy of our system, 
but reduces the number of free parameters down to $3$, which allows to solve the system using a set of three equations out of \cref{eq:Feta0,eq:Fetap0,eq:Infeta,eq:Infetap}.
We call the attention however, that obtaining the same solution for any set is not guaranteed unless relation \cref{eq:deg}, 
$F_{\eta\gamma\gamma}\eta_{\infty} + F_{\eta'\gamma\gamma}\eta'_{\infty} = \frac{3}{2\pi^2}\left( 1 + \frac{8}{9}\delta \right)$, is fulfilled. 
In our case, taking the input values from \cref{tab:chap1mainres}, we obtain $0.89(3)\frac{3}{2\pi^2}$ for the left hand side, whereas the right hand side 
yields $0.85\frac{3}{2\pi^2}$ for $\delta=-0.17$.
Therefore, it seems that neglecting the OZI-violating parameters has not a tremendous impact. Note however that, to reach such agreement, we need to introduce the 
running parameter $\delta$ from \cref{eq:runinf}, which in the FKS scheme should be zero. 

In any case, since the condition \cref{eq:deg} is not exactly fulfilled, every set of equations will yield only marginally-compatible solutions. 
In order to solve the system, we decide to take the result which makes use of $F_{\eta\gamma\gamma}, F_{\eta'\gamma\gamma}$ and $\eta_{\infty}$ alone. 
The reason is motivated in two-fold way. On the one hand, $F_{\eta\gamma\gamma}$ and $F_{\eta'\gamma\gamma}$ have been directly measured to an excellent precision. 
On the other hand, among the asymptotic values, $\eta_{\infty}$ is the one with the most reliable extraction, see \cref{chap:data}. Finally, we expect that the $\eta$ 
parameters are theoretically cleaner, as they are less sensitive to the singlet effects we are neglecting at this stage.
As a result, taking the $F_{\eta\gamma\gamma}, F_{\eta'\gamma\gamma}$ and $\eta_{\infty}$ values from \cref{tab:chap1mainres}, we obtain~\cite{Escribano:2015nra}
\begin{align}
& \frac{F_q}{F_{\pi}} = 1.07(2), \quad \frac{F_s}{F_{\pi}} = 1.29(16), \quad \phi=38.3(1.6)^{\circ}, \label{eq:qfopti} \\
& \frac{F_8}{F_{\pi}} = 1.22(11)   \ \     \frac{F_{0}}{F_{\pi}} = 1.15(5) \quad \theta_8=-21.4(1.9)^{\circ}   \ \    \theta_{0}=-11.2(5.0)^{\circ}, \label{eq:soopti} 
\end{align}
where in the second line we have used \cref{eq:80qs} to translate the result into the octet-singlet basis. As an illustration, had we used $\eta'_{\infty}$ instead of $\eta_{\infty}$, we would have obtained 
$F_q/F_{\pi} = 1.06(1), F_s/F_{\pi} = 1.63(8), \phi=41.1(0.8)^{\circ}$. Had we obviated RG-effects, we would find some deviations in sets containing the $\eta_{\infty}$, while big deviations 
would be found for those containing $\eta'_{\infty}$, as the singlet content is more important for the $\eta'$, see Ref.~\cite{Escribano:2013kba}.
Our result is in line with previous findings~\cite{Leutwyler:1997yr,Feldmann:1998vh,Benayoun:1999au,Escribano:2005qq,Escribano:2013kba} and has competitive errors. 
For comparison, see Fig.~\ref{Fig:MixRes}, Option I. 

\subsection{The $\eta-\eta'$ mixing: $K_2=\Lambda_1 \! = \! 0 , \Lambda_3 \! \neq \! 0$}
\label{sec:MixII}

As illustrated before, the previous approach suffers from the fact that solutions from different sets yield different results which are only marginally compatible. 
This was easy to anticipate given that the degeneracy condition~\labelcref{eq:deg} was only marginally fulfilled for $\Lambda_3=0$. In this second approach, we assume 
that, still, $\Lambda_1=0$, but $\Lambda_3$ is a free parameter, which is fixed as to fulfill \cref{eq:deg}, obtaining~\cite{Escribano:2015nra} $\Lambda_3 = 0.06(4)$. 
Such value may be compared to the result $\Lambda_3=-0.03(2)$ from Ref.~\cite{Benayoun:1999au} obtained from $VP\gamma$ decays. 
They differ in sign, but agree on its small magnitude, even beyond what is expected from the naive $1/N_c$ counting. 
Still, as $\Lambda_1=0$, we stick to the one-angle quark-flavor scheme, whereby any set of three equations can be used with the same result. 
Taking the same inputs as in previous section from \cref{tab:chap1mainres}, we obtain
\begin{align}
 & \frac{F_q}{F_{\pi}} = 1.12(4), \quad \frac{F_s}{F_{\pi}} = 1.52(7), \quad \phi=38.9(1.3)^{\circ}, \label{eq:qfoptii} \\
& \frac{F_8}{F_{\pi}} = 1.40(5)   \ \     \frac{F_{0}}{F_{\pi}} = 1.27(3) \quad \theta_8=-23.6(1.1)^{\circ}   \ \    \theta_{0}=-7.3(3.2)^{\circ}. \label{eq:sooptii}
\end{align}
As an advantage, choosing $\Lambda_3\neq0$, we can obtain analog results for any chosen set of equations, which improves with respect 
to the previous situation. 
Our results are displayed under the label Option II in Fig.~\ref{Fig:MixRes} and show the impact of including the $\Lambda_3$ parameter.

\subsection{The $\eta-\eta'$ mixing: $K_2=0, \Lambda_1,\Lambda_3 \! \neq \! 0$}
\label{sec:MixIII}

The approaches adopted in \cref{sec:MixI,sec:MixII} present, at the formal level, some theoretical inconsistencies. Namely, we found that running effects 
---neglected in the common FKS scheme--- encoded in $\delta$, see \cref{eq:runinf}, were important in our determination. However, these require, formally, the 
presence of the $\Lambda_1$ parameter if the scale-dependency for the asymptotic behavior is to be cancelled ---see \cref{eq:Infeta,eq:Infetap}. 
Similarly, including $\Lambda_3$ requires the presence of $\Lambda_1$ to cancel the scale-dependency in the two photon decays ---see \cref{eq:Feta0,eq:Fetap0}.
Besides, at the phenomenological level, there is further evidence pointing to $\Lambda_1\neq0$ effects. Particularly, our previous results ---and basically every phenomenological 
estimate, see \cref{Fig:MixRes}--- indicate that $F_q>F_{\pi}$ with around $3\sigma$ significance. This, via \cref{eq:FqFs}, implies a non-vanishing positive value for 
$\Lambda_1$, which in our simplified approach was taken to be zero. This in turn, would imply via \cref{eq:sinqs} that $\phi_q\neq\phi_s$, invalidating then our previous 
assumptions and pointing out the necessity of using a general scheme with two different angles and non-zero $\Lambda_{1,3}$ parameters for describing the 
$\eta$ and $\eta'$ decay constants, an approach that we adopt in this section (but still retaining $K_2=0$).\\

\begin{figure}[t]
\centering
  \includegraphics[width=0.8\textwidth]{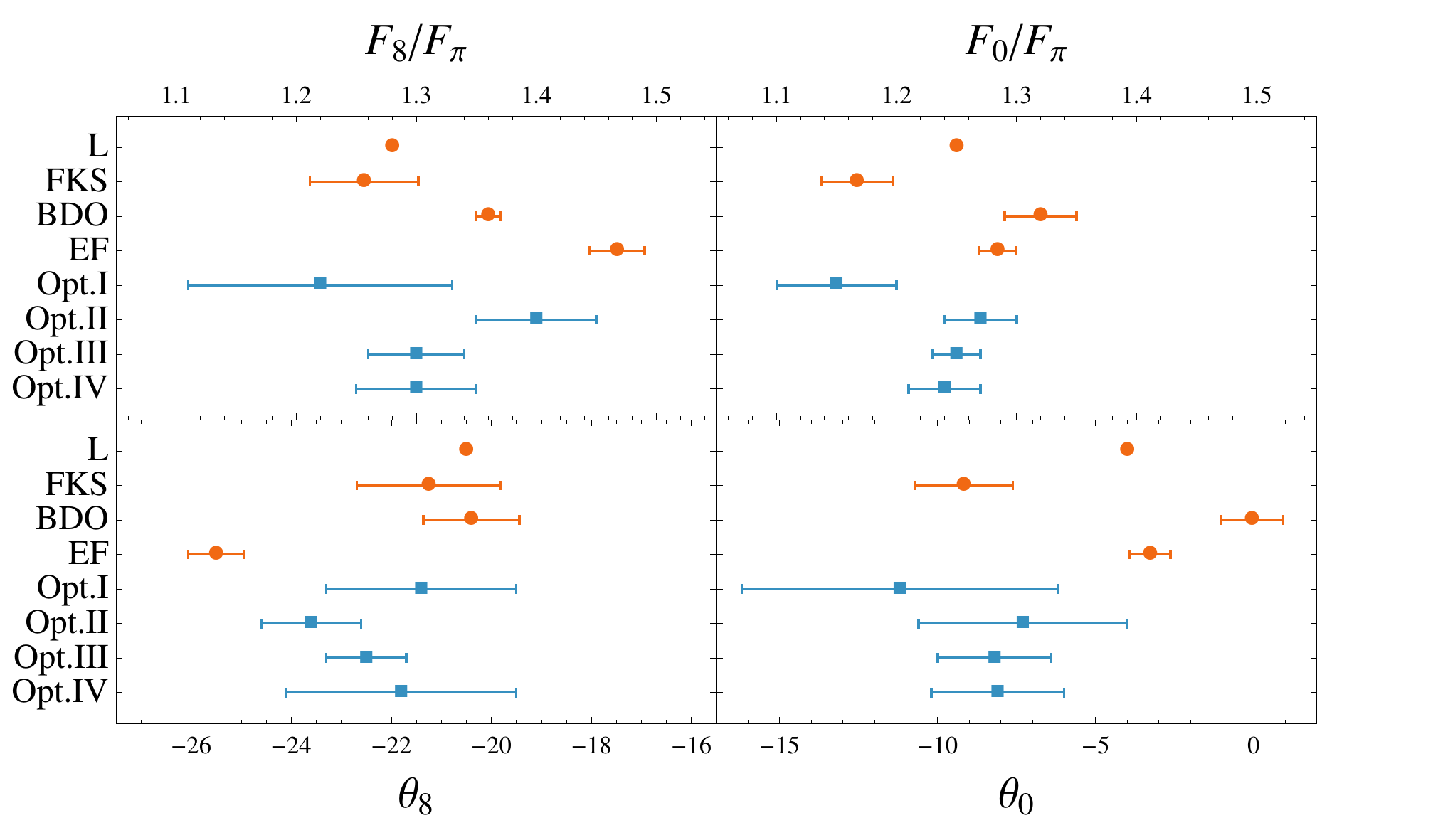}
  \includegraphics[width=0.8\textwidth]{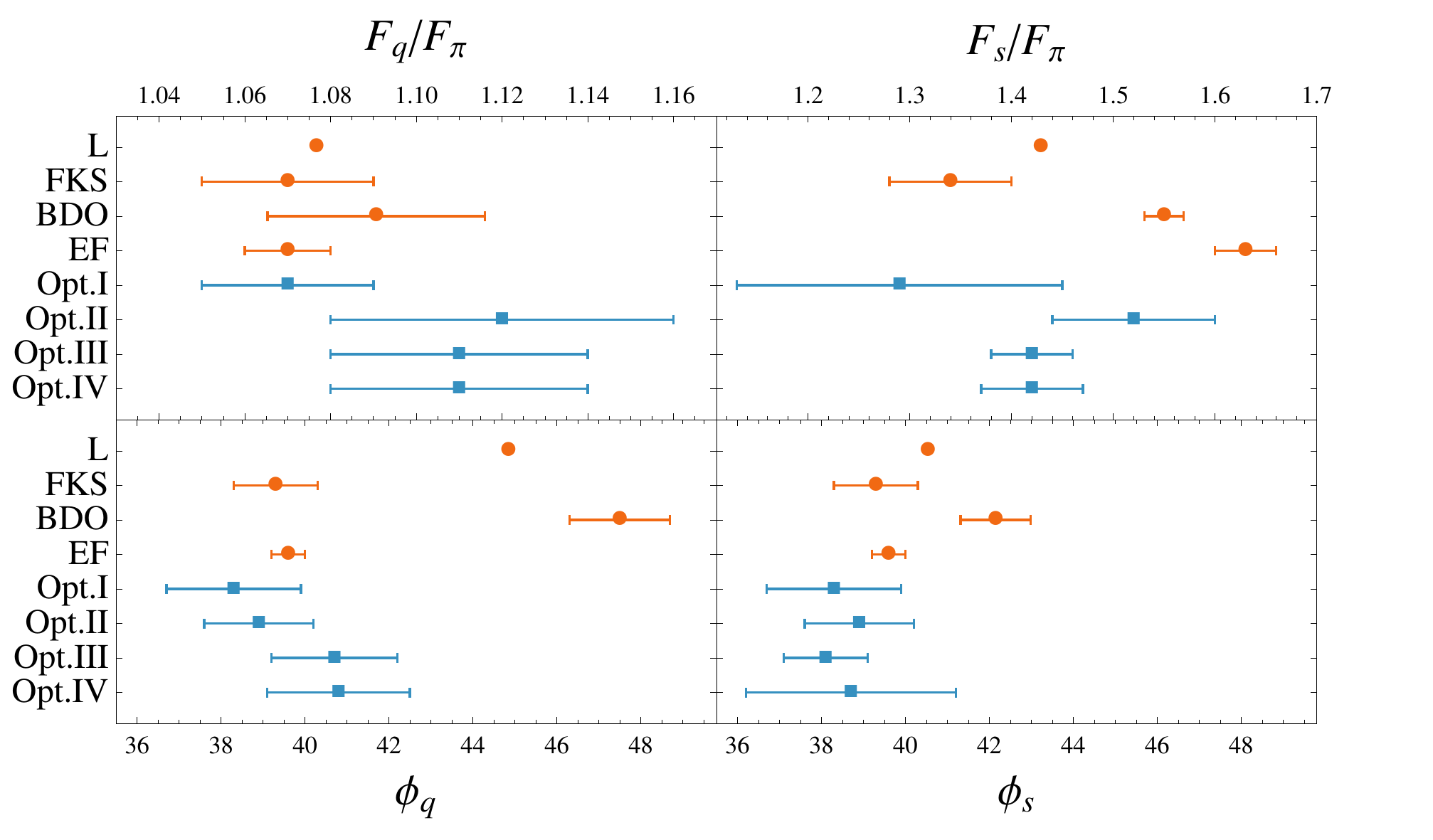}
  \caption{Our mixing parameters \cref{eq:qfopti,eq:qfoptii,eq:qfoptiii,eq:soopti,eq:sooptii,eq:sooptiii} (blue squares) compared to different theoretical results 
   (orange circles), see description in the text. The upper(lower) pannel displays our results in the octet-singlet(quark-flavor) basis. The references stand for  
   L~\cite{Leutwyler:1997yr}, FKS~\cite{Feldmann:1998vh}, BDO~\cite{Benayoun:1999au} EF~\cite{Escribano:2005qq}.}
  \label{Fig:MixRes}
\end{figure}

In order to solve our system, and focusing on the octet-singlet basis, we have at disposal four equations ---\cref{eq:Feta0,eq:Fetap0,eq:Infeta,eq:Infetap}--- and five unknowns 
---$F_8$, $F_0$, $\theta_8$, $\theta_0$ and $\Lambda_3$. In order to cure this situation, we can resort, as in the previous section, to the \cref{eq:deg}, which would provide 
the required constraint to fix $\Lambda_3$, but we still have to face the fact that our system is linear dependent. 
In order to overcome this problem, we notice that NLO \lcpt provides a clean prediction for both, $F_8$ and $F_8F_0\sin(\theta_8 - \theta_0)$ in terms of the well-known value 
for $F_K/F_{\pi}$~\cite{Agashe:2014kda}. Taking either of them as a constraint, one would add an additional equation to the previous system, 
which would provide a unique solution. Taking both, would lead to an overdetermined system, which in general has no solution. For this reason, we adopt a democratic 
procedure\cite{Escribano:2015yup} in which we perform a fit including both $F_8$ and $F_8F_0\sin(\theta_8 - \theta_0)$ 
constraints\footnote{We use preciser relations than those from \cref{sec:singletoctet}: $(F_8/F_{\pi})^2 = 1 +\frac{8}{3}\frac{F_K}{F_{\pi}}(\frac{F_K}{F_{\pi}}-1)$,  
            and $F_8F_0\sin(\theta_8-\theta_0)=-\frac{\sqrt{2}}{3}F_{\pi}^2(\frac{F_K}{F_{\pi}}-1)(4\frac{F_K}{F_{\pi}}+(\Lambda_1\to0))$.  
                   See \cite{Bickert:2015cia,Escribano:2005qq}.}
together with \cref{eq:Feta0,eq:Fetap0,eq:Infeta,eq:Infetap,eq:deg}. 
In addition, we ascribe a $3\%$ theoretical uncertainty for the \lcpt predictions by noticing that $F_K /F_{\pi}$ typically receives $3\%$ corrections from the NNLO\footnote{To see this, consider $F_K/F_{\pi} = 1.198 \simeq 1 +\epsilon+\epsilon^2$. This leads to the estimate for the NNLO correction $\epsilon^2=0.03$. Explicit results in Ref.~\cite{Bijnens:2014lea} leads to similar values too.}. 
Consequently, we add this error in quadrature on top of the one from~\cite{Agashe:2014kda} for our fitting procedure. As in the previous section, we take the inputs 
in \cref{tab:chap1mainres}. We obtain a fit with $\chi^2_{\nu}=0.35$ and the following results for the mixing parameters~\cite{Escribano:2015yup}
\begin{align}
  & \frac{F_8}{F_{\pi}} = 1.30(4),  \ \   \frac{F_{0}}{F_{\pi}} = 1.25(2),     \ \     \theta_8=-22.5(0.8)^{\circ}, \ \   \theta_{0}=-8.2(1.8)^{\circ}, \label{eq:qfoptiii} \\
  & \frac{F_q}{F_{\pi}} = 1.11(3), \quad \frac{F_s}{F_{\pi}} = 1.42(4), \ \ \phi_q=40.7(1.5)^{\circ}, \quad \phi_s=38.1(1.0)^{\circ}. \label{eq:sooptiii} 
\end{align}
These results are labelled as Option III in \cref{Fig:MixRes}, where the impact of including $\Lambda_1\neq0$ can be appreciated. 
In addition, we obtain for the OZI-violating parameters and the state-mixing angle
\begin{equation}
\label{eq:OZIparsiii}
  \Lambda_3=0.05(3), \quad \Lambda_1=0.20(4), \quad   \theta_P =  -15.4(1.0)^{\circ}.
\end{equation}  
Here, $\Lambda_1$ and $\theta_P$ are not directly  fitted parameters, but can be obtained by means of \cref{eq:F8F0}\footnote{Again, we use a preciser relation 
$(F_0/F_{\pi})^2 = (1+\frac{\Lambda_1}{2})^2+ \frac{4}{3}(\frac{F_K}{F_{\pi}}-1)(\frac{F_K}{F_{\pi}}+\frac{\Lambda_1}{2})$ \cite{Bickert:2015cia}.} and \cref{eq:angles}, 
respectively.

\subsection{The $\eta-\eta'$ mixing: $K_2, \Lambda_1,\Lambda_3 \! \neq \! 0$}
\label{sec:MixIV}

Finally, to quantify the impact of a non-zero $K_2$ parameter and to have a fully consistent description at NLO in \lcpt, we include the former in the last step. 
To do so, and given the poor extraction from $\pi^0\to\gamma\gamma$ decays $(K_2=-0.45(58))$, we incorporate this together with the experimental value 
for $F_{\pi\gamma\gamma}$ in our fitting procedure. We obtain a fit with $\chi^2_{\nu}=0.54$ and the following values for the mixing parameters
\begin{align}
  & \frac{F_8}{F_{\pi}} = 1.30(5),  \ \   \frac{F_{0}}{F_{\pi}} = 1.24(3),     \ \     \theta_8=-21.8(2.3)^{\circ}, \ \   \theta_{0}=-8.1(2.1)^{\circ}, \label{eq:qfoptiv} \\
  & \frac{F_q}{F_{\pi}} = 1.11(3), \quad \frac{F_s}{F_{\pi}} = 1.42(5), \ \ \phi_q=40.8(1.7)^{\circ}, \quad \phi_s=38.7(2.5)^{\circ}. \label{eq:sooptiv} 
\end{align}
In addition, we find
\begin{equation}
\label{eq:OZIparsiv}
  \Lambda_3=0.06(3), \  \ K_2 = -0.15(50), \  \ \Lambda_1=0.19(6), \  \   \theta_P =  -14.9(1.9)^{\circ}.
\end{equation}  
The results for the mixing parameters, \cref{eq:qfoptiv,eq:sooptiv,eq:OZIparsiv}, represent the main result from this chapter. 
We remind that we have used in our procedure a renormalization scale $\mu_0=1~\textrm{GeV}$. Consequently, our values should be understood at such scale. This applies to 
the OZI-violating parameters $\Lambda_{1,3}$ and, in the octet-singlet basis, to the singlet decay constants $F_P^0$. Whereas this may be adequate for pQCD studies such as those 
in Refs.~\cite{Agaev:2014wna,Alte:2015dpo},  the \lcpt practitioner may find more helpful the scale-independent $\Lambda_1-2\Lambda_3= 0.07(6)$ quantity.
Our predictions can be compared in \cref{Fig:MixRes}, Option IV, to our previous simplified approaches in order to appreciate the relevance of each parameter and to existing phenomenological determinations. 
Such determinations do not offer in general the values for the OZI-violating parameters, which are assumed to be zero. The exceptions are Ref.~\cite{Benayoun:1999au}, 
$\Lambda_3=-0.03(2)$, $\Lambda_1=0.20(4)$ and Ref.~\cite{Leutwyler:1997yr},  $\Lambda_1-2\Lambda_3=0.25$. \\

In summary, we have performed a new determination for the $\eta-\eta'$ mixing parameters purely based on \lcpt Lagrangian quantities
---to this day, the only consistent framework  to describe the $\eta-\eta'$ system. Our approach fully incorporates the required OZI-violating 
parameters (necessary to render scale-independent results) as well as the $K_2$ LEC, which are neglected in most of the previous phenomenological 
approaches~\cite{Feldmann:1998vh,Feldmann:1998sh,Feldmann:1999uf,Bramon:2000fr,Escribano:2005qq,Escribano:2007cd}. In addition, our approach 
does not rely on a phenomenological model involving further assumptions, as required for instance when using $V\to P\gamma$ transitions 
---find further details in \cref{sec:gvpg}. We note in this respect that previous approaches following the FKS scheme 
should have used $F_q=F_{\pi}$ to be consistent. Finally, we emphasize that our approach makes use of 4 independent quantities alone 
to determine the mixing parameters. This contrasts with previous approaches requiring a larger amount of input in their fits and often with a 
large $\chi^2_{\nu}$ value~\cite{Escribano:2005qq}.

\section{Applications}
\label{sec:appl}

The extraction of the mixing parameters provides an important input to understand the structure of the $\eta-\eta'$, which is still a matter of debate and 
research nowadays due to its complexity ---for the most recent studies, see~\cite{Guo:2015xva,Bickert:2015cia}. However, its interest lies beyond 
unravelling the structure of these pseudoscalars, as these parameters enter in a large variety of phenomenological applications. See for instance those in 
Refs.~\cite{Gilman:1987ax,Feldmann:1999uf,Thomas:2007uy,DiDonato:2011kr,Agaev:2014wna,Alte:2015dpo}, involving processes at low energies, such as 
$p\overline{p}\to \pi^0\eta^{(\prime)}$, mid-energies, such as $B^0\to J/\Psi\eta^{(\prime)}$, or as energetic as $Z\to \eta^{(\prime)}\gamma$ decays.
Consequently, our parameter extraction could be further tested using these processes. We do not pursue here such an ambitious programme, but merely 
describe two selected applications, namely, $V\to P\gamma$ and $P\to V\gamma$ transitions where $P=\eta^{(\prime)}$ and $V=\rho,\omega,\phi$, 
as well as $J/\Psi\to\eta^{(\prime)}\gamma$ decays.

\subsection{Determining the $g_{VP\gamma}$ couplings}
\label{sec:gvpg}

As a first application, we provide 
in this section the $g_{VP\gamma}$ couplings\footnote{The coupling is defined as $\bra{P}J_{\mu}^{EM}\ket{V_{\nu}}|_{(p_{P}-p_V)^2=0} = 
-g_{VP\gamma}\epsilon_{\mu\nu\rho\sigma}p_{P}^{\rho}p_{V}^{\sigma}$~\cite{Ball:1995zv}.} describing the interaction 
of the lowest-lying nonet of vector mesons with the pseudoscalar mesons and a photon. 
As such, they describe $\rho,\omega,\phi\to\eta\gamma$, $\eta'\to \rho(\omega)\gamma$ and $\phi\to\eta'\gamma$ decays, from 
which they can be experimentally extracted.
Alternatively, these parameters can be theoretically related to
the QCD-anomalous Green function $\bra{P} T\left\{J_{\mu}^{EM}(x), J_{\nu}^{a}(0)\right\} \ket{0}$ which, for vanishing virtualities, is given in terms of the 
triangle anomaly. The $g_{VP\gamma}$ couplings appear then when a dispersive representation saturated with the lowest-lying vector 
resonances is adopted~\cite{Ball:1995zv,Escribano:2005qq,Feldmann:1999uf}. The resulting expressions are given in \cref{sec:AppGvp}, which include the OZI violating parameter 
$\Lambda_3$ as appearing in Ref.~\cite{Feldmann:1999uf} and $K_2$ as an additional novelty. 
Our results found for the $g_{VP\gamma}$ couplings are displayed in Tab.~\ref{tab:gVP} together with the experimental values; the  
different outcomes for the methods employed in \cref{sec:MixI,sec:MixII,sec:MixIII,sec:MixIV} are labelled as Option~I, II, III and IV, respectively.
Though the agreement is not excellent, it has to be taken into account that higher resonances and continuum has been neglected in the employed dispersive representation, which implies 
non-negligible modeling associated errors, to some extent common both to the $\eta$ and $\eta'$~\cite{Ball:1995zv}. 
Therefore, it may be more adequate to take the ratio $g_{V\eta\gamma}/g_{V\eta'\gamma}$ instead~\cite{Feldmann:1999uf}, which is displayed in \cref{tab:gVP} as well. 
Actually, the agreement among our predictions and the experiment in these ratios is excellent for the $\rho$ and $\omega$ cases and reasonable for the $\phi$. 
The predictive power for these decays, which are used as inputs in traditional approaches instead, should be considered as an advantage from our approach.
\begin{table}
\footnotesize
\centering
\begin{tabular}{cccccc} \toprule
& Option I & Option II & Option III & Option IV & Experiment\\ \midrule 
$g_{\rho \eta \gamma}$ & $1.50(4)$  & $1.45(2)$  & $1.48(3)$ &  $1.47(5)$ & $1.58(5) $ \\
$g_{\rho \eta^\prime \gamma}$ & $1.18(5)$ & $1.22(3)$ & $1.21(3)$ &   $1.23(8)$ & $1.32(3) $ \\
$g_{\omega \eta \gamma}$ & $0.57(2)$ & $0.56(1)$ & $0.57(1)$ &   $0.56(2)$ & $0.45(2) $ \\
$g_{\omega \eta^\prime \gamma}$ & $0.55(2)$ & $0.56(1)$ & $0.56(1)$ &   $0.56(4)$ & $0.43(2) $ \\
$g_{\phi \eta \gamma}$ & $-0.83(11)$ & $-0.70(4)$ & $-0.78(5)$ &   $-0.72(5)$ & $-0.69(1) $ \\
$g_{\phi \eta^\prime \gamma}$ & $0.98(14)$ & $0.86(7)$ & $0.89(4)$ &   $0.84(5)$ & $0.72(1) $ \\ \midrule
$g_{\rho \eta \gamma}/g_{\rho \eta' \gamma}$ & $1.27(8)$  & $1.22(3)$  & $1.22(4)$ &   $1.19(12)$ & $1.20(5) $ \\
$g_{\omega \eta \gamma}/g_{\omega \eta' \gamma}$ & $1.04(4)$  & $1.00(2)$  & $1.02(2)$ &   $1.00(10)$ & $1.05(7) $ \\
$g_{\phi \eta \gamma}/g_{\phi \eta' \gamma}$ & $-0.85(6)$  & $-0.81(8)$  & $-0.87(7)$ &   $-0.87(7)$ & $-0.96(4) $ \\ \midrule
$R_{J/\psi}$ & $4.74(55)$ & $4.94(46)$ & $5.57(64)$ & $5.66(69)$ & $4.67(20)$ \\ 
\bottomrule
\end{tabular}
\caption{Summary of $g_{VP\gamma}$ couplings together with $R_{J/\Psi}$, see description in the text. 
         Experimental determinations are from Ref.~\cite{Agashe:2014kda}.}
\label{tab:gVP}
\end{table}

\subsection{Charmonium decays: $R_{J/\psi}$}
\label{sec:jpsirad}

It has been argued in Refs.~\cite{Thomas:2007uy,DiDonato:2011kr} that the $\eta-\eta'$ mixing parameters could be used as well to calculate decays in the charmonium region. 
Note that all these processes need to change flavor, which ---neglecting electromagnetic effects--- necessarily happens through OZI violating mechanisms, where the singlet 
sector plays a central role. 
Specially popular, and widely used in phenomenological analyses~\cite{Ball:1995zv,Feldmann:1998vh,Escribano:2005qq} are the $J/\Psi\to\eta^{(\prime)}\gamma$ decays, in 
particular its ratio $R_{J/\Psi}$ defined in \cref{eq:Rjpsi1} below. It is thought that the dominant mechanism underlying these decays is given by an intermediate two gluon 
state as depicted in \cref{fig:jpsi} (see Ref.~\cite{Novikov:1979uy}) which allows to express the ratio as
\begin{equation}
\label{eq:Rjpsi1}
R_{J/\psi} = \frac{BR(J/\psi\rightarrow\eta'\gamma)}{BR(J/\psi\rightarrow\eta\gamma)} =  \left\vert \frac{  \bra{\eta'}  G^{\mu\nu,c}\tilde{G}_{\mu\nu}^c \ket{0}  }{ \bra{\eta}  G^{\mu\nu,c}\tilde{G}_{\mu\nu}^c \ket{0} }  \right\vert^2 
\left( \frac{m_{J/\psi}^2 - m_{\eta'}^2}{m_{J/\psi}^2 - m_{\eta}^2} \right)^3, 
\end{equation}
where the first factor is the matrix element required from the process as outlined in Ref.~\cite{Ball:1995zv} and the second factor is pure phase space. Note that a factorization formalism 
is implicit, assuming as well that everything else but the above matrix elements cancels out in the ratio. 

Remarkably, even though \lcpt does not incorporate gluons as explicit degrees of freedom, it allows to calculate 
Green's functions involving them. This possibility is brought by Ward identities, which in this case via \cref{eq:anom} relate the purely gluonic current in \cref{eq:Rjpsi1} 
to quark currents and their divergencies. Particularly, for each individual flavor $q=u,d,s,...$, \cref{eq:anom}, reads
\begin{equation}
\partial_{\mu}(\overline{q}\gamma^{\mu}\gamma_5 q) = 2m_q \overline{q}i\gamma_5q - \frac{g_s^2}{32\pi^2}\epsilon^{\alpha\beta\mu\nu}G^c_{\alpha\beta}G^c_{\mu\nu}
\equiv2m_q \overline{q}i\gamma_5q + \omega.
\end{equation}
As an interesting academic exercise, we can further explore this relation, which under certain simplifying assumptions, allows to calculate the required 
$\bra{P} G^{\mu\nu,c}\tilde{G}_{\mu\nu}^c \ket{0}$ matrix elements in terms of the mixing parameters~\cite{Ball:1995zv,Feldmann:1998vh,Escribano:2005qq}. 
To show this, note that the divergence of the singlet axial current\footnote{The singlet axial current 
reads $J_{5\mu}^0 = (1/\sqrt{6})\left( \bar{u}\gamma_{\mu}\gamma_5u +\bar{d}\gamma_{\mu}\gamma_5d + \bar{s}\gamma_{\mu}\gamma_5s \right)$; for completness, 
$J_{5\mu}^8 = (1/2\sqrt{3})\left( \bar{u}\gamma_{\mu}\gamma_5u +\bar{d}\gamma_{\mu}\gamma_5d - 2\bar{s}\gamma_{\mu}\gamma_5s \right)$.} in the limit in which 
$m_{u,d}\to0$ reads
\begin{equation}
-\frac{3\alpha_s}{4\pi}G^{\mu\nu,c}\tilde{G}_{\mu\nu}^c \overset{m_{u,d}\to0}{=} \sqrt{3} \left(\sqrt{2} \partial^{\mu}J_{5\mu}^{0}  -(2/\sqrt{3})m_s\bar{s}i\gamma_5s   \right).
\end{equation}
Fortunately, for $m_{u,d}\to0$, the pseudoscalar strange quark current appearing above can be connected to the divergence of the octet axial current which, 
{\textit{in such limit}}, reads $\partial^{\mu}J_{5\mu}^8 = -(2/\sqrt{3})m_s\bar{s}i\gamma_5s$. As a consequence, the following expression has been obtained 
in the literature~\cite{Ball:1995zv,Feldmann:1998vh,Escribano:2005qq}
%
%
%
\begin{equation}
\label{eq:wiJPsi}
-\frac{3\alpha_s}{4\pi}G^{\mu\nu,c}\tilde{G}_{\mu\nu}^c \overset{m_{u,d} \ll m_s}{\simeq} \sqrt{3} \left(  \sqrt{2} \partial^{\mu}J_{5\mu}^{0}  + \partial^{\mu}J_{5\mu}^{8} \right)  
\end{equation}
which holds up to light quark mass corrections or, equivalently, $m_{\pi}^2/m_K^2$ effects~\cite{Feldmann:1999uf}.
\begin{figure}
\centering
   \includegraphics[width=0.5\textwidth]{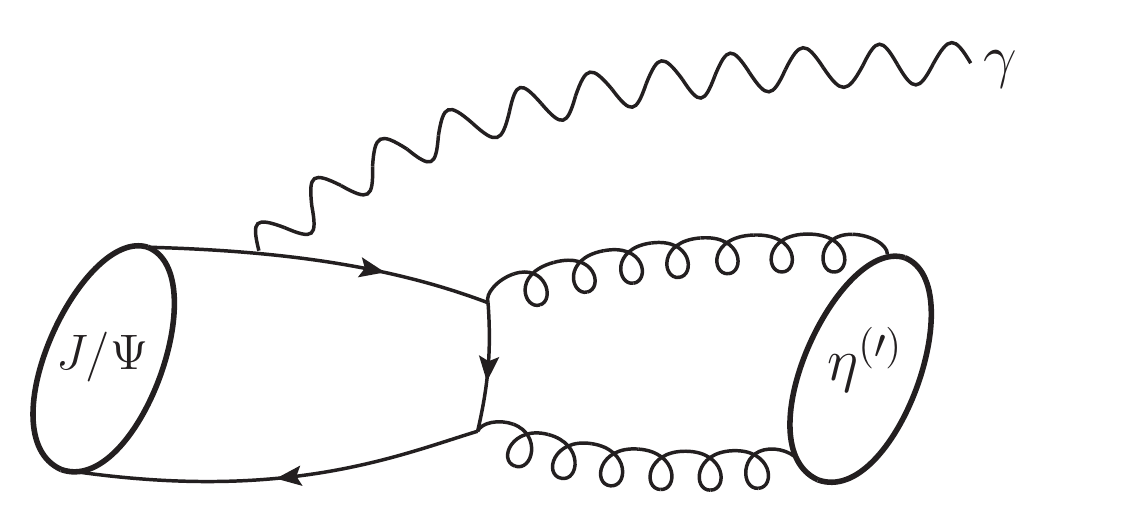}
   \caption{Expected main contribution to $J/\Psi\to\gamma \eta^{(\prime)}$ processes. \label{fig:jpsi}}
\end{figure}
\noindent
The relation above allows to express the $R_{J/\Psi}$ ratio in terms of the axial currents matrix elements as defined in \cref{eq:FP80,eq:defaxialSO}:
\begin{equation}
\label{eq:Rjpsi}
R_{J/\psi}  \simeq \left| \frac{m_{\eta'}^2(F_{8}\sin\theta_8 +\sqrt{2}F_0\cos\theta_0)}{m_{\eta}^2(F_8\cos\theta_8 -\sqrt{2}F_0\sin\theta_0)} \right|^2 
                  \left( \frac{m_{J/\psi}^2 - m_{\eta'}^2}{m_{J/\psi}^2 - m_{\eta}^2} \right)^3 
\end{equation}
%
%
where $\simeq$ stands for $m_{u,d}\neq0$ effects, which will be estimated below. 
To check what is expected in different regimes of the theory as well as the accuracy of the approximation in \cref{eq:wiJPsi}, we take the LO results in \lcpt.  
For the $\hat{m}\to0$ case (i.e. $m_{\pi}^2\to0$) the equality in \cref{eq:wiJPsi} holds exactly, and the gluonic matrix elements read, at LO, 
\begin{equation}
\label{eq:LOapprx}
\sqrt{3}m_P^2(F_P^{8} - \sqrt{2}F_P^0)   \stackrel{\hat{m}\to0}{=}  \bra{P} \omega \ket{0} =   \sqrt{6}F_P^0M_{\tau}^2,
\end{equation}
whereas the ratio itself reads, again at LO, 
\begin{equation}
\label{eq:rationLO}
\left\vert \frac{\bra{\eta'} \omega \ket{0} }{\bra{\eta} \omega \ket{0}}\right\vert^2 = \left\vert \frac{\cos\theta_P}{-\sin\theta_P}\right\vert^2.
\end{equation}
For the special case where $m_s$ effects are negligible as compared to the topological ones, this is, $m_K^2 \ll M_{\tau}^2$, the $\eta$ and $\eta'$ would 
become purely octet and singlet, respectively, with masses and mixing angle at LO
\begin{equation}
m_{\eta}^2 = \frac{4}{3}m_K^2 \left(1-\frac{2}{3}\epsilon\right), \quad m_{\eta'}^2 = M_{\tau}^2\left(1+\frac{2}{3}\epsilon\right), \quad \theta_P =  - \frac{2\sqrt{2}}{3}\epsilon,
\end{equation}
with $\epsilon = m_K^2/M_{\tau}^2$. As a consequence, the $\eta$ would not receive a singlet admixture and would not couple to the gluons, with the ratio  
in \cref{eq:rationLO} diverging as $|-3/(2\sqrt{2})\epsilon^{-1}|^2$. 

An opposite scenario would be that in which the large-$N_c$ limit represents an excellent approximation, 
whereby $M_{\tau}^2\to0$ and $FM_{\tau}^2 = 6\tau/F \sim 1/\sqrt{N_c}\to0$, but $m_s>0$ (i.e. $M_{\tau}^2\ll m_K^2$). 
In such a case, the $\eta$ would be become a massive $\eta_s$ meson, 
whereas the $\eta'$ would become a massless $\eta_q$, with masses and mixing angle at LO
\begin{equation}
m_{\eta}^2 = 2m_K^2(1+\frac{1}{6}\tilde{\epsilon}), \ \ m_{\eta'}^2 = \frac{2}{3}M_{\tau}^2(1-\frac{1}{6}\tilde{\epsilon}), \ \ \theta_P = \pi - \theta_{ideal} + \frac{1}{3\sqrt{2}}\tilde{\epsilon},
\end{equation}
with now $\tilde{\epsilon}=M_{\tau}^2/m_K^2$. In this case, both matrix elements would vanish as $M_{\tau}^2\to0$, but its ratio in \cref{eq:rationLO} would be kept 
fixed at $|-\sqrt{2}(1-(1/2)\tilde{\epsilon})|^2$, with 2 its limiting value. Consequently, as far as LO results are concerned, a result $R_{J/\Psi}>2$ would directly point 
towards $M_{\tau}>m_s$. 

Finally, but still at LO, we discuss the accuracy of the approximation of neglecting the light quark masses in \cref{eq:wiJPsi}. From the LO results in 
Ref.~\cite{Bickert:2015cia}, for which $M_{\tau}=0.82$~GeV and $\theta_P=-19.6^{\circ}$, we obtain for the left hand side of \cref{eq:LOapprx} $0.60$ and $1.58$ for the $\eta$ and $\eta'$. 
For the right hand side, the results read $0.55$ and $1.55$, respectively. As a consequence, we obtain that the equality \cref{eq:wiJPsi} holds at around $5\%$ 
precision for the matrix elements, implying a $10\%$ systematic uncertainty for the $R_{J/\Psi}$ result.

After this discussion, we proceed to our determination. From \cref{eq:Rjpsi} and our mixing parameters determination from 
\cref{sec:MixI,sec:MixII,sec:MixIII,sec:MixIV}, we obtain the results quoted in the last row from 
\cref{tab:gVP}. We find a difference of $1.4\sigma$  among our final result for the mixing parameters prediction (Option IV) and experiment. 
Yet this is not large, it would be interesting to have a preciser theoretical and experimental prediction, as this process could be sensitive to 
non-standard phenomena such as gluonium admixtures or $c\bar{c}$ content in the $\eta'$. However, to confirm such eventual discrepancy may require a 
more detailed analysis, including the light-quark mass effects neglected above, that could be around a $10\%$ effect and would involve additional 
$\Lambda_i$ OZI-violating parameters. In addition, it would be interesting to retain additional OZI-suppressed contributions to the hard process non considered 
in \cref{fig:jpsi} and which may be non-negligible in the light of $\psi(2S)$ decays ---see discussions in~\cite{Gerard:2013gya}. 
Finally, it has to be mentioned that previous analysis did not include the RG effects which would appear in such process, necessary 
to render the amplitude scale-independent. A similar argument to that above \cref{eq:runinf} would 
imply $(1-\delta_{\textrm{RG}}(m_{J/\Psi}^2))=(1-0.05)$. Re-evaluating then \cref{eq:Rjpsi} including such factor in the $F_0$ terms, 
we obtain $R_{J/\Psi}=4.99(61)$, a non-negligible effect that shifts our value closer to the experimental one, and suggests the relevance of a more refined analysis.

\subsection{Light- and strange-quark transition form factors}
\label{sec:lqtff}

As explained in Sec.~\ref{sec:qfbasis}, under the assumption that large-$N_c$ OZI-violating effects are negligible, the $\eta$ and $\eta'$ Fock states may be described 
through the use of a single angle in terms of the light and strange quarks wave functions $\Psi_q, \Psi_s$, common to the $\eta$ and $\eta'$. These define the meson distribution 
amplitudes $\phi_{\eta}^q=\phi_{\eta'}^q\equiv\phi_q$ and $\phi_{\eta}^s=\phi_{\eta'}^s\equiv\phi_s$, \cref{eq:DAdef}, which are used to calculate the $\eta$ and $\eta'$ TFFs. 
Such distribution amplitudes can be used to obtain the unphysical ---i.e., non measurable--- light- and strange-quark TFF, $F_{q\gamma^*\gamma}(Q^2)$ and $F_{s\gamma^*\gamma}(Q^2)$, respectively, in terms of which the physical $\eta$ and $\eta'$ TFF can be expressed as
\begin{align}
 F_{\eta\gamma^*\gamma}(Q^2) = & \  \cos\phi F_{q\gamma^*\gamma}(Q^2) -  \sin\phi F_{s\gamma^*\gamma}(Q^2) , \\
  F_{\eta'\gamma^*\gamma}(Q^2) =  & \ \sin\phi F_{q\gamma^*\gamma}(Q^2) +  \cos\phi F_{s\gamma^*\gamma}(Q^2) . 
\end{align}
The light- and strange-quark TFFs are related to the physical ones via rotation
\begin{align}
 F_{q\gamma^*\gamma}(Q^2) = & \  \cos\phi F_{\eta\gamma^*\gamma}(Q^2) +  \sin\phi F_{\eta'\gamma^*\gamma}(Q^2) , \label{eq:Fq}\\
  F_{s\gamma^*\gamma}(Q^2) =  & \ -\sin\phi F_{\eta\gamma^*\gamma}(Q^2) +  \cos\phi F_{\eta'\gamma^*\gamma}(Q^2) . \label{eq:Fs} 
\end{align}
Our mixing parameters extraction would allow to find such a decomposition, which represents an interesting theoretical result. 
In order to reconstruct them, we take our averaged result $\phi\equiv(\phi_q + \phi_s)/2 = 39.5(1.1)^{\circ}$ from the mixing angles 
obtained in \cref{sec:MixIII}\footnote{This should not be a bad approximation given our results in the previous section; we note however that $\Lambda_1\neq0$ 
implies that this is not a strict result but an approximate one.}
together with our fits from the TFFs in \cref{chap:data}.
\begin{figure}
\centering
   \includegraphics[width=0.495\textwidth]{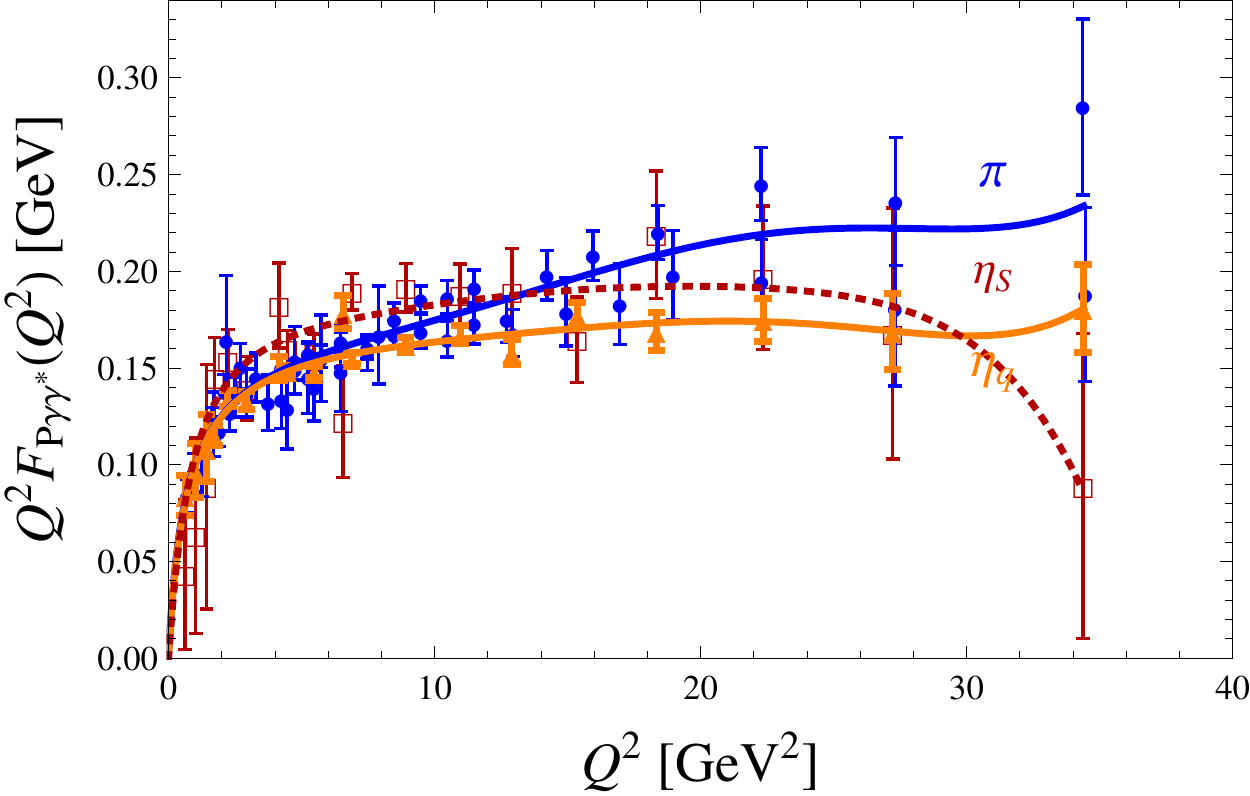}
   \includegraphics[width=0.495\textwidth]{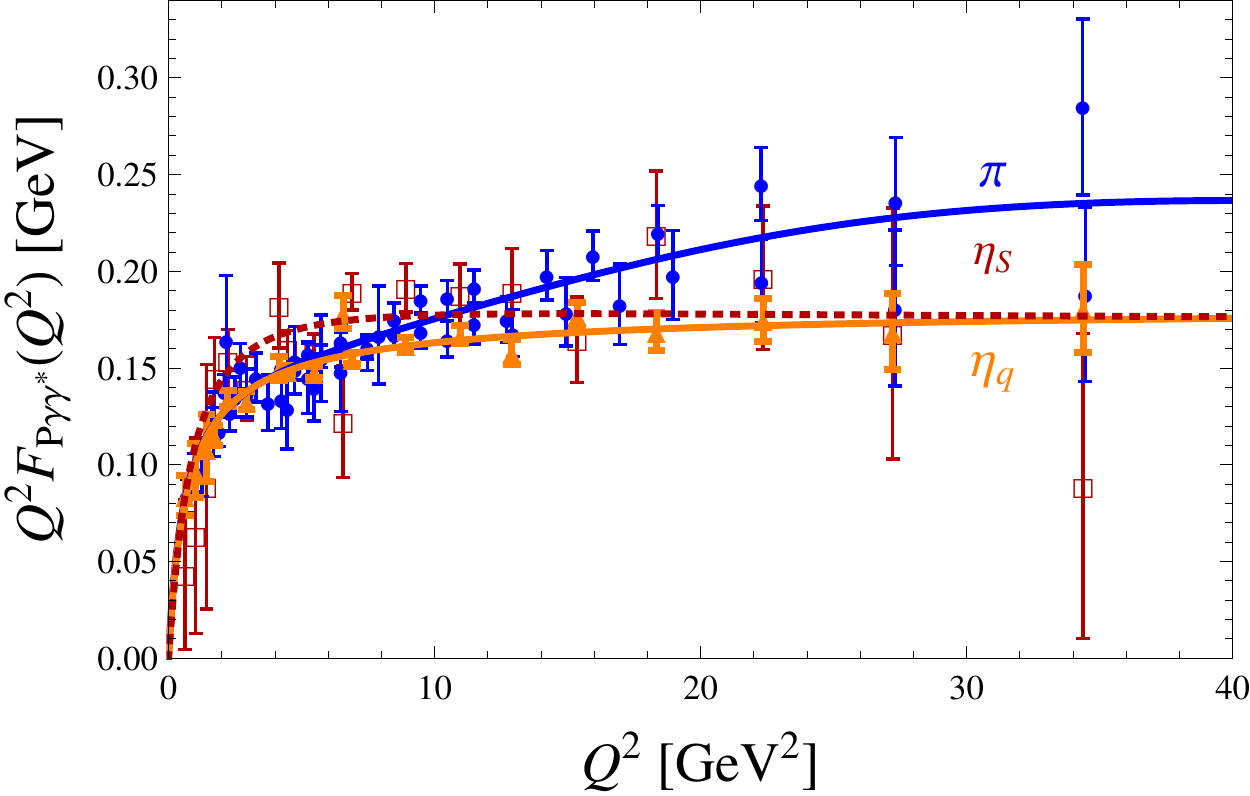}
   \caption{The light(strange)-quark TFF in orange(dotted-red) together with the $\pi^0$ TFF (blue). The left plot shows the $P^N_N$-based description, whereas the right one 
                  represents the $P^N_1$ one. The former TFFs have been multiplied by a charge factor $3/5$ and $3/\sqrt{2}$, 
                 respectively (see details in the text). When possible, the $\eta$ and $\eta'$ TFF data points have been combined to 
                 extract what would be the light- and strange-quark TFF data as orange triangles and open-red squares, respectively. The data for the $\pi^0$ appears as blue points. }
   \label{fig:qfTFF}
\end{figure}
In addition,  as a consequence of assuming a mild large-$N_c$ OZI violating effects, which seems a reasonable estimation according to our results, it is theoretically 
expected that the $\pi^0$ distribution amplitude $\phi_{\pi}$ should be the same as that from the light-quarks $\phi_q$. This is easy to understand as, in this limit, the $U(3)_F$ 
symmetry would be recovered, guaranteeing then the equality of all distribution amplitudes ---symmetry breaking effects should be accounted though for the strange quark, which 
does not represent a problem for the arguments above. Consequently, the resulting TFF should be, 
up to a charge factor 5/3, equivalent ($F_{q\gamma^*\gamma}(Q^2) = (5/3)F_{\pi\gamma^*\gamma}(Q^2)$). 
For this reason, we plot in Fig.~\ref{fig:qfTFF}, the results for the $\pi^0, \eta_q$ and $\eta_s$ TFF obtained from \cref{eq:Fq,eq:Fs} and normalized to the $\pi^0$ charge. 
This amounts to multiply the light- and strange-quark TFF by the charge factors $3/5$ and $3/\sqrt{2}$, respectively. We find that actually the light-quark and the $\pi^0$ TFFs 
match each other up to the $Q^2\sim 6$~GeV$^2$ scale, 
where the controversial Belle-\babar discrepancy manifests~\cite{Uehara:2012ag,Aubert:2009mc}. Provided $\phi_q\simeq\phi_{\pi}$, our approach supports Belle data against \babar and 
strongly calls for a new preciser measurement at Belle II. In addition, the results above 
show a behavior beyond the simplest VMD ($P^0_1$ approximant) approach and should warn therefore against oversimplified descriptions. 
Finally, we give the resulting (dimensionful, i.e. $m_P=1$ in \cref{eq:series}) slope for these TFFs
\begin{align}
 b_{\eta_q}= & \ 1.67(3)~\textrm{GeV}^{-2}  =(0.774(6)~\textrm{GeV})^{-2} , \\
 b_{\eta_s}= & \ 0.57(^{+0.17}_{-0.19})~\textrm{GeV}^{-2} =(1.43(^{+0.37}_{-0.20})~\textrm{GeV})^{-2},
\end{align}
which has been obtained from our values in \cref{tab:chap1mainres}. These could be compared with the results for the $\pi^0$, $\eta$, and $\eta'$ results from 
\cref{chap:data}\footnote{To obtain them, the results from \cref{tab:chap1mainres} should be multiplied by $m_P^{-2}$.},  
\begin{align}
b_{\pi} &=1.78(12)~\textrm{GeV}^{-2} =(0.750(26)~\textrm{GeV})^{-2},  \\
b_{\eta}&=1.916(39)~\textrm{GeV}^{-2}=(0.722(7)~\textrm{GeV})^{-2},  \\
b_{\eta'}&=1.42(3)~\textrm{GeV}^{-2}  =(0.874(13)~\textrm{GeV})^{-2},
\end{align}
which shows again the expected similarity among the $\pi^0$  and the light-quark quantities.

\section{Conclusions and outlook}
\label{sec:mixingconcl}

In this chapter, we have presented a new and alternative determination for the $\eta-\eta'$ mixing parameters using information on 
the TFFs exclusively. As an advantage, our formulation allows for a straightforward connection to the quantities arising in the \lcpt Lagrangian ---up to day, 
the only consistent framework to describe the $\eta-\eta'$ system--- and avoids thereby the use of models and approximations as those taken in studies using 
$V\to P\gamma$ and $P\to V\gamma$ processes or $J/\Psi$ decays. Moreover, besides implementing the full NLO \lcpt expressions including the relevant OZI-violating 
parameters,  we have been able to provide a determination for them. Even if we find small values for them, their role is not negligible and plays a crucial role 
in the TFFs asymptotic behavior ---the role of the LEC $K_2$ is by contrast negligible. 
We remark that including them is necessary to achieve formally  a consistent picture. This is a disadvantage from previous approaches, in which these parameters were 
kept finite for some quantities and vanishing in others. To illustrate their impact, we used a sequential approach in which the different 
OZI-violating effects and finally $K_2$ were sequentially included one by one. Remarkably, we achieve a competitive prediction 
with respect to existing approaches, that required a large amount of inputs in their fits and usually obtained a 
large $\chi^2_{\nu}$ value, highlighting possible model-dependencies. This put us in a perfect position to test the mixing-scheme in different observables. 

Possible venues to improve and extend our work would be a thorough and detailed calculation of the RG-equation for the singlet axial current, 
including higher orders. In addition, it would 
be interesting to see if ongoing studies of the $\eta-\eta'$ provide additional insights which may help in extracting the mixing 
parameters~\cite{Guo:2015xva,Bickert:2015cia,Bickert:2016fgy}. Lattice studies such as~\cite{Michael:2013vba} would help in this point as well ---note however that they obtain the pseudoscalar, 
rather than the axial current matrix element. A final point of interest would be the application of our results to the calculation of additional charmonium and weak decays 
in lines of Refs.~\cite{DiDonato:2011kr,Thomas:2007uy} with a proper account of OZI-violating effects.

\chapter{Pseudoscalar to lepton pair decays}
\label{chap:PLL}
\minitoc

\section{Introduction}

The psedusocalar decays into lepton pairs, \PtoLL, are a beautiful place to keep track of the evolution of our understanding of QCD, which is behind the mechanism driving these 
processes. Its pioneering study was initiated by Drell~\cite{Drell:1959} back in 1959, well before the time where the pseudoscalar decays into photons were properly understood on basis 
of the Adler~\cite{Adler:1969gk}-Bell-Jackiw~\cite{Bell:1969ts} (ABJ) anomaly. Still, he was able to set a lower bound for the $\pi^0\rightarrow e^+e^-$ decay. 
Further studies (some of them rather qualitative) appeared in the 60's with the advent of VMD ideas~\cite{Berman:1960zz,Geffen:1965zz,Sehgal:1966wr,Young:1967zzb,Quigg:1968zz} 
which were just being develeoped at that time. 
Later on, the development of perturbative QCD stimulated different approaches in the 80's. Among them, quark loop models based on duality 
ideas~\cite{Ametller:1983ec,Bergstrom:1983ay,Scadron:1983nt,Pich:1983zk,Margolis:1992gf,deRafael:2011ga} and phenomenological 
models based on the novel understanding of exclusive reactions in pQCD~\cite{Babu:1982yz} ---which were improved through the use of 
data~\cite{Dorokhov:2007bd,Dorokhov:2008cd,Dorokhov:2009xs}. More recently, the development of \cpt, the low-energy effective field theory of QCD, 
provided an alternative approach to study these decays~\cite{Savage:1992ac,GomezDumm:1998gw}, which in addition may be complemented with large-$N_c$ 
and resonant ideas~\cite{Knecht:1999gb,Husek:2015wta}.
The motivation for this continuous study has been undoubtedly bound to the different experimental anomalies appearing in these processes along the years, 
stimulating a continuous revision and speculation about new-physics effects~\cite{Herczeg:1977gy,Soni:1974aw,Bergstrom:1982zq,Kahn:2007ru,Chang:2008np}.

In this chapter, we apply all the machinery developed for reconstructing the TFFs and benefit from our novel ideas, gaining on precision and obtaining, for the first time, a 
reliable systematic error estimation, taking special care of the $\eta$ and $\eta'$ cases. 
In this way, we want to update the status of these decays to the standards of precision met nowadays ---required for testing the low-energy frontier of 
the SM~\cite{Masjuan:2015lca,Masjuan:2015cjl}. The calculation details of these processes together with their relevant features are outlined in \cref{sec:calc}. 
The systematic error assessment is described in 
\cref{sec:CAsSys}, including a careful description of some particular features ---previously overlooked--- present for the $\eta$ and $\eta'$ but not for the $\pi^0$. 
Our results, discussed in \cref{sec:results}, show interesting features when compared to \cpt as we describe in \cref{sec:pllcpt}. 
Finally, we discuss new physics implications in \cref{sec:np}.

\section{The process: basic properties and concepts}
\label{sec:calc}

The leading order\footnote{An additional but subleading tree-level $Z^0$ boson electroweak contribution exists too, cf. \cref{sec:np}.} QED contribution to \PtoLL decays is 
mediated through an intermediate two-photon state as sketched in \cref{fig:ptoll}. 
\begin{figure}[t]
\centering
  \includegraphics[width=0.5\textwidth]{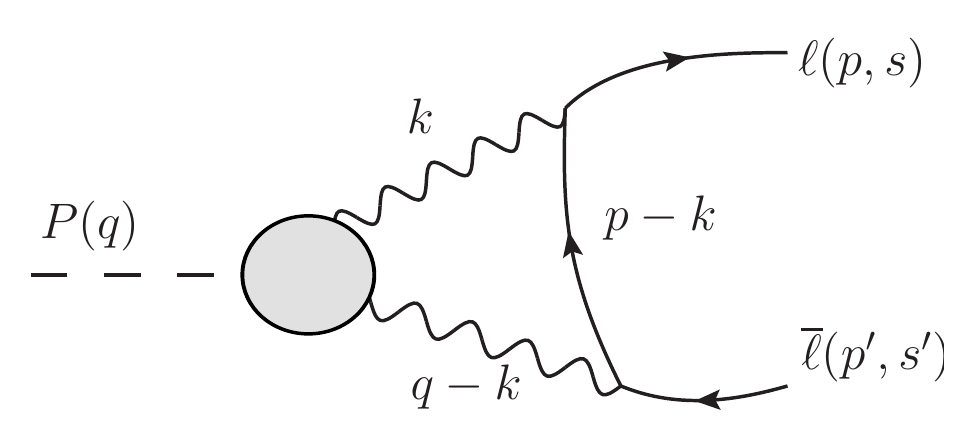}
  \caption{The leading order contribution to \PtoLL processes. The shadowed blob stands for the QCD dynamics in the $P\to\gamma^*\gamma^*$ transition encoded in 
                $F_{P\gamma^*\gamma^*}(k^2,(q-k)^2)$.\label{fig:ptoll}}
\end{figure}
The gray blob appearing there stands for the hadronic effects encoded in the $P\to\gamma^*\gamma^*$ transition. For real photons, such process is theoretically well
known in terms of the ABJ anomaly, and can be obtained as well in the odd-parity  sector of \cpt, see \cref{sec:tffcpt}. 
For deeply virtual photons, the limits $\lim_{Q^2\to\infty} F_{P\gamma^*\gamma}(Q^2)$~\cite{Brodsky:1981rp} and 
$\lim_{Q^2\to\infty} F_{P\gamma^*\gamma^*}(Q^2,Q^2)$~\cite{Novikov:1979uy} (see \cref{sec:tffpqcd}) are known as well. 
However, the interpolation in between these two regimes is a theoretically unknown territory, what has been amended through wise and different modeling procedures, 
explaining the large amount of studies on these processes. Parametrizing such interaction in terms of the most general 
TFF, $F_{P\gamma^*\gamma^*}(q_1^2,q_2^2)$, we obtain for the matrix element
\begin{align}
\label{eq:feynrul}
i\mathcal{M} = \int \frac{d^4k}{(2\pi)^4}& (-ie^2 F_{P\gamma^*\gamma^*}(k^2,(k-q)^2))\epsilon^{\mu\nu\rho\sigma}k_{\mu}(q-k)_{\rho}\frac{-ig_{\nu\nu'}}{k^2}\frac{-ig_{\sigma\sigma'}}{(q-k)^2} \nonumber \\
                                      & \times \overline{u}_{p,s}(-ie\gamma^{\nu'})i\frac{(\slashed{p}-\slashed{k})+m_{\ell}}{(p-k)^2-m_{\ell}^2}(-ie\gamma^{\sigma'})v_{p's'} \nonumber \\
    =\int \frac{d^4k}{(2\pi)^4}& e^4\epsilon^{\mu\nu\rho\sigma}k_{\mu}q_{\rho} \frac{[\overline{u}_{p,s} \gamma_{\nu} ((\slashed{p}-\slashed{k})+m_{\ell}) \gamma_{\sigma}v_{p's'}]}{k^2(q-k)^2((p-k)^2-m_{\ell}^2)}F_{P\gamma^*\gamma^*}(k^2,(k-q)^2),
\end{align}
where $k$ is the momentum running through the loop and must be integrated over all energies. The definitions for the different elements follow the 
conventions\footnote{Here it may worth to stress that in our convention $\epsilon^{0123}=+1$.} in~\cite{Peskin:1995ev} and can be found in \cref{app:conv}. 
At this stage of the calculation, it is convenient to evaluate the spinor contractions. This can be done using the pseudoscalar projector defined in Eq.~(A16) from 
Ref.~\cite{Martin:1970ai}. We recall it here adapted to our conventions ---which amounts to shift the antisymmetric tensor sign with respect to \cite{Martin:1970ai}--- for completeness, 
\begin{equation}
\label{eq:proj}
v_{p',s'}\overline{u}_{p,s}\vert_{\textrm{out},\mathcal{P}}  = \frac{1}{2\sqrt{2q^2}}\left[ -2m_{\ell}\slashed{q}\gamma_5 -i\epsilon_{\alpha\beta\gamma\delta}\gamma^{\alpha}\gamma^{\beta}p^{\gamma}p'^{\delta} + q^2\gamma_5 \right].
\end{equation} 
The subindex out means that such equality holds for the final state particles, while subindex $\mathcal{P}$ means that it is in a pseudoscalar state. Using standard trace 
techniques together with \cref{eq:proj}, we find that the spinorial part in square brackets from \cref{eq:feynrul} yields 
$-i(2\sqrt{2}m_{\ell}/m_P)\epsilon_{\nu\sigma\alpha\beta}q^{\alpha}k^{\beta}$. Inserting back into \cref{eq:feynrul} 
and using $\epsilon^{\nu\sigma\mu\rho}\epsilon_{\nu\sigma\alpha\beta}=-2(\delta^{\mu}_{\alpha}\delta^{\rho}_{\beta} - \delta^{\mu}_{\beta}\delta^{\rho}_{\alpha})$, we obtain the final result
\begin{equation}
\label{eq:amp1}
i\mathcal{M} = 2\sqrt{2}m_{\ell}m_P\alpha^2F_{P\gamma\gamma} \frac{2i}{\pi^2q^2}  \int d^4k \frac{[k^2q^2-(k\cdot q)^2] \tilde{F}_{P\gamma^*\gamma^*}(k^2,(k-q)^2)}{k^2(q-k)^2[(p-k)^2-m_{\ell}^2]},
\end{equation}
where the $\sqrt{2}m_P$ term can be traced back to the effective pseudoscalar $\overline{u}\gamma_5v$ interaction\footnote{To see this, note that, from \cref{eq:proj},  
$\textrm{Tr}(\overline{u}\gamma_5v) = \sqrt{2}m_P$. In addition, this allows to effectively express 
$i\mathcal{M} = -i(\bar{u}i\gamma_5v) 2m_{\ell}\alpha^2F_{P\gamma\gamma}\mathcal{A}(q^2)$, with $\mathcal{A}(q^2)$ defined in \cref{eq:loopamp}} and $m_{\ell}$ 
to the helicity flip. $F_{P\gamma\gamma}\equiv F_{P\gamma\gamma}(0,0)$ and so $\tilde{F}_{P\gamma^*\gamma^*}(k^2,(k-q)^2)$ is the normalized TFF, 
$\tilde{F}_{P\gamma^*\gamma^*}(0,0)=1$. The decay width reads then (see \cref{sec:crossdecay})
\begin{equation}
\label{eq:dwidth}
\Gamma(P\rightarrow\bar{\ell}\ell) = \frac{1}{16\pi m_P}\beta_{\ell} \vert \mathcal{M} \vert^2,
\end{equation}
with $\beta_{\ell} = \sqrt{1-4m_{\ell}^2/m_P^2}$ the lepton velocity. It is customary in the literature to express \cref{eq:dwidth} in terms of the 
$\Gamma(P\rightarrow\gamma\gamma)$\footnote{The two photon decay-width reads 
$\Gamma(P\rightarrow\gamma\gamma)=\frac{e^4m_P^3}{64\pi}\vert F_{P\gamma\gamma}\vert^2$.} result, so the normalization for the 
TFF dependency disappears, which is the reason that it was factored out in \cref{eq:amp1}. In such a way, the final result reads
\begin{equation}
\label{eq:br}
\frac{\textrm{BR}(P\rightarrow\bar{\ell}\ell)}{\textrm{BR}(P\rightarrow\gamma\gamma)} = 2\left( \frac{\alpha m_{\ell}}{\pi m_P} \right)^2 \beta_{\ell} \vert \mathcal{A}(q^2) \vert^2.
\end{equation}
The prefactor\footnote{ The prefactor in \cref{eq:br} is $\mathcal{O}(10^{-10})$ for the $\pi^0\rightarrow e^+e^-$, $\mathcal{O}(10^{-11}(10^{-7}))$ for the 
$\eta\rightarrow e^+e^-(\mu^+\mu^-)$ and $\mathcal{O}(10^{-12}(10^{-7}))$ for the $\eta'\rightarrow e^+e^-(\mu^+\mu^-)$.} in \cref{eq:br} already predicts tiny BRs 
for these processes, which are known as rare decays. 
This is due to the electromagnetic $\alpha^2$ and the helicity flip suppression $m_{\ell}^2/m_P^2$ factors with respect to the $P\rightarrow\gamma\gamma$ decay.
The last parameter, $\mathcal{A}(q^2)$, is related to the loop amplitude and encode the QCD dynamics encapsulated in the TFF,
\begin{equation}
\label{eq:loopamp}
\mathcal{A}(q^2) = \frac{2i}{\pi^2q^2} \int d^4k \frac{\left(k^2q^2-(k\cdot q)^2\right) \tilde{F}_{P\gamma^*\gamma^*}(k^2,(q-k)^2)}{k^2(q-k)^2\left((p-k)^2-m_{\ell}^2\right)}.
\end{equation}
The formulae in \cref{eq:br,eq:loopamp} represent the main standard results necessary to calculate the BRs. 
At this point, it may seem hopeless to say anything about \cref{eq:loopamp} without any information on the TFF, which is actually required  
to render the ---otherwise divergent--- loop integral finite. 
However, it is still possible to derive some important general results. Among them, the unitary bound obtained by 
Drell~\cite{Drell:1959}, the result for a constant TFF (of relevance for \cpt) and the relevant regimes in which a precise 
TFF determination is required. The latter is an essential prerequisite for any proper discussion on systematic errors and how to reconstruct the TFF.

\subsection{The unitary bound}
\label{sec:Cutcosky}

To derive the imaginary part associated to these processes, we use the Cutcosky rules, 
relating the imaginary part of the diagram to its discontinuities~\cite{Peskin:1995ev}. 
%
%
The latter are computed replacing the propagators which can be put on-shell as $\frac{1}{p^2-m^2+i\epsilon} \rightarrow -2\pi i \delta(p^2-m^2)\theta(p^0)$. 
For the $\pi^0$ ---being the lightest hadronic particle--- the only possible intermediate state appearing in the loop is the two photon 
one. Following Cutcosky and replacing the photon propagators in \cref{eq:loopamp}, 
one obtains\footnote{We use polar coordinates $dk^4 = d\Omega_3 \ dk^0 \frac{1}{2}\mathbf{k}d\mathbf{k}^2$ and specialize 
to the pseudoscalar rest frame, where $\vec{q}=(m_P,\vec{0})$ and $\vec{p} = m_P/2(1,\vec{\beta_{\ell}})$. 
To perform integration over $d\mathbf{k}^2$ we use $\delta(k^2-m_i^2) = \delta((k^0)^2-\mathbf{k}^2-m_i^2)$.}

\begin{align}
\operatorname{Im}\mathcal{A}_{\gamma\gamma} = \ & \frac{(-2\pi i)^2}{\pi^2q^2} \int d^4k \frac{(q^2k^2 - (q\cdot k)^2)  \tilde{F}_{P\gamma\gamma}(k^2,(q-k)^2) }{((p-k)^2-m^2)} \delta(k^2 )\delta((q-k)^2), \nonumber \\
                        = \ & \frac{-2}{m_P^2} \int  \ d\Omega_3 \ dk^0 \frac{m_P^2(k^0)^3  \tilde{F}_{P\gamma\gamma}(0,0) }{   m_Pk^0(1 - \beta_{\ell} \cos\theta))  }  \frac{1}{2m_P}\delta(k^0- \frac{m_P}{2}), \nonumber \\
                        = \ & \frac{\pi}{4} \int  \ d\Omega_3 \frac{1}{ \beta_{\ell} \cos\theta -1 } = \frac{\pi}{2\beta_{\ell}} \ln \left( \frac{1-\beta_{\ell}}{1+\beta_{\ell}}  \right). \label{eq:ImGG}
\end{align}
Remarkably, this observation allowed Drell~\cite{Drell:1959} to put already a lower bound in 1959, which is known as the unitary bound,
\begin{equation}
|\mathcal{A}(m_{\pi}^2)|^2 \geq (\operatorname{Im} \mathcal{A}_{\gamma\gamma}(m_{\pi}^2))^2 = \left( \frac{\pi}{2\beta_{\ell}} \ln \left( \frac{1-\beta_{\ell}}{1+\beta_{\ell}}  \right) \right)^2 = (-17.52)^2.
\end{equation}
Quite often, this bound has been extended to the heavier $\eta, \eta'$ and $K_L$ pseudoscalar states.
This generalization is however incorrect, as all of these particles will have intermediate $\pi^+\pi^-\gamma$ states in addition, cf. \cref{fig:UB}.
This is specially important for the $\eta'$, where such $\pi^+\pi^-$ state becomes resonant at the $\rho$ peak, 
besides the additional $\omega$ resonance. 
This feature is carefully illustrated for the $\eta$ and $\eta'$ in \cref{sec:erroreta} in order to assess the systematic error. 
We find small corrections for the $\eta$, but large deviations for the $\eta'$. As a further illustration, we derive in \cref{app:im} 
the additional contributions to the imaginary part that a narrow-width vector meson would produce.
\\

Repeatedly, this result has been used in the literature for estimating the whole amplitude using Cauchy's integral formula, which is often 
referred to as a dispersion relation. This consists in reconstructing the original function \cref{eq:loopamp} from its $\gamma\gamma$ discontinuity 
above $q^2=0$. 
As the imaginary part, \cref{eq:ImGG}, does not fall rapidly enough at infinity ---which is related to the divergent character of \cref{eq:loopamp} for a constant TFF---  
a subtraction is required, so the final result reads~\cite{Bergstrom:1983ay,Ametller:1984uk,Dorokhov:2007bd}
\begin{equation}
  \operatorname{Re}\mathcal{A}(q^2) =   \mathcal{A}(0) + \frac{q^2}{\pi}  \int_0^{\infty}  ds \frac{\operatorname{Im}\mathcal{A}_{\gamma\gamma}(s) }{ s(s-q^2) }.
\end{equation}
Still, the value for $\mathcal{A}(0)$ must be calculated from \cref{eq:loopamp}, which represents though a simpler calculation. 
The result from the dispersive integral leads exactly to the terms in brackets in \cref{eq:cnstTFF}. 
We note here that such calculations are approximate. For a general pseudoscalar mass the additional contributions to the imaginary 
part coming from the TFF must be specified ---actually these would allow to write an unsubtracted dispersion relation, cf. \cref{app:im}.
Consequently, such calculations are approximate as they would neglect all kinds of $m_{P,\ell}/\Lambda$ corrections, where $\Lambda$  
is some TFF characteristic scale~\cite{Bergstrom:1983ay,Ametller:1984uk,Dorokhov:2009xs}.

\subsection{Results for a constant form factor}
\label{sec:cnstTFF}

Before continuing, it will be useful in view of the next discussion and \cref{sec:pllcpt}, to estimate the result 
which is obtained when taking a constant (WZW) TFF. 
Obviously, the result will include some divergent term ---to be cancelled once the TFF is switched on--- which 
needs regularization. Taking $\tilde{F}_{P\gamma^*\gamma^*}(k^2,(k-q)^2)=1$, the loop integral \cref{eq:loopamp} can be expressed using 
dimensional regularization 
in terms of known scalar integrals
\begin{align}
\mathcal{A}^{\textrm{\tiny WZW}}(q^2) &=   2\frac{q^{\mu}q^{\nu}}{q^2}C_{\mu\nu}(q^2,m_{\ell}^2,m_{\ell}^2;0,0,m_{\ell}^2)  -2B_0(m_{\ell}^2;0,m_{\ell}^2) \nonumber \\
        &= \frac{1}{2} \left( q^2C_0(q^2,m_{\ell}^2,m_{\ell}^2;0,0,m_{\ell}^2) - 3B_0(m_{\ell}^2;0,m_{\ell}^2) +1\right). \label{eq:cffscalar}
\end{align}
Note here that if we were to use some cut-off in our integrals for the regularization procedure ---which is particularly useful for deriving the approximate formula--- 
the peculiarities of dimensional regularization must be accounted for carefully. As an example, from the first line in \cref{eq:cffscalar}, the divergent 
part arises from
\begin{equation}
  2\frac{4}{d}\textrm{Div}[C_{00}]   -2\textrm{Div}[B_0]  = \frac{2}{d}\Delta_{\epsilon}   -2\Delta_{\epsilon}
                      = -\frac{3}{2}\Delta_{\epsilon} + \frac{1}{4}, \label{eq:dreg}
\end{equation}
where we have used $d= 4-\epsilon $ and $\Delta_{\epsilon} = \frac{2}{\epsilon} -\gamma_E + \ln4\pi$. The additional finite extra-term 
which is found should be subtracted from \cref{eq:cffscalar} if not using dimensional regularization. Performing the calculation for the scalar 
functions $C_0$ and $B_0$, we find, in dimensional regularization, 
\begin{equation}
\mathcal{A}^{\textrm{\tiny WZW}}(q^2) = \frac{i\pi}{2\beta_{\ell}}L + \frac{1}{\beta_{\ell}}\left[  \frac{1}{4}L^2 + \frac{\pi^2}{12} +\textrm{Li}_2\left( \frac{\beta_{\ell}-1}{1+\beta_{\ell}} \right)  \right] 
 -\frac{5}{2} +\frac{3}{2}\ln\left( \frac{m_{\ell}^2}{\mu^2} \right) , \label{eq:cnstTFF}
\end{equation}
%
%
%
where $L=\ln\left( \frac{1-\beta_{\ell}}{1+\beta_{\ell}}\right)$,  $\beta_{\ell} = \sqrt{1-\frac{4m_{\ell}}{q^2}}$  
is the lepton velocity and $\textrm{Li}_2(x)$ is the dilogarithm function\footnote{The dilogarithm or Spence's function is defined as $\textrm{Li}_2(x) = -\int_0^x dt \frac{\ln(1-t)}{t}$.}. 
If we were using a cut-off regularization $\mu^2\rightarrow\infty$, from \cref{eq:cffscalar}, and accounting for the last piece in \cref{eq:dreg} , we would find similar results 
but replacing the last terms in \cref{eq:cnstTFF} by $ -\frac{5}{4} -\frac{3}{2}\ln(1+ \frac{\mu^2}{m_{\ell}^2} ) $.

\subsection{Approximate results and main properties}

Before providing any input for the TFF, it is very convenient to analyze the loop-integral. This allows to identify the relevant scales involved in the problem, which is 
extremely important in order to achieve the most appropriate TFF description. For this task, it is very convenient to carry out an approximate calculation in terms of 
$m_{P,\ell}^2/\Lambda^2$, where $\Lambda$ is some characteristic scale encoded in the form factor. Following~\cite{Knecht:1999gb}, we take
\begin{equation}
\mathcal{A}(q^2) = \mathcal{A}^{\textrm{\tiny WZW}}(q^2) + \frac{2i}{\pi^2q^2} \! \int \! d^4k \frac{\left(k^2q^2-(k\cdot q)^2\right)(F_{P\gamma^*\gamma^*}(k^2,(q-k)^2) -1 ) }{k^2(q-k)^2\left((p-k)^2-m_{\ell}^2\right)} ,
\end{equation}
where we have added and subtracted a constant term ---precisely, that in \cref{eq:cnstTFF}. 
The remaining integral is essentially zero at scales $k^2\sim m^2_{P}, m^2_{\ell} $ below $ \Lambda^2$, as the TFF remains constant. 
Above, all the terms $\mathcal{O}(p^2,q^2,m_{\ell}^2)$ can be neglected. 
At such scales, the leading term from the tensor $k^{\mu}k^{\nu}q_{\mu}q_{\nu}$ part is given by $k^{\mu}k^{\nu} \sim (1/d)k^2g^{\mu\nu}$, 
as additional terms are $m_P^2/\Lambda^2$ suppressed. We are left then with 
\begin{align}
\mathcal{A}(q^2) & \simeq \mathcal{A}^{\textrm{\tiny WZW}}(q^2) + \frac{2i}{\pi^2}(1-\frac{1}{d}) \int d^4k \frac{F_{P\gamma^*\gamma^*}(k^2,k^2) -1  }{(k^2)^2} , \nonumber \\
                            & = \mathcal{A}^{\textrm{\tiny WZW}}(q^2) -3 \int_0^{\mu} dQ \frac{F_{P\gamma^*\gamma^*}(Q^2,Q^2) -1  }{Q}. \label{eq:approx1}
\end{align}
The first line corresponds, essentially, to the result Eq.~(12) in \cite{Knecht:1999gb}, whereas in the second one, we have Wick-rotated and introduced a cut-off regularization. 
The obtained integral is still divergent for $\mu\rightarrow\infty$, which is expected as it must cancel the divergency in 
$\mathcal{A}^{\textrm{\tiny WZW}}(q^2)$, see \cref{eq:cnstTFF}. In order to remove it, we identify the origin of the UV divergent term in \cref{eq:cnstTFF}, 
subtract\footnote{That amounts to remove the $ -\frac{3}{2}\ln(1+ \mu^2/m_{\ell}^2 ) $ term from $\mathcal{A}^{\textrm{\tiny WZW}}(q^2)$. Note that we are using a cut-off regularization, 
so the comments below \cref{eq:cnstTFF} apply.} there, and plug into \cref{eq:approx1}, obtaining
\begin{multline}
\mathcal{A}^{\textrm{app}}(q^2) = \frac{i\pi}{2\beta_{\ell}}L +  \frac{1}{\beta_{\ell}}  \left[  \frac{1}{4}L^2 + \frac{\pi^2}{12} +\textrm{Li}_2\left( \frac{\beta_{\ell}-1}{1+\beta_{\ell}} \right)  \right]   -  \frac{5}{4}     \\
 + \int_0^{\infty}dQ\frac{3}{Q}\left(  \frac{m_{\ell}^2}{Q^2+m_{\ell}^2} - F_{P\gamma^*\gamma^*}(Q^2,Q^2)  \right). \label{eq:approx2}
\end{multline}
%
%
%
%
%
This kind of approximation, obtained in many different ways, has been widely used in the literature, see explicitly in 
Refs.~\cite{,Knecht:1999gb,Dorokhov:2007bd} and implicit in most of the quoted references.
Exceptions are the full calculation in Ref.~\cite{Silagadze:2006rt}, and those including partial corrections in Refs.~\cite{Babu:1982yz,Dorokhov:2008cd,Dorokhov:2009xs}. 
While these are relevant to the precision we are aiming, specially for the $\eta$ and $\eta'$ cases, the approximation in \cref{eq:approx2} is
enough, at least for the $\pi^0$, to understand the relevant dynamics in this process. 
To illustrate this, we plot in \cref{fig:ker} the integrand of \cref{eq:approx2}, $\mathcal{K}(Q^2)$, for the electron case.
\begin{figure}[t]
  \includegraphics[width=0.5\textwidth]{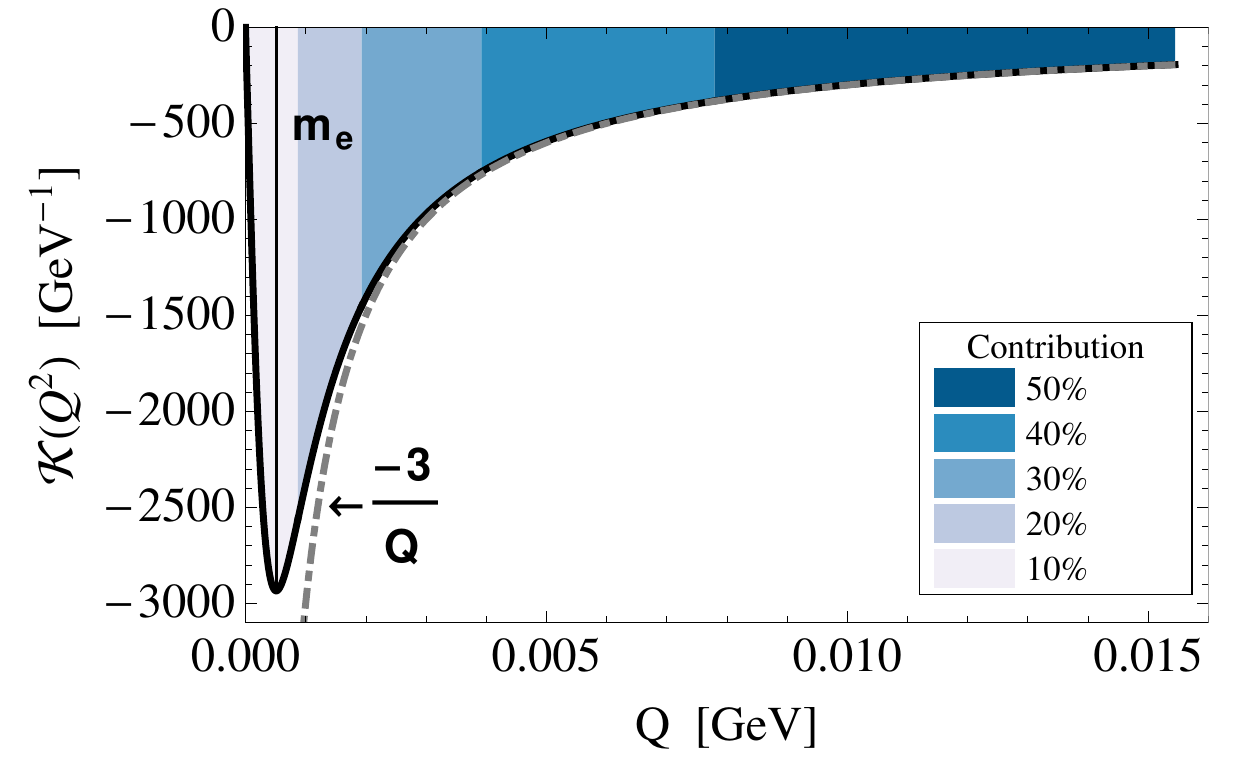}
  \includegraphics[width=0.5\textwidth]{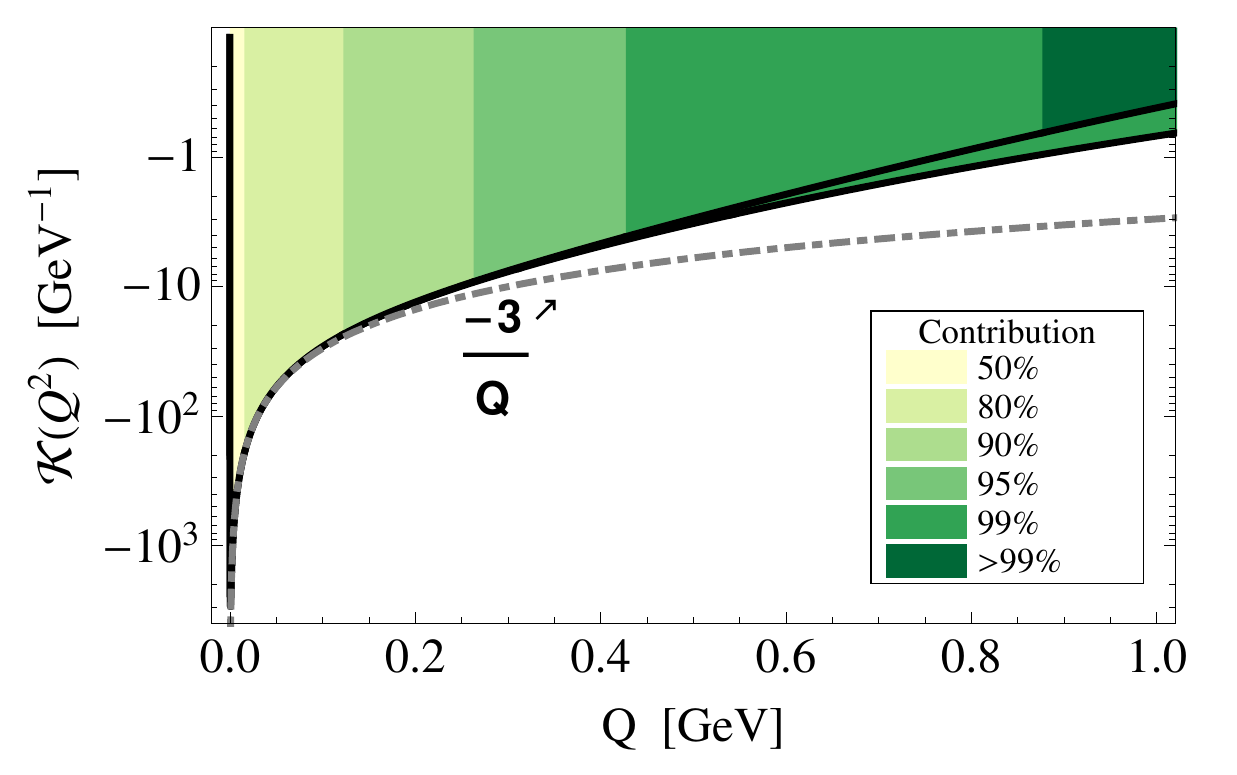}
\caption{The integrand in \cref{eq:approx2}, $\mathcal{K}(Q^2)$, for $\ell=e$ and partial contributions to the integral. The left figure stands for the low-energies, whereas the 
right one, in logarithmic-scale, stands for the high-energies. The upper(lower) black line stand from our factorization(OPE) models (see \cref{sec:c01rec}) and the 
dot-dashed gray line for a constant TFF.\label{fig:ker}}
\end{figure}
As one can see, it involves the space-like symmetric $(Q_1^2=Q_2^2)$ kinematics. In addition, the integrand is peaked at very low-energies close to the lepton mass, 
where the TFF essentially remains constant. 
The TFF effects become visible and specially relevant in the $(0.1-0.4)$~GeV region, where the slope parameter is roughly enough to describe the TFF; 
the effects from additional parameters appear roughly above this region ---where the two black lines in \cref{fig:ker} separate--- 
and represent a minor contribution to the integral. The high-energy tail plays though a non-negligible role too.
Given the sensitivity to the double virtual regime, this challenging process would represent the first experimental probe to the TFF double-vitrtual kinematics. 
%
%
%
%
From the features enumerated above, any serious approach developed to deal with this process should implement:
\begin{itemize}
  \item The appropriate space-like low-energy behavior. Particularly, the slope parameter should be described as precise as possible to obtain the most accurate 
        description below $1$~GeV.
  \item A proper implementation (not modeling) of the doubly-virtual behavior.
  \item A minimum implementation for the high-energy behavior (i.e. the correct $Q^2$ behavior discussed in \cref{sec:tffpqcd}).
\end{itemize}
However, previous approaches have often adopted either 
quark models~\cite{Ametller:1983ec,Bergstrom:1983ay,Scadron:1983nt,Pich:1983zk,Margolis:1992gf,deRafael:2011ga} 
or simplified VMD models saturated with the lowest-lying resonance~\cite{Quigg:1968zz,Bergstrom:1983ay,Babu:1982yz,Knecht:2014sea,Dorokhov:2009xs}. 
As such models cannot properly reproduce the data, one may doubt 
about the accuracy of their results, this is, their non-assessed systematic errors. To supply this, some approaches have adopted a VMD-like approach where the 
effective mass is obtained from a fitting procedure to high-energy (mostly above $2$~GeV$^2$) space-like data~\cite{Dorokhov:2008cd,Dorokhov:2009xs}. We have 
seen in \cref{chap:data} that this can be understood as the first element of 
a PA, which systematic error is certainly large. Besides, these approaches face the problem that no double-virtual data is available so far, for which some model must be 
assumed\footnote{Note that those parameterizations using a factorized form, such as Ref.~\cite{Dorokhov:2009xs}, imply an important unaccounted error, as such model 
violates the OPE expansion.}~\cite{Dorokhov:2007bd,Dorokhov:2008cd}. The associated error of this procedure or how the data would be incorporated 
into their descriptions  is not clear.
A possible alternative to circumvent these problems is provided by \cpt~\cite{Savage:1992ac}. In such framework, every pseudoscalar decay depends, at LO, on the same common 
counterterm (see \cref{sec:pllcpt}). Obtaining this from a particular channel, one may predict the others. We only note for the moment that the NLO effects cannot be neglected 
as will be discussed in \cref{sec:pllcpt}. Therefore, this approach is not feasible at the required precision. Last but not least, (as explained) most of the calculations 
employed so far rely on numerical approximations for the calculation in \cref{eq:loopamp}. If this is not a bad approximation for the $\pi^0\to e^+e^-$ decay given the strong 
$m_e\ll m_{\pi} \ll \Lambda$ hierarchy ($\Lambda$ represents the TFF scale), these approximations, which would simplify the loop calculation, are not appropriate for the 
heavier $\eta$ and $\eta'$ cases, where the induced error may become larger than the statistic and systematic ones.

\section{A rational description for $F_{P\gamma^*\gamma^*}(Q_1^2,Q_2^2)$}
\label{sec:CAsSys}

In view of the problems raised in the preceding section, CAs seem an ideal and robust framework  to deal with these shortcomings. First, they are able to systematically 
incorporate the appropriate low-energy expansion, not only for the single- but for the double-virtual case. Second, they are able to accommodate what is known from the high-energies. 
Third, they provide a method to obtain a systematic error.
We analyze therefore the \PtoLL processes in the light of CAs with the idea of achieving a preciser and more accurate prediction for these decays, including a systematic error 
and performing a precise numerical evaluation. This, together with the most recent evaluation for the radiative corrections in Refs.~\cite{Vasko:2011pi,Husek:2014tna,Husek:2015sma}, 
would promote their SM prediction to the standards of precision met nowadays, and would provide a reliable quantity to compare with the available or upcoming experimental results. 
\\

\subsection{Reconstructing the transition from factor}
\label{sec:c01rec}

The reconstruction of a general function from CAs was discussed in great detail in \cref{chap:CA}. The lowest approximant at our disposal corresponds to the $C^0_1(Q_1^2,Q_2^2)$, 
which, given the low-energy TFF expansion,
\begin{equation}
\tilde{F}_{P\gamma^*\gamma^*}(Q_1^2,Q_2^2) = 1 - \frac{b_P(Q_1^2+Q_2^2)}{m_P^2} + \frac{a_{P;1,1}Q_1^2Q_2^2}{m_P^4} +\frac{c_P(Q_1^4+Q_2^4)}{m_P^4} + ... ,
\end{equation}
can be reconstructed, fully-based on its low-energy expansion, as 
\begin{equation}
\label{eq:c01low}
C^0_1(Q_1^2,Q_2^2) =  \frac{1}{1+\frac{b_P}{m_P^2}(Q_1^2+Q_2^2) + \frac{2b_P^2-a_{P;1,1}}{m_P^4} Q_1^2Q_2^2},
\end{equation}
where all the single virtual parameters have already been determined in \cref{chap:data} (see \cref{tab:chap1mainres}). 
It remains then to assign a numerical value to the double-virtual parameter $a_{P;1,1}$ ---not determined so far due to the absence of double virtual experimental data.
In order to make a well-educated guess for this parameter, we consider the two extreme regimes relevant for our calculation. On the one hand, at the very low-energies 
involved in our calculation, \cpt should provide a reliable estimate for the TFF behavior. It turns out that, according to the study in Ref.~\cite{Bijnens:2012hf}, 
the chiral leading logs suggest that a factorization approach ($F_{P\gamma^*\gamma^*}(Q_1^2,Q_2^2)\simeq F_{P\gamma^*\gamma}(Q_1^2)\times F_{P\gamma^*\gamma}(Q_2^2)$) 
should provide a good approximation\footnote{Actually, this approximation is supported as well from the dispersive study in~\cite{Xiao:2015uva}.}, 
implying that $a_{P;1,1}\simeq b_P^2$ ---corrections appearing an order higher (even two in the chiral limit) than expected.
On the other hand, at the very high-energies relevant for the integrand tail, the OPE expansion (i.e., that $F_{P\gamma^*\gamma^*}(Q^2,Q^2) \sim Q^{-2}$) requires 
$a_{P;1,1} = 2b_P^2$ if a two-point approximation is employed (see \cref{sec:caope}), suggesting that corrections to the chiral leading logs should push the factorized 
value upwards. For these reasons, we choose to take the  
$b_P^2<a_{P;1,1}<2b_P^2$\footnote{Note that values above $2b_P^2$ would imply in addition a pole in the SL region.} band as a compromise between the low energies 
and the appropriate high-energy behavior~\cite{Masjuan:2015lca,Masjuan:2015cjl}. If the real value ---to be extracted from the experiment or lattice results--- is eventually observed to lie 
within this band, success is guaranteed.

In order to improve this description, we should move on along the $C^N_{N+1}$ sequence and construct larger approximants.
This would allow then to implement both, the low- and the high-energy behavior at the same time, and would make the preceding discussion unnecessary.
However, the next approximant, the $C^1_2$, already contains too many double-virtual parameters, further complicating its reconstruction and numerical evaluation, 
for which we omit its discussion here\footnote{For an extended discussion for the $C^1_2$ approximant we refer to \cref{chap:gm2}.}.

\subsection{Systematic error I: the $\pi^0$}
\label{sec:errorpi0}

Given the short length of our sequence, including only a single element, it is extremely important 
to check on the systematics. For these reasons, we come back once more to our recurrent logarithmic and Regge models defined in \cref{chap:CA} for the most general 
double-virtual case,
\begin{align}
F^{\textrm{Regge}}_{\pi^0\gamma^*\gamma^*}(Q_1^2,Q_2^2) &= \frac{aF_{P\gamma\gamma}}{Q_1^2-Q_2^2}
                                         \frac{\left[ \psi^{(0)}\left(\frac{M^2+Q_1^2}{a}\right) -\psi^{(0)}\left(\frac{M^2+Q_2^2}{a}\right) \right]}{\psi^{(1)}\left(\frac{M^2}{a}\right)}, \\
F^{\textrm{log}}_{\pi^0\gamma^*\gamma^*}(Q_1^2,Q_2^2) &= \frac{F_{P\gamma\gamma}M^2}{Q_1^2-Q_2^2}\ln\left( \frac{M^2+Q_1^2}{M^2+Q_2^2} \right).
\end{align}
We note that for the Regge and logarithmic models the condition $b_P^2<a_{P;1,1}<2b_P^2$ is satisfied. In particular, for the Regge model, 
$a_{P;1,1} = [2\psi^{(1)}(\frac{M^2}{a})\psi^{(3)}(\frac{M^2}{a})]/[3(\psi^{(2)}(\frac{M^2}{a})^2]b_P^2 = 1.13b_P^2$, whereas for the logarithmic 
one $a_{P;1,1} = (4/3)b_P^2$. Performing the numerical integration in \cref{eq:loopamp}, 
we obtain\footnote{We use $F_{P\gamma^*\gamma^*}(Q_1^2,Q_2^2)=\frac{1}{\psi^{(1)}(M^2/a)}\sum_{m=0}^{\infty} \frac{\Lambda^4}{(Q_1^2+(M^2+na))(Q_2^2+(M^2+na))} $~\cite{RuizArriola:2006jge} 
for the Regge model; for the logarithmic model, we use \cref{eq:logflat}.}
%
%
%
\begin{gather}
  \mathcal{A}(m_{\pi}^2) = 9.73-17.52i \qquad \textrm{BR}(\pi^0\rightarrow e^+e^-) = 6.14\times10^{-8},\\
  \mathcal{A}(m_{\pi}^2) = 8.78-17.52i \qquad \textrm{BR}(\pi^0\rightarrow e^+e^-) = 5.87\times10^{-8},
\end{gather}
for the Regge and logarithmic models respectively. This is to be compared with their corresponding $C^0_1$ reconstruction, which for the chosen  $b_P^2<a_{P;1,1}<2b_P^2$ band yields, 
\begin{gather}
  \mathcal{A}(m_{\pi}^2) = (9.63 \! \div \! 10.09) \! - \! 17.52i \ \ \textrm{BR}(\pi^0 \!  \! \rightarrow  \!  e^+ \! e^-) = (6.08 \! \div \! 6.22) \!  \! \times \!  \! 10^{-8},\\
  \mathcal{A}(m_{\pi}^2) = (8.78 \! \div \!   9.24) \!  - \! 17.52i \ \  \textrm{BR}(\pi^0 \!  \! \rightarrow  \! e^+ \! e^-) = (5.87 \! \div \! 6.00) \!  \! \times \!  \! 10^{-8},
\end{gather}
where the first(second) value corresponds to $a_{P;1,1}=2b_P^2(b_P^2)$, i.e., the value implied by OPE(factorization). 
As a curiosity, we find that, for the logarithmic model, constraining the OPE $(a_{P;1,1}=2b_P^2)$ seems to be the better choice. This is just an accident which 
can be understood from the fact that, for $Q_1^2=Q_2^2$, that model is parametrically equivalent to such approximant\footnote{This would not be the 
case for the heavier $\eta$ and $\eta'$, as the behavior for $Q_1^2\neq Q_2^2$ becomes relevant too.}, see \cref{eq:appelleq}. 
Indeed, this observation does not apply to the Regge model. In general, whether the result is closer to the OPE or the factorization choice 
will depend on the pseudoscalar masses, the double-virtual low-energy behavior and how the TFF approaches the asymptotic regime. 
It seems hard to us to judge on a better choice with a single approximant at hand. 
Consequently, we take the given band as the best (more conservative) error estimation one can do at this point.
As a further comment, we find that additional sources of error beyond the double-virtual reconstruction are masked within this band.

In principle, it seems that these results would apply for the $\eta$ and $\eta'$ cases. However, such extrapolation cannot be strictly performed. For the $\eta$ and $\eta'$, the 
approximation in \cref{eq:approx2} is not appropriate anymore; it is easy to see that the loop-integral in \cref{eq:loopamp} does not involve space-like arguments for the TFF alone, 
but time-like ones in the $-m_P^2 \leq Q^2 \leq 0$ region too. Whereas this does not represent a problem for the $\pi^0$,  
it poses a problem for the $\eta$ and $\eta'$ cases, as such region includes the $\pi\pi$ threshold for the $\eta$ and reaches the $\rho$ and $\omega$ resonances for the  
$\eta'$. It has yet to be seen if our approximants have the ability to reproduce the corresponding real and imaginary parts required 
in these processes. In the following section, we discuss that this is actually possible provided that we deal with Stieltjes functions, a unique feature which 
cannot be reproduced in traditional approaches.

\subsection{Systematic error II: the $\eta$ and $\eta'$}
\label{sec:erroreta}

The $\eta$ and $\eta^{\prime}$ masses are large enough to yield intermediate hadronic states in the $P \to \bar{\ell} \ell$ processes as sketched in \cref{fig:UB}, which implies 
an additional imaginary part beyond that of the $\gamma\gamma$ contribution. As we will show, this diminishes the imaginary part, invalidating then the 
unitary bound.
\begin{figure}[t]
\centering
   \includegraphics[width=0.8\textwidth]{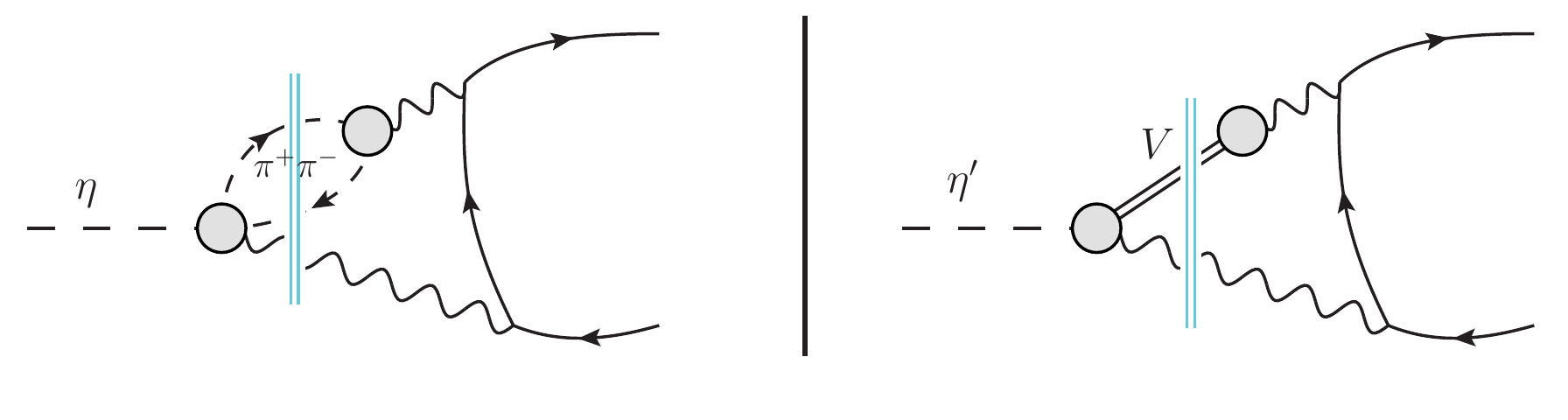}
   \caption{Intermediate hadronic states invalidating the unitary bound for the $\eta,(\eta^{\prime})$, left(right).}
   \label{fig:UB}
\end{figure}
This effect has never been considered before when calculating these decays and must be taken into account when evaluating the systematic error. Indeed, when the 
intermediate state becomes resonant, this effect becomes the dominant source of error. 

To quantitatively study this effect, we take a toy-model for the TFF that includes both, a 
two-pion production threshold and a vector resonance. The model is conceived in such a way that the time-like region contains all the required features of the physical TFF up 
to the $\eta^{\prime}$ mass. The first ingredient in our toy-model is factorization, which as explained before seems a reasonable choice at low-energies 
and does not spoil our discussion. The second ingredient is the use of vector meson dominance ideas~\cite{Landsberg:1986fd} allowing to express the (normalized) 
single-virtual TFF as 
\begin{equation}
\label{eq:VMD}
\tilde{F}_{P\gamma^*\gamma}(s) = c_{P\rho} G_{\rho}(s) + c_{P\omega} G_{\omega}(s) + c_{P\phi} G_{\phi}(s),
\end{equation}
where $G_V(s)$ are the different resonance contributions weighted by the dimensionless couplings $c_{PV}$  obtained from a quark-model, 
$c_{\eta(\eta^{\prime})\rho}=9/8(9/14)$, $c_{\eta(\eta^{\prime})\omega}=1/8(1/14)$, $c_{\eta(\eta^{\prime})\phi}=-2/8(4/14)$~\cite{Hanhart:2013vba}, and $G_V(0)=1$. 
In order to incorporate the $\pi\pi$ intermediate branch cut in \cref{fig:UB}, fulfilling unitarity and analyticity, we take for the $\rho$ contribution, $G_{\rho}(s)$, a 
model based on Refs.~\cite{GomezDumm:2000fz,Dumm:2013zh}
\begin{equation}
\label{eq:rho}
G_{\rho}(s) = \frac{M_{\rho}^2}{   M_{\rho}^2 - s +\frac{s M_{\rho}^2}{96\pi^2F_{\pi}^2}\left( \ln\left(\frac{m_{\pi}^2}{\mu^2}\right)  +\frac{8m_{\pi}^2}{s} -\frac{5}{3} - \sigma(s)^3 \ln\left(\frac{\sigma(s)-1}{\sigma(s)+1}\right)   \right)   }
\end{equation}
with $\sigma(s) = \sqrt{1-4m_{\pi}^2/s}$, and the parameters $M_{\rho}=0.815$ GeV, $F_{\pi}=0.115$ GeV, $\mu=0.775$ GeV, and $m_{\pi}=0.139$ GeV, chosen to reproduce the 
pole position $s_{\rho}=(M - i \Gamma/2)^2$ with $M =  0.764$ GeV and $\Gamma = 0.144$ GeV from\cite{Masjuan:2013jha}, while for the (narrow-width) $\omega,\phi$ resonances, 
we take\footnote{We explored further refined models with an improved threshold behavior for the $\omega$ and $\phi$ resonances. Given their narrow width they led to very similar 
results and we decided to take the ones in \cref{eq:omegaphi} for not obscuring our study and deviating the attention from our main concern, an estimation of a systematic error.}
\begin{equation}
\label{eq:omegaphi}
 G_{\omega,\phi} = \frac{M_{\omega,\phi}^2 + M_{\omega,\phi}\Gamma_{\omega,\phi}(s_{th}/M^2_{\omega,\phi})^{3/2}}{M_{\omega,\phi}^2 - s + M_{\omega,\phi}\Gamma_{\omega,\phi} ((s_{th}-s)/M^2_{\omega,\phi})^{3/2}},
\end{equation}
with parameters fixed from PDG masses and widths~\cite{Agashe:2014kda}. 
This choice makes our model very similar to the dispersive approach formulated in~\cite{Hanhart:2013vba}. 

To evaluate the BR, we calculate the loop amplitude in \cref{eq:loopamp} with the TFF from \cref{eq:VMD} as an input ---this parametrization already implements 
the desired threshold and resonance effects displayed in \cref{fig:UB}. It is convenient for the integration procedure to employ a Cauchy integral representation  for the TFF,
%
\begin{equation}
\label{eq:cauchy}
\tilde{F}_{P\gamma^*\gamma^*}(q_1^2,q_2^2) = \int_{s_{th}}^{\infty} dM_1^2  \int_{s_{th}}^{\infty} dM_2^2  \frac{\operatorname{Im}\tilde{F}_{P\gamma^*\gamma}(M_1^2)}{q_1^2 - M_1^2 -i\epsilon} \frac{\operatorname{Im}\tilde{F}_{P\gamma^*\gamma}(M_2^2)}{q_2^2 - M_2^2 -i\epsilon}.   
\end{equation}
The loop integral in \cref{eq:loopamp} can be expressed then, after changing the integration order, as
\begin{align}
\label{eq:loopdisp}
\mathcal{A}(q^2)   = & \frac{1}{\pi^2}  \int_{s_{th}}^{\infty} dM_1^2  \int_{s_{th}}^{\infty} dM_2^2  \operatorname{Im}\tilde{F}_{P\gamma^*\gamma}(M_1^2)  \operatorname{Im}\tilde{F}_{P\gamma^*\gamma}(M_2^2) \nonumber \\
 &  \times   \left(2i \int \frac{d^4k}{\pi^2} \frac{ (q^2k^2 - (q k)^2)}{ q^2 k^2(q-k)^2((p-k)^2-m_{\ell}^2)}\frac{1}{k^2-M_1^2}\frac{1}{(q-k)^2-M_2^2}\right) \nonumber \\
 \equiv   &   \frac{2}{\pi^2} \int_{s_{th}}^{\infty} dM_1^2  \int_{s_{th}}^{M_1^2} dM_2^2\,  \operatorname{Im}\tilde{F}_{P\gamma^*\gamma}(M_1^2)  \operatorname{Im}\tilde{F}_{P\gamma^*\gamma}(M_2^2)   \! \times \!    K(M_1^2,M_2^2).
\end{align}
This procedure results in an easy evaluation of the loop amplitude, denoted as $K(M_1^2,M_2^2)$, through standard one-loop techniques~\cite{'tHooft:1978xw} or a numerical 
evaluation using \texttt{FeynCalc}~\cite{Mertig:1990an} and \texttt{LoopTools}~\cite{Hahn:1998yk}. 
Now, the threshold effects are clear and easier to handle. To illustrate them, we plot the imaginary part of the integrand in \cref{eq:loopdisp} in terms of 
$\operatorname{Im}\tilde{F}_{P\gamma^*\gamma}(M_V^2)$ and $\operatorname{Im}K(M_V^2)$ ---containing both $\gamma \gamma$ and vector contributions--- when 
dispersing only one virtuality in \cref{eq:loopdisp} for simplicity (i.e., we consider a $q^2$-independent narrow width approximation for the second virtuality). 
The resulting plot is shown in \cref{fig:imdisp} as a solid-black (dashed-purple) line for the $\eta (\eta^{\prime})$ in terms of the dispersive variable $M_V$ once the $\int d^4k$ 
integration has been performed to give $K(M_V^2)$ in the last line of \cref{eq:loopdisp}. These lines have to be convoluted with $\operatorname{Im}\tilde{F}_{P\gamma^*\gamma}(M_V^2)$ 
(bluish area in \cref{fig:imdisp}) in order to obtain $\operatorname{Im}\mathcal{A}(q^2)$. 
For $M_V>m_P$, the imaginary part corresponds to the $\gamma\gamma$ contribution, which diminishes as soon as $M_V<m_P$. Consequently, for resonances heavier 
than the pseudoscalar mass, there will be a a slight modification whenever the resonance tail (in our case at $2m_{\pi}$) appears below $m_P$. On the other hand, for resonances 
lighter than the pseudoscalar mass, the shift will be considerable. All in all, as unitarity implies, the imaginary part will be shifted whenever an intermediate hadronic 
channel appears below $m_P$.
\begin{figure}[t]
\centering
\includegraphics[width=0.6\textwidth]{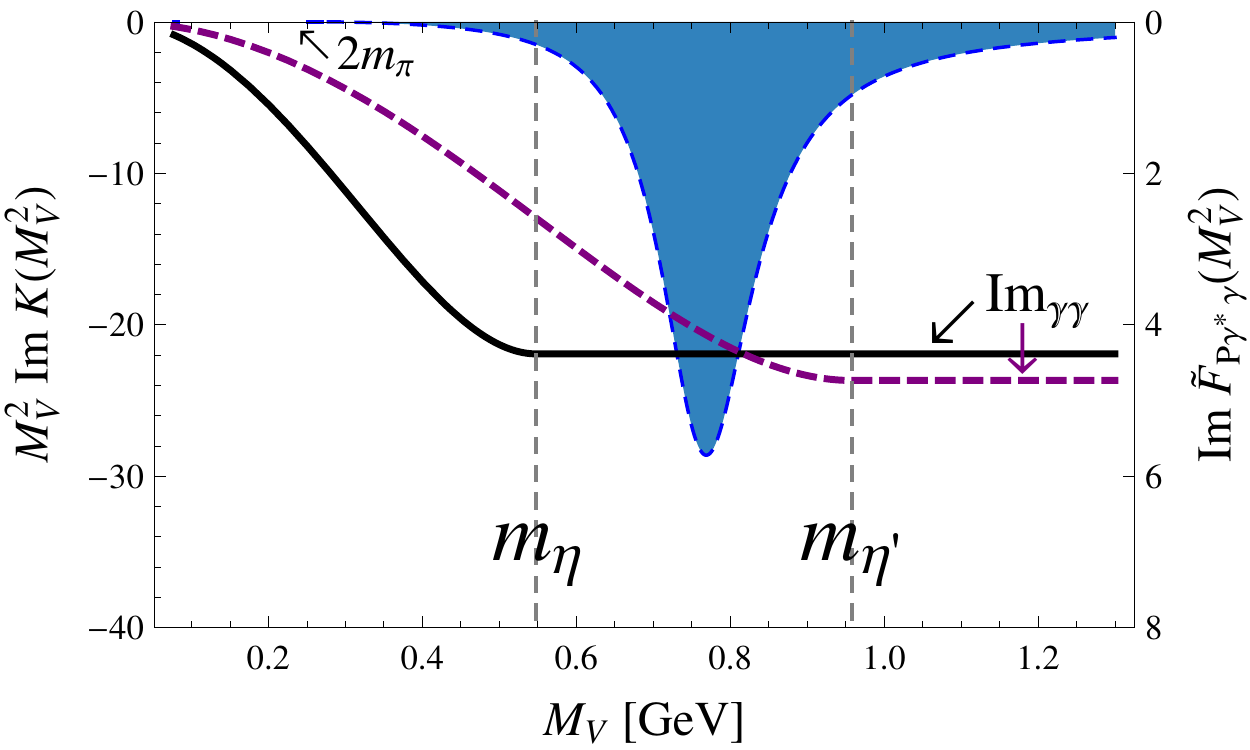}
\caption{The imaginary part for integrand \cref{eq:loopdisp} expressed in terms of $M_V^2\operatorname{Im} K(M_V^2)$ (black and dashed-purple lines for the $\eta$ and $\eta'$, respectively) 
which has then to be convoluted with $\operatorname{Im} \tilde{F}_{P\gamma^*\gamma}(M_V^2)$; in the figure 
$\operatorname{Im} \tilde{F}_{P\gamma^*\gamma}(M_V^2) = \operatorname{Im} G_{\rho}(M_V^2)$ is shown ---the $\omega$ and $\phi$ resonances woud produce sharp peaks on top.}
\label{fig:imdisp}
\end{figure}
For completeness, we illustrate in \cref{tab:im} the numerical shift in the imaginary part with respect to the $\gamma\gamma$ contribution in our toy-model \cref{eq:VMD}, showing 
the break of the unitary bound.
\begin{table}[t]
\small
\centering
\begin{tabular}{cccc}\toprule
 &  & $\gamma\gamma$ & Total \\ \hline
\multirow{2}{*}{$\operatorname{Im}\mathcal{A}_{\eta\rightarrow \ell\ell}(m_{\eta}^2)$}    & $ee$         & $-21.920$ & $-21.805$ \\ 
                                                                                                                   & $\mu\mu$ & $-5.468$ & $-5.441$ \\ \midrule
\multirow{2}{*}{$\operatorname{Im}\mathcal{A}_{\eta^{\prime}\rightarrow \ell\ell}(m_{\eta^{\prime}}^2)$}   & $ee$         & $-23.675$ & $-19.251$ \\ 
                                                                                                                   & $\mu\mu$ & $-7.060$ & $-5.733$ \\ \bottomrule
\end{tabular}
\caption{Imaginary part of $\mathcal{A}(q^2)$ (Total) compared to the imaginary part calculated from the $\gamma\gamma$ channel alone. The hadronic contributions lower the total value of the imaginary part with respect to the $\gamma\gamma$ contribution, invalidating the unitary bound.}
\label{tab:im}
\end{table}

Given that our model is a Stieltjes function, it is well known that the $C^N_{N+1}(Q_1^2,Q_2^2)$ sequence is guaranteed to converge in the whole complex plane, except along the 
cut~\cite{Baker}, where zeros and poles of our CA will clutter to reproduce the discontinuity~\cite{Baker,Masjuan:2009wy}, see \cref{fig:stielt}. Such poles will be responsible 
for effectively generating an imaginary part in our integral mimicking the cut contribution due to Cuachy's integral theorem ---even if the approximation for 
$\tilde{F}_{P\gamma^*\gamma^*}(q_1^2,q_2^2)$ does not converge above the cut. As an illustration, we collect the results for both, BR and $\mathcal{A}(m_P^2)$, from our simplest 
approximant, the $C^0_1(Q_1^2,Q_2^2)$, in \cref{tab:c01res} and compare its results with the toy model. The comparison of the BRs reveals a systematic error induced by the fact 
that we have truncated the CA sequence. For the $\eta$, such error is almost negligible (the role of the vector resonances is very mild there), whereas for the $\eta^{\prime}$ 
it goes almost up to $20\%$. These percentages will be used as an estimate of our systematic error in our final results for the $C^0_1(Q_1^2,Q_2^2)$ element.

We would like to remark at this point that using a VMD model with the $\rho$ mass ---which was standard in the past for performing this calculation--- instead of the more sophisticated model in \cref{eq:VMD}, we would have found BR$(\eta\rightarrow e e) = 5.30\times10^{-9}$, which implies a larger systematic uncertainty compared to our result in \cref{tab:c01res}. A VMD fit to 
generated space-like data in the $(0-15)$~GeV$^2$ does not improve on the result either. In such case, we would have obtained BR$(\eta\rightarrow e e) =5.26\times10^{-9}$. 
These numbers illustrate the potential large systematic error coming from the usage of VMD data-fitting procedures from high-energies for processes which are low-energy dominated, even if the quality of the fit is good enough.\\ 

\begin{table}[t]
\centering
\scriptsize
\begin{tabular}{c   @{\hspace{3pt}}c   @{\hspace{3pt}}c  @{\hspace{3pt}}c } \toprule
 $\textrm{BR}(P\rightarrow\ell\ell)$ & Toy-model & $C^0_1$ & Error ($\%$) \\ \midrule	
  $(\eta\rightarrow e e)  \times  10^{-9}$ & $5.4095$ & $5.4179$ & $0.16$ \\
  $(\eta\rightarrow \mu\mu)  \times 10^{-6}$ & $4.49361$ & $4.52701$ & $0.74$ \\  \midrule
  $(\eta^{\prime}\rightarrow e e)  \times  10^{-10}$ & $1.70507$ & $1.88331$ & $9$ \\
  $(\eta^{\prime}\rightarrow \mu\mu)  \times 10^{-7}$ & $1.1953$ & $1.46089$ & $18$ \\ \bottomrule
\end{tabular}
\begin{tabular}{c  @{\hspace{3pt}}c  @{\hspace{5pt}}c} \toprule
 $\mathcal{A}(m_P^2)$ & Toy-model & $C^0_1$\\ \midrule	
  $(\eta\rightarrow e e)$ & $31.4-21.8i$ & $31.4-21.9i$\\
  $(\eta\rightarrow \mu\mu)$ & $-1.09-5.44i$ & $-1.05-5.47i$\\  \midrule
  $(\eta^{\prime}\rightarrow e e)$ & $46.4-19.2i$ & $48.7-20.5i$\\
  $(\eta^{\prime}\rightarrow \mu\mu)$ & $3.09-5.73i$ & $3.82-6.10i$\\ \bottomrule
\end{tabular}
\caption{Comparison between our toy-model result and the simplest $C^0_1(Q_1^2,Q_2^2)$ approximation for each channel. The Error column represents the relative deviation between the model and the approximation. Left part collects the BR, whereas the right part contains the loop amplitude $\mathcal{A}(m_P^2)$. }
\label{tab:c01res}
\end{table} 
%
%

As we have said, the convergence of the CA sequence to our toy model is guaranteed~\cite{Baker,Masjuan:2009wy,Alabiso:1974vk}, and we show it by constructing the higher elements of 
the $C^N_M(Q_1^2,Q_2^2)$ sequence, calculating with them the amplitude in \cref{eq:loopamp}, $\mathcal{A}(m_P^2)^{\textrm{CA}}$ 
for short, and studying the relative distance in the complex plane, defined as $|1-\mathcal{A}(m_P^2)^{\textrm{CA}}/\mathcal{A}(m_P^2)|$. The results are 
shown in \cref{fig:pacut}, where, for simplicity, we employ only the $G_{\rho}$ contribution in \cref{eq:VMD} (without $\omega, \phi$ contributions).

The results in  \cref{fig:pacut} show the ability of our approximants to systematically account for the TFF to arbitrary precision since the relative distance decreases when the order of 
the CA increases, even in the presence of  the non trivial behavior of the branch cut from the intermediate hadronic states. Note the \textit{a priori} irregular convergence for the $\eta$ 
case in \cref{fig:pacut} (top panel). This is just an accident due to the appearance of effective poles for the particular chosen TFF close to the $\eta$ mass; whenever 
some pole is located close to the $\eta$ mass, it leads to a bad determination. This is compensated in higher approximants with a nearby zero to this pole, alleviating this effect and 
making it negligible as $N\rightarrow\infty$ as shown in \cref{fig:pacut} (bottom right panel), where the poles and zeros for different approximants are plotted. 

We find then that, for the $C^N_{N+1}(Q_1^2,Q_2^2)$ element, the systematic error can be accounted for by the difference in the BR with respect to the $C^{N-1}_{N}(Q_1^2,Q_2^2)$ 
result. As in our case study, we only reach the $C^0_1(Q_1^2,Q_2^2)$ approximant, this procedure does not apply and we take as the systematic error for the BR the one which is 
displayed in the fourth column in \cref{tab:c01res}. This possibly overestimates the systematic error, see comments in \cref{sec:results}, but we opt for this to remain on the 
conservative side. As soon as experimental data on the doubly virtual TFF becomes available, we will be able to extend our CA sequence and reduce the systematic error.
\begin{figure}[t]
\centering
   \begin{minipage}[b]{0.49\textwidth}
       \centering
       \includegraphics[width=\textwidth]{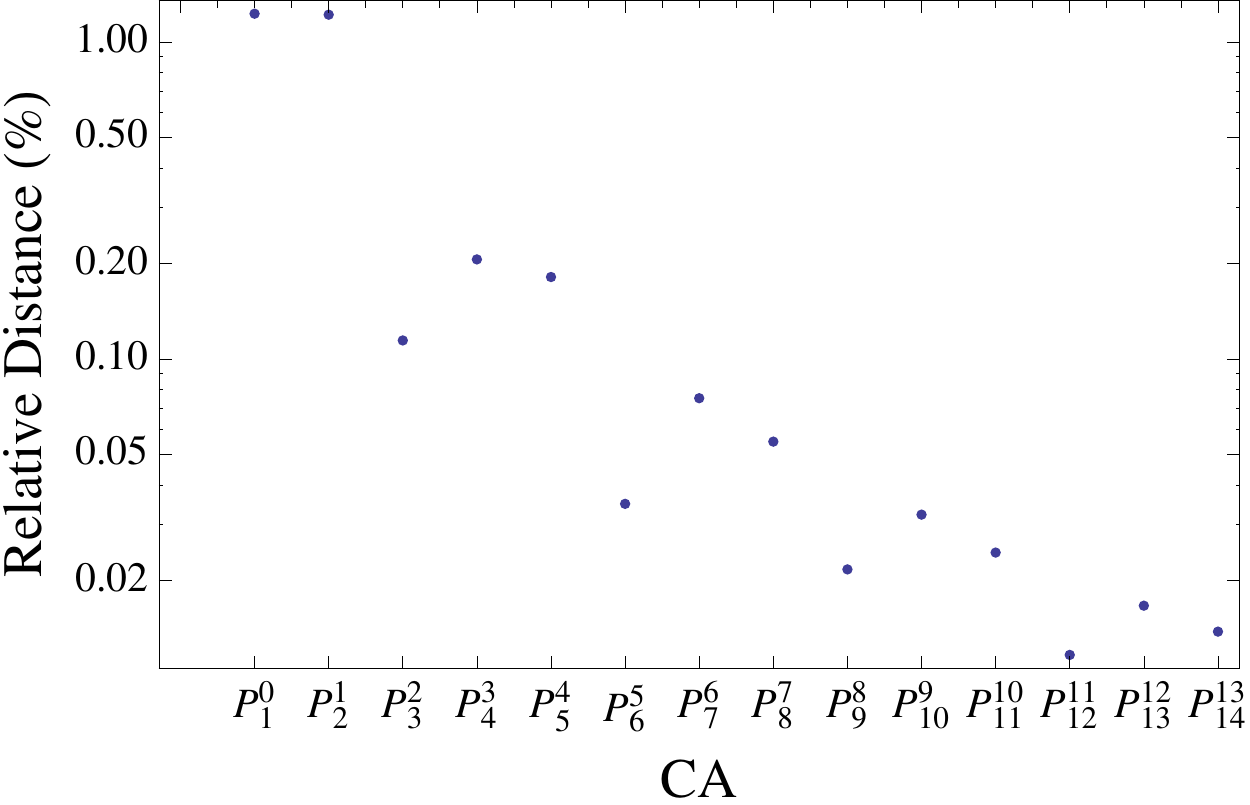}   
   \end{minipage} 
      \hspace{0.04\textwidth}
   \begin{minipage}[b][4.2cm][c]{0.45\textwidth}
       \caption{Top (bottom left) panel represent the $C^N_M$ pattern of convergence to $|\mathcal{A}(m_P^2)|$ for the toy-model \cref{eq:VMD} for $\eta(\eta^{\prime})$. Bottom right figure represents the poles (open circles) and zeros (slashes) from our approximants where $[N/M]$ stands for $C^N_M$.  \label{fig:pacut}}
   \end{minipage}
   \includegraphics[width=0.49\textwidth]{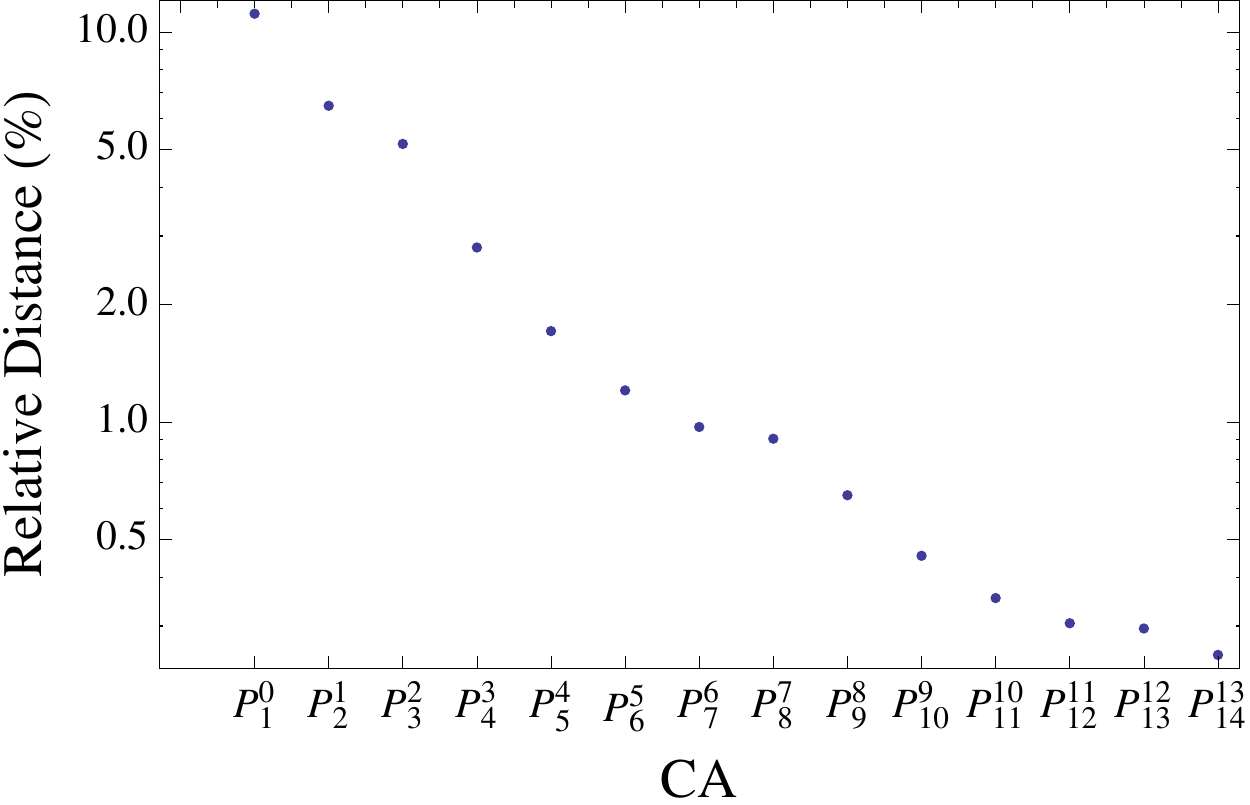}
   \includegraphics[width=0.49\textwidth]{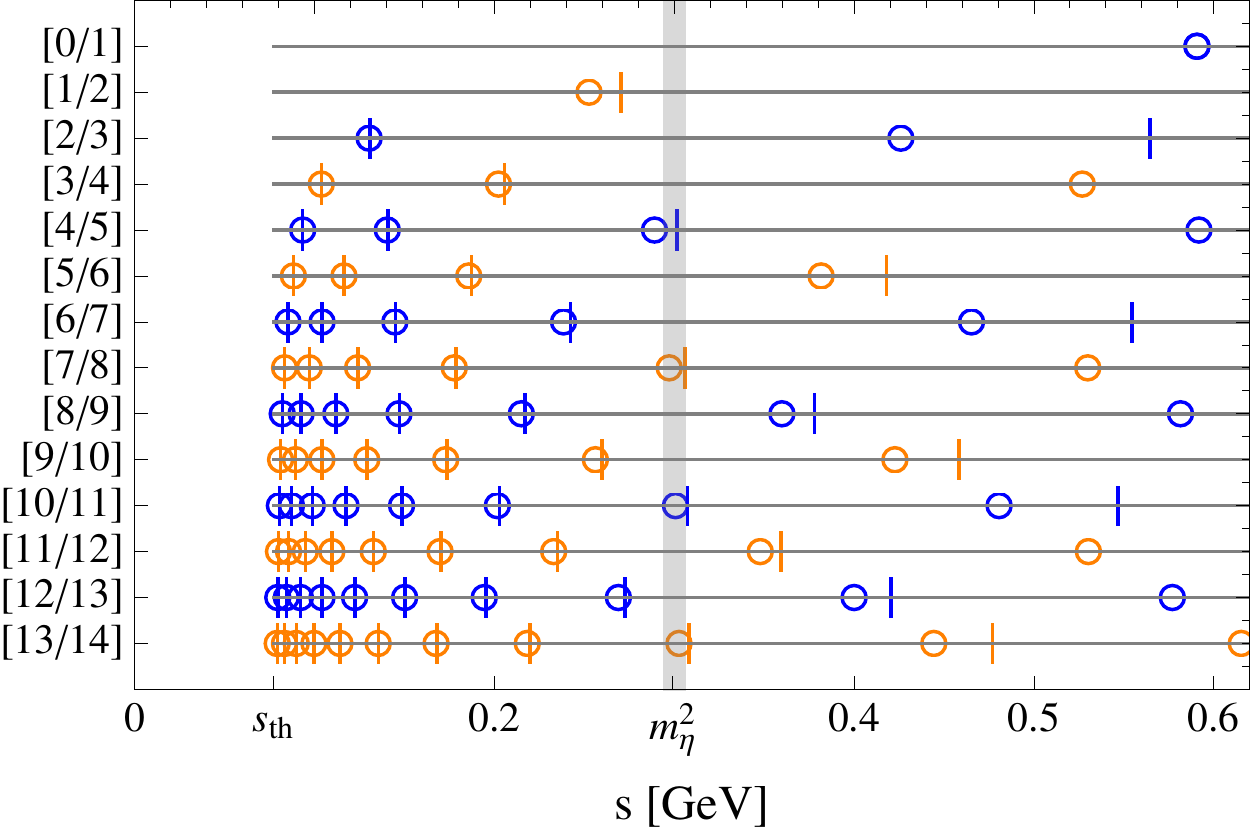}
\end{figure}

\section{Final results}
\label{sec:results}

Having carefully revised all the systematic errors which must be accounted for the $\pi^0, \eta$ and $\eta'$, we are finally in the position to give our final results for these decays.
As it has been explained, given the uncertainty on the TFF double-virtual behavior, we restrict ourselves to the lowest approximant ($C^0_1$) and take the double-virtual 
parameter, $a_{P;1,1}$, in the $b_P^2 \leq a_{P;1,1} \leq 2b_P^2$  range. This choice represents a compromise between the low- and high-energy regimes. Particularly, we 
showed in \cref{sec:errorpi0} that this range provides a band wide enough to cover the full systematic error for the $C^0_1$ for the $\pi^0$ case. 
Regretfully, this did not apply to the $\eta$ and $\eta'$, as they feature additional intermediate states implying a slower convergence. For this reason, we evaluated 
in \cref{tab:c01res} an additional systematic error that should be included on top of the previous band (see \cref{tab:c01res}).
Our final results are obtained through a precise numerical evaluation of \cref{eq:loopamp} ---involving no approximations--- 
using \texttt{FeynCalc}~\cite{Mertig:1990an}\footnote{For the factorization limit, this is possible using partial fraction decomposition and involves one, two and three point 
scalar functions. For the OPE limit, the TFF $\sim \left( k^2 +(q-k)^2 -M^2\right)^{-1}$ can be expressed as $\sim \left((k-q/2)^2 -(M^2/2 -q^2/4)\right)^{-1}/2$ and 
its integration involes one, two, three and four point functions.} and 
\texttt{LoopTools}~\cite{Hahn:1998yk}\footnote{In addition, we checked the results by performing an analytical calculation for the relevant scalar integrals using 
the techniques in Ref.~\cite{'tHooft:1978xw}.}.\\

The results for the $\mathcal{A}(m_P^2)$ amplitude in \cref{eq:loopamp} are displayed in \cref{tab:mainamp} in form of a range  
associated to $a_{P;1,1}=(2b_P^2\div b_P^2)\equiv(\textrm{OPE}\div\textrm{Fact})$.
\begin{table}[t]
\centering
\footnotesize
\begin{tabular}{c  r@{\hspace{2pt}}r@{\hspace{2pt}}c@{}c @{\hspace{2pt}}l    r@{}l    r@{\hspace{2pt}}r@{\hspace{2pt}}c} \toprule
Process & \multicolumn{5}{c}{$\mathcal{A}(m_P^2)$} & \multicolumn{2}{c}{$\mathcal{A}^{Z^0}(m_P^2)$}        &  \multicolumn{3}{c}{$\mathcal{A}^{\textrm{app}}(m_P^2) $}  \\ \midrule
$\pi^0\rightarrow e^+e^-$      &  $(10.00$&$\div$&$10.46)$&$(12) $&$-17.52i$           & $-$&$0.05$      &   $(9.84$&$\div$&$10.30)-17.52i$ \\ \midrule
$\eta\rightarrow e^+e^-$       &  $(30.95$&$\div$&$31.51)$&$(11) $&$-21.92(0)i$        & $-$&$0.03$      & $(27.53$&$\div$&$28.00)-21.92i$   \\ 
$\eta\rightarrow \mu^+\mu^-$   &  $-(1.52$&$\div$&$0.99)$&$(5)   $&$-5.47(0)i$         & $-$&$0.03$      & $-(2.33$&$\div$&$1.87)-5.47i$   \\ \midrule
$\eta^{\prime}\rightarrow e^+e^-$      &  $(47.4$&$\div$&$48.2)$&$(5)  $&$-21.0(5)i$     & & $0.03$      & $(35.20$&$\div$&$35.66)-23.68i$  \\ 
$\eta^{\prime}\rightarrow \mu^+\mu^-$  &  $(2.95$&$\div$&$3.65)$&$(19)  $&$-6.3(17)i$       & & $0.03$   & $-(0.66$&$\div$&$0.20)-7.06i$  \\ \bottomrule
\end{tabular}
\caption{Our results for $a_{P;11}\in(2b_P^2\div b_P^2)\equiv(\textrm{OPE}\div\textrm{Factorization})$. The error is statistical alone. We quote the $Z^0$ boson contribution $\mathcal{A}^{Z^0}(m_P^2)$ separately and show the approximated $\mathcal{A}^{\textrm{app}}(m_P^2)$ result, \cref{eq:approx2}, for comparison.}
\label{tab:mainamp}
\end{table}
%
We include the statistical and systematic error associated to the $b_P$ parameter determination obtained in \cref{chap:data}, \cref{tab:chap1mainres} ---required for the TFF 
reconstruction. In addition, we include  the $Z^0$ boson contribution separately (see details in \cref{sec:np}) and the result that would have been obtained from the approximate 
formula \cref{eq:approx2} in the third and fourth columns, respectively.  We note that employing the approximate result implies non-negligible errors, especially for the 
$\eta'$, as can be seen from the last row in \cref{tab:mainamp} ---the agreement observed for the last row in \cref{tab:mainres}, which does not involve $\mathcal{A}(m_P^2)$ but 
$|\mathcal{A}(m_P^2)|$, is just accidental.\\

The results for the BR, including the  ---often neglected--- $Z^0$ boson contribution, are given in \cref{tab:mainres}.
There, we include three different sources of errors on top of the $a_{P;1,1}$ range. The first one is associated to the 
experimental measurement for $\textrm{BR}(P\rightarrow\gamma\gamma)$ and has been frequently ignored; the second is that arising from the $b_P$ parameter; 
the third one, which applies for the $\eta$ and $\eta'$ alone, corresponds to the additional systematic error associated to the resonant region, see \cref{sec:erroreta}. 
In addition, we show in the third and fourth columns the result that would have been obtained if not including the $Z^0$ boson and using the approximated 
formula \cref{eq:approx2}, respectively.\\

\begin{table}[t]
\footnotesize
\centering
\begin{tabular}{c @{\hspace{6pt}}l@{\hspace{2pt}}c@{\hspace{2pt}}r@{}c @{}c@{}l @{}l @{\hspace{4pt}}r@{\hspace{2pt}}c@{\hspace{2pt}}l   @{\hspace{4pt}}r@{\hspace{2pt}}c@{\hspace{2pt}}l} \toprule
Process & \multicolumn{7}{c}{BR}       & \multicolumn{3}{c}{BR w/$Z^0$}    &        \multicolumn{3}{c}{BR app} \\ \midrule
$\pi^0\rightarrow e^+e^-$  &  $(6.20$&$\div$&$6.35)$&$(0)$&$(4)$&$(0) $&$ \times10^{-8}$  &$(6.22$&$\div$&$6.36) \!\times\!10^{-8}$ & $(6.17$&$\div$&$6.31) \!\times\!10^{-8}$  \\ \midrule
$\eta\rightarrow e^+e^-$  & $(5.31$&$\div$&$5.44)$&$(3)$&$(2)$&$(1) $&$ \times10^{-9}$ & $(5.32$&$ \div $&$5.45) \!\times\!10^{-9}$ & $(4.58 $&$\div$&$ 4.68) \!\times\!10^{-9}$  \\ 
$\eta\rightarrow \mu^+\mu^-$ &  $(4.72$&$\div$&$4.52)$&$(2)$&$(3)$&$(4) $&$ \times10^{-6}$  & $(4.70$&$ \div $&$4.51) \!\times\!10^{-6}$& $(5.16$&$ \div$&$4.88) \!\times\!10^{-6}$  \\ \midrule
$\eta^{\prime}\rightarrow e^+e^-$ & $(1.82$&$\div$&$1.87)$&$(7)$&$(2)$&$(16) $&$ \times10^{-10}$ & $(1.82$&$ \div $&$1.87) \!\times\!10^{-10}$ & $(1.22$&$ \div$&$ 1.24) \!\times\!10^{-10}$ \\ 
$\eta^{\prime}\rightarrow \mu^+\mu^-$  & $(1.36$&$\div$&$1.49)$&$(5)$&$(3)$&$(25) $&$ \times10^{-7}$ & $(1.35$&$ \div$&$ 1.48) \!\times\!10^{-7}$ & $(1.42$&$ \div$&$ 1.41) \!\times\!10^{-7}$  \\ \bottomrule
\end{tabular}
\caption{Our results for $a_{P;11}\in(2b_P^2\div b_P^2)\equiv(\textrm{OPE}\div\textrm{Factorization})$.  The errors refer respectively to those from BR$(P\rightarrow\gamma\gamma)$,  $b_P$ and the systematic one. We compare to the results either neglecting the $Z^0$ boson contribution (BR w/$Z^0$) or using the approximation in \cref{eq:approx2}. 
See details in the text.}
\label{tab:mainres}
\end{table}

\cref{tab:mainamp,tab:mainres} represent the main results from this chapter~\cite{Masjuan:2015lca,Masjuan:2015cjl}. They provide un updated calculation of the SM values 
for \PtoLL decays.As a novelty, they are the first ones making full use of the available data for the relevant TFFs. Moreover, we are the first ones implementing an 
appropriate low-energy description, which is crucial for these processes, as well as an appropriate double-virtual description accounting for the low- 
and high-energy effects, reflected in the given band. Furthermore, we are able to estimate, for the first time, a systematic error, which is by no means negligible 
for the $\eta$ and, specially, the $\eta'$, and has been previously overlooked. In addition, our calculation does not involve numerical approximations when calculating the loop 
integral, which for the $\eta$ and $\eta'$ becomes a large effect as can be inferred from \cref{tab:mainamp,tab:mainres}. As a result, we find that the lepton mass corrections 
neglected in \cite{Dorokhov:2009xs} are by no means negligible for the muonic channels at the precision we are aiming.\\

From the quoted results, we find that the main source of error for the $\pi^0$ and $\eta$  is the double-virtual description, which could be dramatically improved by constructing the 
$C^1_2$ approximant. This would be possible if having double-virtual experimental data or some additional constraints. Still, we emphasize that the current value is already 
below the experimental uncertainties and improves previous estimates, see \cref{tab:finalexp}.
Concerning the $\eta'$, the major source of errors comes from the systematic uncertainty associated to threshold and resonance effects (we note though that this error 
is likely to be overestimated, as the model employed does not provide a realistic SL description) that could be partially improved, again, if reaching the $C^1_2$ 
approximant. Investigations in this respect are undergoing\footnote{For further details on the $C^1_2$, see \cref{chap:gm2}.}. 
Still it would be desirable to have an alternative approach to systematically implement not only the low- and high-energy behaviors, but the information about the time-like region, 
such as physical resonances and threshold discontinuities.\\

Finally, we compare to the experimental available results in \cref{tab:finalexp}. For the ease of comparison, we take the middle value from \cref{tab:mainres} and include the
(OPE$\div$Fact) range as an additional source of error (see comments in \cref{tab:finalexp}).
\begin{table}[t]
\footnotesize
\centering
\begin{tabular}{cccc} \toprule
BR & This Work & Exp. & Previous SM~\cite{Dorokhov:2009jd} \\ \midrule
$\textrm{BR}(\pi^0\to e^+e^-)\times10^{8}$     & $6.28(7)(4)[8]$  & $7.48(38)$~\cite{Abouzaid:2006kk} &  $6.23(12)\to6.26$ \\ \midrule
$\textrm{BR}(\eta\to e^+e^-)\times10^{9}$     & $5.38(6)(4)[7]$  & $\leq2.3\times10^3$~\cite{Agakishiev:2013fwl} &  $4.53(9)\to5.19$ \\ 
$\textrm{BR}(\eta\to \mu^+\mu^-)\times10^{6}$     & $4.62(10)(5)[11]$  & $5.8(8)$~\cite{Agashe:2014kda,Abegg:1994wx} &  $5.35(27)\to4.76$ \\ \midrule
$\textrm{BR}(\eta'\to e^+e^-)\times10^{10}$     & $1.85(2)(18)[18]$  & $\leq56$~\cite{Akhmetshin:2014hxv,Achasov:2015mek} &  $1.182(14)\to1.83$ \\ 
$\textrm{BR}(\eta'\to \mu^+\mu^-)\times10^{7}$     & $1.42(7)(26)[27]$  & --- &  $1.364(10)\to1.24$ \\ \bottomrule
\end{tabular}
\caption{Our final results for the BRs as compared to the available experimental measurements. The first error gives the chosen (OPE$\div$Fact) band; the second 
         one is the combined error from \cref{tab:mainres}; the third error, in brackets, is the combination of them. For completeness, we give the commonly 
         quoted as SM values: the first value with the error uses an approximate calculation, whereas the second one implements certain corrections.\label{tab:finalexp}}
\end{table}
\begin{itemize}
\item For $\pi^0\to e^+e^-$, the most recent result ---dominating the current PDG~\cite{Agashe:2014kda} value--- comes from KTeV Collaboration~\cite{Abouzaid:2006kk} 
and implies a $3\sigma$ deviation from our theoretical result. 
Such value is extracted from $\textrm{BR}(\pi^0\to e^+ e^-, x_D>0.95)=6.44(25)(22)\times10^{-8}$~\cite{Abouzaid:2006kk}, where $x_D=m_{e^+e^-}^2/m_{\pi}^2$ and 
the first(second) error is statistical(systematic). Accounting for the radiative corrections (RC) in \cite{Bergstrom:1982wk} and extrapolating to $x_D=1$, they obtain 
$\textrm{BR}(\pi^0\to e^+e^-)=7.48(29)(25)\times10^{-8}$. 
As a result of the discrepancy, the authors in Ref.~\cite{Vasko:2011pi} have performed a full two-loop evaluation of the RC with the Bremsstrahlung diagrams evaluated in the 
soft-photon approximation. There, the authors noticed that the previous estimate~\cite{Bergstrom:1982wk} neglected a class of subleading diagrams, which due to partial 
cancelations among the leading ones, turned out to be dominant and reduced the size of the RC from $14\%$~\cite{Bergstrom:1982wk} down to $6\%$~\cite{Vasko:2011pi}. 
Finally, the authors in~\cite{Husek:2014tna} have performed the exact calculation for the Bremsstrahlung diagrams, confirming the goodness of the soft photon approximation
and closing the exact full two-loop evaluation of the RC. The work from Refs.~\cite{Vasko:2011pi,Husek:2014tna} suggest then 
$\textrm{BR}(\pi^0\to e^+e^-)=6.87(36)\times10^{-8}$, $1.5\sigma$ away from our result.
Still, this discrepancy is hard to be explained within QCD, as it would require 
an extremely damped TFF at very low-energies, implying an unexpected TFF behavior as well as a slow convergence for the OPE expansion\footnote{More details in \cref{chap:gm2}.}. 
In light of this result, it is tempting to discuss about new physics scenarios, a debate to which we come back in \cref{sec:np}.\\

\item For the $\eta\to\mu^+\mu^-$, we observe an interesting discrepancy with respect to the experimental result, which corresponds to a $2\sigma$ deviation  
---note that this discrepancy would disappear if we would have used the approximated result \cref{eq:approx2} instead. Still, the experimental accuracy prevents 
us from drawing any conclusion. For this, a new preciser experiment would be desired. In this respect, there exists the possibility 
that such decay could be measured at the LHCb~\cite{Huong:2016gob} Collaboration. Amusingly, if the discrepancy were to be explained on 
QCD grounds, this time we would require a flatter TFF, contrary to the $\pi^0$ case, which represents an intriguing situation. A similar situation is found 
when looking at possible new physics scenarios, a discussion to which we come back in \cref{sec:np}.\\

\item Finally, we turn our attention to the $\eta'$ decays. At present, only a recent upper bound exists for the $\eta'\to e^+e^-$ channel from 
VEPP-2000 at Novosibirsk~\cite{Akhmetshin:2014hxv,Achasov:2015mek}, which improves the previous one by two orders of magnitude~\cite{Agashe:2014kda}, but is 
still two orders of magnitude above our prediction. In the future, it may be possible as well to find the first signal for the $\eta'\to \mu^+\mu^-$ channel at LHCb~\cite{Huong:2016gob}. 
\end{itemize}

\section{Implications for \cpt}
\label{sec:pllcpt}

At LO, the \cpt prediction involves two different contributions. The first one is obtained when replacing the LO \cpt result for the TFF, this is, 
the constant WZW term (left diagram in \cref{fig:pllcpt}). The second one is the counterterm required to regularize the 
divergent integral and is obtained from the following lagrangian~\cite{Savage:1992ac,GomezDumm:1998gw,Knecht:1999gb}
\begin{equation}
\label{eq:ChPTct}
\frac{3i\alpha^2}{32\pi^2} \bar{\ell}\gamma^{\mu}\gamma_5\ell \left[  \chi_1\operatorname{tr}\! \left(\! \mathcal{Q}^2\!\left\{U^{\dagger},\partial_{\mu}U \right\} \!\right)  + 
\chi_2\operatorname{tr}\! \left(\! \mathcal{Q}U^{\dagger}\mathcal{Q}\partial_{\mu}U  -   \mathcal{Q}\partial_{\mu}U^{\dagger}\mathcal{Q}U \!\right) \right],
\end{equation}
where $\mathcal{Q}$ stands for the charge matrix. Following the definitions in \cref{app:conv}, the leading term, depicted in the right diagram from \cref{fig:pllcpt}, 
yields
\begin{equation}
i\mathcal{M} = 2\sqrt{2}m_{\ell}m_P\alpha^2 F_{P\gamma\gamma}\chi(\mu),
\end{equation}
where $\chi(\mu) \equiv -(\chi_1(\mu) + \chi_2(\mu))/4$ is the (scale-dependent) counterterm and the TFF result is to be taken from the LO piece in \cref{sec:tffcpt}.
\begin{figure}[t]
\centering
   \includegraphics[width=0.8\textwidth]{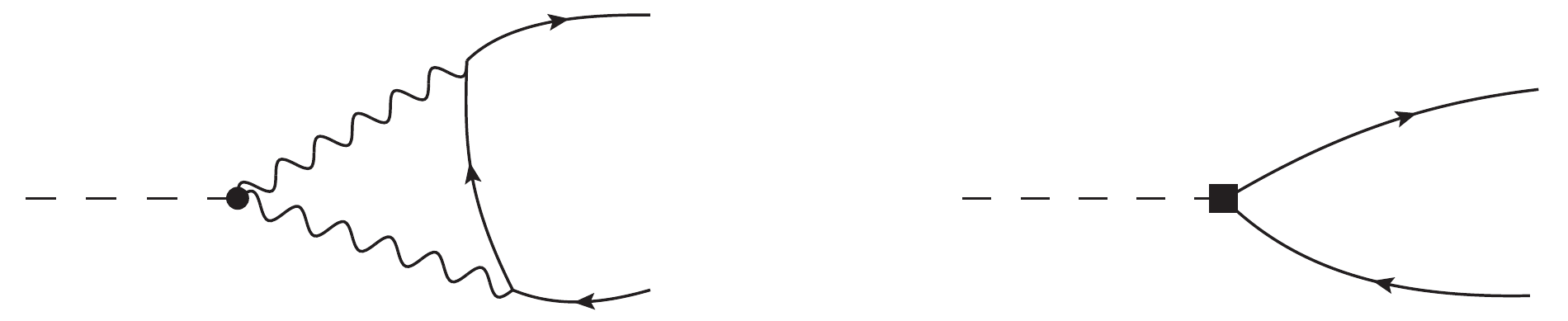}
\caption{The LO contributions to the \PtoLL process in \cpt. The diagram on the left stands for the WZW (constant) TFF. The one on the right is the required counterterm.\label{fig:pllcpt}}
\end{figure}
Recalling our result for a constant TFF, \cref{eq:cnstTFF}, the LO \cpt prediction reads
\begin{equation}
\mathcal{A}^{\textrm{LO}}(q^2) = \frac{i\pi}{2\beta_{\ell}}L + \frac{1}{\beta_{\ell}}\left[  \frac{1}{4}L^2 + \frac{\pi^2}{12} +\textrm{Li}_2\left( \frac{\beta_{\ell}-1}{1+\beta_{\ell}} \right)  \right] 
 -\frac{5}{2} +\frac{3}{2}\ln\left( \frac{m_{\ell}^2}{\mu^2} \right) + \chi(\mu). \label{eq:PllChPT}
\end{equation}
As we see, there exists at this order a single available term, $\chi(\mu)$, to determine all the \PtoLL processes, including $\pi^0,\eta, \eta'$ as well as  $\ell=e,\mu$ final states. 
This approach benefits from being rigorous, model-independent  and very predictive at the leading order. Regretfully, we find that large corrections are expected to arise 
at higher orders, requiring a NLO calculation with its consequent loss of predictiveness as the number of counterterms increases.\\

To illustrate this, we obtain for each particular decay the associated counterterm $\chi(\mu)$ which is required to reproduce our results from \cref{tab:mainamp}. This is, we 
subtract \cref{eq:PllChPT} from our results. The obtained values are shown in the first row from \cref{tab:chpt}. The large $\chi(\mu)$ variations which arise when comparing 
different channels indicates the relevance of NLO corrections and the danger of fixing some counterterm from a particular channel to predict the others.
\begin{table}[t]
\centering
\scriptsize
\begin{tabular}{cc@{\hspace{7pt}}c@{\hspace{7pt}}c@{\hspace{7pt}}l@{\hspace{7pt}}c} \toprule
                                   & $\pi^0\rightarrow e^+e^-$ & $\eta\rightarrow e^+e^-$ & $\eta\rightarrow \mu^+\mu^-$ & \multicolumn{1}{c}{$\eta^{\prime}\rightarrow e^+e^-$} & $\eta^{\prime}\rightarrow \mu^+\mu^-$ \\ \midrule
    $\chi(\mu)$             & $(2.53\!\div\!2.99)$ & $(5.90\!\div\!6.46)$ & $(3.29\!\div\!3.82)$ & $(14.2\!\div\!14.9) + 2.52i$ & $(5.61\!\div\!6.31)+0.75i$ \\ 
$\chi(\mu)_{m_{\pi}}$ &  $(2.53\!\div\!2.99)$ & $(2.66\!\div\!3.12)$ & $-$ & $(2.16\!\div\!2.62)$ & $-$ \\  
    $\chi(\mu)_{UV}$ &  $(2.53\!\div\!2.99)$ & $(5.50\!\div\!6.05)$ & $(3.11\!\div\!3.64)$ & $(16.8\!\div\!17.7) + 7.09i$ & $(6.56\!\div\!7.35)+2.12i$ \\  \bottomrule
\end{tabular}
\caption{Our equivalent $\chi$PT counter-term $\chi(\mu)$ together with its equal-mass version, $\chi(\mu)_{m_{\pi}}$, and the $U(3)_F$-symmetric TFFs version, $\chi(\mu)_{UV}$. Results for 
$\mu=0.77$~GeV.}
\label{tab:chpt}
\end{table}

In the following, we investigate the sources of these differences and identify which are the most relevant effects. On the one hand, there is a clear difference among each of 
the pseudoscalars which should arise from $U(3)_F$-breaking effects. As a first step, we take all the masses to be equal to the $\pi^0$ and calculate again 
the results from  \cref{tab:mainamp}. Subtracting \cref{eq:PllChPT}, we obtain the results in the $\chi(\mu)_{m_{\pi}}$ row  from \cref{tab:chpt}, which represent a large effect. 
All the remaining differences in this row arise from $U(3)$-breaking effects in the TFFs alone (i.e., $b_{\pi}\neq b_{\eta}\neq b_{\eta'}$). These are less pronounced as may be inferred 
from the $\chi(\mu)_{UV}$ row in \cref{tab:chpt}, where we recalculate $\chi(\mu)$ for the case in which all the TFFs are equal to that of the $\pi^0$, but the pseudoscalar masses 
are the physical ones. On the other hand,  there is a notorious impact among the different leptonic channels, which is clear when comparing the $\eta$ and $\eta'$ electronic channels 
against the muonic ones.\\

All these effects will be generated at higher orders in \cpt as one obtains a $q^2$-dependent TFF introducing some hadronic scale. This will generate 
additional $m_{P,\ell}^2/\Lambda^2$ corrections explaining the observed differences, which  will be further commented in \cref{sec:cptcorr}. As an illustration, this 
is the only way to generate an imaginary part for the $\eta'$, to be associated with the pion loop in \cref{fig:UB}, left.

\subsection{The $\pi^0$-exchange contribution to the $2S$ hyperfine- \\ splitting in the muonic hydrogen}
\label{sec:lambshift}

The results collected in \cref{tab:chpt}, first row, are also relevant for calculating the $\pi^0$ {\textit{pole}}-contribution to the $2S$ hyperfine-splitting in the muonic 
hydrogen~\cite{Huong:2015naj,Hagelstein:2015egb,Zhou:2015bea} ($\Delta E_{\textrm{HFS}}^{\pi}$). Such calculation can be performed within $\chi$PT, which involves 
again \cref{eq:PllChPT}. However, the kinematics of the process involves a vanishingly small $Q^2$ space-like momentum for the $\pi^0$, since its contribution to the $2S$ 
hyperfine-splitting appears in the \textit{t}-channel. As such, it is $\mathcal{A}(Q^2\simeq0)$ instead of $\mathcal{A}(m_{\pi}^2)$ which is relevant now~\cite{Huong:2015naj}, 
shifting the values obtained in \cref{tab:chpt}. To illustrate this, we recalculate $\mathcal{A}(0)$ from \cref{eq:loopamp} taking the limit $Q^2 \to 0$, and obtain the 
new subtraction constant which should be used in \cref{eq:PllChPT} to reproduce our results. We obtain
\begin{equation}
\chi^{ee}_{\pi^0}(\mu)=(2.37\div2.83)
\label{eq:HFSe}
\end{equation}
for $\ell=e$, which is smaller than its counterpart collected in \cref{tab:chpt}. However, for the $2S$ hyperfine-splitting in muonic hydrogen what is needed is the coupling 
to muons ($\ell=\mu$). In that case, we obtain
%
\begin{equation}
\chi^{\mu\mu}_{\pi^0}(\mu)=(2.18\div2.63),
\label{eq:HFSmu}
\end{equation}
which is even lower than \cref{eq:HFSe}. Note that the shift is of the order of the uncertainties quoted in \cref{tab:chpt} and arises again from the full $q^2$ and 
$m_{\ell}^2$ dependence in \cref{eq:loopamp}, which is not accounted for at LO in $\chi$PT. 
To close the discussion, we note that it was pointed out for the first time in Ref.~\cite{Hagelstein:2015egb} that, to obtain the $\pi^0$ contribution to 
$\Delta E_{\textrm{HFS}}^{\pi}$, it was necessary to account for the full $q^2$-dependency of $\mathcal{A}(q^2)$ given its non-analytic behavior at $q^2=0$. Consequently, and 
for the sake of completeness, we quote what would be obtained in such case using our exact $\mathcal{A}(q^2)$ numerical result. 
This can be calculated through Eq.~(37) and Eq.~(39) in Ref.~\cite{Huong:2015naj} and leads\footnote{In our approach, we take the $\Lambda_{\pi}\to\infty$ limit 
in Ref.~\cite{Huong:2015naj}, which corresponds with the treatment in~\cite{Hagelstein:2015egb} and corresponds to the $\pi^0$-pole.} 
\begin{equation}
	\Delta E_{\textrm{HFS}}^{\pi} = -(0.13\div0.12)~\mu\textrm{eV},
\end{equation}
where the uncertainties from the slope and TFF normalization can be neglected against the dominating one, that arises from the chosen (OPE$\div$Fact) range chosen for the 
double-virtual parameter.  
We note that the connection between the $\chi^{ee}_{\pi^0}(\mu)$ in 
\cref{eq:HFSe} and $\chi^{\mu\mu}_{\pi^0}(\mu)$ in \cref{eq:HFSmu} and that extracted from the experimental results is non-trivial as it is TFF dependent. In quoting 
our results, we implicitly assume that there is no new-physics contribution. However, if the current discrepancies among theory and experiment persists, indicating new physics 
contribution ---which we will discuss in \cref{sec:np}--- the connection between the experimental $\chi(\mu)$ and that in \cref{eq:HFSe,eq:HFSmu} will 
depend on the particular new-physics scenario and will have to be reanalyzed.

The results above are illustrative as well regarding $(g-2)_{\mu}$ hadronic contributions, which in $\chi$PT involve $\chi(\mu)$ together with an additional counterterm, $C(\mu)$, as 
an input~\cite{RamseyMusolf:2002cy}. If we were able to determine $C(\mu)$ somehow, from $(g-2)_{e}$ for example, and $\chi(\mu)$ would be taken from the experimental 
$\pi^0\rightarrow e^+e^-$  result, extrapolating up to the $\mu$ case may imply a non-negligible error as illustrated above; similar effects may arise for $C(\mu)$ itself too.

\subsection{Corrections}
\label{sec:cptcorr}

As discussed above, the precision which is reached at the LO in \cpt for processes involving a $P\bar{\ell}\ell$ vertex may not be enough ---a feature which 
manifests when comparing the same process for a different $\ell=e,\mu$ channel. This suggests to look at the next to lading order. 
In this respect, \cpt would yield a power series expansion for the TFF\footnote{For simplicity, we have assumed a single scale for the TFF inspired in typical VMD 
models. Note that logarithmic terms coming from loops are of course present too. However, they are subleading as compared to the power expansion and may be 
Taylor expanded for the $\pi^0$ and $\eta$ cases.} 
\begin{equation}
\label{eq:cpttff}
\tilde{F}_{P\gamma^*\gamma^*}(q_1^2,q_2^2) =  \underbrace{\phantom{\frac{1}{\Lambda^2}} \! \! \! \! \! \! \! \! \! 1}_{\textrm{LO}}
                                             +   \underbrace{ \frac{1}{\Lambda^2}(q_1^2 + q_2^2) }_{\textrm{NLO}} + 
                                              \underbrace{ \frac{1}{\Lambda^4}(q_1^4 + q_2^4) +   \frac{1}{\Lambda^4}(q_1^2q_2^2)  }_{\textrm{NNLO}} + \mathcal{O}\left(\frac{q^6}{\Lambda^6}\right)  .
\end{equation}
Then, we could calculate the result of \cref{eq:loopamp} for the TFF in \cref{eq:cpttff}, 
\begin{align}
\mathcal{A}(q^2,m_{\ell}^2) &= \frac{2i}{\pi^2q^2}  \int  d^4k \frac{\left(k^2q^2-(k\cdot q)^2\right)}{k^2(q-k)^2\left((p-k)^2-m_{\ell}^2\right)}  \left[ 1 + \frac{(...)}{\Lambda^2}+ \frac{(...)}{\Lambda^4} + ...  \right] \nonumber \\
 \label{eq:CptExp}                           &\equiv   \mathcal{A}^{\textrm{LO}}(q^2,m_{\ell}^2)   +    \mathcal{A}^{\textrm{NLO}}(q^2,m_{\ell}^2)  +  \mathcal{A}^{\textrm{NNLO}}(q^2,m_{\ell}^2) + ... \ ,
\end{align}
where $\mathcal{A}^{\textrm{LO}}(q^2)$ has been given in \cref{eq:PllChPT} and 
\begin{align}
%
%
\mathcal{A}^{\textrm{NLO}}(q^2,m_{\ell}^2)  \! &=  \frac{1}{3\Lambda^2}(q^2-10m_{\ell}^2)\left( 1 -L_{\ell}  \right) +\frac{1}{9\Lambda^2}(4m_{\ell}^2 - q^2) , \\
\mathcal{A}^{\textrm{NNLO}}(q^2,m_{\ell}^2)  \! &= \!  \! \left[\frac{126m_{\ell}^4 \! -  \! q^4  \! -  \! 8m_{\ell}^2q^2 }{12\Lambda^4}L_{\ell}  
 + \frac{26m_{\ell}^2q^2  \! +  \! 7q^4  \!-  \! 702m_{\ell}^4}{72\Lambda^4} \right]  \!    . 
\end{align}
where $L_{\ell} = \ln(m_{\ell}^2/\Lambda^2)$.
We notice that the LO leading logs $L_{\ell}$ correspond ---not surprisingly as they arise from a power-like expansion as well--- 
to the corrections found in~\cite{Dorokhov:2008cd,Dorokhov:2009xs} if $\Lambda$ is taken as the VMD scale. We adopt then a more modest approach and retain the leading 
logs alone, which represents a good approximation. This would produce a straightforward generalization to higher orders as well as a tool to estimate the convergence of the 
chiral expansion. Of particular relevance is the difference $\mathcal{A}(q^2,m_{\mu}^2)-\mathcal{A}(q^2,m_{e}^2)$, where one expects a better convergence for 
\cref{eq:CptExp} due to partial cancellations. Taking into account the smallness of the lepton masses, we find that such a shift is given, to a reasonable accuracy, as
\begin{multline}
\mathcal{A}(q^2,m_{e}^2)-\mathcal{A}(q^2,m_{\mu}^2) = \mathcal{A}^{\textrm{LO}}(q^2,m_{e}^2)-\mathcal{A}^{\textrm{LO}}(q^2,m_{\mu}^2) \\
  + \frac{q^2}{3\Lambda^2}\left( 1+ \frac{q^2}{4\Lambda^2} \right)  \ln\left( \frac{m_{\mu}^2}{m_e^2} \right) 
   + \frac{10m_{\mu}^2}{3\Lambda^2}\ln\left( \frac{\Lambda^2}{m_{\mu}^2} \right). \label{eq:NNLOcorr}
\end{multline} 
Whereas our theoretical results for the leptonic and muonic channels in \cref{tab:mainamp} could not be reproduced at LO with an unique counterterm, the 
observed differences in \cref{tab:chpt} \cref{sec:lambshift} can be easily accounted for, to a good approximation, taking into account the additional  terms in \cref{eq:NNLOcorr} 
---an exception is the $\eta'$ case, for which the pion loops cannot be neglected in order to extract an imaginary part.
The expansion above, \cref{eq:NNLOcorr}, proves extremely useful to relate different leptonic channels, which is not only relevant in the cases discussed above 
but  for \cpt studies on lepton flavor violation in $K_L\to \bar{\ell}\ell$ decays~\cite{Crivellin:2016vjc}.

\section{Implications for new physics contributions}
\label{sec:np}

\subsection{Generic new physics scenarios}

Given the current puzzles existing in the low-energy precision frontier of particle physics--- 
specifically, the long standing discrepancy among the electron and muon anomalous magnetic moments~\cite{Agashe:2014kda,Jegerlehner:2009ry}, and the most 
recent proton radius puzzle coming from the different values obtained from electronic- and muonic-hydrogen experiments~\cite{Antognini:1900ns}, 
together with $\mathcal{R}_K$ and $\mathcal{R}_{D^{(*)}}$~\cite{Alonso:2015sja} from $B$-decays--- where lepton universality seems to fail 
contrary to what is expected in the standard model, it would be very interesting to study whether similar puzzles appear in the processes discussed here as well.
Having updated the SM values for \PtoLL decays with careful account of systematic errors, we discuss possible new physics (NP) contributions, specially given the 
current discrepancies in the two existing measured decays. As it is explained in \cref{sec:fierz}, 
any additional contribution ---such as leptoquark-like--- will always manifest, after Fierz-rearrangement, only through effective 
pseudoscalar $(\mathcal{P})$ and axial $(A)$ contributions which, given the existing well-motivated models~\cite{Davoudiasl:2012ag,Davoudiasl:2012qa,Chang:2008np}, 
are conveniently expressed using the effective Lagrangian
\begin{equation}\nonumber
 \label{eq:NPL}
\mathcal{L} = \frac{g}{4m_W} \sum_{f} m_Ac^{A}_{f} \left(\overline{f}\slashed{A}\gamma_5f \right) + 2 m_f c^{\mathcal{P}}_{f} \left(\overline{f}i\gamma_5 f \right)\mathcal{P}, 
\end{equation}
where $g, m_W$ are the standard electroweak parameters, and $c^{A,\mathcal{P}}_{f}$ are dimensionless couplings to the fermions $f=\{u,d,s,e,\mu\}$.
These interactions yield additional tree-level contributions as shown in \cref{fig:np}.
\begin{figure}[t]
   \includegraphics[width=\textwidth]{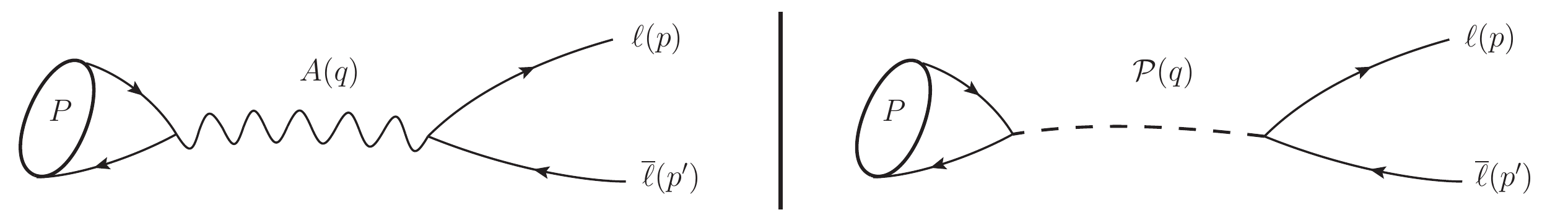}
   \caption{Left(right): additional tree level contributions from an axial(pseudoscalar) field. The $P$ 
                 stands for the pseudoscalar meson; $A(\mathcal{P})$ stands for the axial(pseudoscalar) field with momentum $q$; $\ell (\bar{\ell})$ for the (anti)lepton with momentum 
                 $p (p')$.\label{fig:np}}
\end{figure}
Their corresponding amplitudes (see \cref{app:conv}) read
\begin{align}
i\mathcal{M} = &\frac{igc^A_{\ell}m_A}{4m_W}[\overline{u}_{p,s}\gamma_{\mu}\gamma_5v_{p',s'}] \frac{-i\left(g_{\mu\nu} - \frac{q_{\mu}q_{\nu}}{m_A^2}\right)}{m_P^2-m_A^2}   \frac{igm_A}{4m_W} \overbrace{\sum_q \bra{0} c^A_q\overline{q}\gamma^{\mu}\gamma_5q\ket{P(q)}}^{\bra{0} J_{\mu5}^{\textrm{NP}}\ket{P(q)}}, \label{eq:npAX}\\
i\mathcal{M} = &\frac{i g c_{\ell}^{\mathcal{P}}}{2m_W} m_{\ell} [\overline{u}_{p,s}i\gamma_5 v_{p',s'}] \frac{i}{m_P^2-m_{\mathcal{P}}^2} \frac{i g}{2m_W} 
\overbrace{ \sum_q \bra{0}c^{\mathcal{P}}_qm_q\overline{q}i\gamma_5q\ket{P(q)} }^{\bra{0} \mathcal{P}^{\textrm{NP}} \ket{P(q)} }, \label{eq:npPS}
\end{align}
for the axial and pseudoscalar contribution, respectively.
In order to relate the hadronic matrix element $\bra{0} J_{\mu5}^{\textrm{NP}}\ket{P(q)}$ in \cref{eq:npAX} to the pseudoscalar decay constants
\begin{equation}
\label{eq:axdef}
  \bra{0} J_{\mu5}^a \ket{P(q)} \equiv i q_{\mu}F_P^a, \qquad J_{\mu5}^a = \overline{q} \gamma_{\mu}\gamma_5 \frac{\lambda^a}{2} q, \quad q=(u,d,s)^T,
\end{equation}
we re-express $J_{\mu5}^{\textrm{NP}}$ ---as defined in \cref{eq:npAX}--- in terms of the 
$U(3)_F$ axial current in \cref{eq:axdef}\footnote{In the flavor basis $\lambda^8$ and $\lambda^0$ can be traded for $\lambda^q=\textrm{diag}(1,1,0)$ and 
$\lambda^s=\textrm{diag}(0,0,\sqrt{2})$, see \cref{chap:mixing}. An analogous procedure aplies then.}. 
For that, we use the relation $\textrm{Tr}(\lambda^a\lambda^b)=2\delta^{ab}$,  whereby we obtain
\begin{align}
\bra{0} J^{\textrm{NP}}_{\mu5} \ket{P}   = & \bra{0} \sum_a \textrm{Tr}(J^{\textrm{NP}}_{\mu5}\lambda^a)J_{\mu5}^a \ket{P} \nonumber \\
= & \sum_a \textrm{Tr}(\textrm{diag}(c^A_u,c^A_d,c^A_s)\lambda^a) \bra{0}J_{\mu5}^a\ket{P}. \label{eq:axdec}
\end{align}
Then, from \cref{eq:axdec,eq:axdef,eq:axdec} and the equations of motion\footnote{For the spinors, these imply $\overline{u}_{p,s} \slashed{p} = - \slashed{p}'v_{p',s'}=m_{\ell}$, 
see see \cref{app:conv}.}, \cref{eq:npAX} can be expressed as
\begin{align}
i\mathcal{M} = & -ic_{\ell}^A\frac{g^2}{16m_W^2} 2m_{\ell}(\overline{u}_{p,s}i\gamma_5v_{p',s'})\sum_a\textrm{Tr}(J^{\textrm{NP}}_{\mu5}\lambda^a)F_P^a \nonumber \\
                     = & -ic_{\ell}^A\frac{G_F}{\sqrt{2}}m_{\ell}(\overline{u}_{p,s}i\gamma_5v_{p',s'})\sum_a\textrm{Tr}(J^{\textrm{NP}}_{\mu5}\lambda^a)F_P^a \nonumber \\
                     = & c^A_{\ell}m_{\ell}m_PG_F\sum_a\textrm{Tr}(J^{\textrm{NP}}_{\mu5}\lambda^a)F_P^a ,
\end{align}
where in the second and third lines we have used $G_F=g^2/(4\sqrt{2}m_W^2)$ and the projector in \cref{eq:proj}, respectively. 
This produces an effective additional contribution to the $\mathcal{A}(q^2)$ loop amplitude in \cref{eq:loopamp},
\begin{equation}
\label{eq:axfinal}
\mathcal{A}(q^2) \rightarrow \mathcal{A}(q^2) + \frac{\sqrt{2}G_F}{4\alpha^2F_{P\gamma\gamma}} c^A_{\ell}\sum_a\textrm{Tr}(J^{\textrm{NP}}_{\mu5}\lambda^a)F_P^a.
\end{equation}
As an example, the $Z^0$ boson contribution is obtained after taking $c^Z_{u}=-c^Z_{d,s,e,\mu}=1$, leading for $P=\{\pi^0,\eta,\eta'\}$ 
\begin{equation}\nonumber
\mathcal{A}(q^2) \rightarrow \mathcal{A}(q^2) - \frac{2\sqrt{2}G_FF_{\pi}}{4\alpha^2F_{P\gamma\gamma}}\left\{1, \frac{F_{\eta}^8}{\sqrt{3}F_{\pi}} - \frac{F_{\eta}^0}{\sqrt{6}F_{\pi}} ,  \frac{F_{\eta^{\prime}}^8}{\sqrt{3}F_{\pi}} - \frac{F_{\eta^{\prime}}^0}{\sqrt{6}F_{\pi}}\right\} .
\end{equation}
Alternatively, we could have used the flavor basis instead, then
\begin{equation}\nonumber
\mathcal{A}(q^2) \rightarrow \mathcal{A}(q^2) - \frac{2\sqrt{2}G_FF_{\pi}}{4\alpha^2F_{P\gamma\gamma}}\left\{1, -\frac{F_{\eta}^s}{\sqrt{2}F_{\pi}} , -\frac{F_{\eta^{\prime}}^s}{\sqrt{2}F_{\pi}}\right\} .
\end{equation}
For the pseudoscalar contribution, the $\bra{0} \mathcal{P}^{\textrm{NP}} \ket{P(q)} $ hadronic matrix element determination in \cref{eq:npPS} is more involved whenever 
the singlet component appears, which is the case for the $\eta$ and $\eta^{\prime}$. To illustrate this, we outline its LO calculation in \cpt, which amounts to retain  
the leading term from the LO lagrangian\footnote{This is, $\frac{F^2}{4}(\chi^{\dagger}U + U^{\dagger}\chi).$} arising from the 
interaction between the pseudoscalar field $P$ and the pseudoscalar current  $\mathcal{P}$ defined in \cpt from the building block $\chi \equiv 2 Bi \mathcal{P}$. 
Then, in the presence of new physics of pseudoscalar type, $\chi \rightarrow2Bi \mathcal{P}^{\textrm{NP}}$. For the $\pi^0$ such term corresponds to 
\begin{equation}\nonumber
 F B\hat{m}(c_u^{\mathcal{P}} - c_d^{\mathcal{P}})\mathcal{P}^{\textrm{NP}}\pi^0, 
\end{equation}
from which the matrix element reads ($2B \hat{m} =  m_\pi^2$)
\begin{equation} \nonumber
\label{eq:pspion}
\bra{0} \mathcal{P}^{\textrm{NP}}\ket{\pi^0}=  F B\hat{m}(c_u^{\mathcal{P}} - c_d^{\mathcal{P}}) = \frac{F_{\pi}}{2}m_{\pi}^2(c_u^{\mathcal{P}} - c_d^{\mathcal{P}}).
\end{equation}
where the LO results $F_{\pi}=F$ and $m_{\pi}=2B\hat{m}$ have been used, see \cref{eq:cptmass}. For the $\eta$ and $\eta'$, it gets more involved. 
The analogous  LO term in the effective lagrangian contributing to the matrix element reads now
\begin{equation}
\label{eq:psmxel1}
F B \mathcal{P}^{\textrm{NP}}\left(\frac{1}{\sqrt{3}}\left( \hat{m}(c^{\mathcal{P}}_u+c^{\mathcal{P}}_d) - 2c^{\mathcal{P}}_s m_s \right)\eta_8 + \sqrt{\frac{2}{3}}\left( \hat{m}(c^{\mathcal{P}}_u+c^{\mathcal{P}}_d) + c^{\mathcal{P}}_s m_s \right)\eta_1 \right).
\end{equation}
After relabeling, introducing $g_8\equiv (c^{\mathcal{P}}_u+c^{\mathcal{P}}_d-2c^{\mathcal{P}}_s)/\sqrt{3}$ and $g_0\equiv\sqrt{2}(c^{\mathcal{P}}_u+c^{\mathcal{P}}_d+c^{\mathcal{P}}_s)/\sqrt{3}$, together with the definitions in \cref{chap:mixing}, \cref{eq:psmxel1} reads
\begin{equation}\nonumber
\frac{F_0}{2}\left( \eta_8(g_8M_8^2+g_0M_{80}^2) + \eta_1(g_8M_{80}^2+g_0M_0^2) \right)\mathcal{P}^{\textrm{NP}},
\end{equation}
Finally, using the $\eta-\eta^{\prime}$ masses, mixing and decay constants at LO\footnote{At LO, $\eta_{8(0)}=\eta(\eta')\cos\theta_P \pm \eta'(\eta)\sin\theta_P$, 
$\theta_P=-19.6^{\circ}$ and the decay constants read $F_{\eta}^8 = F_0 \cos\theta_P, F_{\eta'}^8 = F_0 \sin\theta_P, F_{\eta}^0 = -F_0\sin\theta_P, F_{\eta'}^0 = F_0 \cos\theta_P$. 
In addition $\tan\theta_P = M_{80}^2(M_0^2\!+\!M_{\tau}^2\!-\!M_{\eta}^2)^{-1}= M_{80}^2(M_{\eta'}^2\!-\!M_8^2)^{-1} = (M_8^2\!-\!M_{\eta}^2)/M_{80}^2 = (M_{\eta'}^2\!-\!M_0^2\!-\!M_{\tau}^2)/M_{80}^2$.}, 
we obtain for the matrix element
\begin{equation}
\label{eq:cptps}
\bra{0} \mathcal{P}^{\textrm{NP}} \ket{\eta(\eta^{\prime})}= \sum_a \frac{1}{2}F_{\eta(\eta^{\prime})}^a g_a m_{\eta(\eta^{\prime})}^2\left(1-\delta^{a0}\frac{M_{\tau}^2}{M_{\eta(\eta^{\prime})}^2} \right),
\end{equation}
where $g_a$ has been defined above and $M_{\tau}^2=6\tau/F^2$ is the topological mass term ---see \cref{chap:mixing}. After some algebra, we have obtained 
a relation which is very similar 
to the $\pi^0$ result ---except for the singlet $a=0$ term--- and resembling that of the axial current matrix element. The natural question is how to find a 
general result valid at all orders in an easy way, for which is convenient to recall the 
Ward identity \cref{eq:anom}
%
\begin{equation}
\partial_{\mu}(\overline{q}\gamma^{\mu}\gamma_5 q) = 2m_q \overline{q}i\gamma_5q - \frac{g_s^2}{32\pi^2}\epsilon^{\alpha\beta\mu\nu}G^c_{\alpha\beta}G^c_{\mu\nu}
\equiv2m_q \overline{q}i\gamma_5q + \omega,
\end{equation}
which for the $U(3)_F$ axial current, \cref{eq:axdef}, reads
\begin{equation}\nonumber
\partial^{\mu}J_{\mu5}^{a} = \left\{ \mathcal{P}^a,\mathcal{M} \right\} + \delta^{a0}\sqrt{N_F/2}~\omega; \quad \mathcal{P}^a = \overline{q}i\gamma_5\frac{\lambda^a}{2}q, \ \  q=(u,d,s)^T,
\end{equation}
where $\mathcal{M}=\textrm{diag}(\hat{m},\hat{m},m_s)$ is the quark mass matrix. In such a way, the pseudoscalar current can be expressed in terms of the axial 
current and the winding number density $\omega$. Then, using the same algebra as previously, the matrix element can be expressed as  
\begin{align}\nonumber
\bra{0} \mathcal{P}^{\textrm{NP}}\ket{P(p)} = &\frac{1}{2} \sum_a\textrm{Tr}(\mathcal{P}^{\textrm{NP}}\lambda^a) \bra{0}\partial^{\mu}J_{5\mu}^a -\delta^{a0}\sqrt{3/2}~\omega \ket{P(p)} \\
 = & \frac{m_P^2}{2} \sum_a\textrm{Tr}(\textrm{diag}(c_u^{\mathcal{P}},c_d^{\mathcal{P}},c_s^{\mathcal{P}})\lambda^a)F_{P}^a(1-\Delta\delta^{0a}) \label{eq:singapp}
\end{align}
where $\Delta=\bra{0}\sqrt{6}\omega \ket{P}/m_P^2F_P^0$. Still, $\Delta$ needs to be determined. A nice solution can be borrowed from~\cite{Escribano:2005qq}. 
Neglecting the $u$ and $d$ quark masses $\hat{m}$ ---which roughly amounts to take $m_{\pi}^2/m_{K,\eta,\eta'}^2 \rightarrow 0$--- we obtain from the octet and singlet Ward identities,
\begin{equation}\nonumber
\sqrt{3/2}~\omega = \partial^{\mu}J_{\mu5}^0 + \frac{1}{\sqrt{2}}\partial^{\mu}J_{\mu5}^8 .
\end{equation}
Plugging this relation into \cref{eq:singapp}, we obtain $\Delta=1+F_P^8/(\sqrt{2}F_P^0)$, so the pseudoscalar contribution to $P\rightarrow \bar{\ell}\ell$ can be finally expressed as
\begin{align}
i\mathcal{M} = & -i [\overline{u}_{p,s}i\gamma_5 v_{p',s'}]\frac{g^2}{8m_W^2} \frac{m_P^2m_{\ell}c^{\mathcal{P}}_{\ell}}{m_P^2-m_{\mathcal{P}}^2} 
\sum_a\textrm{Tr}(\mathcal{P}^{\textrm{NP}}\lambda^a)F_{P}^a(1 - \delta^{0a} - \frac{\delta^{0a}F_P^8}{F_P^0\sqrt{2}})), \nonumber  \\
= & -i [\overline{u}_{p,s}i\gamma_5 v_{p',s'}]\frac{G_F}{\sqrt{2}} \frac{m_P^2m_{\ell}c^{\mathcal{P}}_{\ell}}{m_P^2-m_{\mathcal{P}}^2} 
\sum_a\textrm{Tr}(\mathcal{P}^{\textrm{NP}}\lambda^a)F_{P}^a(1-\delta^{0a}(1+\frac{F_P^8}{F_P^0\sqrt{2}})),  \nonumber  \\
= & G_F \frac{m_P^3m_{\ell}c^{\mathcal{P}}_{\ell}}{m_P^2-m_{\mathcal{P}}^2} 
\sum_a\textrm{Tr}(\mathcal{P}^{\textrm{NP}}\lambda^a)F_{P}^a(1-\delta^{0a}(1+\frac{F_P^8}{F_P^0\sqrt{2}})).
\end{align}
This induces an additional contribution to the $\mathcal{A}(q^2)$ loop amplitude in \cref{eq:loopamp},
\begin{equation}
\mathcal{A}(q^2) \rightarrow \mathcal{A}(q^2) + \frac{\sqrt{2}G_Fm_P^2 c^{\textrm{NP}}_{\ell}}{4\alpha^2F_{P\gamma\gamma}(m_P^2-m^2_{\mathcal{P}})}\sum_a\textrm{Tr}(\mathcal{P}^{\textrm{NP}}\lambda^a)F_{P}^a   (1-\delta^{0a}(1+\frac{F_P^8}{F_P^0\sqrt{2}}) ).
\end{equation}
We note that the approximation taken for calculating the $\bra{0}\omega\ket{P}$ matrix element has been used with great success in 
$J/\Psi\rightarrow\gamma\eta(\eta^{\prime})$ decays~\cite{Escribano:2005qq} and has been checked in \cref{sec:jpsirad}. 
Actually, at LO in $\chi$PT\footnote{At this order, the $\eta$ and $\eta^{\prime}$ masses 
are~\cite{Bickert:2015cia} $M_{\eta}^2=0.244\textrm{GeV}^2$ $M_{\eta^{\prime}}^2=0.917\textrm{GeV}^2$ and 
$M_{0}^2=0.673\textrm{GeV}^2$.}, the difference between \cref{eq:cptps} and \cref{eq:singapp} is of $8\%(1\%)$ for the $\eta(\eta^{\prime})$, enough for our study.

In the flavor basis, neglecting the $u$ and $d$ quark masses, only the strange part contributes. Using an analogous procedure, we find
\begin{equation}
\mathcal{A}(q^2) \rightarrow \mathcal{A}(q^2) + \frac{\sqrt{2}G_Fm_P^2 c^{\textrm{NP}}_{\ell}}{4\alpha^2F_{P\gamma\gamma}(m_P^2-m^2_{\mathcal{P}})}\sum_a\textrm{Tr}(\mathcal{P}^{\textrm{NP}}\lambda^a)F_{P}^a(1-\frac{F_P^q}{\sqrt{2}F_P^s})(1-\delta^{aq}).
\end{equation}

\subsection{Implications for new physics}

All in all, both contributions may be summarized to yield an additional term modifying \cref{eq:loopamp} as
\begin{equation}
   \label{eq:NP}
   \mathcal{A}(q^2) \rightarrow  \mathcal{A}(q^2) + \frac{\sqrt{2} G_F F_{\pi}}{4\alpha_{em}^2 F_{P\gamma\gamma}}(\lambda^A_P+\lambda^{\mathcal{P}}_P) , 
\end{equation}
where $G_F$ is the Fermi coupling constant, and $F_{\pi}\simeq92$~MeV is the pion decay constant. The $\lambda$-terms depend on the pseudoscalar meson structure, 
which for the $\eta$ and $\eta^{\prime}$ involve the mixing parameters. In the flavor-mixing scheme, they read\footnote{By definition, $F_{\pi^0}^8=F_{\pi^0}^0\equiv0$ and 
$F_{\pi^0}^3\equiv F_{\pi}$. From \cref{chap:mixing}, $F_{\eta(\eta^{\prime})}^q=0.84(0.72)F_{\pi}$, $F_{\eta(\eta^{\prime})}^s=-0.90(1.14)F_{\pi}$ 
and $F_{\eta(\eta^{\prime})}^3\equiv0$.}
\begin{align}
\label{eq:NPA}
\lambda^{A}_{P} & =  c^A_{\ell} \left[    \frac{F_{P}^3}{F_{\pi}} \left( c^{A}_u - c^{A}_d \right)   +   \frac{F_{P}^q}{F_{\pi}} \left( c^{A}_u + c^{A}_d \right)    +    \frac{F_{P}^s}{F_{\pi}} \sqrt{2}c^{A}_s \right], \\ 
\label{eq:NPP}
\lambda^{\mathcal{P}}_{P} & =   \frac{c^{\mathcal{P}}_{\ell}}{1- \frac{m_{\mathcal{P}}^2}{m_{P}^2}} 
 \left[    \frac{F_{P}^3}{F_{\pi}} \left( c^{\mathcal{P}}_u - c^{\mathcal{P}}_d \right)   +   \frac{F_{P}^q}{F_{\pi}} \left( -c^{\mathcal{P}}_s \right)    +    \frac{F_{P}^s}{F_{\pi}} \sqrt{2}c^{\mathcal{P}}_s \right].  
\end{align}
Taking the result from the mixing parameters in \cref{chap:mixing} to numerically calculate \cref{eq:NPA,eq:NPP} \cref{eq:NP} yields
\begin{align}\nonumber
   & \mathcal{A}(m_{\pi^0}^2) + 0.026 \left(c^A_{\ell}(c^A_u -c^A_d) + c^{\mathcal{P}}_{\ell}(c^{\mathcal{P}}_u -c^{\mathcal{P}}_d)(1-m_{\mathcal{P}}^2/m_P^2)^{-1}  \right) , \\\nonumber
   & \mathcal{A}(m_{\eta}^2) + 0.026  \left(0.84c^A_{\ell}(c^A_u +c^A_d) - 1.27c^A_{\ell}c^A_s  -2.11c^{\mathcal{P}}_{\ell}c^{\mathcal{P}}_{s}(1-m_{\mathcal{P}}^2/m_P^2)^{-1} \right) , \\\nonumber
   & \mathcal{A}(m_{\eta^{\prime}}^2) + 0.021 \left(0.72c^A_{\ell}(c^A_u +c^A_d) + 1.61c^A_{\ell}c^A_s  +0.89c^{\mathcal{P}}_{\ell}c^{\mathcal{P}}_{s}(1-m_{\mathcal{P}}^2/m_P^2)^{-1} \right) . 
\end{align}
To discuss the sensitivity of each particular channel to NP, it is convenient to cast a very approximate result for $\mathcal{A}(m_P^2)$, namely 
\begin{equation}
\label{eq:AmpApp}
\mathcal{A}(m_P^2) \simeq i\pi \left[ \ln\left(\frac{m_{\ell}}{m_P}\right) \right] + \left[ \ln^2\left(\frac{m_{\ell}}{m_P}\right)  - 3\ln\left(\frac{\Lambda}{m_{\ell}}\right) +\delta_{\textrm{NP}} \right],
\end{equation}
where $\Lambda$ is some effective hadronic scale characterizing the TFF and $\delta_{\textrm{NP}}$ is the NP contribution in \cref{eq:NP}. 
From \cref{eq:AmpApp}, we see that, as the lepton mass gets lighter, the amplitude will be dominated by the $\ln(m_{\ell}/m_P)$ terms, which become large and 
make the NP contribution harder to see. Indeed, for $\ell = e$, the relative NP contribution to the BR is approximately given by 
$ 2 \delta_{\textrm{NP}}(\ln^2(\frac{m_{e}}{m_P}) + \pi^2)^{-1}$. If we are aiming to find contributions from NP, it is therefore much easier to look for the $\ell=\mu$ 
channel as the NP part is insensitive to $m_{\ell}$ (see \cref{eq:NP}). 

With respect to $m_P$, from the logarithmic scaling, we infer that there is no big difference in the SM in choosing either $\pi^0,\eta$, or $\eta^{\prime}$ as their masses are of same order. Furthermore, the NP axial contribution does not depend on $m_P$, see \cref{eq:NPA}, meaning that is equally likely to appear in any case. This contrasts with the pseudoscalar NP contribution, which strongly depends on $m_P$ (cf. \cref{eq:NPP}) and gets bigger as $m_P$ and $m_{\mathcal{P}}$ (the mass of the new pseudoscalar particle) approach each other. Still, this is \textit{a priori} irrelevant unless there is a well-motivated NP scale which is close to either the $\pi^0, \eta$, or $\eta^{\prime}$ masses.

From this discussion, we conclude that $\eta(\eta^{\prime})\rightarrow \mu^+\mu^-$ decays are the best candidates to look for NP effects (as the $\pi^0$ cannot decay into muons). For illustrating the statements above, we give the approximate NP contribution to the branching ratio for each particular process,
\begin{align}\nonumber
   & BR(\pi^0\rightarrow e^+e^-)\left( 1 +0.001\left[c^A_{\ell}(c^A_u -c^A_d) +  c^{\mathcal{P}}_{\ell}\frac{c^{\mathcal{P}}_u -c^{\mathcal{P}}_d}{1-m_{\mathcal{P}}^2/m_P^2} \right]\right) , \\\nonumber
   & BR(\eta\rightarrow^{\mu^+\mu^-}_{e^+e^-}) \left(1 +\left(^{-0.002}_{+0.001}\right) \left[ 0.84c^A_{\ell}(c^A_u +c^A_d) - 1.27c^A_{\ell}c^A_s  -\frac{2.11c^{\mathcal{P}}_{\ell}c^{\mathcal{P}}_{s}}{1-m_{\mathcal{P}}^2/m_P^2} \right]\right) , \\\nonumber
   & BR(\eta'\rightarrow^{\mu^+\mu^-}_{e^+e^-}) \left(1 +\left(^{+0.003}_{+0.001}\right)\left[ 0.72c^A_{\ell}(c^A_u +c^A_d) + 1.61c^A_{\ell}c^A_s  + \frac{0.89c^{\mathcal{P}}_{\ell}c^{\mathcal{P}}_{s}}{1-m_{\mathcal{P}}^2/m_P^2} \right]\right) . 
\end{align}
We see that, as stated above, the $\ell = e$ channel has the same sensitivity for every pseudoscalar. For $\ell=\mu$, we find it two(three) times more sensitive than 
the $\ell=e$ channel  for the $\eta(\eta^{\prime})$. These numbers imply, together with the experimental precision reached for the $\pi^0(\eta)$ decay (we do not consider the 
central value, but the obtained precision), bounds for the $c^{A}$ parameters of the order of $7(8)$. As an example, for the $Z^0$ boson $(c^A_{\ell}=c^A_{d,s}=-c^A_u\equiv1)$, 
the $c^A_f$ combination is $-2(-1.27)[1.61]$ for $\pi^0(\eta)[\eta^{\prime}]$.

Interesting enough, a typical $Z^0$-like contribution has opposite sign for $\pi^0\rightarrow e^+e^-$ than for $\eta\rightarrow\mu^+\mu^-$, contrary to experimental implications. This would suggest either different couplings (necessarily $SU(2)_F$ breaking), or lepton flavor violating (LFV) models, which would couple different to distinct generation of quarks, leptons, or both. Moreover, in order to avoid $(g-2)_{\mu}$ problems, we would need, either some balance from an additional vector-like contribution\footnote{The dominant Schwinger-like contribution for a vector(scalar)-like coupling has positive sign whereas the axial(pseudoscalar) one has opposite sign, providing a fine tuning cancelation.} or, again, LFV models in which the coupling to the muon is suppressed.

For a pseudoscalar contribution, as in Ref.~\cite{Chang:2008np}, the effective couplings may become even larger as the new particle mass approaches the $\pi^0,\eta,\eta^{\prime}$ masses, meaning that would be visible for one of the pseudoscalars alone. Finally, we comment on the existing correlations given the pseudoscalar structure. We see for instance that $\pi^0\rightarrow e^+e^-$ and $\eta\rightarrow \mu^+\mu^-$ are, in general, anti-correlated unless there is a pseudoscalar particle $\mathcal{P}$ with $m_{\pi^0}<m_{\mathcal{P}}<m_{\eta}$ (or a different structure for distinct generations). Again, $(g-2)_{\mu}$ would play an important constraint for the pseudoscalar case as well.

To conclude, there is still the chance to look for NP contributions, specially in the $\ell=\mu$ channel, and a variety of phenomenology is possible depending on which kind of interaction is chosen. Still, our study suggests to go beyond simple scenarios; this seems nevertheless the standard in high energy physics nowadays, and scenarios of this kind have been and are still studied at present. In this discussion, we have omitted a detailed discussion of available physical constraints for these scenarios. This constitutes a field of study by itself. To mention some constraints, $(g-2)_{\mu}$ and low-energy parity violating would provide tight bounds. For additional discussion along these lines, see Refs.~\cite{Davoudiasl:2012ag,Davoudiasl:2012qa,Chang:2008np,Carlson:2012pc,Karshenboim:2014tka}.

\section{Conclusions and outlook}

In this chapter, we have reviewed as a first application of CAs the status of pseudoscalar decays into lepton pair \PtoLL processes.
We have shown that the main problem in these processes is to obtain a precise and reliable determination for the TFFs, not only at the high 
energies, but ---especially--- at the low-energies. 
This feature, which has been known since long, has been ignored due to the lack of ability to incorporate these two regimes at once 
in a single theory, systematically, precisely and model independently ---the perfect scenario to test and apply our acquired knowledge.

Thanks to our method we have been able, for the first time, to provide a systematic error for these processes. This was specially 
important regarding the $\eta$ and $\eta'$, where previously unaccounted systematic errors associated to the existence of 
threshold production lead to unrealistic underestimated errors. Still, thanks to the precise achieved description, we have been able 
to improve on the precision in most of results, even after the inclusion of previously unaccounted errors. 
In addition, we have carried out a precise numerical evaluation and avoided approximations commonly employed in the literature. 
Such an error cannot be neglected at all when dealing with the $\eta$ and $\eta'$, which would induce a very large systematic error. 
For completeness, we have included the $Z^0$ boson contribution as well.

From our results, we have confirmed the present experimental discrepancies in the $\pi^0\to e^+e^-$ and $\eta\to\mu^+\mu^-$ decays ---the latter 
often obviated in the literature because of the approximations employed in the loop integral among others. In light of this situation we have 
discussed the possible implications of new physics. We find that 
that appropriate scenarios to describe the discrepancy most likely require light new-physics degrees of freedom of lepton-flavor violating nature. 
Finally, we have shown that previous \cpt-based calculations at LO imply non-negligible errors and should be avoided. For this reason, we have 
provided a simple formula which provides, in a simple way, the required corrections.

For the moment, we have only employed the simplest $C^0_1$ approximant due to the absence of double-virtual data. Reaching the 
$C^1_2$ approximant would greatly reduce the obtained uncertainty and evidence the performance of the method. This is an ongoing 
effort which we will briefly discuss in the next chapter. An additional line of thought to be followed is developing a modified approach  
for the $\eta'$ in which the time-like features could be easily implemented too.

\chapter{The muon $(g-2)$: pseudoscalar-pole contribution}
\label{chap:gm2}
\minitoc

\section{Introduction}


The anomalous magnetic moment of fermions, proportional to $(g-2)$\footnote{In particular, given a fermion $\ell$, $\boldsymbol{\mu}_m=g_{\ell}\frac{e\mathcal{Q}}{2m_{\ell}}\boldsymbol{S}$, whereby   $\boldsymbol{\mu}_{\textrm{anom}}=(g_{\ell}-2)\frac{e\mathcal{Q}}{2m_{\ell}}\boldsymbol{S}$.}, 
has been a path of effort and triumphs in theoretical and experimental particle physics. First, back in 1928, the new 
relativistic Dirac theory for {\textit{elementary}} spin-$1/2$ fermions $\ell$ predicted $g_{\ell}=2$, in contrast to the classical expectation $g_{\ell}=1$~\cite{Dirac:1928hu}. The 
$g_e$ measurement in 1934~\cite{Kinsler} confirmed the  Dirac theory of electrons. Nevertheless, subsequent preciser measurements were 
performed~\cite{Nafe:1947zz,Nagle:1947zz,Breit:1947zzb,Foley:1948zz} finding slight deviations from $g_e=2$. This could be soon explained after the great effort from Tomonaga, 
Feynmann, Schwinger and Dyson in the development of the renormalization of Quantum Field Theories (QFT), culminating with the Schwinger prediction of $g_{\ell}$ at 
NLO~\cite{Schwinger:1948iu}, 
\begin{equation}
\label{eq:schw}
  a_{\ell} \equiv \frac{g_{\ell}-2}{2} = \frac{\alpha}{2\pi},
\end{equation}
which established QED ---the very first QFT--- as a serious microscopic theory of the electromagnetic interactions. 
Since then, experiments and theory have evolved, and still, $(g-2)$ continues to be one of the finest tests of our understanding of particle physics.  
At present, both $e$ and $\mu$ anomalous magnetic moments have been measured; their most recent results read
\begin{align}
  a_{e}^{\textrm{exp}} = & \    115965218.073(28) \times 10^{-11}, \\
  a_{\mu}^{\textrm{exp}} = & \ 116592091(63) \times 10^{-11}.
\end{align}
The first one is the result from~\cite{Hanneke:2008tm}, whereas the second one is the updated value~\cite{PDG:amm} from~\cite{Bennett:2006fi}, after the new 
muon-to-proton magnetic ratio determination~\cite{Mohr:2012tt}.
Regardless the precise determination for $a_{e}$, it is $a_{\mu}$ on which we focus from now on. This is due to its higher sensitivity to new physics in the naive scaling  
$\delta a_{\ell} = \mathcal{C} m_{\ell}^2/\Lambda^2_{NP}$ with $\mathcal{C}\sim\mathcal{O}\left( \frac{\alpha}{\pi} \right)$~\cite{Jegerlehner:2009ry}, 
which make heavy leptons more interesting (unfortunately, precise experiments are not yet accessible 
for the heavier $\tau$ lepton). Given the current precision, $a_{\mu(e)}$ is sensitive to $\mathcal{O}(200(50)~\textrm{GeV})$ physics\footnote{There are however some exceptions 
violating this scaling~\cite{Giudice:2012ms} and would make $a_e$ very interesting as well for testing new physics 
scenarios~\cite{Giudice:2012ms,Knecht:2014sea}.}, which 
is complementary to the LHC.
As an example, it could help in distinguishing among SUSY models~\cite{vonWeitershausen:2010zr}. Alternatively, for models with extra-dimensions, it would be sensitive 
to Kaluza-Klein gravitons~\cite{Kim:2001rc} despite of constraints from electroweak precision observables~\cite{Beneke:2012ie}. This contrasts with Littlest Higgs models which 
have little influence on $(g_{\mu}-2)$~\cite{Blanke:2007db}. Finally, it is well suited for testing Dark Photons~\cite{Okun:1982xi,Holdom:1985ag} scenarios. 
However, before searching for new physics, it is necessary 
to provide a robust theoretical prediction within the SM at the same level of precision as the experimental one. This is very pressing given the expected precision in the forthcoming 
muon $(g-2)$ experiments at Fermilab~\cite{LeeRoberts:2011zz} and J-PARC~\cite{Mibe:2010zz} around $16\times10^{-11}$. Below, we review the current status and motivate the needs 
for improving the current estimation for the hadronic light-by-light pseudoscalar pole contribution, to which this chapter is devoted.

\section{Standard Model contributions to $a_{\mu}$}

\subsection{QED}

In the SM, the major contribution to $a_{\mu}$ arises from QED corrections including $e,\mu$ and $\tau$ leptons alone. At one loop, the only diagram is the 
Schwinger term, Fig.~\ref{fig:g2QED} left, which was calculated by Schwinger in 1948~\cite{Schwinger:1948iu}. Then, at two loops, there are 9 diagrams contributing 
to $a_{\mu}$,  among which we find the so-called vacuum polarization, see Fig.~\ref{fig:g2QED} center ---the full calculation was carried out by Petermann and Sommerfield 
in 1957~\cite{Petermann:1957hs,Sommerfield:1957zz,Sommerfield:AP}. At three-loops, there are 72 diagrams, including the light-by-light
one, Fig~\ref{fig:g2QED} right, and their 
calculation required a huge effort taking almost 40 years~\cite{Laporta:1996mq,Samuel:1990qf,Li:1992xf,Laporta:1993ju,Laporta:1992pa,Czarnecki:1998rc}. Up to this order, 
analytic calculations are tractable, whereas 
\begin{figure}[t]
\centering
  \includegraphics[width=0.8\textwidth]{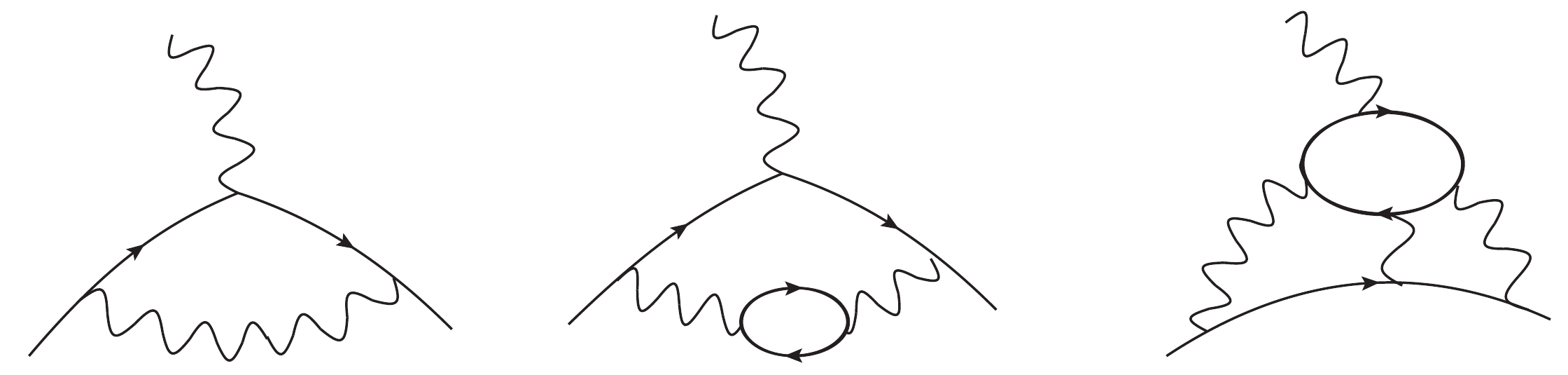}
  \caption{Representatives contribution to $a_{\ell}$. First, the one-loop Schwinger contribution. Second, one of the two-loops contributions: the vacuum polarization. Third, one 
           of the three-loops contribution: the light-by-light. In all these diragrams the loops contain charged leptons alone, i.e., $\ell=e,\mu,\tau$.}
  \label{fig:g2QED}
\end{figure}
at higher orders the number of diagrams as well as the calculational complexity increases. Nonetheless, a great effort has been done from the group of Kinoshita and collaborators 
to numerically compute the four- and five-loop contributions~\cite{Aoyama:2012wk}.
The up-to-date result is
%
\begin{align}
a_{\mu}^{\textrm{QED}}= & \ \frac{1}{2} \left(\frac{\alpha}{\pi}\right) + 0.765857425(17) \left(\frac{\alpha}{\pi}\right)^2 + 24.05050996(32) \left(\frac{\alpha}{\pi}\right)^3 \nonumber \\
 & \ + 130.8796(63) \left(\frac{\alpha}{\pi}\right)^4 +  753.29(1.04) \left(\frac{\alpha}{\pi}\right)^5 \nonumber \\
 = & \  116 \ 584 \ 718.951(80)\times10^{-11}. \label{eq:g2QED}
\end{align}
In the calculation we used the most precise determination from $\alpha^{-1}=137.035 \ 999 \ 049(90)$~\cite{PDG:amm,Aoyama:2012wk} from Rb-atom~\cite{Bouchendira:2010es} combined with the Rydberg constant and 
$m_{\textrm{Rb}}/m_e$ in~\cite{Mohr:2012tt}. The errors are dominated from the $\alpha$ determination in the Schwinger term and, to a lesser extent, the computational error at 
four-loops. 
Likewise, for the electron~\cite{Aoyama:2012wj} $a_{e}^{\textrm{QED}}=115 \ 965 \ 218.007(77)\times10^{-11}$.

\subsection{Electroweak}

The next sizable contributions to $a_{\mu}$ are the hadronic ones, and part of them are indeed one of the main objects of study in this thesis. However, due to their complexity, we leave 
their discussion for the last part of this section. Then, the last piece remaining in the SM are the electroweak contributions ---find the one-loop contributions in \cref{fig:EW}. 
They have been analytically computed at one- and two-loops, in Ref.~\cite{Fujikawa:1972fe} and Refs.~\cite{Czarnecki:1995sz,Knecht:2002hr,Czarnecki:2002nt}, respectively. Remarkably, such calculation 
was the first at two-loop that was performed within the electroweak sector of the SM. The last full re-evaluation after the Higgs discovery obtained~\cite{Gnendiger:2013pva}
\begin{equation}
\label{eq:g2EW}
 a_{\mu}^{\textrm{EW}} = \left(194.80(1) - 41.23(1.0) \right)\times10^{-11} = 153.6(1)\times10^{-11},
\end{equation}
where the first and second terms represent the one- and two-loop contributions. The error is dominated in this case by hadronic uncertainties. 
Similarly, for the electron $a_e^{\textrm{EW}} = 0.00297(5)\times10^{-11}$~\cite{Knecht:2014sea}.

\begin{figure}[t]
\centering
  \includegraphics[width=0.7\textwidth]{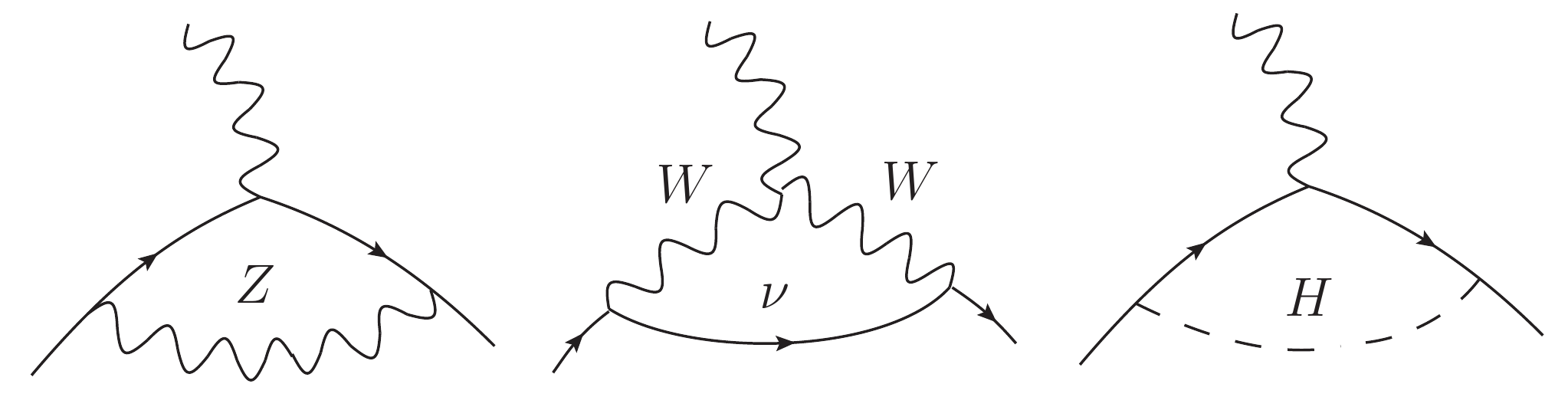}
  \caption{Electroweak contributions to $a_{\ell}$ at one loop.\label{fig:EW}}
\end{figure}

\subsection{QCD}

As anticipated, we finally discuss the QCD or hadronic contributions. In contrast to the previous cases, these cannot be perturbatively calculated in a combined $\alpha$ and $\alpha_s$ 
expansion, as the latter becomes non-perturbative at low-energies, which turns out to be the most relevant region in $a_{\mu}$ calculations. Therefore, we must relegate to a 
perturbative expansion in $\alpha$ together with some machinery dealing with the hadronic interactions in its non-perturbative regime.

\subsubsection{HVP}
   
At order $\mathcal{O}(\alpha^2)$, the only hadronic contribution is the hadronic vacuum polarization (HVP), which is shown in 
Fig.~\ref{fig:g2Had} left. Fortunately for this case, data comes to our rescue. The reason being that the HVP is an analytic 
function of which the imaginary part is related to the process $e^+e^-\rightarrow \textrm{hadrons}$ by virtue of 
the optical theorem. Then, a dispersive representation allows to express such contribution as an integral over the mentioned 
cross-section~\cite{Jegerlehner:2009ry}. This procedure allows to include all the hadronic effects in a data-driven approach using 
the available exclusive processes at low-energies and a matching to the pQCD prediction at the high-energies, obtaining~\cite{Davier:2010nc}
\begin{equation}
  \label{eq:HVPLOee}
  a_{\mu}^{\textrm{HVP-LO};e^+e^-}=6923(42)\times10^{-11}.
\end{equation}
Alternatively, it is possible to use $\tau\rightarrow\nu+\textrm{hadrons}$ data after correcting for isospin effects, which yields~\cite{Davier:2010nc}
$7015(47)\times10^{-11}$ instead
and shows some tension with the $e^+e^-$-based calculation at the level of $1.5\sigma$.
At higher orders $\mathcal{O}(\alpha^3, \alpha^4)$, there appear corrections to the HVP ---see the second and third diagrams in Fig.~\ref{fig:g2Had}.
Again, all the hadronic information can be 
obtained from data. 
From the update~\cite{Hagiwara:2011af} 
of Ref.~\cite{Hagiwara:2003da},
\begin{equation}
  \label{eq:HVPNLO}
  a_{\mu}^{\textrm{HVP-NLO}}=-98.4(7)\times10^{-11}.
\end{equation}
\begin{figure}[t]
\centering
\includegraphics[width=\textwidth]{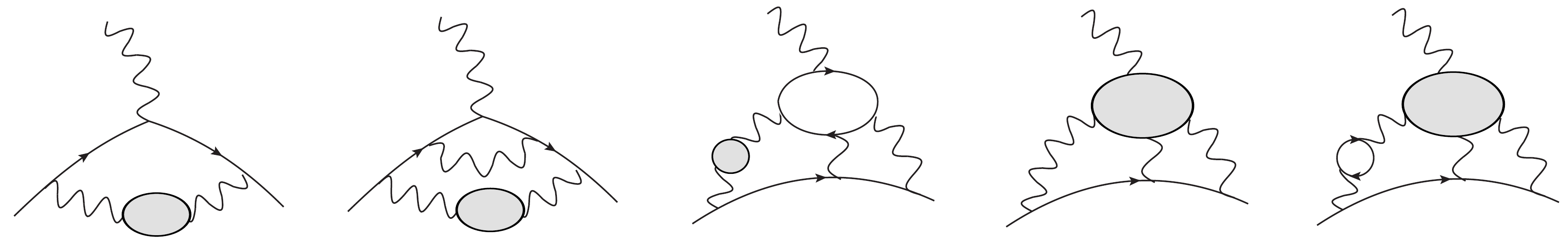}
\caption{Hadronic contributions to $a_{\mu}$: The first, second and third stand for the LO and representative NLO and NNLO hadronic vacuum polarization. Fourth and fifth represent 
the LO and representative NLO hadronic light-by-light contribution. The blobs represents hadronic physics whereas loops represent the leptons.}
\label{fig:g2Had}
\end{figure}
Finally, the very recent result of Ref.~\cite{Kurz:2014wya} obtains 
\begin{equation}
  \label{eq:HVPNNLO}
  a_{\mu}^{\textrm{HVP-NNLO}}=12.4(1)\times10^{-11}.
\end{equation}
%
Putting together the information from \cref{eq:HVPLOee,eq:HVPNLO,eq:HVPNNLO}, we obtain for the HVP contributions
\begin{equation}
  \label{eq:HVP}
  a_{\mu}^{\textrm{HVP-}e^+e^-}=6837(42)\times10^{-11}, 
\end{equation}
of which the error is dominated by the $\sigma(e^+e^-\to\pi^+\pi^-)$ experimental data uncertainty. The present accuracy is of the order of the current 
$(g_{\mu}-2)$ experiment, but three times larger than the projected ones. 
This situation is planned to be solved with a more precise and more extensive experimental programme. It is 
important to remark at this point that there are alternative determinations obtaining slightly different results~\cite{Jegerlehner:2009ry,Benayoun:2011mm,Benayoun:2015gxa}. 
Moreover, some experimental discrepancies exist ---see Ref.~\cite{Ablikim:2015orh}--- which illustrates that a closer work among the theoretical and experimental community 
 to agree on a common procedure and database is required.
In addition, there are alternative ideas to attack this problem, such the space-like approach in Ref.~\cite{Calame:2015fva} or the effort from the lattice community: MILC Collaboration~\cite{Aubin:2006xv}, RBC-UKQCD~\cite{Boyle:2011hu}, Mainz~\cite{DellaMorte:2011aa} and  ETM Collaboration~\cite{Feng:2011zk}. While their results are promising, 
additional work is required in order to reduce the error. Hopefully, in the near future the different approaches may converge to a very precise and robust determination for the HVP. 
Similarly, for the electron~\cite{Nomura:2012sb,Kurz:2014wya}
\begin{equation}
 a_e^{\textrm{HVP}} = (0.1866(11) - 0.02234(14)+0.0028(1) )\times10^{-11} = 0.16706(11)\times10^{-11}.
\end{equation}

\subsubsection{HLbL}

The last relevant hadronic contribution, starting at $\mathcal{O}(\alpha^3)$, is the hadronic light-by-light (HLbL), ---fourth diagram 
in Fig.~\ref{fig:g2Had}. Regretfully, this process cannot be directly related to a measurable cross section. 
Being a function of the four incoming-momenta ---though the external one may be set to zero for our purposes--- the underlying hadronic function depends 
on many invariants and is much more complicated than the HVP, which depends on a single quantity. 
This implies that mixed regions involving low- and high-energies at the same time appear, involving both non-perturbative and perturbative input.\\

\begin{figure}[t]
   \includegraphics[width=\textwidth]{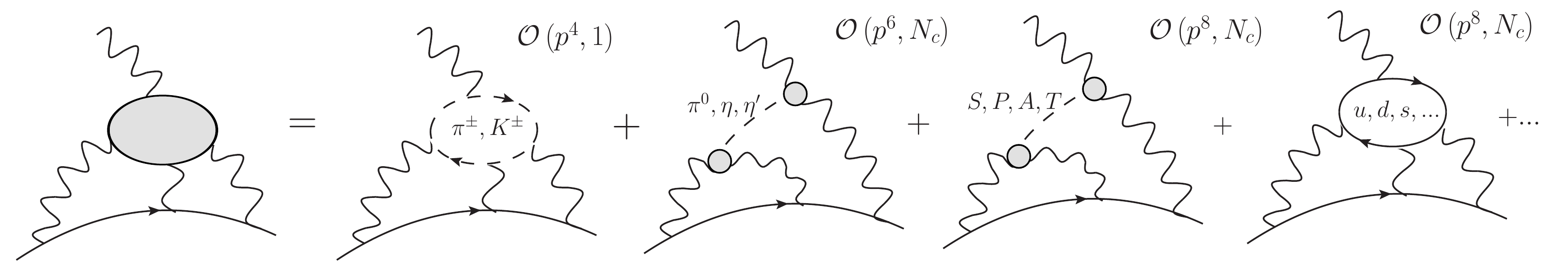}
   \caption{Chiral and large-$N_c$ decomposition of the HLbL, see Ref.~\cite{deRafael:1993za}.}
\label{fig:LNCRafael}
\end{figure}

With no data at rescue, one needs some theoretical expansion parameter. Given the size of $\alpha_s$, the only perturbative parameters 
at hand are the chiral expansion in terms of small momenta (anticipating the impact of low energies in this quantity), 
and the large number of colors, $N_c$. This observation allowed to a first decomposition of the leading terms 
in Ref.~\cite{deRafael:1993za}, which are shown in Fig.~\ref{fig:LNCRafael}. In such expansion, the leading contributions are the charged pion and kaon loops, and the pseudo-Goldstone bosons $\pi^0$, $\eta$ and $\eta'$ exchanges. Numerically however, it is the latter that dominates.\\

Still, the diagrams in Fig.~\ref{fig:LNCRafael} cannot be calculated from first principles in QCD. Therefore, different approaches have been used. As an example, there 
exist Extended Nambu-Jona-Lasinio models, effective theories such as Hidden Gauge Symmetry, or large-$N_c$ models where the minimum amount of resonances required to fulfill 
the high-energy behavior are included (see Refs. in~\cite{Prades:2009tw}). The fact that all these approaches do not actually calculate the same quantities as shown in 
Fig.~\ref{fig:LNCRafael} and the ambiguities when including the known constraints explain the range of different results and the lack of agreement within the community. An attempt 
to reconcile all these approaches lead to the estimate $a_{\mu}^{\textrm{HLbL-LO}} = 105(26)\times10^{-11}$~\cite{Prades:2009tw}. A more recent evaluation for this quantity appeared 
in Ref.~\cite{Jegerlehner:2009ry}, which most recent update~\cite{Jegerlehner:2015stw} reads
\begin{equation}
\label{eq:g2HLBLLO}
a_{\mu}^{\textrm{HLbL-LO}} = 102(39)\times10^{-11},
\end{equation}
Nevertheless, this result may neglect important theoretical uncertainties from the models. Therefore, its error should not be taken on the same 
foot as in the HVP case, where this is associated to the data uncertainties alone. Similarly, for the electron 
$a_{e}^{\textrm{HLbL-LO}}= 3.9(1.3)\times10^{-14}$~\cite{Jegerlehner:2009ry}.
Data-based approaches would help in solving this situation, and such is the concern of this thesis chapter. As an example, dispersive 
approaches have been proposed both in Mainz~\cite{Pauk:2014rfa} and Bern~\cite{Colangelo:2014dfa,Colangelo:2014pva}, though they are limited 
in the energy range of applicability~\cite{Masjuan:2014rea}. Note at this respect that the pseudo-Goldstone boson exchanges were already calculated as an euclidean two-loop integral 
in Ref.~\cite{Knecht:2001qf}. In addition, there 
are ongoing promising proposals in the Lattice community aiming for this calculation~\cite{Blum:2014oka,Green:2015sra,Blum:2015gfa} as well as approaches from 
Dyson-Schwinger equations~\cite{Goecke:2010if}. Finally, NLO corrections have been estimated~\cite{Colangelo:2014qya},
\begin{equation}
\label{eq:g2HLBLNLO}
a_{\mu}^{\textrm{HLbL-NLO}}= 3(2)\times10^{-11}.
\end{equation}

Putting all the pieces ---\cref{eq:g2QED,eq:g2EW,eq:HVP,eq:g2HLBLLO,eq:g2HLBLNLO}--- together, we obtain the full SM contribution
\begin{align}
a_{e}^{\textrm{th}} = & \ 115965218.181(77)\times10^{-11}, \\
a_{\mu}^{\textrm{th}} = & \ 116591815(57)\times10^{-11}, 
\end{align}
where the errors are totally dominated from the $\alpha$ determination for the first, and from QCD errors for the second. Comparison to experiment gives
\begin{align}
a_{e}^{\textrm{exp}}-a_{e}^{\textrm{th}} = & \ -0.108(82)\times10^{-11}, \\
a_{\mu}^{\textrm{exp}}-a_{\mu}^{\textrm{th}} = & \ 276(85)\times10^{-11}.
\end{align}
For the electron, there is a nice agreement between theory and experiment and the current $\alpha$ 
determination is the limiting factor when aiming for precision. By contrast, for the muon, there is a significant $3.2\sigma$ discrepancy, which suggests the possibility 
that NP effects are present in this quantity~\cite{Jegerlehner:2009ry}. In order to establish whether the discrepancy is here to stay or if it is a statistical fluctuation, two 
experiments have been proposed at Fermilab~\cite{LeeRoberts:2011zz} and J-PARC~\cite{Mibe:2010zz} 
with a precision around $16\times10^{-11}$ for $a_{\mu}$. However, this would not be 
significant if the theoretical uncertainty, fully dominated by the LO HVP and HLbL contributions, is not improved accordingly. Whereas the first contribution is expected to be
improved with the forthcoming new data, such as those from $e^+e^-\to\pi^+\pi^-$ cross section measurements, more work is required to improve the error on the LO HLbL contribution.\\

It is the subject of this work to improve on the current precision of the HLbL. In particular, we focus on the dominant contribution among those 
depicted in \cref{fig:LNCRafael}: the pseudoscalar-pole, which is required at the $10\%$ accuracy level according to future experiments. 
At this level of precision, one needs to carefully account for all possible source of errors, specially the systematic ones, and avoid model dependencies. This contrasts to previous 
determinations, for which the current experimental error did not require such standards of precision and had different concerns, such as the sign problem~\cite{Knecht:2001qf} 
or the full-HLbL tensor high-energy behavior~\cite{Melnikov:2003xd,Nyffeler:2009tw,Jegerlehner:2009ry}. 
In \cref{sec:alblgen}, the formalism to calculate the most general HLbL contribution is introduced. Then, we focus on the pseudoscalar pole contribution in 
\cref{sec:psex}, where the meaning of the former ---not to be confused with alternative ``off-shell'' approaches--- is carefully outlined. 
Such contribution requires a precise model-independent description for the pseudoscalar TFFs, which we implement once more through the techniques of CAs. 
These are introduced ---and their systematic error carefully discussed--- and successfully employed for extracting a precise determination for the pseudoscalar pole 
contribution in \cref{sec:rationalTFF}. In \cref{sec:bpole}, we outline how our description can be implemented into previous approaches. Finally, we combine our value with 
the additional contributions to give an estimate for the full HLbL in \cref{sec:HLBLfinalres} and give the conclusions and outlook in \cref{sec:gm2conc}.

\section{Generic HLbL  contribution to $a_{\mu}$}
\label{sec:alblgen}

\begin{figure}[t]
\centering
  \includegraphics[width=\textwidth]{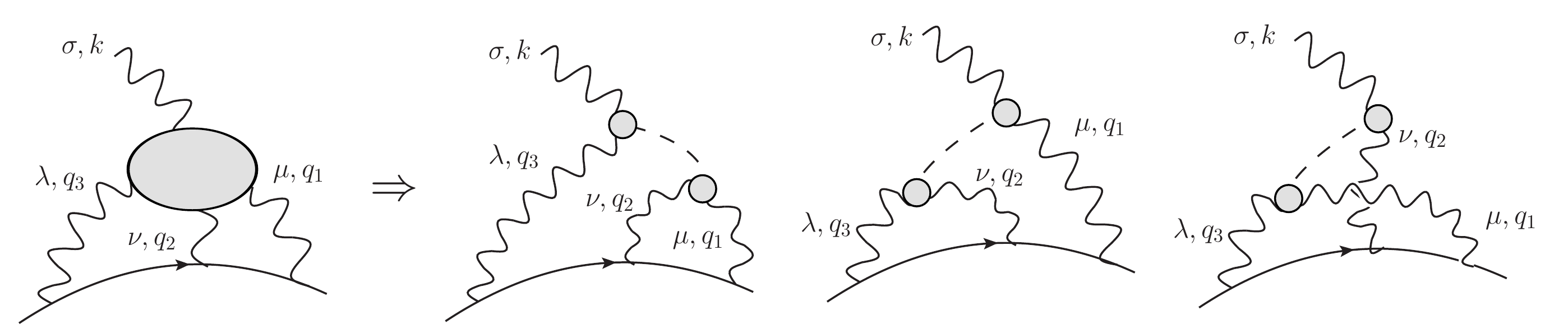}
  \caption{The pseudoscalar-exchange contribution to $(g-2)$. All the momenta, except for the external photon $(k)$, are outgoing from the TFF vertex.
           We take $q_1,q_2,k$ independent, then $q_3=(k-q_1-q_2)$.}
  \label{fig:pspole}
\end{figure}

The most general vertex describing the fermion-photon interaction
\begin{equation}
\bra{\ell^-(p')} (ie)j^{\mu}(0)\ket{\ell^-(p)}\equiv  -ie\overline{u}(p') \Gamma^{\mu}(p',p)  u(p) 
\end{equation} 
can be parametrized relying on $C,P$, and $T$ invariance as~\cite{Peskin:1995ev} 
\begin{equation}
\label{eq:emvertex}
\Gamma^{\mu}(p',p) = \gamma^{\mu} F_1(k^2) + i\frac{\sigma^{\mu\nu}k_{\nu}}{2m_{\ell}}F_2(k^2),
\end{equation}
where $k=p'-p$ and $F_{1,2}(k^2)$ are the Dirac and Pauli form factors. The former is fixed by gauge-invariance at $k^2=0$, which via Ward-identities constrains 
$F_1(0)=1$. The latter, which vanishes at tree-level, is not constrained by any symmetry. From this parametrization, it is possible to calculate the electromagnetic interactions 
at the classical level. A comparison to classical equations allows then to identify the gyromagnetic ratio $g_{\ell}=2(F_1(0)+F_2(0))= 2 +2F_2(0)$~\cite{Peskin:1995ev}, thus 
$a_{\ell}=F_2(0)$, which is our object of study.\\

In our case of study, following Ref.~\cite{Knecht:2001qf}, the HLbL diagram (left diagram in Fig.~\ref{fig:pspole}), gives the following contribution to the electromagnetic vertex 
\begin{align}
-ie\Gamma_{\rho}(p',p)  = & \  
  \int \frac{d^4q_1}{(2\pi)^4}  \int \frac{d^4q_2}{(2\pi)^4} \frac{(-i)^3}{q_1^2 q_2^2 (k-q_1-q_2)^2} \nonumber\\ 
 & \  \times \frac{i}{(p'-q_1)^2 -m_{\ell}^2} \frac{i}{(p'-q_1-q_2)^2 -m_{\ell}^2} \nonumber \\
& \  \times (-ie)^3\gamma^{\mu}(\slashed{p}'- \slashed{q}_1 +m_{\ell})\gamma^{\nu}(\slashed{p}'- \slashed{q}_1 - \slashed{q}_2 +m_{\ell})\gamma^{\lambda} \nonumber \\
& \  \times (ie)^4\Pi_{\mu\nu\lambda\rho}(q_1,q_2,k-q_1-q_2), \label{eq:HLbLgen}
\end{align}
%
where $\Pi_{\mu\nu\lambda\rho}(q_1,q_2,k-q_1-q_2)$ denotes the HLbL tensor for light quarks $q=u,d,s$, defined in terms of the QCD Green's function
\begin{align}
\Pi_{\mu\nu\lambda\rho}(q_1,q_2,q_3) & = \int d^4x_1 \int d^4x_2 \int d^4x_3 e^{i(q_1\cdot x_1  +  q_2\cdot x_2   +   q_3\cdot x_3)}  \nonumber \\
& \ \times\bra{0} T  j_{\mu}(x_1)j_{\nu}(x_2)j_{\lambda}(x_3)j_{\rho}(0)  \ket{0}, \label{eq:hlbltensor}
\end{align}
in which $j_{\mu}= \frac{2}{3}\overline{u}\gamma_{\mu}u - \frac{1}{3}\overline{d}\gamma_{\mu}d - \frac{1}{3}\overline{s}\gamma_{\mu}s $ stands for the electromagnetic current 
and $\ket{0}$ represents the QCD vacuum. In addition, the Ward identities  
$\{ q_1^{\mu}; q_2^{\nu}; q_3^{\lambda}; k^{\rho} \} \Pi_{\mu\nu\lambda\rho}(q_1,q_2,q_3)=0$ allow to rewrite the HLbL tensor as\footnote{To see this, take 
$0=\partial/\partial k^{\rho}(k^{\sigma}\Pi_{\mu\nu\lambda\sigma}(q_1,q_2,q_3)) = \delta^{\sigma}_{\rho}\Pi_{\mu\nu\lambda\sigma}(q_1,q_2,q_3) + 
k^{\sigma}(\partial/\partial k^{\rho}) \Pi_{\mu\nu\lambda\sigma}(q_1,q_2,q_3)$, from which previous identity follows.}
\begin{equation}
\label{eq:redHLbL}
 \Pi_{\mu\nu\lambda\rho}(q_1,q_2,q_3) = -k^{\sigma}(\partial/\partial k^{\rho}) \Pi_{\mu\nu\lambda\sigma}(q_1,q_2,q_3).
\end{equation}
%
Inserting this back into \cref{eq:HLbLgen} results in an expression of the kind  
$\Gamma_{\rho}(p',p) = k^{\sigma}\int ... \ (\partial/\partial k^{\rho}) \Pi_{\mu\nu\lambda\sigma}(q_1,q_2,q_3)\equiv k^{\sigma}\Gamma_{\rho\sigma}$, 
that allows to use the trace technique described in Ref.~\cite{Knecht:2001qf}, 
\begin{equation}
F_2(0) = \frac{1}{48m_{\ell}}\textrm{tr}\left( (\slashed{p}+m_{\ell})[\gamma^{\rho},\gamma^{\sigma}](\slashed{p}+m_{\ell})\Gamma_{\rho\sigma}(p,p) \right),
\end{equation}
which allows to take the limit $k\rightarrow0$ afterwards without introducing any kinematical singularity. Then, our desired $a_{\mu}^{\textrm{HLbL}}$ contribution is given as
\begin{align}
a_{\mu}^{\textrm{HLbL}} = 
 &  -ie^6  \int  \frac{d^4q_1}{(2\pi)^4}  \int  \frac{d^4q_2}{(2\pi)^4}  \frac{1}{q_1^2 q_2^2 (q_1+q_2)^2} \frac{1}{(p-q_1-q_2)^2 -m_{\ell}^2} \frac{1}{48m_{\ell}}  \nonumber \\
&  \times \! \frac{1}{(p-q_1)^2 -m_{\ell}^2} \textrm{tr} \left( (\slashed{p}+m_{\ell})[\gamma^{\rho},\gamma^{\sigma}](\slashed{p}+m_{\ell})  \gamma^{\mu}(\slashed{p}- \slashed{q}_1 \! +m_{\ell})  \right. \nonumber \\ 
& \left.  \gamma^{\nu}(\slashed{p}- \slashed{q}_1 \! - \slashed{q}_2 \! +m_{\ell})\gamma^{\lambda} \right) \!
 \frac{\partial}{\partial k^{\rho}}\Pi_{\mu\nu\lambda\sigma}(q_1,q_2,k-q_1-q_2) \Big\vert_{k\rightarrow 0}.  \label{eq:HLbLint}
\end{align}
At this point, an input for the HLbL tensor $\Pi_{\mu\nu\lambda\rho}(q_1,q_2,q_3)$ is required. 
As previously stated, the most relevant features for this quantity can be classified according to a combined chiral and large-$N_c$ counting. In the following, we extract from the 
HLbL tensor expression, \cref{eq:hlbltensor}, what has been phenomenologically found to be the dominant contribution, the pseudoscalar-pole. We stress that such a piece can 
be model-independently defined in contrast to other approaches~\cite{Melnikov:2003xd,Nyffeler:2009tw,Jegerlehner:2009ry} in terms of the pseudoscalar TFFs, which is essential in 
avoiding additional sources of systematic uncertainties. Then, we use the framework of CAs in order to describe the TFFs, providing a critical revision of systematic errors.
We remark that the chiral large-$N_c$ counting is used to identify the most relevant contributions alone ---our pseudoscalar-pole description involves however 
no chiral, large-$N_c$, or any other approximation, and aspires to give a full theoretical description for this quantity, which can serve as well as an input in dispersive approaches 
that evaluate further contributions beyond the pseudoscalar-pole.

\section{The pseudoscalar-pole contribution}
\label{sec:psex}

\subsection{The pole approximation to the HLbL}
\label{sec:polology}

As quoted by Weinberg~\cite{Weinberg:1995mt} (see chapter 10.2), ``often the $S$-matrix for a physical process can be well approximated by the construction of a single-pole''. 
To understand this, we follow Weinberg and discuss the particular case of the HLbL tensor Green's function\footnote{Usually $x_4=0$ is taken 
together with four-momentum conservation. By retaining this, we explicitly obtain the momentum conservation, $(2\pi)^4\delta^{(4)}(\sum_i p_i)$ function, in \cref{eq:WGreenHLBL}.}
\begin{align}
\Pi^{\mu\nu\rho\sigma}(p_1,p_2,p_3,p_4)  = & \left(\prod_i^4 \int d^4x_i\right) e^{ip_1\cdot x_1}e^{ip_2\cdot x_2}e^{-ip_3\cdot x_3}e^{-ip_4\cdot x_4} \nonumber \\ 
& \times \bra{0} T\{ j^{\mu}(x_1) j^{\nu}(x_2) j^{\rho}(x_3) j^{\sigma}(x_4) \} \ket{0} \label{eq:WGreenHLBL}.
\end{align}
Inserting intermediate particle states in \cref{eq:WGreenHLBL}, we obtain
\begin{align}
\Pi^{\mu\nu\rho\sigma}(p_1,p_2,p_3,p_4)  = &  \int d^4x_2 d^4x_4 e^{ip_2\cdot x_2}e^{-ip_4\cdot x_4} \frac{i}{q^2-m_P^2+i\epsilon}  \nonumber \\ 
& \times \bra{0} T\{ j^{\mu}(0) j^{\nu}(x_2)  \} \ket{P(q)} \bra{P(q)} T\{ j^{\rho}(0) j^{\sigma}(x_4) \} \ket{0} \nonumber \\
& \times (2\pi)^4\delta^{(4)}(p_1+p_2-p_3-p_4) + \textrm{OT},
\end{align}
where $q=p_1+p_2=p_3+p_4$, $P$ refers to intermediate (on-shell) pseudoscalar states, in our case, $P=\pi^0,\eta$ and $\eta'$ and OT refers to crossed channels 
(i.e. different time-orderings) and additional (multi)particle states not necessarily of pseudoscalar nature. Identifying the above matrix elements 
with the $S$-matrix for a pseudoscalar to electromagnetic current transition\footnote{ Note that coupling then to the photons would 
require an additional $(ie)^2$ factor, which has actually been accounted for in \cref{eq:HLbLgen}.},
\begin{align}
\label{eq:GreenToTFF}
\int d^4x e^{iq\cdot x} \bra{0} T\{j^{\mu}(x)j^{\nu}(0) \} \ket{P} = i\mathcal{M}^{\mu\nu}_{P\rightarrow\gamma^*\gamma^*}, 
\end{align}
allows to express the behavior around the pseudoscalar poles for the HLbL Green's function (momentum conservation is now implied) as
\begin{equation}
\label{eq:GtoGG}
\Pi^{\mu\nu\rho\sigma}(p_1,p_2,p_3) = i\mathcal{M}^{\mu\nu}_{P\rightarrow\gamma^*\gamma^*} \frac{i}{(p_1+p_2)^2-m_P^2} i\mathcal{M}^{\rho\sigma}_{\gamma^*\gamma^*\rightarrow P} + \textrm{crossed}.
\end{equation} 
Accounting for the additional crossed channels, we find what is known as the pseudoscalar-pole contribution to  
the HLbL tensor. For the kinematics\footnote{Remember that the vector currents are defined to have $q_1,q_2,q_3,-k$ outgoing momenta 
($k=q_1+q_2+q_3$).} described in \cref{fig:pspole}~\cite{Knecht:2001qf}, this is given as
\begin{align}
\Pi_{\mu\nu\lambda\rho}^{P-\textrm{pole}}(q_1,q_2,q_3) = & \  i\frac{F_{P\gamma^*\gamma^*}(q_1^2,q_2^2)F_{P\gamma^*\gamma^*}(q_3^2,k^2)}{(q_1+q_2)^2-m_P^2}
               \epsilon_{\mu\nu\alpha\beta}q_1^{\alpha}q_2^{\beta}\epsilon_{\lambda\rho\sigma\tau}q_3^{\sigma}k^{\tau}  \nonumber \\
 & +  i\frac{F_{P\gamma^*\gamma^*}(q_1^2,k^2)F_{P\gamma^*\gamma^*}(q_3^2,q_2^2)}{(q_2+q_3)^2-m_P^2} 
            \epsilon_{\mu\rho\alpha\beta}q_1^{\alpha}k^{\beta}\epsilon_{\nu\lambda\sigma\tau}q_2^{\sigma}q_3^{\tau}  \nonumber \\
 & +  i\frac{F_{P\gamma^*\gamma^*}(q_1^2,q_3^2)F_{P\gamma^*\gamma^*}(k^2,q_2^2)}{(q_1+q_3)^2-m_P^2} 
            \epsilon_{\mu\lambda\alpha\beta}q_1^{\alpha}q_3^{\beta}\epsilon_{\nu\rho\sigma\tau}q_2^{\sigma}k^{\tau},  \label{eq:polehlbl}
\end{align}
where the different terms correspond to the $s$, $t$ and $u$ channels 
depicted in \cref{fig:ppoleHLBL}.
\begin{figure}[t]
\centering
\includegraphics[width=0.8\textwidth]{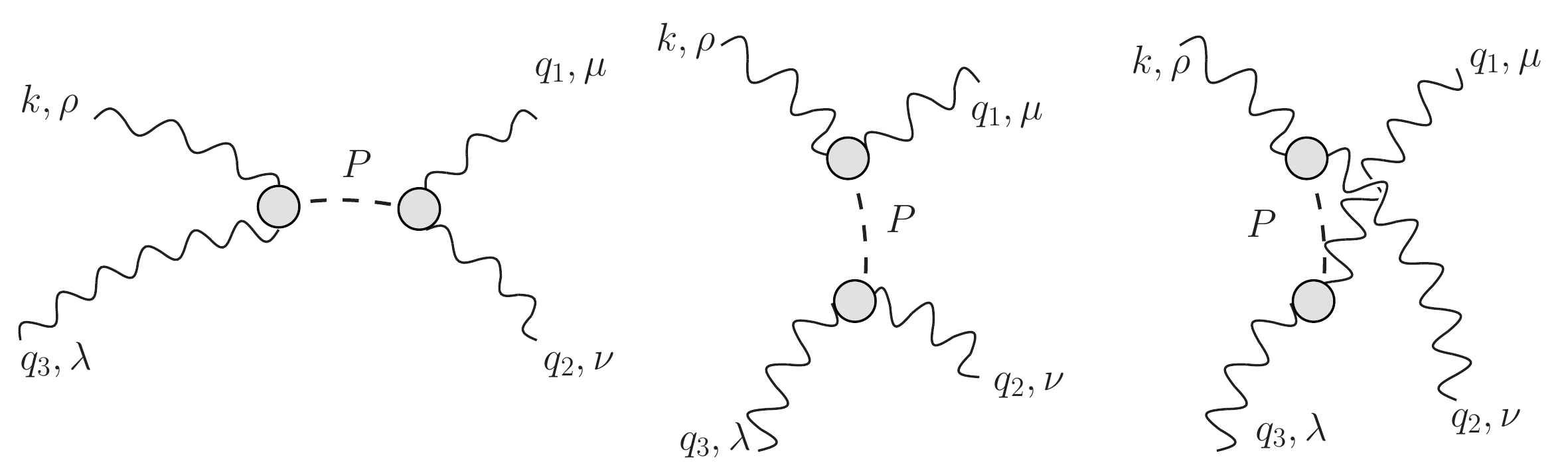}
\caption{The $s,t$ and $u$ channel pseudoscalar-pole contribution to the HLbL tensor. \label{fig:ppoleHLBL} }
\end{figure}
The procedure outlined above allows then to extract a contribution to the HLbL tensor which is defined in terms of  a physical measurable quantity, the 
pseudoscalar TFFs.

In the preceding discussion we have retained a particular (exclusive) contribution among all the intermediate states.
At present, there is an ongoing effort to improve the charged pion loop~\cite{Colangelo:2014dfa,Colangelo:2014pva} contribution, 
which should be the most relevant multiparticle contribution to $a_{\mu}^{\textrm{HLbL}}$; the relevance of pQCD may be 
estimated, in a model-dependent way, from the OPE expansion, a discussion which we postpone to \cref{sec:bpole}. Still, 
as we have outlined, the pseudoscalar-pole contribution is a model-independent and properly defined contribution in QFT,  
associated to an isolated pole in the $S$ matrix, which we proceed to discuss.

\subsection{Master formula and main properties}

In this section, we provide the calculation details following Refs.~\cite{Knecht:2001qf,Jegerlehner:2009ry}. 
Moreover, we discuss the relevant kinematical regions of the integral in view of the obtained kernel functions.
Plugging the pole-contribution, \cref{eq:polehlbl}, into the master formula \cref{eq:HLbLint}, 
one obtains the result, see Ref.~\cite{Knecht:2001qf}
\begin{align}
a_{\mu}^{\textrm{HLbL};P} = & \ -e^6\int \frac{d^4q_1}{(2\pi)^4}\frac{d^4q_2}{(2\pi)^4} \frac{1}{q_1^2 q_2^2 (q_1+q_2)^2 [(p+q_1)^2-m_{\ell}^2] [(p-q_2)^2-m_{\ell}^2]} \nonumber \\ 
 & \ \left[ \frac{F_{P\gamma^*\gamma^*}(q_1^2,(q_1+q_2)^2) F_{P\gamma^*\gamma^*}(q_2^2,0)}{q_2^2 - m_{\pi}^2} T_1(q_1,q_2;p) \right. \nonumber \\
 & \ \left. + \frac{F_{P\gamma^*\gamma^*}(q_1^2,q_2^2) F_{P\gamma^*\gamma^*}((q_1+q_2)^2,0)}{(q_1+q_2)^2 - m_{\pi}^2} T_2(q_1,q_2;p) \right], \label{eq:ppole}
\end{align}
where
\begin{align}
  T_{1}(q_1,q_2;p) = & \ \frac{16}{3}(p\cdot q_1)(p\cdot q_2)(q_1\cdot q_2) - \frac{16}{3}(p\cdot q_2)^2q_1^2  \nonumber \\
                     & \ - \frac{8}{3}(p\cdot q_1)(q_1\cdot q_2)q_2^2 +8(p\cdot q_2)q_1^2q_2^2  \nonumber \\ 
                     & \ - \frac{16}{3}(p\cdot q_2)(q_1\cdot q_2)^2 + \frac{16}{3}m_{\ell}^2q_1^2q_2^2 - \frac{16}{3}m_{\ell}^2(q_1\cdot q_2)^2, \label{eq:T1} 
\end{align}
\begin{align}
  T_{2}(q_1,q_2;p) = & \ \frac{16}{3}(p\cdot q_1)(p\cdot q_2)(q_1\cdot q_2) - \frac{16}{3}(p\cdot q_1)^2q_2^2 \nonumber \\
                     & \ +\frac{8}{3}(p\cdot q_1)(q_1\cdot q_2)q_2^2 + \frac{8}{3}(p\cdot q_1)q_1^2q_2^2 \nonumber \\ 
                     & \ + \frac{8}{3}m_{\ell}^2q_1^2q_2^2 - \frac{8}{3}m_{\ell}^2(q_1\cdot q_2)^2. \label{eq:T2}
\end{align}
In deriving~\cref{eq:ppole,eq:T1,eq:T2}, the change of variables $q_2\rightarrow q_2 -q_1$, then $q_1\rightarrow-q_1$ has been used. The second and third 
graphs in Fig.~\ref{fig:pspole} give the same contribution proportional to $T_1(q_1,q_2;p)$, whereas the fourth diagram in Fig.~\ref{fig:pspole} provides the 
term proportional to $T_2(q_1,q_2;p)$. In the last, the symmetry property $q_1 \leftrightarrow -q_2$ has been used.
To further simplify the integral~\cref{eq:ppole}, we use the Wick rotation and employ the techniques of Gegenbauer polynomials~\cite{Knecht:2001qf,Jegerlehner:2009ry}. 
Following the approach in Ref.~\cite{Jegerlehner:2009ry},
\begin{align}
a_{\ell}^{\textrm{HLbL};P} = & \  \frac{-2\pi}{3}\left(\frac{\alpha}{\pi}\right)^3 \int_0^{\infty}dQ_1dQ_2 \int_{-1}^{+1}dt \sqrt{1-t^2} Q_1^3Q_2^3  \nonumber \\
                            & \ \times  \left[  \frac{F_1 I_1(Q_1,Q_2,t)}{Q_2^2+m_{P}^2}  +    \frac{F_2 I_2(Q_1,Q_2,t)}{Q_3^2+m_{P}^2}  \right], \label{eq:hlblpex}
\end{align}
where $Q_3^2 = Q_1^2+Q_2^2+2Q_1Q_2t$ and 
\begin{align}
F_1 &= F_{P\gamma^*\gamma^*}(Q_1^2,Q_3^2)F_{P\gamma^*\gamma^*}(Q_2^2,0), \\ F_2 &= F_{P\gamma^*\gamma^*}(Q_1^2,Q_2^2)F_{P\gamma^*\gamma^*}(Q_3^2,0),
\end{align}
and
\begin{align}
I_1(Q_1,Q_2,t) = & \ \frac{1}{m_{\ell}^2Q_3^2}\left[ -\frac{4m_{\ell}^2t}{Q_1Q_2} - (1-R_{m_1})\left( \frac{2Q_1t}{Q_2}-4(1-t^2) \right) \right. \nonumber \\
  & \ + \left. (1-R_{m_1})^2\frac{Q_1t}{Q_2} + 8X(Q_1,Q_2,t)\left( Q_2^2 -2m_{\ell}^2)(1-t^2) \right) \right], \\
I_2(Q_1,Q_2,t) = & \ \frac{1}{m_{\ell}^2Q_3^2}\left[ -2(1-R_{m_1})\left(\frac{Q_1t}{Q_2}+1\right) -2(1-R_{m_2})\left(\frac{Q_2t}{Q_1}+1\right)  \right. \nonumber \\
               & \ - \left. 4X(Q_1,Q_2,t)\left(Q_3^2 +2m_{\ell}^2(1-t^2) \right)\right].
\end{align}
In the last term, the $Q_1\leftrightarrow - Q_2$ invariance has been used again to make $I_2(Q_1,Q_2,t)$ symmetric. The above expressions employ the functions 
arising from angular integration,
\begin{align}
X(Q_1,Q_2,t) = & \ \frac{1}{Q_1Q_2\sqrt{1-t^2}}\arctan\left(\frac{z\sqrt{1-t^2}}{1-zt}\right),  \\
z = & \ \frac{Q_1Q_2}{4m_{\ell}^2}(1-R_{m_1})(1-R_{m_2}), \\
R_{m_i}= & \ \sqrt{1+4m_{\ell}^2/Q_i^2}.
\end{align}

Having defined all the required equations, it is interesting before embarking on the TFF description and performing the numerical calculation, to discuss 
the main aspects of the integrand in \cref{eq:hlblpex} and study the main features which are required in order to provide a very precise estimation of this quantity 
---for a thorough study see Ref.~\cite{Nyffeler:2016gnb}. 
First we plot, up to an overall $(\alpha/\pi)^3$ factor, the two terms in \cref{eq:hlblpex} for $P=\pi^0$ and a constant TFF in Fig.~\ref{fig:g2kernel}.
\begin{figure}[t]
\centering
   \includegraphics[width=0.44\textwidth]{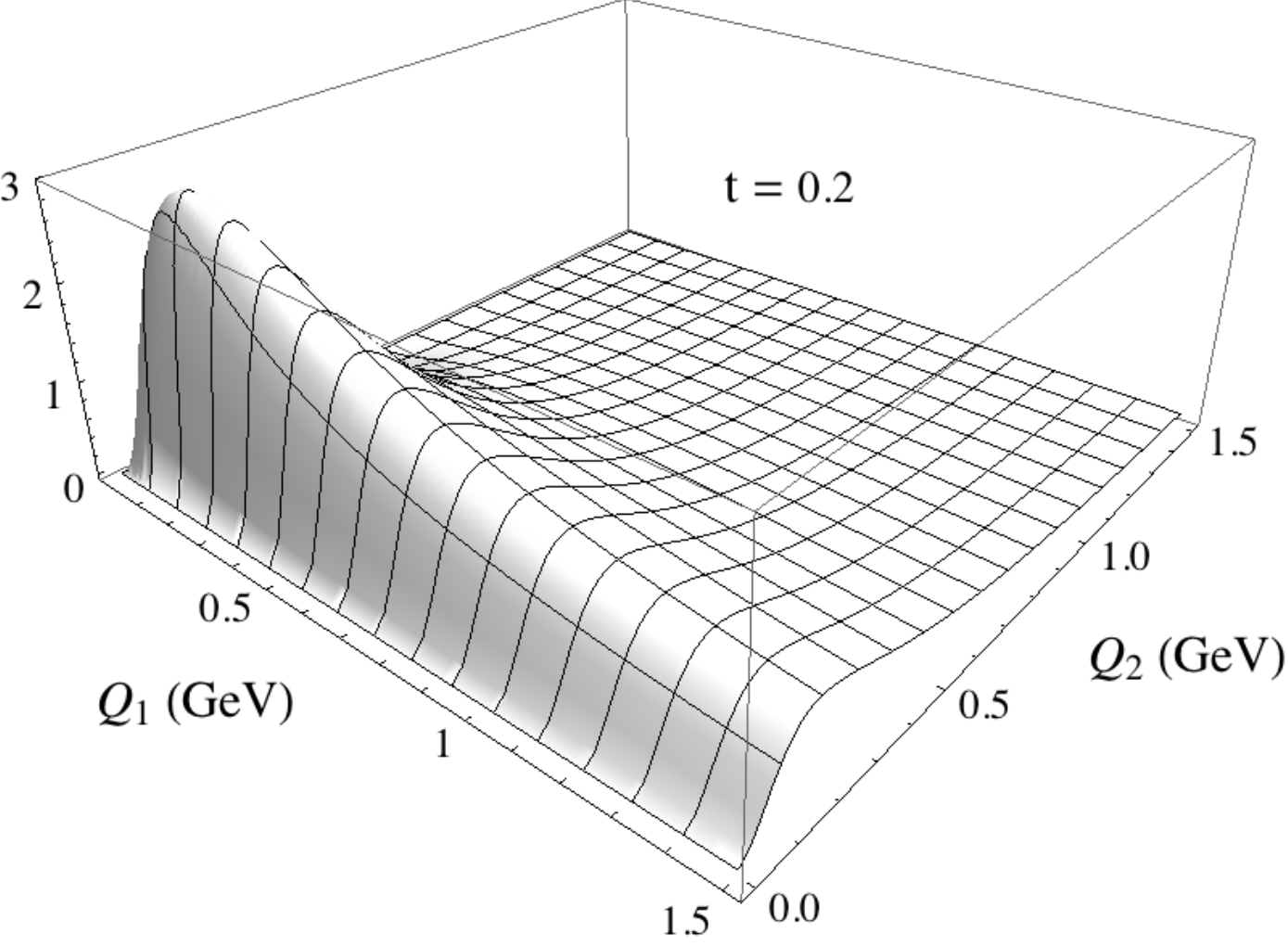}
   \includegraphics[width=0.44\textwidth]{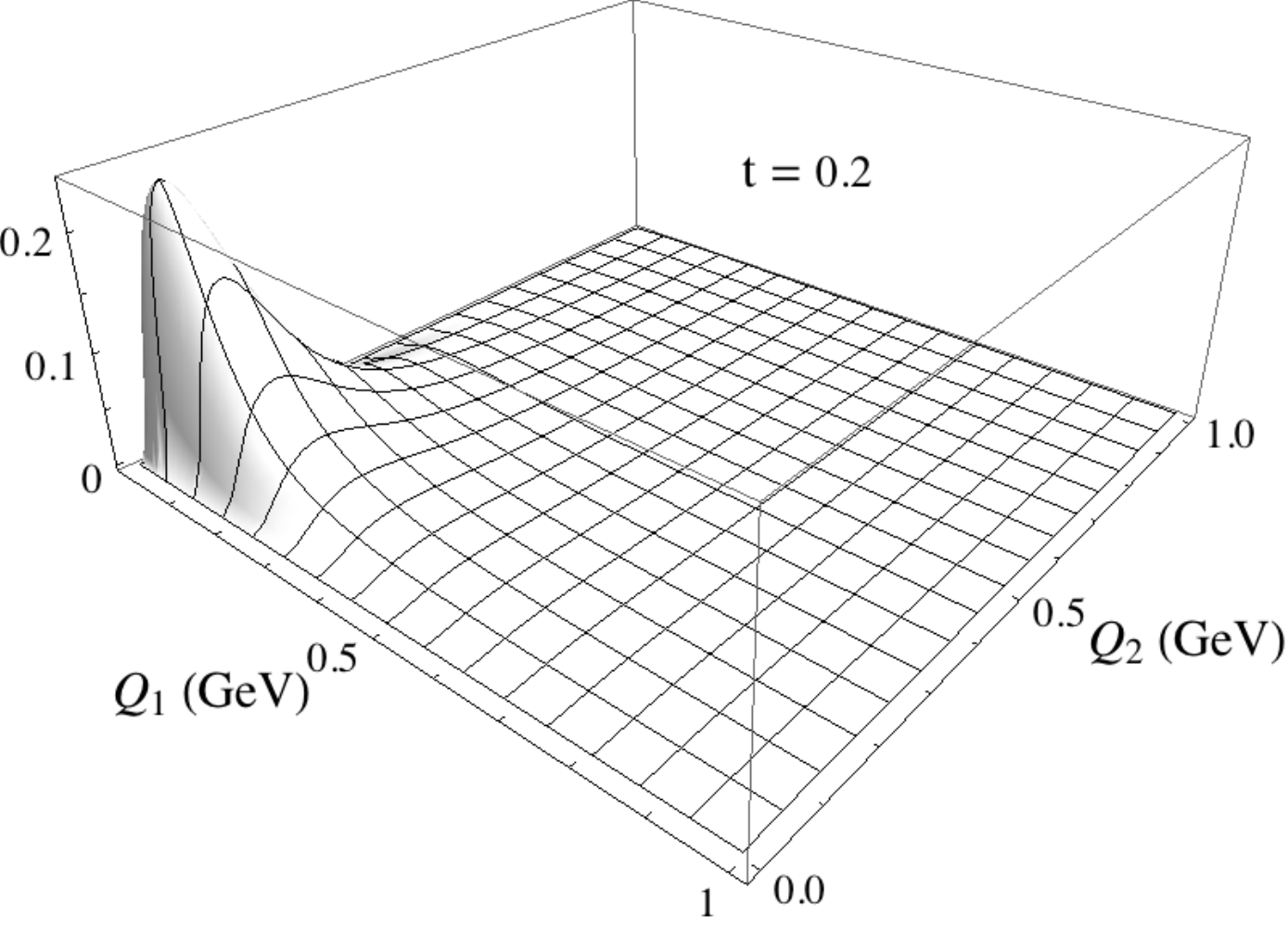}
   \caption{Integrands in \cref{eq:hlblpex} for $P=\pi^0$ and a constant TFF and $t=0.2$ (we omit the overall $(\alpha/\pi)^3$ factor). Left and right stand for the first 
            ($\sim F_1$) and second ($\sim F_2$) terms in the integrand in \cref{eq:hlblpex}, respectively.}
   \label{fig:g2kernel}
\end{figure}
For plotting, we choose $t=0.2$, though a similar shape appears for different $t$ values. As one can see, both integrands peak at very low-energies  and the 
first one features a non-negligible tail extending up to moderate energies. 
From this observation it is clear that any approach aiming for a precise determination must 
provide an extremely precise description for the TFF below $1$~GeV~\cite{Masjuan:2017tvw}. Unfortunately, there is no data available in this region  
for the single virtual TFF, see \cref{fig:datakernel}  ---there is the notorious exception of the $\eta'$ data from L3, which however has never been included in previous analyses. 
In practice, this means that previous calculations have required a model for the TFF together 
with an extrapolation down to $Q^2\sim0$. The error that such extrapolation may induce was a systematic source of error not accounted for. 
\begin{figure}
\centering
   \includegraphics[width=\textwidth]{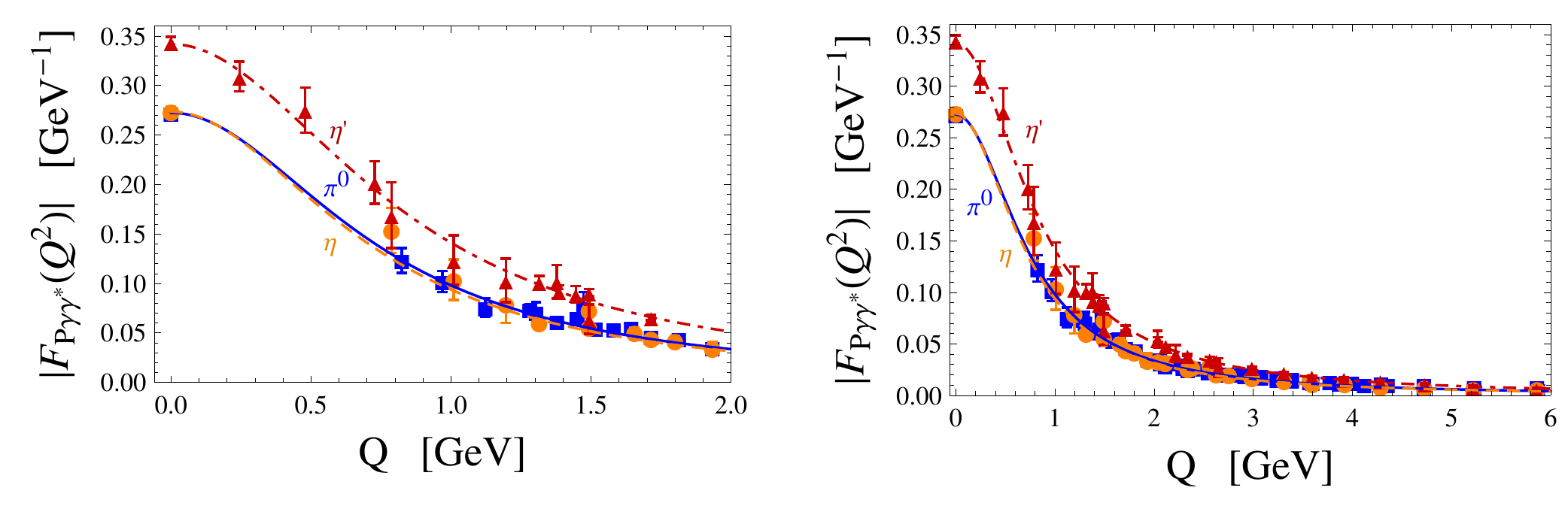}
   \caption{The single virtual $\pi^0,\eta$ and $\eta'$ TFF description from PAs from {\cref{chap:data}} as blue, dashed orange and red dash-dotted lines,  
   together with the available space-like data  at low-energies 
   from CELLO~\cite{Behrend:1990sr}, CLEO~\cite{Gronberg:1997fj}, L3~\cite{Acciarri:1997yx} and the normalization extracted from~\cite{Agashe:2014kda} in 
   blue squares, orange circles and red triangles for the $\pi^0,\eta$ and $\eta'$, respectively.}
   \label{fig:datakernel}
\end{figure}
However, a precise low-energy description ---which in principle would be provided from \cpt--- is not enough. 
At high energies, the mentioned tail for the first integral cannot be neglected; the behavior of such is given (after angular $t$ integration) for large $Q_1$ in \cref{eq:F1asQ1}, 
for large $Q_2$ in \cref{eq:F1asQ2} and for large $Q_1=Q_2\equiv Q$ in \cref{eq:F1asQ1} below~\cite{Knecht:2001qf,Nyffeler:2016gnb}   
\begin{align}
 \frac{8\pi^2Q_2^3(1-t^2)^{3/2}}{3m_{\mu}^2(m_{P}^2+Q_2^2)}\frac{1}{Q_1}\left( R_{m_2}(2m_{\mu}^2-Q_2^2)+Q_2^2 \right) + \mathcal{O}(Q_1^{-2}), \label{eq:F1asQ1} \\
 \frac{8\pi Q_1^2 t(1-t^2)^{3/2}}{3m_{\mu}^2Q_2^2}\left( Q_1^2(R_{m_1}-1) -2m_{\mu}^2 \right)+ \mathcal{O}(Q_2^{-3}), \label{eq:F1asQ2} \\
 \frac{8\pi m_{\mu}^2\sqrt{1-t^2}(3-t)(1-t)}{3Q^2} + \mathcal{O}(Q^{-4}). \label{eq:F1asQ1Q2}
\end{align}
As one can see, the resulting integral diverges for a constant TFF. Consequently, any parametrization for the TFF must incorporate, beyond a precise 
TFF description at low-energies, the appropriate high-energy behavior which is necessary to render the integral finite. 
On turn, the second (subleading) term falls much faster. Its high-energy behavior is given for large $Q_{1,2}$ values in \cref{eq:F2asQ12} and for large 
$Q_1=Q_2\equiv Q$ in \cref{eq:F2asQ1Q2} below~\cite{Knecht:2001qf,Nyffeler:2016gnb} 
\begin{align}
\frac{8\pi Q_{2,1}^3(1-t^2)^{3/2}}{9m_{\mu}^2 Q_{1,2}^3} \left( (R_{m_{2,1}}-1)(m_{\mu}^2+Q_{2,1}^2) -2m_{\mu}^2  \right) + \mathcal{O}(Q_{1,2}^{-4}), \label{eq:F2asQ12} \\
\frac{4m_{\mu}^4\pi(1-t)^{3/2}(2-t)}{9\sqrt{1+t}Q^4} \label{eq:F2asQ1Q2}.
\end{align}
From the equations above, it can be observed that such an integral yields a finite result even for a constant TFF.

\subsection{A rational description for $F_{P\gamma^*\gamma^*}(Q_1^2,Q_2^2)$}
\label{sec:rationalTFF}

In the previous subsection, we found that any description for $F_{P\gamma^*\gamma^*}(Q_1^2,Q_2^2)$ aiming to obtain a precise determination for the pseudoscalar-pole 
contribution \alblp requires:
\begin{itemize}
\item A full-energy description in the whole space-like region; the time-like features do not directly play a role in this calculation, as one has performed 
a Wick rotation to Euclidean space.
\item An extremely precise description at energies below $1$~GeV, with special emphasis on the region $Q^2\sim0$.
\item An appropriate high-energy behavior providing a convergent integral. Actually, given the non-negligible tail from the integrand, 
      implementing the correct $Q^2$ power-like behavior becomes relevant. 
\end{itemize}
This means that any phenomenological approach should:
\begin{itemize}
\item Reproduce the available data in the space-like region. In addition, 
having a finite-data set, the approach should guarantee that the extrapolation down to $Q^2=0$ and $Q^2=\infty$ converges to 
the original function; in this sense, the method should provide an error estimation for the extrapolation procedure. 
\item Have the ability to be systematically improved to meet the eventual required precision.
\end{itemize}
For the moment, we can distinguish three different approaches calculating the pseudoscalar-pole contribution. 
First, there are those which provide a theoretical model for the TFF~\cite{Bijnens:1995xf,Bijnens:2001cq}. These may entail large systematic 
uncertainties inherent to the models, which do not correspond to the full QCD theory and are hard to estimate. 
Second, there are those phenomenological data-based parametrizations, typically inspired 
on large-$N_c$ and VMD ideas, see for instance~\cite{Hayakawa:1997rq,Knecht:2001qf,Roig:2014uja}. 
Whereas these kind of models may reproduce the fitted data, it is unclear how precise their extrapolations to 
$Q^2=0$ and $Q^2=\infty$ are and their associated errors ---the reason for which 
very precise low-energy data is highly desired in these approaches. In addition, strictly speaking, they should include the whole 
family of vector resonances; their truncation, arbitrary choice of resonances, the connection among the required large-$N_c$ inputs 
with the real world, and the unavoidable large-$N_c$ corrections must play a role, of which the systematic error is missing and 
hard to account for. Finally, we find the (recently proposed) dispersive 
reconstruction of the TFFs~\cite{Colangelo:2014dfa,Colangelo:2014pva}. 
Whereas this approach would provide in principle 
an exact numerical calculation, in the real world some approximations must be taken. As an example, only the lowest-lying 
thresholds are employed and the dispersive integrals must 
be cut at some value. A serious disadvantage arising from this feature is the inability to extend up to arbitrary large $Q^2$ values.  
Estimating the ($Q^2$-dependent) systematic error and eventually improving their result ---if the available experimental data eventually 
requires it--- is a weakness of the method, while they may profit from a large amount of time-like data as compared to other approaches. 
\\

To amend for these shortcomings and determine this calculation to the required precision nowadays ---which is beyond the reach of previous 
studies--- we propose to use a rational approach description based on Canterbury approximants. 
As we have seen, our approach provides a corpus to extract and implement not only the relevant low-energy behavior but the high-energy one 
as well from a data-based procedure. The uniqueness of the method resides in the convergence checks and provides a 
tool to safely extrapolate to the regions where no data is available. On top, its sequential implementation allows to account for 
a systematic error in an easy way. Given the available limited information ---specially on the doubly-virtual TFF--- we restrict our studies 
to the $C^0_1(Q_1^2,Q_2^2)$ and $C^1_2(Q_1^2,Q_2^2)$ elements of the $C^N_{N+1}(Q_1^2,Q_2^2)$ sequence; different sequences turn out 
not to obey the high-energy behavior, for which they are not considered.

\subsubsection{The first element: $C^0_1(Q_1^2,Q_2^2)$}

Given the low-energy expansion for the TFF 
\begin{equation}
\label{eq:tffexpch6}
  F_{P\gamma^*\gamma^*}(Q_1^2,Q_2^2) =   F_{P\gamma\gamma}(1 -\frac{b_P(Q_1^2+Q_2^2)}{m_P^2} + \frac{a_{P;1,1}Q_1^2Q_2^2}{m_P^4} + \frac{c_P(Q_1^4+Q_2^4)}{m_P^4} + ...),
\end{equation}
the lowest CA we can construct in terms of its low-energy parameters defined above is given as 
\begin{equation}
\label{eq:c01low}
C^0_1(Q_1^2,Q_2^2) =  \frac{F_{P\gamma\gamma}}{1+\frac{b_P}{m_P^2}(Q_1^2+Q_2^2) + \frac{2b_P^2-a_{P;1,1}}{m_P^4} Q_1^2Q_2^2}.
\end{equation}
All the necessary single-virtual parameters ---$F_{P\gamma\gamma}$ and $b_P$--- have been determined so far in \cref{chap:data} and summarized \cref{tab:chap1mainres}. 
It remains however the determination for the double-virtual parameter $a_{P;1,1}$. 
Even if this is not available at the moment, the possibility of having access to the doubly-virtual TFF in the near future would allow for such an extraction 
(efforts are being made in BESIII). Still, for the moment, we are compelled to judge on a theoretical reasonable estimate. 
On one side, given the low-energy dominance of the process we are looking for, we may find guidance in $\chi$PT. From the work in Ref.~\cite{Bijnens:2012hf}, it seems 
that $\chi$PT favors a factorized behavior, namely, that $F_{P\gamma^*\gamma^*}(Q_1^2,Q_2^2)\sim F_{P\gamma^*\gamma}(Q_1^2)F_{P\gamma^*\gamma}(Q_2^2)$, implying
$a_{P;1,1} = b_P^2$. Actually, such behavior was obtained as well for the $\eta$ case in a dispersive 
analysis~\cite{Xiao:2015uva}. On the other side, we may find help from the high energies, where the two point approximant with the OPE behavior (see \cref{eq:OPElim}) 
built-in\footnote{With the OPE behavior we mean that $F_{P\gamma^*\gamma^*}(Q^2,Q^2)$ behaves as $Q^{-2}$ as $Q^2\to\infty$.} implies, as a first approximation, 
that $a_{P;1,1} = 2b_P^2$ ---moreover, this upper value avoids for poles in the SL region. 
This suggests that the high-energy corrections should drive the value which is obtained from the low-energies in some region in between $b_P^2\leq a_{P;1,1} \leq 2b_P^2$. 
Then, we will take this range as a theoretical estimate where the real value ---to be determined from data--- is likely to be found. Of course, experimental data will have the 
last word on this choice.

\subsubsection{Second element: $C^1_2(Q_1^2,Q_2^2)$}

The next CA within the chosen sequence is parametrically given as 
\begin{equation}
  C^1_2(Q_1^2,Q_2^2) =  
\textstyle\frac{F_{P\gamma\gamma}(1+ \alpha_1(Q_1^2+Q_2^2) + \alpha_{1,1}Q_1^2Q_2^2)}{1+\beta_1(Q_1^2+Q_2^2) +\beta_2(Q_1^4+Q_2^4) + \beta_{1,1} Q_1^2Q_2^2 + \beta_{2,1}Q_1^2Q_2^2(Q_1^2+Q_2^2) + \beta_{2,2}Q_1^4Q_2^4}. 
\label{eq:c12def}
\end{equation} 
The single virtual parameters can be related to the low-energy expansion of the TFF, see \cref{eq:series}, as
\begin{equation}
\label{eq:leplow}
\alpha_1 = \frac{-b_P^3 + 2b_Pc_P -d_P}{m_P^2(b_P^2 -c_P)}, \beta_1 =  \frac{b_Pc_P-d_P}{m_P^2(b_P^2-c_P)}, \beta_2 =  \frac{c_P^2-b_Pd_P}{m_P^4(b_P^2-c_P)}.
\end{equation}
Alternatively, we could have employed a combined low- and high-energy expansion (see \cref{sec:PAnpoint}) which enforces the BL behavior 
($F_{P\gamma^*\gamma}(Q^2)\sim P_{\infty}Q^{-2}$). Sacrificing $d_P$ in favor of $P_{\infty}$, we would obtain
\begin{equation}
\label{eq:lepbl}
\alpha_1 = \frac{P_{\infty}(b_P^2-c_P)}{F_{P\gamma\gamma}-b_P P_{\infty}}, \beta_1 =  \frac{b_PF_{P\gamma\gamma} - c_PP_{\infty}}{F_{P\gamma\gamma} -b_P P_{\infty} }, \beta_2 =  \frac{F_{P\gamma\gamma}(b_P^2-c_P)}{F_{P\gamma\gamma}-b_P P_{\infty}}.
\end{equation}
The additional $\alpha_{i,j}$ and $\beta_{i,j}$ parameters are connected to the double virtual series expansion, see \cref{chap:CA}. 
Given our lack of experimental or theoretical information for the doubly-virtual TFF, it is hard to express them in terms of the low-energy expansion. 
As a first start, we proceed analogous to the previous section and fix the value from $\beta_{1,1}$ from the low-energy parameter $a_{P;1,1}$, 
which leads to the constraint
\begin{equation}
\beta_{1,1} = -a_{P;1,1} +\alpha_{1,1} -2\alpha_1\beta_1 + 2\beta_1^2
\end{equation}
To fix the additional remaining parameters we are doomed to use some high-energy constraints, even 
though this may come at cost of the low-energy description. Nicely, the OPE behavior allows to set $\beta_{2,2}=0$. Moreover, if we do not only 
fix the power-like behavior as for the $C^0_1(Q_1^2,Q_2^2)$ case, but constrain its leading coefficient from \cref{eq:OPElim}, we find
\begin{equation}
   \alpha_{1,1} = (2/3)\beta_{2,1}P_{\infty}/F_{P\gamma\gamma},
\end{equation}
where $P_{\infty}$ is the BL TFF asymptotic behavior. For the $\pi^0$, $\pi^0_{\infty}=2F_{\pi}$, whereas for the $\eta$ and $\eta'$ this depends on the mixing parameters and was 
determined in \cref{chap:data}. Still, there is an additional undetermined parameter, $\beta_{2,1}$. Therefore, we make use of the higher order terms in the OPE expansion, 
which for the $\pi^0$ reads~\cite{Novikov:1983jt,Jegerlehner:2009ry}
\begin{equation}
\label{eq:tffope}
F_{\pi\gamma^*\gamma^*}(Q^2,Q^2) = (2/3)F_{\pi}\left( \frac{1}{Q^2} -\frac{8}{9}\frac{\delta^2}{Q^4} + ... \right). 
\end{equation}
The $\delta$ parameter has been estimated using sum rules, obtaining $\delta^2=0.20(2)$ \cite{Novikov:1983jt,Jegerlehner:2009ry}. 
To extend this value to the $\eta$ and $\eta'$ cases, we replace $2F_{\pi}\to\eta_{\infty}^{(\prime)}$ and apply an additional $30\%$ uncertainty due to 
$SU(3)_F$ breaking and large-$N_c$ corrections 
(note that this is enough for all the low- and high-energy parameters analyzed so far). 
This provides the remaining constraint 
\begin{equation}
\beta_{2,1} = \frac{9F_{P\gamma\gamma}(a_{1;PP}P_{\infty} +2\alpha_1(3F_{P\gamma\gamma}+P_{\infty}\beta_1) -2P_{\infty}(\beta_1^2+\beta_2) )}{2P_{\infty}(3P_{\infty}-8F_{P\gamma\gamma}\delta^2)}. 
\end{equation}
Finally, we could employ our previous estimation $b_P^2\leq a_{P;1,1} \leq 2b_P^2$. However, the appearance of poles for certain $a_{P;1,1}$ values restrict the chosen range. 
To see this, take $a_{P;1,1} = \lambda b_P^2$ and the phenomenological observation that $c_p\sim b_P^{2}$ and $d_P\sim b_P^{3}$; avoiding the appearance of poles requires then
\begin{equation}
\frac{9(-3b_P F_{P\gamma\gamma}^2 + b_P^2 F_{P\gamma\gamma} P_{\infty} \lambda )}{2P_{\infty}(3P_{\infty} - 8F_{P\gamma\gamma}\delta^2)} + \mathcal{O}(\epsilon) >0
\longrightarrow \lambda>\frac{3F_{P\gamma\gamma}}{b_P P_{\infty}} + \mathcal{O}(\epsilon),
\end{equation}
where we used that, phenomenologically, $(3P_{\infty} - 8F_{P\gamma\gamma}\delta^2)>0$. It turns out that $\lambda\sim 2$ in our cases, 
supporting our assumption that high-energy QCD properties should drive up the factorization value closer to the OPE choice, $a_{P;1,1} = 2b_P^2$ 
and naturally providing an $a_{P;1,1}$ lower bound. In addition, from Pad\'e theory, it is not expected to find complex-conjugated poles\footnote{The PA reality condition 
forces the approximant to have either real or pair of complex-conjugated poles.} in the SL region; this provides in practice an upper bound for $a_{P;1,1}$ above $2b_P^2$ and a 
(more generous) band for $a_{P;1,1}$ solely based in Pad\'e theory criteria.

\subsubsection{Systematic errors}

Before we present our final results, it is necessary to set up some procedure allowing to determine the systematic error to be associated to a certain element within 
our chosen $C^N_{N+1}$ sequence. Actually, given the length of our sequence, consisting of two elements alone, it is extremely important to check on the expected 
convergence. For this purpose, we come back again to two models which have been widely-employed along this work. These are the 
Regge~\cite{RuizArriola:2006jge,Arriola:2010aq} and the proposed doubly-virtual logarithmic models defined in \cref{sec:examples}, 
\begin{align}
F^{\textrm{Regge}}_{P\gamma^*\gamma^*}(Q_1^2,Q_2^2) &= \frac{aF_{P\gamma\gamma}}{Q_1^2-Q_2^2}
                                         \frac{\left[ \psi^{(0)}\left(\frac{M^2+Q_1^2}{a}\right) -\psi^{(0)}\left(\frac{M^2+Q_2^2}{a}\right) \right]}{\psi^{(1)}\left(\frac{M^2}{a}\right)}, \\
\label{eq:logmod}
F^{\textrm{log}}_{P\gamma^*\gamma^*}(Q_1^2,Q_2^2) &= \frac{F_{P\gamma\gamma}M^2}{Q_1^2-Q_2^2}\ln\left( \frac{1+Q_1^2/M^2}{1+Q_2^2/M^2} \right).
\end{align}
Recall that such models incorporate a well-defined high-energy behavior for the doubly-virtual TFF, whereas the single-virtual one behave as $\ln(Q^2)Q^{-2}$,  not
fulfilling the BL behavior, but convergent enough in order to perform the integral \cref{eq:hlblpex}. The relevance of these models is that an infinite sequence of CAs 
is required to describe the underlying function, which makes them an ideal laboratory to test convergence properties.\\

To test the performance of our approximants, we calculate the HLbL contribution, \cref{eq:hlblpex}, for the specific (dominant) case of the $\pi^0$. We show our results together 
with the exact outcome from the model in 
Table~\ref{tab:regge}. To test the accuracy of different assumptions, we show the result from different strategies:
matching all the doubly-virtual low-energy parameters (LE), setting the OPE $Q^{-2}$ power-like behavior as well as including its leading $c/Q^{2}$ 
coefficient\footnote{Note that, for the logarithmic model \cref{eq:logmod}, the $C^N_{N+1}$ approximants with the OPE behavior built-in already reproduce the 
whole $Q_1^2=Q_2^2$ regime, see \cref{eq:appelleq}, so the entries OPE:$~Q^{-2}$ and OPE:$c/Q^2$ in \cref{tab:regge} are equivalent.} and, finally, using a factorized form (Fact), 
which has been a common approach.
\begin{table}
\small
\centering
\begin{tabular}{ccccc} \toprule
  & \multicolumn{4}{c}{Regge model}  \\
        & $C^0_1$   &  $C^1_2$ &  $C^2_3$ &  $C^3_4$   \\ \midrule
     LE & $66.0$   & $71.9$   &  $72.8$     &  $73.1$    \\
OPE$:\sim Q^{-2}$ & $77.4$  & $73.4$   &  $73.3$   &  $73.3$    \\
OPE$:c/Q^2$ &     -    & $73.1$   &  $73.3$   &  $73.3$   \\ \midrule
Fact & $65.1$   & $68.8$   &  $69.0$   &  $69.1$    \\  
Fit$^{\textrm{OPE}}$  & $80.2$   & $75.1$   &  $73.7$   &  $73.4$    \\ \midrule
Exact & \multicolumn{4}{c}{$73.3$}  \\ \bottomrule
\end{tabular}
\begin{tabular}{cccc} \toprule
   \multicolumn{4}{c}{Logarithmic model} \\ 
   $C^0_1$   &  $C^1_2$ &  $C^2_3$ &  $C^3_4$ \\ \midrule
   $87.9$   & $97.6$   &  $99.7$  &  $100.5$   \\
   \multirow{ 2}{*}{$99.5$}  & \multirow{ 2}{*}{$101.2$}   &  \multirow{ 2}{*}{$101.4$}   &  \multirow{ 2}{*}{$101.5$}   \\
          &    &   &  \\  \midrule
   $85.2$   & $92.4$   &  $93.6$  &  $94.0$   \\ 
   $113.1$   & $104.0$   &  $102.2$  &  $101.8$   \\ \midrule
 \multicolumn{4}{c}{$101.5$} \\ \bottomrule
\end{tabular}
\caption{Result for $a_{\mu}^{\textrm{HLbL};\pi}$, \cref{eq:hlblpex}, for different approximants compared to the exact result. The first columns stand for the Regge model, 
whereas the last four columns stand for the logarithmic one. See details in the text.}
\label{tab:regge}
\end{table}
We find that, whereas the factorization result does not converge to the model value ---a feature to be expected as the original 
models do not factorize--- the LE approach approximates the exact value even if the proper 
doubly-virtual high-energy behavior is not built-in. This requires however the use of a large sequence and may not be the 
best choice in our case. Certainly, any of the OPE choices seems to provide the best convergence pattern, which can be understood 
as the OPE becomes relevant already at a low scale; still, the first element could involve a large systematic error. 
Finally, we observe that the difference among the $C^N_{N+1}$ and $C^{N-1}_{N}$ elements is enough to give the size of the systematic error. 
Summarizing, we conclude from this study that CAs provide a reliable systematic approach to perform the desired calculation, which systematic error 
estimation can be accounted for from the difference of a given element with the previous one and convergence rate is improved when implementing the OPE 
even if this is not necessary.
In our discussion above, it cannot be overemphasized the relevance of having employed the low-energy TFF expansion in \cref{eq:tffexpch6} when reconstructing 
the approximants ---as the framework requires--- rather than fitting the rational functions to data themselves. As an illustration, we show in the Fit$^{\textrm{OPE}}$ 
row of \cref{tab:regge} what would have been obtained if fitting the $C^N_{N+1}(Q_1^2,Q_2^2)$ rational functions, with the OPE power-like behavior implemented, 
to a $16\times16$ grid of equally-spaced double-virtual data ranging from $0\leq Q_{1,2}^2 \leq 35~\textrm{GeV}^2$. The convergence obtained is slower, and illustrates 
the difference and the power of CAs with respect to standard fitting approaches ---we stress in addition that, in these fits, no assumption of factorization either in numerator 
or denominator has ben employed, which differs form traditional rational approaches, for which we foresee a yet slower convergence.

\subsection{Results for the pseudoscalar-pole contribution}
\label{sec:poleresults}

Having discussed the construction of the approximants and the associated systematic errors, we are in the position to give our final results 
for the pseudoscalar-pole contribution to \albl. For this, we take our results from \cref{tab:chap1mainres} for $F_{P\gamma\gamma}, b_P,c_P,d_P$ and 
$P_{\infty}$, which are required for reconstructing the CAs. The pseudoscalar and lepton masses are taken from~\cite{Agashe:2014kda}.
For the $\eta$ and $\eta'$ cases, we reconstruct the single-virtual parameters using the low-energy constraints, see \cref{eq:leplow}. For the 
$\pi^0$ case, there is no reliable extraction for the $d_{\pi}$ parameter so far. Consequently, we use the BL behavior in order to determine the 
$C^1_2(Q_1^2,Q_2^2)$ approximant single-virtual parameters, cf. \cref{eq:lepbl}.

\begin{table}
\small
\centering
\begin{tabular}{ccc} \toprule
$a_{\mu}^{\textrm{HLbL};P}$ & Fact ($a_{P;1,1}=b_P^2$) & OPE ($a_{P;1,1}=2b_P^2$) \\ \midrule
$\pi^0$ & $54.0(1.1)_F(2.5)_{b_{\pi}}[2.7]_t$  &   $64.9(1.4)_F(2.8)_{b_{\pi}}[3.1]_t$  \\
$\eta$    &     $13.0(0.4)_F(0.4)_{b_{\eta}}[0.6]_t$    &    $17.0(0.6)_F(0.4)_{b_{\eta}}[0.7]_t$    \\
$\eta'$    &    $12.0(0.4)_F(0.3)_{b_{\eta'}}[0.5]_t$    &    $16.0(0.5)_F(0.3)_{b_{\eta'}}[0.6]_t$   \\ \midrule
Total & $79.0[2.8]_t$ & $97.9[3.2]_t$  \\ \bottomrule
\end{tabular}
\caption{Result for $a_{\mu}^{\textrm{HLbL};P}$, \cref{eq:hlblpex}, for the $C^0_1$ approximants for different $a_{P;1,1}$ values in units of $10^{-11}$. See 
description in the text.}
\label{tab:c01pole}
\end{table}

We show the results from the $C^0_1(Q_1^2,Q_2^2)$ approximant for the different pseudoscalars in \cref{tab:c01pole}~\cite{Masjuan:2017tvw}. 
There, we display the results for the double-virtual parameter $b_P^2\leq a_{P;1,1}\leq 2b_P^2$, labelled as 
Fact and OPE, respectively. The errors are separated into those arising from the TFF normalization, $(\cdot)_F$ and those coming 
from the slope, $(\cdot)_{b_P}$ and are symmetrized. The total error, $[\cdot]_t$, is the combination in quadrature of both of them. 
The sum of the $\pi^0$, $\eta$ and $\eta'$ contributions from our $C^0_1(Q_1^2,Q_2^2)$ approximant considering 
our $a_{P;1,1}$ range and adding errors in quadrature reads
\begin{equation}
\label{eq:c01pole}
  a_{\mu}^{\textrm{HLbL};P;C^0_1} = (79.0\div97.9)(3.2)\times10^{-11}.
\end{equation}

For the $C^1_2(Q_1^2,Q_2^2)$ approximation, we estimate our results for the chosen $a_{P;1,1}$ range in which no space-like poles appear, 
$a_{P;1,1}^{\textrm{min}}\leq a_{P;1,1}\leq a_{P;1,1}^{\textrm{max}}$, as previously explained\footnote{This leads, for the minimum, $a_{\pi;1,1}^{\textrm{min}}=1.92b_{\pi}^2$, 
$a_{\eta;1,1}^{\textrm{min}}=1.84b_{\eta}^2$ and $a_{\eta';1,1}^{\textrm{min}}=1.32b_{\eta'}^2$, whereas for the maximum $a_{\pi;1,1}^{\textrm{max}}=2.07b_{\pi}^2$, 
$a_{\eta;1,1}^{\textrm{max}}=2.33b_{\eta}^2$ and $a_{\eta';1,1}^{\textrm{max}}=3.41b_{\eta'}^2$.}. 
For the OPE parameter $\delta$, see \cref{eq:tffope}, we take $\delta^2=0.20(2)$ from Refs.~\cite{Novikov:1983jt,Jegerlehner:2009ry}, and apply 
the mentioned $30\%$ correction to the $\eta$ and $\eta'$ to account for the symmetry breaking effects.  
Again, we decompose the different sources of errors into the single-virtual terms, $(\cdot)_L$, the 
uncertainty on $\delta$, $(\cdot)_{\delta}$, and add them in quadrature to obtain the total error, which is given as $[\cdot]_t$ and symmetrized. 
The numerical values are given in \cref{tab:c12pole}.
\begin{table}
\small
\centering
\begin{tabular}{ccc} \toprule
$a_{\mu}^{\textrm{HLbL};P}$ & $a_{P;1,1}^{\textrm{min}}$ & $a_{P;1,1}^{\textrm{max}}$ \\ \midrule
$\pi^0$ & $63.9(1.3)_L(0)_{\delta}[1.3]_t$  &   $62.9(1.2)_L(0.3)_{\delta}[1.2]_t$  \\
$\eta$    &     $16.6(0.8)_L(0)_{\delta}[1.0]_t$    &    $16.2(0.8)_L(0.5)_{\delta}[0.9]_t$     \\
$\eta'$    &    $14.7(0.7)_L(0)_{\delta}[0.7]_t$    &    $14.3(0.5)_L(0.5)_{\delta}[0.7]_t$  \\ \midrule
Total & $95.2[1.7]_t$ & $93.4[1.7]_t$  \\ \bottomrule
\end{tabular}
\caption{Result for $a_{\mu}^{\textrm{HLbL};P}$, \cref{eq:hlblpex}, for the $C^1_2(Q_1^2,Q_2^2)$ approximants for different $a_{P;1,1}$ values in units of $10^{-11}$. See 
description in the text.}
\label{tab:c12pole}
\end{table}
%
%
%
The sum of the different pseudoscalars $C^1_2(Q_1^2,Q_2^2)$ results for the given band reads
\begin{equation}
\label{eq:c12pole}
  a_{\mu}^{\textrm{HLbL};P;C^1_2} = (93.4\div95.2)(1.7)\times10^{-11}.
\end{equation}
Comparing with the OPE result from the previous element in \cref{eq:c01pole}, $97.9(3.2)\times10^{-11}$,  
we obtain the systematic error, leading to the final result for the $\pi^0$, $\eta$ and $\eta'$ pseudoscalar-pole contributions to the HLbL
%
%
\begin{equation}
\label{eq:finalpole}
  a_{\mu}^{\textrm{HLbL};P} = (93.4\div95.2)(1.7)_{\textrm{stat}}(4.5)_{\textrm{sys}}[4.8]_t\times10^{-11},
\end{equation}
where the first error includes both statistic and systematic errors from the CA reconstruction, the second one is the systematic error associated with the $C^1_2$ element, 
and the last one is a combination in quadrature of the formers and is dominated by systematics. Our result may be compared to that in Ref.~\cite{Knecht:2001qf}, 
$a_{\mu}^{\textrm{HLbL};P} = (58(10) + 13(1)+12(1)=83(12))\times10^{-11}$, 
and the more recent result from Ref.~\cite{Roig:2014uja}, $a_{\mu}^{\textrm{HLbL};P} = (57.5(6) + 14.4(2.6)+10.8(0.9)=82.7(2.8))\times10^{-11}$ ---note that the latter 
does not use any data to parameterize the $\eta$ and $\eta'$ TFFs beyond the information which is included in the mixing parameters and represents a major drawback at the 
required precision. Our approach represents a clear improvement over previous estimates since
\begin{itemize}
\item It is the only one making full use of data for the $\eta$ and $\eta'$, which must be carefully described given the required $10\%$ precision. 
As an example, Ref.~\cite{Knecht:2001qf} did use the CLEO slope~\cite{Gronberg:1997fj} only, whereas Ref.~\cite{Roig:2014uja} did not directly use 
any data, but a prediction from their framework based on the $\pi^0$ TFF, suffering from (unaccounted) $SU(3)_F$-breaking and large-$N_c$ corrections. 
Both of these approaches {\textit{cannot}} reproduce the experimental data accounted for in our approach, which reflects a relevant systematic source of error. 
Furthermore, it is the only one which is fully-data driven.
\item It incorporates the appropriate low-energy behavior encoded in the low-energy parameters (which previous 
approaches cannot guarantee  and is crucial for this calculation) together with the high energies.  
This comes out naturally by construction in our framework. Note that incorporating the high-energy behavior for the $\eta$ and $\eta'$ 
mesons is required at the desired $10\%$ precision, as it can be observed when comparing the two columns in \cref{tab:c01pole}, which 
contrasts with the ---commonly employed--- factorization approached.
\item Finally, we are the first to provide a systematic error, which is by no means negligible. 
If the approaches in Refs.~\cite{Knecht:2001qf,Roig:2014uja} were reconstructed in the spirit of Pad\'e type 
approximants, see \cref{sec:patype,sec:ptype}, one would expect a systematic error larger than our. 
\end{itemize}
%
Last, we comment on the error sources. We find that the statistical error in each channel is similar regarding the single-virtual 
parameters, the $\delta$ parameter and our ignorance on the $a_{P;1,1}$ parameter, which are required to improve if a better precision is desired. 
A first measurement of the double virtuality would drastically improve on the $a_{P;1,1}$-induced error and would represent an important milestone. 
Such measurements would be possible in the future, at least for the $\pi^0$, at BESIII~\cite{Adlarson:2014hka}. Actually, 
this would possibly allow to trade $\delta$ for a low-energy parameter, which would be very interesting. 
To improve the single-virtual parameters errors for the $\pi^0$ would require, looking at \cref{tab:c01pole,tab:c12pole}, an improved determination 
for the $b_{\pi}$, $c_{\pi}$ and ---given the relevance of the low energies--- 
to eventually employ $d_{\pi}$ instead of the asymptotic value. This would be possible with new low-energy data which are expected in the 
near future from the BESIII~\cite{Adlarson:2014hka} experiment in the SL region, and from the NA62~\cite{Hoecker:2016lxt} and 
A2~\cite{Marc:private} collaborations in the low-energy TL region. 
In addition, it is expected that further low-energy SL data in the $(0.01-0.4)~\textrm{GeV}^2$ range would be provided by the KLOE-2~\cite{Babusci:2011bg} 
and and $GlueX$~\cite{Gan:2015nyc} collaborations. For the $\eta$ and $\eta'$, this would require new precise measurements from their two-photons 
decays, which would be possible at the $GlueX$ experiment \cite{Dudek:2012vr}.
Regarding the systematic error, we find that this is similar to the statistical one for the $\pi^0$ and $\eta$ cases and 
larger for the $\eta'$, which points out the relevance of the high-energies for the latter due to its mass. 
In this respect, it would be very interesting to have precise high-energy data for the $\eta'$, which would be possible in Belle II 
experiment. Still, to pin down the systematic errors would be possible only if higher approximants could be constructed, demanding 
the determination of additional parameters, where double-virtual measurements cannot be avoided. Actually, it is the 
systematic error which dominates the final number as this source is taken to be fully correlated among the pseudoscalars.  
This is natural to expect if one assumes a similar convergence pattern for the different channels.


\subsection{Cross-checks I: The light-quark transition form factor}

Given the precision we are aiming for in our calculation, every possible cross-check poses a valuable result, which is specially 
important for the dominant $\pi^0$ contribution. Actually, much has been discussed given the differences between 
\babar~\cite{Aubert:2009mc} and Belle~\cite{Uehara:2012ag} results regarding the $\pi^0$ TFF. We note in this respect that the disagreement\footnote{Actually, 
the global difference is not statistically significant ---around $0.8\sigma$.} 
arises mainly from the region at $(8-13)~\textrm{GeV}^2$\footnote{The apparent rising from \babar data at high-energies 
is much less important and we checked this to be irrelevant for $(g_{\mu}-2)$ ---see Ref.~\cite{Nyffeler:2009uw} as well.}; it was checked in 
Ref.~\cite{Masjuan:2012wy} that still, removing either Belle or \babar from the data sets produced compatible results for the LEPs extraction, 
clearing up any possible inconsistency or additional errors. Nevertheless, given such disagreement at intermediate energies, a second test 
would be welcome. In this respect, we mentioned in \cref{sec:lqtff} that, to a reasonable accuracy and up to an overall charge factor, the light 
quark and the $\pi^0$ TFFs should be very similar (see \cref{fig:qfTFF} at this respect). This offers the opportunity to calculate again the 
$a_{\mu}^{\textrm{HLbL};\pi^0}$ contribution employing the light-quark TFF instead of the $\pi^0$ one. The obtained results are labelled as 
LQ I and shown in \cref{tab:gm2LQC01,tab:gm2LQC12} for the $C^0_1(Q_1^2,Q_2^2)$ and $C^1_2(Q_1^2,Q_2^2)$ approximants. 
\begin{table}
\small
\centering
\begin{tabular}{ccc} \toprule
$a_{\mu}^{\textrm{HLbL};P}$ & Fact ($a_{P;1,1}=b_P^2$) & OPE ($a_{P;1,1}=2b_P^2$) \\ \midrule
LQ I & $50.4(1.3)_F(0.5)_{b_{\pi}}[1.4]_t$  &   $60.4(1.5)_F(0.6)_{b_{\pi}}[1.6]_t$  \\ 
LQ II & $56.2(1.5)_F(0.6)_{b_{\pi}}[1.6]_t$  &   $67.4(1.7)_F(0.7)_{b_{\pi}}[1.8]_t$  \\ \bottomrule
\end{tabular}
\caption{Result for $a_{\mu}^{\textrm{HLbL};\pi^0}$ from $C^0_1$ using the light quark TFF in $10^{-11}$ units. See details in the text.\label{tab:gm2LQC01}}
\end{table}
\begin{table}
\small
\centering
\begin{tabular}{ccc} \toprule
$a_{\mu}^{\textrm{HLbL};P}$ & $a_{P;1,1}^{\textrm{min}}$ & $a_{P;1,1}^{\textrm{max}}$ \\ \midrule
LQ I & $57.1(1.6)_L(0)_{\delta}[1.6]_t$  &   $57.1(2.0)_L(1.1)_{\delta}[2.3]_t$  \\
LQ II    &     $63.7(1.8)_L(0)_{\delta}[1.8]_t$    &    $63.7(2.2)_L(1.2)_{\delta}[2.5]_t$     \\ \bottomrule
\end{tabular}
\caption{Result for $a_{\mu}^{\textrm{HLbL};\pi^0}$ from $C^1_2$ using the light quark TFF in $10^{-11}$ units. See details in the text.\label{tab:gm2LQC12}}
\end{table}
The results are close, but are not compatible with those from \cref{tab:c01pole,tab:c12pole}. However, as we said, such equivalence cannot be exact as 
$\Lambda_1>0$ was found, which particularly implies $F_q>F_{\pi}$. At large energies, the BL behavior enhancement is roughly compensated through 
the singlet axial current running effects, see \cref{eq:runinf}; at low-energies however, there exist no compensation, producing a lower value for the TFF 
normalization, which in turn is the most relevant parameter. This effect can be corrected by normalizing the light quark TFF to the $\pi^0$ one. In this way, the 
results labelled as LQ II in \cref{tab:gm2LQC01,tab:gm2LQC12} are obtained, which agreement to the $\pi^0$ TFF results is embarrassingly good. 
We conclude therefore that new data, like that expected from BESIII can improve in precision but is unlikely to shift much the obtained central results in 
the previous section. This closes the discussion regarding the single-virtual part, but leaves the double-virtual part unanswered, to which we proceed below.

\subsection{Cross-checks II: $\pi^0\to e^+e^-$ implications on $(g_{\mu}-2)$}

As it has been discussed, the lack of experimental double-virtual data for the TFFs represents one of the major problems for reconstructing our 
approximants, which requires then some additional theoretical inputs often motivated from the high-energy regime ---where we have better control on QCD. 
Still, we discussed in \cref{chap:PLL} that \PtoLL decays may provide indirect experimental evidence of this behavior, as they involve an integral 
---with similar weights to that in \albl--- over the double-virtual TFF. Consequently, one may constrain some parameter of the approximant requiring this to 
reproduce the observed BRs. \\

As we pointed out in \cref{chap:PLL}, there is an interesting discrepancy in the $\pi^0\to e^+e^-$ channel measured by KTeV Collaboration~\cite{Abouzaid:2006kk} ---though this 
is reduced when taking into account the latest RC~\cite{Vasko:2011pi,Husek:2014tna}. We discuss in this section the impact that such measurement has in the \albl contribution 
to $(g_{\mu}-2)$.  For this purpose, we require our $C^0_1$ and $C^1_2$ approximants to reproduce the RC-corrected value 
$\textrm{BR}(\pi^0\to e^+e^-) =6.87(36)\times10^{-8}$. 

For the lowest $C^0_1(Q_1^2,Q_2^2)$ approximant, there is only one parameter to be tuned, this is $a_{\pi;1,1}$. We find that reproducing KTeV value requires 
then $a_{\pi;1,1} = -(32\div4)b_{\pi}^2$, where the first number stands for reproducing the central value and the second that which is $1\sigma$ 
below\footnote{This calculation and the one below have been preformed employing the approximate methods in Ref.~\cite{Dorokhov:2009xs}, which are accuerate enough 
for the $\pi^0\to e^+e^-$ decay}. 
One may argue that this approximant does not obey the OPE and should not be trusted then. However, we emphasize once more at this point 
that it is the low-energy behavior of the approximant the one which is responsible for reproducing the experimental value, well before the OPE comes 
into play. Consequently, this should not greatly change the conclusions with respect to those obtained with higher elements implementing the OPE.

For the next approximant, the $C^1_2(Q_1^2,Q_2^2)$, we have two free parameters, that associated to $a_{\pi;1,1}$, and, in addition, that we associated to $\delta^2$, 
see \cref{eq:tffope}. We notice that reproducing KTeV results requires that $\delta^2\gtrsim 10~\textrm{GeV}^2$ together with $a_{\pi;1,1} = -(39\div4)b_{\pi}^2$ 
---very similar to the $C^0_1$ results as we anticipated. \\

\begin{figure}
\centering
   \includegraphics[width=0.48\textwidth]{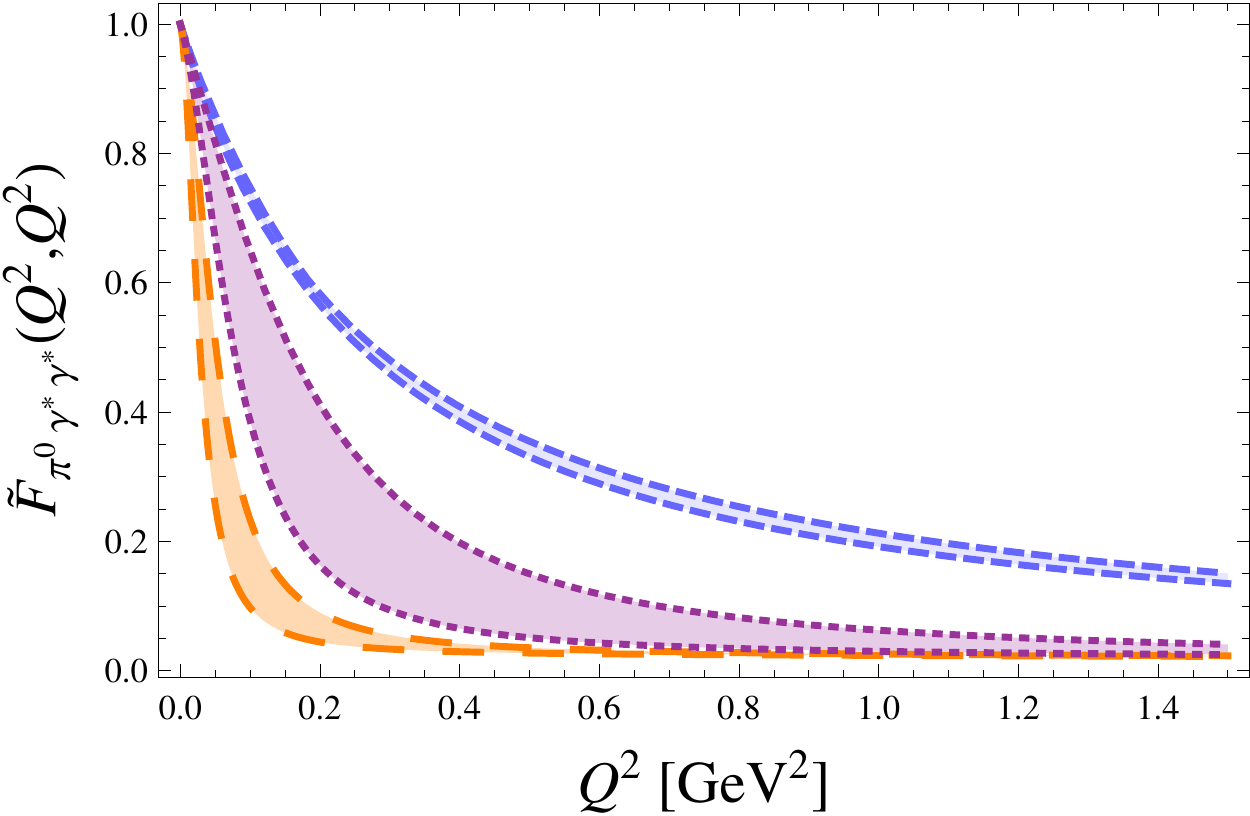}
   \includegraphics[width=0.48\textwidth]{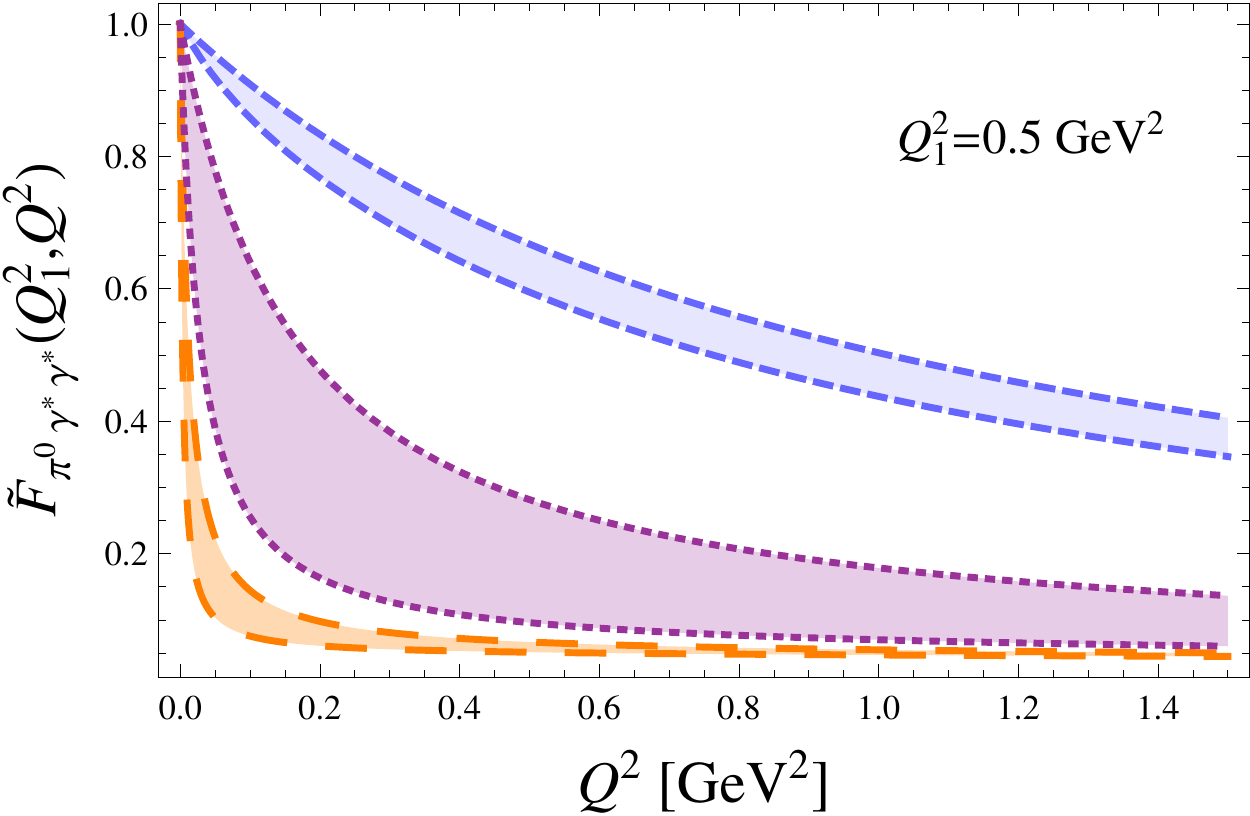}
   \caption{Left pannel: normalized TFF assuming $Q_1^2=Q_2^2\equiv Q^2$. Right pannel: normalized TFF assuming $Q_1^2=0.5~\textrm{GeV}^2$.
   Upper (blue) band shows our $C^1_2(Q_1^2,Q_2^2)$ estimation with $1.92b_{\pi}^2 \leq a_{\pi;1,1} \leq 2b_{\pi}^2$. Middle (purple) band reproduces KTeV within $1\sigma$ 
   when latest RC are included . Lower (orange) band reproduces KTeV measurement within $1\sigma$ when the latest RC are not included.  See details in the text.\label{fig:gm2KTeV} }
\end{figure}

All in all, we find that a strongly decreasing TFF is required. In addition, from the $\delta^2$ value obtained for the $C^1_2$ approximant, the OPE convergence 
should be rather slow (cf. \cref{eq:tffope}), a feature not observed so far. These results are shown as a purple band in \cref{fig:gm2KTeV}.  While the very low-energy 
behavior is not extremely different from the approximant in our previous section (blue band), it is clear that an experimental measurement above $0.2~\textrm{GeV}^2$ 
would clearly distinguish both scenarios without requiring a high precision (around $(30-50)\%$). \\

Translating the previous numbers into the the $\pi^0$-pole contribution, we obtain, for the $C^1_2$ approximant, 
$a_{\mu}^{\textrm{HLbL};\pi^0-\textrm{KTeV}} = 36(7)\times10^{-11}$. Not surprisingly, the same result would have been 
obtained for the $C^0_1$ approximant, which shows the potentiality of CAs to investigate the role of data in $(g_{\mu}-2)$. 
The present result represents a large deviation when comparing to \cref{tab:c12pole} and is larger than the projected experimental uncertainties. 
In this respect, it would be very  interesting to have an experimental analyisis on double-virtual data. As it is shown in \cref{fig:gm2KTeV}, this would not require a 
tremendous precision. Moreover, it does not necessarily involve a measurement for $Q_1^2=Q_2^2$. Keeping a photon virtuality finite, but different from zero, would 
provide an interesting result, see \cref{fig:gm2KTeV} right. Had we use the KTeV experimental result without the latest RC would accentuate 
the differences indicated above. As an example, we show in \cref{fig:gm2KTeV} the TFF that such value would imply as an orange band.

\section{Beyond pole approximation}
\label{sec:bpole}

So far, we have calculated the $\pi^0,\eta,$ and $\eta'$ pole contributions to \albl. However, it is clear that such contributions cannot account for all the QCD 
properties alone, and particularly the high-energy behavior, which can be described in terms of quarks and gluons. 
Indeed, when deriving the pole-contribution from \cref{eq:WGreenHLBL}, we dismissed any multiparticle state ($3\pi,...$), resonances ($\pi(1300),...$), 
and $q\bar{q}$ continuum. 
In order to effectively account for these additional QCD effects in the overall \albl calculation, it has been customary to analyze the high-energy behavior of QCD, 
which is given by the OPE.
In this line, it was pointed out for the first time by Melnikov and Vainshtein~\cite{Melnikov:2003xd}, that the pion-pole contribution cannot account for the HLbL high-energy QCD 
behavior which is obtained from the OPE in some particular kinematical limit, a feature which should be fixed. We refer to this approach as MV.
Later on, the author(s) in~\cite{Nyffeler:2009tw,Jegerlehner:2009ry} pointed to similar features arising when studying the $\langle VVP \rangle$ Green's function 
high-energy behavior~\cite{Knecht:2001xc}. 
Their approach to solve this is referred to as JN from now on. These approaches are the ones employed when calculating the current values for \albl.
In the following sections, we briefly describe these approaches, what they may physically stand for, their pros and cons. 
We note that taking these approaches one is providing a model with no clear connection to 
physical observables. The separation of the different contributions is hard to perform and one incurs in potential double-counting problems.

\subsection{\albl \textit{\`a la} Melnikov-Vainshtein }
\label{sec:MV}

To obtain the relevant kinematics for the HLbL tensor in the \albl scattering, we need to take the case for which one of the vector currents attaches to a real photon. 
This means, shifting from our previous general HLbL tensor, \cref{eq:hlbltensor}, to the following Green's function
\begin{equation}
\label{eq:greenMV}
\int d^4x d^4y e^{-iq_1x}e^{-iq_2y} \bra{0} T\{ j_{\mu_1}(x) j_{\mu_2}(y) j_{\mu_3}(0) \} \ket{\gamma},
\end{equation}
where, for the calculation, we can take the limit of vanishing photon momenta, then $q_1+q_2+q_3=0$.
In the space-like region, this allows for two relevant different regimes. The first, is that in which 
$Q_1^2\sim Q_2^2\sim Q_3^2$. The second, is that for which one of the photon momenta is much smaller, 
as an example $Q_1^2\sim Q_2^2\gg Q_3^2$. In this particular kinematic regime, the OPE for these two highly 
virtual photons can be easily performed. Following Ref.~\cite{Melnikov:2003xd}, \cref{eq:greenMV} reduces to\footnote{In Ref.~\cite{Melnikov:2003xd},  $j_5^{\rho}(z)$ is defined as 
$\overline{q}(z)\mathcal{Q}^2\gamma^{\rho}\gamma_5q(z)$, where $q$ stands for the light quarks and $\mathcal{Q}$ and for the charge operator.} 
\begin{equation}
\label{eq:greenMVope}
\int d^4z e^{-i(q_1+q_2)z}4\frac{(q_1-q_2)_{\delta}}{(q_1-q_2)^2} \epsilon_{\phantom{\rho\delta}\mu_1\mu_2}^{\rho\delta}  \bra{0} T\{ j_{5\rho}(z) j_{\mu_3}(0) \} \ket{\gamma}.
\end{equation}
The required matrix element, connected to the famous triangle-amplitude, is well-known~\cite{Melnikov:2003xd} and is related to the 
Adler~\cite{Adler:1969gk}-Bell-Jackiw~\cite{Bell:1969ts} anomaly for non-singlet currents. Thanks to this and non-renormalizability theorems for the anomaly, 
the authors claim that they are able to relate the behavior at $Q_3^2\sim0$ with that at $Q_3^2\to\infty$, obtaining that no suppression ---beyond that of the 
pseudoscalar propagator--- is required for the vertex involving the external photon. This observation leads them to the conclusion that no TFF 
should be employed at the external vertex, since otherwise this would introduce an additional suppresion.
As the authors point out, constraining such behavior is a modelization for the required HLbL function, including effects beyond the pseudo-Goldstone 
bosons poles; labeling this contribution as the ``pion-pole'' is just an abuse of language then. 
Somehow, in analogy to PAs, they are modeling some general QCD Green's function using some low- and high-energy constraints. 
After this, all different contributions get entangled and cannot be separated, which makes hard to tell what is included in their model and what is not. 
We only emphasize here that such derivation was obtained in a particular kinematical limit for the HLbL function and its implementation is certainly 
model-dependent. This point will be better understood in the next section when dealing with the JN approach.\\

Implementing then our approach into their method is straightforward; it reduces to set the vertex with the external photon to $F_{P\gamma\gamma}$. 
Following the same procedure as in \cref{sec:psex}, we quote our results for the $C^0_1$ and $C^1_2$ approximants in \cref{tab:c01MV,tab:c12MV}, respectively. 
\begin{table}
\small
\centering
\begin{tabular}{ccc} \toprule
$a_{\mu}^{\textrm{HLbL};P}$ & Fact & OPE \\ \midrule
$\pi^0$  &   $66.3(1.4)_F(2.6)_{b_{\pi}}[3.0]_t$  &   $84.5(1.8)_F(2.9)_{b_{\pi}}[3.4]_t$    \\
$\eta$   &     $19.6(7)_F(5)_{b_{\eta}}[8]_t$    &    $28.9(1.0)_F(0.6)_{b_{\eta}}[1.1]_t$    \\
$\eta'$   &    $19.4(6)_F(4)_{b_{\eta'}}[7]_t$    &    $30.4(1.0)_F(0.5)_{b_{\eta'}}[1.1]_t$   \\ \midrule
Total & $105.3[3.2]_t$ & $143.8[3.7]_t$  \\ \bottomrule
\end{tabular}
\caption{Result for $a_{\mu}^{\textrm{HLbL};P}$, \cref{eq:hlblpex}, for the $C^0_1$ approximants for different $a_{P;1,1}$ values in units of $10^{-11}$. See 
description in the text. }
\label{tab:c01MV}
\end{table}
%
\begin{table}
\small
\centering
\begin{tabular}{ccc} \toprule
$a_{\mu}^{\textrm{HLbL};P}$ & $a_{P;1,1}^{\textrm{min}}$ & $a_{P;1,1}^{\textrm{max}}$ \\ \midrule
$\pi^0$ & $82.7(1.7)_L(0)_{\delta}[1.7]_t$  &   $80.8(1.3)_L(0.5)_{\delta}[1.4]_t$  \\
$\eta$    &     $27.8(1.3)_L(0)_{\delta}[1.3]_t$    &    $27.0(1.4)_L(0.8)_{\delta}[1.6]_t$     \\
$\eta'$    &    $26.8(1.1)_L(0)_{\delta}[1.1]_t$    &    $25.8(0.7)_L(0.9)_{\delta}[1.1]_t$  \\ \midrule
Total & $137.3[2.4]_t$ & $133.6[2.4]_t$  \\ \bottomrule
\end{tabular}
\caption{Result for $a_{\mu}^{\textrm{HLbL};P}$, \cref{eq:hlblpex}, for the $C^1_2$ approximants for different $a_{P;1,1}$ values 
in units of $10^{-11}$. See description in the text.}
\label{tab:c12MV}
\end{table}
%
Accounting for the errors in exactly the same way as in \cref{sec:psex}, we obtain
\begin{equation}
\label{eq:finalMV}
  a_{\mu}^{\textrm{HLbL};P-\textrm{MV}} = (134\div137)(2)_{\textrm{sys}}(10)_{\textrm{stat}}[10]_t\times10^{-11}.
\end{equation}
Again, it is the systematic error (specially relevant for the $\eta'$) which dominates the full error.
Our value can be compared to the original one in~\cite{Melnikov:2003xd}, $a_{\mu}^{\textrm{HLbL};P-\textrm{MV}}=(76.5+18+18)\times10^{-11}\to114(10)\times10^{-11}$.
%
We find that including the $\eta$ and $\eta'$ high-energy behavior is once more very important to the precision we are aiming for, 
whereas this was not implemented in Ref.~\cite{Melnikov:2003xd}. Moreover, it must be emphasized again that our approach allows 
for the proper implementation of the low-energies at the same time. 
Finally, our method allows to estimate for a systematic error concerning the pseudoscalars TFF description.
%
The question still remains on the systematic error in the modellization which has been done when taking the external vertex as 
constant and possible corrections particular to the singlet component. Actually, this procedure to calculate \albl has been 
criticized in Ref.~\cite{Jegerlehner:2009ry}, see Sections 5.1.1, 5.2.1 and 6.2 therein, a debate which is left to the authors.

\subsection{\albl \textit{\`a la} Jegerlehner Nyffeler }
\label{sec:JN}

A relevant quantity for the \albl calculation, but simpler than the HLbL Green's function, \cref{eq:WGreenHLBL}, is the $\langle VVP \rangle$ 
Green's function discussed in ~\cite{Knecht:2001xc}\footnote{Note the relative minus sign 
with respect to~\cite{Knecht:2001xc} arising from our antisymmetric tensor conventions.}
\begin{align}
  \left(\Pi_{VVP}\right)^{\mu\nu}_c & \equiv \int d^4x \int d^4y \ e^{i(px+qy)} \bra{0} T\{ j^{\mu}(x) j^{\nu}(y) \mathcal{P}^c(0) \} \ket{0} \nonumber \\
  & \equiv -2\operatorname{tr}(\mathcal{Q}^2\lambda^c)\epsilon^{\mu\nu\alpha\beta}p_{\alpha}q_{\beta}\mathcal{H}_V(p^2,q^2;(p+q)^2).
\end{align}
Note that we have already specialized to the electromagnetic current as compared to~\cite{Knecht:2001xc}; $\mathcal{P}^c=\overline{q}i\gamma_5\frac{\lambda^c}{2}q$ 
stands for the pseudoscalar source and $\mathcal{Q}$ is the charge operator. 
The OPE expansion for $\langle VVP \rangle$ was obtained in~\cite{Knecht:2001xc} and reads\footnote{$\langle\overline{\psi}\psi\rangle_0$ refers to the quark condensate
in the chiral limit~\cite{Gasser:1984gg}.}
\begin{align}
\label{eq:OPEi}
\lim_{\lambda\to\infty} \mathcal{H}_V((\lambda p)^2,(\lambda q)^2;(\lambda q +\lambda p)^2) = & \ -\frac{\langle \overline{\psi}\psi \rangle_0}{2\lambda^4}  \frac{p^2+q^2+(p+q)^2}{p^2q^2(p+q)^2},  \\
\lim_{\lambda\to\infty} \mathcal{H}_V((\lambda p)^2,(q-\lambda p)^2;q^2) = & \ \frac{-1}{\lambda^2}  \langle \overline{\psi}\psi \rangle_0 \frac{1}{p^2q^2}, \label{eq:OPEii} \\
\lim_{\lambda\to\infty} \mathcal{H}_V((\lambda p)^2,q^2;(q+\lambda p)^2) = & \ \frac{1}{\lambda^2}\frac{1}{p^2}  \Pi_{VT}(q^2).  \label{eq:OPEiii}
\end{align}
The limiting behaviors for the $\Pi_{VT}(q^2)$ function read~\cite{Knecht:2001xc,Jegerlehner:2009ry}
\begin{equation}
\lim_{\lambda\to\infty}\Pi_{VT}((\lambda q)^2)  = -\frac{1}{\lambda^2}\frac{\langle\overline{\psi}\psi\rangle_0}{q^2} + \mathcal{O}(\lambda^{-4}), \ \ 
\Pi_{VT}(0) = -\frac{\langle \overline{\psi}\psi\rangle_0}{2}\chi,
\end{equation}
where $\chi$ is the quark condensate magnetic susceptibility~\cite{Jegerlehner:2009ry}.
%
It is relevant at this point to note the connection (in the chiral and large-$N_c$ limit) to the Goldstone bosons ($\pi^0,\eta,\eta'$) pole contributions. 
From the LSZ-reduction formalism, see Ref.~\cite{Peskin:1995ev}, we know that such function behaves as 
\begin{equation}
  \left(\Pi_{VVP}\right)^{\mu\nu}_c = \frac{i\sqrt{Z_P}}{(p+q)^2}\int d^4x e^{iqx} \bra{0} T\{ j^{\mu}(x)j^{\nu}(0) \} \ket{P(p+q)} + ...
\end{equation}
where the ellipses refer to terms which are non-singular at $(p+q)^2=0$ and the residue stands for the matrix element 
$\sqrt{Z_P^c}=\bra{0}\mathcal{P}^c(0)\ket{P}=-\frac{\langle\overline{\psi}\psi\rangle_0}{F}=F B$\footnote{$F$ is the decay constant in the chiral limit~\cite{Gasser:1984gg}; see 
\cref{sec:cptintro} for the origin of $B$, which is to be taken in the chiral limit too.}. 
This allows to connect with the pseudoscalar-pole contribution 
\begin{align}
  \lim_{(p+q)^2\to0}  (p+q)^2\left(\Pi_{VVP}\right)^{\mu\nu}_c &\equiv -2\mathrm{tr}(\mathcal{Q}^2\lambda^c)\epsilon^{\mu\nu\alpha\beta}p_{\alpha}q_{\beta}\mathcal{H}^{P}_V(p^2,q^2)  \\
   & = i\sqrt{Z_P^c}\int d^4x e^{iqx} \bra{0} T\{ j^{\mu}(x)j^{\nu}(0) \} \ket{P(p+q)}. \nonumber
\end{align}
Taking the last line in terms of the matrix element for the pseudoscalar to photons transition, \cref{eq:GreenToTFF}, and using the TFF 
definition, see \cref{eq:tffdef}, 
we obtain the desired connection\footnote{Note that here we omit the $(ie)^2$ coupling 
arising from the photons. Therefore, the TFF here defined has a relative minus sign with respect to those in previous chapters.}
\begin{equation}
F_{P\gamma^*\gamma^*}(p^2,q^2) = -\frac{F}{\langle\overline{\psi}\psi\rangle_0}  2\operatorname{tr}(\mathcal{Q}^2\lambda^c)  
\mathcal{H}^{P}_V(p^2,q^2). 
\label{eq:GreenToFF}
\end{equation}
%
The previous analysis could have been performed analogous to the procedure in \cref{sec:polology}.
It is tempting at this point to follow Weinberg again, as we did to calculate the pseudoscalar-pole contributions, and create a general off-shell pseudoscalar 
function which we approximate with the Goldstone boson ($P$) pole. 
Up to some irrelevant factors in \cref{eq:GreenToFF}, 
\begin{equation}
\label{eq:OffShell}
  \mathcal{H}_V(p^2,q^2,(p+q)^2) \simeq \frac{F_{\mathcal{P}^*\gamma^*\gamma^*}(p^2,q^2;(p+q)^2)}{(p+q)^2} 
   \to \frac{F_{P\gamma^*\gamma^*}(p^2,q^2)}{(p+q)^2}.
\end{equation}
an approximation to the pole-contribution ---expected to dominate at low-energies--- to such function.  
The first term has been named the off-shell pseudoscalar form factor. Care must be taken in not to associate this to any physical TFF, a connection 
which is only possible in the on-shell case via the LSZ-formalism~\cite{Peskin:1995ev}. 
\\

The relevant point here is that $F_{\mathcal{P}^*\gamma^*\gamma^*}(p^2,q^2)/(p+q)^2$ should obey then the OPE expansion \cref{eq:OPEi,eq:OPEii,eq:OPEiii}, 
this is~\cite{Nyffeler:2009tw,Jegerlehner:2009ry}, 
%
\begin{align}
\label{eq:fpggOPE1}
\lim_{\lambda\to\infty} F_{\mathcal{P}^*\gamma^*\gamma^*} ((\lambda p)^2,(\lambda q)^2;(\lambda q +\lambda p)^2) = & \operatorname{tr}(\mathcal{Q}^2\lambda^c)
          \ \frac{F}{\lambda^2}\frac{p^2+q^2+(p+q)^2}{p^2q^2},  \\
\label{eq:fpggOPE2}
\lim_{\lambda\to\infty} F_{\mathcal{P}^*\gamma^*\gamma^*} ((\lambda p)^2,(q-\lambda p)^2;q^2) = & 2\operatorname{tr}(\mathcal{Q}^2\lambda^c)
           \ \frac{F}{\lambda^2}\frac{1}{p^2}, \\
\label{eq:fpggOPE3}
\lim_{\lambda\to\infty} F_{\mathcal{P}^*\gamma^*\gamma^*} ((\lambda p)^2,0;(\lambda p)^2) = & \operatorname{tr}(\mathcal{Q}^2\lambda^c)
           \ F\chi.
\end{align}
Identifying $F_{\mathcal{P}^*\gamma^*\gamma^*}(p^2,q^2)$ with $F_{P\gamma^*\gamma^*}(p^2,q^2)$, we find that the first condition, \cref{eq:fpggOPE1}, 
is accounted from $F_{P\gamma^*\gamma^*}(p^2,q^2)$ up to an overall constant, the second, \cref{eq:fpggOPE2}, 
is trivially satisfied from $F_{P\gamma^*\gamma^*}(p^2,q^2)$ as well. However, the last one, \cref{eq:fpggOPE3}, is not satisfied as it would contradict the BL 
limit. Consequently, the pseudoscalar pole approximation cannot be accurate at high-energies, as it contradicts the $\langle VVP \rangle$ OPE behavior. 
This feature took the authors in~\cite{Nyffeler:2009tw,Jegerlehner:2009ry} to redefine an off-shell TFF for the $\pi^0$ which effectively 
accounts for \cref{eq:fpggOPE3}. 
We emphasize again that this is a modelization which goes beyond the pion-pole, including then additional QCD contributions ---implying similar problems to that in 
previous section. In the following, we describe how these high-energy constraints may be accounted for in our formalism in a similar but systematic manner.

\subsubsection{$C^0_1(Q_1^2,Q_2^2)$ implementation}

First, we start from the lowest element, the $C^0_1$, and construct the approximant for what has been defined as the pseudoscalar off-shell form factor, 
\cref{eq:OffShell}, based on the constraints in  \cref{eq:fpggOPE1,eq:fpggOPE2,eq:fpggOPE3}. 
As said, the conditions in \cref{eq:fpggOPE1,eq:fpggOPE2} were already satisfied within the pole approximation $F_{P\gamma^*\gamma^*}(p^2,q^2)$, but not the condition 
in \cref{eq:fpggOPE3}. This is possible to achieve if we would modify the $C^0_1$ approximant by adding some polynomial depending on the pseudoscalar virtuality 
in the numerator. However, such a piece would spoil the first condition, \cref{eq:fpggOPE1}. It seems hopeless then to obtain this with the lowest approximant. 
However, we can work instead the piece appearing in the loop integral arising from the external vertex and reverse the relation in \cref{eq:OffShell}
\begin{equation}
\frac{F_{P\gamma^*\gamma^*}(Q^2,0)}{Q^2+m_P^2} \rightarrow  \frac{F_{\mathcal{P}^*\gamma^*\gamma}(Q^2,0;Q^2)}{Q^2+m_P^2} \simeq  \mathcal{H}_V(Q^2,0,Q^2),
\end{equation}
where we have switched off the chiral limit ($m_{P}\neq0$). We could think of this as a $C^0_1$-type approximant
\begin{equation}
\mathcal{H}_V(p^2,q^2,(p+q)^2) \simeq \frac{F_{P\gamma\gamma}}{1-\frac{b_P}{m_P^2}(p^2+q^2) }\frac{1}{(p+q)^2-m_P^2}.
\end{equation}
If we insist in constraining the third condition \cref{eq:fpggOPE3}, we are forced to remove then the term in the denominator proportional to $(p+q)^2(p^2+q^2)$
\begin{equation}
\label{eq:c01jn}
\mathcal{H}_V(p^2,q^2,(p+q)^2) \simeq -\frac{F_{P\gamma\gamma}}{m_P^2 - b_P(p^2+q^2) - (p+q^2)},
\end{equation}
a valid procedure from our approach which does not require preserving the pole structure. In this way, we are able to implement the power-like behavior in 
\cref{eq:fpggOPE1,eq:fpggOPE2,eq:fpggOPE3}. 
If we would use \cref{eq:c01jn} to predict the value for $\chi$ upon comparing to \cref{eq:OPEiii}, we 
would find $\chi= -F_{P\gamma\gamma}/(F_0(1+b_P)\operatorname{tr}(Q^2\lambda^c)) =-8.9$ 
for the $\pi^0$ case. 
The \alblp results using \cref{eq:c01jn} for the external vertex and our previous $C^0_1$ description in \cref{eq:c01low} with the OPE built-in ($a_{P;1,1}=2b_P^2$) are 
given in \cref{tab:lbljnc01}.
\begin{table}
\small
\centering
\begin{tabular}{cc} \toprule
$a_{\mu}^{\textrm{HLbL};P-\textrm{JN}}$ &  \\ \midrule
$\pi^0$  &   $82.5(1.8)_F(3.1)_{b_{\pi}}[3.5]_t$   \\
$\eta$   &     $22.8(0.8)_F(0.6)_{b_{\eta}}[1.0]_t$   \\
$\eta'$   &    $20.6(0.7)_F(0.5)_{b_{\eta'}}[0.8]_t$    \\ \midrule
Total & $125.9[3.7]_t$  \\ \bottomrule
\end{tabular}
\caption{Result for $a_{\mu}^{\textrm{HLbL};P-\textrm{JN}}$, \cref{eq:hlblpex}, using the $C^0_1$ approximant defined in \cref{eq:c01jn} at the external vertex in units of $10^{-11}$. 
See description in the text.\label{tab:lbljnc01}}
\end{table}

\subsubsection{$C^1_2(Q_1^2,Q_2^2)$ implementation}

In this second element, there is more freedom to implement the high-energy conditions, \cref{eq:fpggOPE1,eq:fpggOPE2,eq:fpggOPE3}. 
Once more, we proceed \textit{\`a la} Pad\'e type and fix the pion pole for the denominator. However, in contrast to the previous element, 
we can incorporate in general additional $(Q_1+Q_2)^2$ and $(Q_1+Q_2)^2(Q_1^2 + Q_2^2)$ terms (in accordance with our systematic expansion) 
in the numerator of \cref{eq:c12def} which do not spoil the high-energy behavior. The second of these terms can be related to \cref{eq:fpggOPE3}, 
whereas the first one can be related, after taking $Q_1^2=Q_2^2=0$, to the low-energy chiral expansion for the $\langle VVP \rangle$ function, see Eq.~(5) in 
Ref.~\cite{Moussallam:1994xp} (very similar results would be obtained from \cite{Roig:2014uja}). This is very 
important as the low-energies play a major role in the \albl integrand, \cref{eq:hlblpex}. 
Fulfilling \cref{eq:fpggOPE3} requires adding to the numerator $\sim (Q_1^2 + Q_2^2)((Q_1+Q_2)^2+m_P^2)(-F_0\chi \operatorname{tr}(Q^2\lambda^c))$,
whereas fulfilling the low-energy chiral expansion requires ---note that Ref.~\cite{Moussallam:1994xp} works in the chiral limit--- 
\begin{equation}
\label{eq:psvirt}
 \mathcal{H}_V(0,0,Q^2) = \mathrm{Tr}(\hat{Q}^2\lambda^c)\frac{N_c}{8\pi^2F}\frac{1}{-Q^2}\bigg( 1 + \underbrace{-\frac{(16\pi)^2}{N_c}t_1}_{0.32\pm0.10\pm0.12} Q^2 \bigg) 
\end{equation}
where $t_1$ is a LEC. The numerical value $0.32(10)(12)$ has been obtained from the estimate in~\cite{Moussallam:1994xp}, i.e., $t_1\simeq -F^2/(64M_V^4)$, where 
$M_V=0.77$~GeV and $F=F_{\pi}$ is used. The $\pm0.10$ error has been obtained taking the difference among $F$ and $F_{\pi}$~\cite{Ecker:2013pba} and the 
half-width rule for $M_{\rho}$~\cite{Arriola:2012vk}.
This should account for the $\pi^0$ given its small mass. For extending to the $\eta$ and, ---given its singlet nature--- 
specially the $\eta'$, we assume an additional $30\%$ correction for symmetry breaking effects, which should be enough for describing the differences. 
This leads the $\pm0.12$ error in \cref{eq:psvirt}. All in all, we require adding to the numerator in \cref{eq:c12def}
\begin{equation}
\label{eq:c12jnmod}
0.32((Q_1+Q_2)^2+m_P^2)  - \beta_2(Q_1^2 + Q_2^2)((Q_1+Q_2)^2+m_P^2) F_{\pi}\chi \operatorname{tr}(Q^2\lambda^c).
\end{equation}
Note that additional $m_P^2$ terms have been included in order to recover the pole contribution as $(Q_1+Q_2)^2\to -m_P^2$. 
In addition, we take from~\cite{Jegerlehner:2009ry} 
$\chi=-3.3(1.1)$ together with an additional $30\%$ error for the $\eta$ and $\eta'$ accounting for symmetry breaking effects.
The obtained results are given in \cref{tab:c12JN}.
\begin{table}
\small
\centering
\begin{tabular}{ccc} \toprule
$a_{\mu}^{\textrm{HLbL};P-\textrm{JN}}$ & $a_{P;1,1}^{\textrm{min}}$ & $a_{P;1,1}^{\textrm{max}}$ \\ \midrule
$\pi^0$ & $78.7$  &   $69.4(1.5)_L(0.6)_{\delta}(0.4)_{\chi}(1.8)_{\textrm{Low}}[2.5]_t$  \\
$\eta$    &     $32.9$    &    $22.2(2.7)_L(1.7)_{\delta}(0.8)_{\chi}(2.2)_{\textrm{Low}}[3.9]_t$     \\
$\eta'$    &    $41.2$    &    $25.6(0.9)_L(2.1)_{\delta}(1.8)_{\chi}(4.0)_{\textrm{Low}}[5.0]_t$  \\ \midrule
Total & $152.8$ & $120.6[6.8]_t$  \\ \bottomrule
\end{tabular}
\caption{Result for $a_{\mu}^{\textrm{HLbL};P-\textrm{JN}}$ for the $C^1_2$ approximants modified as in \cref{eq:c12jnmod} for different $a_{P;1,1}$
 values in units of $10^{-11}$. See description in the text.}
\label{tab:c12JN}
\end{table}
%
We note that the large difference among the values in the ``min'' and ``OPE'' columns are due to the fact that the first one does not 
obey the OPE constraint \cref{eq:fpggOPE3}\footnote{For $a_{P;1,1}=a_{P;1,1}^{\textrm{min}}$, the $\alpha_{1,1}$ and $\beta_{2,1}$ 
parameters in \cref{eq:c12def} go to $0$.}. Consequently, this value should be thought as a limiting value, 
and shows the necessity of having a determination for the doubly-virtual coefficients. 
The displayed errors are those which have been already defined in \cref{sec:poleresults} together with that from the magnetic suseptibility $\chi$ parameter, 
$(\cdot)_{\chi}$, and the low-energy behavior in \cref{eq:psvirt}, $(\cdot)_{\textrm{Low}}$. 
We find again non-negligible differences with respect to the pole contributions. Note however that these effects come mainly from the low-energies. 
Actually, if we would have retained the OPE condition but switching off the low-energy constraint (i.e., $t_1=0$), we would have found for the $a_{P;1,1}^{\textrm{max}}$ column in \cref{tab:c12JN}, 
$64.1, 18.2, 18.5$ for the $\pi^0$, $\eta$ and $\eta'$, respectively, in units of $10^{-11}$.

Furtheremore, if we would have applied the modified form factor \cref{eq:OffShell} at the external vertex alone, similar to the previous section, we would have found 
for the ``OPE'' column in \cref{tab:c12JN},  $67.0, 20.3, 18.1$ for the $\pi^0,\eta,\eta'$, respectively, in units of $10^{-11}$ and a weak dependence on $\chi$. 
Our result should be compared with that in~\cite{Jegerlehner:2009ry}, 
$a_{\mu}^{\textrm{HLbL};P-\textrm{JN}} = (72(12)+14.5(4.8) + 12.5(4.2) = 99(16) )\times10^{-11}$ and points out 
the necessity to implement the high-energy behavior not only for the $\pi^0$ but for the $\eta$ and $\eta'$ mesons as well.

In the light of previous results, we think that implementing our framework in this approach would require a minimal information 
on the double-virtual TFF in order to narrow down the errors. Similarly, a comparison between the $C^1_2$ and $C^0_1$ approximation 
is difficult given the slight different procedures. Estimating a reliable systematic error would require thus reproducing a 
higher element, say $C^2_3$, which requires again, among others, double-virtual information. For all these reasons, we do not 
consider this number for updating the full \albl contribution. 

We remark that the difference in errors with respect to Ref.~\cite{Jegerlehner:2009ry} is related to their approximation style, 
which resembles a CA-type approximation which avoids some of the problems encountered here. The additional error that this may 
induce is unknown. Still, the systematic errors pointed out above, signal a potentially large unaccounted systematic error.

\section{Final results for \albl}
\label{sec:HLBLfinalres}

Having discussed the results for the pseudoscalar-pole contribution, we give the final results for the total \albl. 
For this, we need to incorporate, in addition to the former, the additional contributions outlined in \cref{fig:LNCRafael}. 
These are, the charged pseudoscalar loops~\cite{Bijnens:1995xf,Bijnens:2001cq,Jegerlehner:2009ry}, 
higher resonances exchanges (we consider the axials~\cite{Pauk:2014rta,Jegerlehner:2015stw}, but not the 
scalars~\cite{Prades:2009tw,Jegerlehner:2009ry} and tensors~\cite{Pauk:2014rta} as they are partially accounted for by 
the $\pi^+\pi^-$ loop and we may incur in a double-counting problem) and the quark 
loop~\cite{Bijnens:1995xf,Bijnens:2001cq,Jegerlehner:2009ry,Masjuan:2012qn}, which seems necessary 
in this approach to account for the high-energy behavior. Taking the central value from \cref{eq:finalpole} leads to 
\begin{align}
a_{\mu}^{\textrm{HLbL}} &= ( 94.4(4.8) -19(13)_{P\textrm{loop}} +6.4(2.0)_{\textrm{axial}} +21(3)_{Q\textrm{loop}} )\times10^{-11} \nonumber \\
&= 102.8(14)\times10^{-11},
\end{align}
where errors have been added in quadrature as they are taken independent from each other. 
The total error is fully dominated by the charged pseudoscalar loop contribution. 
In this respect, it is pressing to improve such error as well as to determine a reliable systematic error for it, 
which would be possible in dispersive analysis for the $\pi^+\pi^-$ contribution. From our point of view, this would 
set up the foundations to have a reliable precise determination for the \albl. 
\\

Finally, we address the impact of our study to existing alternative approaches. First, respecting the MV approach~\cite{Melnikov:2003xd} 
from \cref{sec:MV}, the authors argue that only pseudoscalar and axial contributions should be accounted. Taking our result from 
\cref{sec:MV} and updating the axial contribution~\cite{Pauk:2014rta,Jegerlehner:2015stw}, we obtain
\begin{align}
a_{\mu}^{\textrm{HLbL}} &=   (136(9) -0(10)_{P\textrm{loop}} + 6.4(2.0)_{\textrm{axial}})\times10^{-11} \\ \nonumber 
&= 142(21)\times10^{-11},
\end{align}
where the second number is a theoretical error they estimate for the pseudoscalar loop contribution. In order to compare with 
their result, $136(25)\times10^{-11}$, errors have been added linearly as well. 
In addition, previous estimation was used in the Glasgow consensus~\cite{Prades:2009tw} to obtain the \albl. 
Substituting for this new value and updating the axial vector contribution 
as well~\cite{Pauk:2014rta,Jegerlehner:2015stw}, we obtain
\begin{align}
a_{\mu}^{\textrm{HLbL}} &= (136(11))-19(19)_{P\textrm{loop}}+6.4(2.0)_{\textrm{axial}}-7(7)_{\textrm{scalar}}+2.3_{c})\times10^{-11}\nonumber\\ 
& = 119(23)\times10^{-11},
\end{align}
where the last contribution is from the $c$ quark. In the result above, errors have been combined in quadrature; the result should be compared 
against $105(26)\times10^{-11}$~\cite{Prades:2009tw}.

\section{Conclusions and outlook}
\label{sec:gm2conc}

In this chapter, we have updated the pseudoscalar pole contribution to the $(g_{\mu}-2)$ hadronic light-by-light, 
where the key quantities are, once more, the pseudoscalar TFFs. 
Such calculation requires a precise error in order to meet future experiments criteria ---at the order of $10\%$---  
which cannot be easily obtained using model 
approximations to QCD. Moreover, the phenomenological approaches employed so far, in which experimental data is 
used to reduce the model-dependence, lack the presence of experimental data at low energies. Unfortunately, this 
turns out to be the most relevant region in the present calculation and their accuracy relies therefore on 
extrapolations. Moreover, their choices and particular ansatz for the fitting functions may incur in 
additional theoretical errors which are, so far, unquantified. 

For these reasons, we advocate the use of Canterbury approximants in order to reconstruct the pseudoscalar TFFs, 
which have been introduced and worked out in previous chapters. These allow to implement both, the low energies and 
the high ones, which play a relevant role as well in the calculation and have been often disregarded for the $\eta$ 
and $\eta'$. To demonstrate their performance, we have made 
use of two different theoretical models for the TFF which have proven useful before. This has allowed to illustrate 
the convergence of the approach and how the systematic error can be obtained.
The reconstruction of the first two elements of the chosen CA sequence has been illustrated then, requiring full 
use of the available information on pseudoscalar TFFs. This has allowed to obtain a precise determination meeting the 
future experimental criteria for the pseudoscalar pole contribution to \albl, including a precise determination for 
the systematic error which, globally, turns out to dominate the full calculation and represents one of the main 
advances with respect to previous approaches.   

We have been very careful in order to illustrate what the pseudoscalar pole contribution means and why we advocate 
such calculation. Still, our approach can be incorporated into alternative approaches including a pion pole, such as 
the MV or JN approaches. 

Finally, we have employed the existing determinations for the additional contributions to the \albl in order to estimate 
the full number. We find that the dominating error at the moment is the pseudoscalar loop contribution, which is expected to be 
improved in the near future from ongoing dispersive approaches, and would set up the foundations to achieve a precise and 
model-independent calculation for the \albl.

In addition, we have shown that our results could be improved in the near future given the intensive experimental 
activity regarding $\gamma\gamma$ physics, which has received a strong incentive from the future $(g_{\mu}-2)$ experiments. 
In particular, BESIII, NA62 and A2 future results regarding the $\pi^0$ TFF are relevant ---also future experiments at KLOE-2 and 
$GlueX$ collaborations are expected to provide valuable information on this. For the $\eta$ and $\eta'$, the $GlueX$ Collaboration is likely to improve 
the two-photon decays and TFFs. More important, it is possible that, in the future, the BESIII Collaboration provides the first measurement on the double-virtual 
$\pi^0$ TFF, which is specially relevant for this calculation.

\chapter{Conclusions and outlook}

In the present thesis, I have studied the lightest pseudoscalars, $\pi^0$, $\eta$ and $\eta'$, transition form factors (TFFs). 
The objective was to achieve a precise space-like low-energy description and simultaneously to incorporate the high energies 
with a realistic estimate of the the systematic error. 
These features are crucial to provide a precise determination for the hadronic light-by-light (HLbL) contribution to 
the muon anomalous magnetic moment, $(g_{\mu}-2)$, and have proven useful for further applications. 
For this purpose, we used the theory of Pad\'e approximants to describe the 
single-virtual TFFs. The relevant feature of the approach was to provide a mathematical well-defined framework where the 
previous requirements can in principle be systematically implemented to arbitrary precision. 
One of the main features was the systematic implementation. The resulting pattern allowed to check the performance and the systematic errors.
 \\

The central quantities required in our approach were the low-energy parameters appearing in the TFFs series expansion, guaranteeing the 
appropriate description at low-energies. Determining such parameters without any theoretical prejudice was achieved through a data-fitting procedure to the 
existing space-like data from $e^+e^-$ colliders using Pad\'e approximants. Remarkably, the high-energy data was fundamental in order to achieve 
a precise description free of large systematic errors. As an outcome, we anticipated that the resulting parameterization would provide an excellent description 
for the low-energy time-like data ---unlike previous vector meson dominance descriptions--- at least, below production thresholds. This hypothesis was checked for 
the time-like data for the $\eta$ meson at the A2 Collaboration at MAMI. Our parameterization was found to provide an excellent description of data, including those above threshold. 
Furthermore, our results were corroborated in dispersive approaches, all in all, confirming the power and reliability of the method. The success of the method in the low-energy 
time-like region could have been anticipated given the $P$-wave nature of the discontinuity, which softens out the non-analiticities and allows to understand the 
latest results for the $\eta'$ as well. For these reasons, we included in a second stage these data-sets into our approach, obtaining the most precise determination for the 
$\eta$ and $\eta'$ low-energy parameters. 
In the near future, the upcoming experimental results will provide valuable information and will help to improve our results. Most importantly, there is an ongoing 
analysis of the $\pi^0$ TFF at low space-like energies at BESIII ---further in the future, even lower energies will be accessed at KLOE-2 and $GlueX$ 
collaborations. In addition, low-energy time-like data from the Dalitz decay are expected to appear from NA62 and A2 collaborations, which will definitely improve 
our low-energy parameters determination. Additional data is expected for the $\eta$ and $\eta'$ too.
\\

Beyond the single-virtual TFF studies, we discussed how to implement the most general double-virtual case, which is a prerequisite for 
the calculations developed in this thesis. This required to introduce, for the first time in this context, the notion of Canterbury approximants, which 
serve as a generalization of Pad\'e approximants to the bivariate case. The current lack of any data did not allow to extract the required low-energy 
parameters belonging to the double-virtual TFF series expansion. Nonetheless, we provided a careful analysis based on pseudo-data showing 
the potential of future double-virtual measurements to extract the required parameters.
At present, there is an ongoing effort at BESIII to measure the $\pi^0$ double-virtual TFF. The framework provided in this work would serve as an 
important analysis tool for the experimentalists at BESIII as well as in extracting the required parameters. 
A further opportunity would be the investigation of the existing $V\to P\gamma^*$ processes, in which the approximants are constructed {\textit{\`a la}} Pad\'e type, 
for narrow vector mesons $V$. 
\\

The phenomenology related to the physics of TFFs is very rich and is not restricted to $(g_{\mu}-2)$ physics. To start with, 
the connection of the low- and high-energy behaviors of the $\eta$ and $\eta'$ TFFs made possible to study the $\eta-\eta'$ mixing.
For this purpose, we carefully discussed the relevance of using a two-angle formalism for describing the decay constants as well as accounting for 
the peculiarities of the singlet content, which, even if $N_c$-suppressed, produce non-negligible effects in the asymptotic behaviors. 
As an advantage with respect to traditional approaches, the adopted formalism benefits from using inputs which are well-defined in 
large-$N_c$ chiral perturbation theory ---our best tool so far to describe the $\eta$ and $\eta'$. 
The equations involved in our approach resulted in a degenerate system of equations. Remarkably, this could be used to obtain an additional 
OZI-violating parameter often ignored. As a result, our framework consistently incorporated all the chiral corrections and OZI-violating parameters 
involved at NLO in large-$N_c$ chiral perturbation theory, which have been commonly neglected in most of the phenomenological studies. 
The obtained results were competitive in comparison to existing determinations despite 
the small amount of required input and are of relevance for the study of exclusive 
processes involving the $\eta$ and $\eta'$, which require an accurate input for the mixing parameters.
\\

As a first test of the double-virtual implementation, we discussed the application of our approach to the rare $P\to\bar{\ell}\ell$ decays. 
The involved calculation not only required a precise TFF description at low space-like energies, but a reasonable description of the 
high-energies, providing an excellent ground to test our description. 
As a further advantage of our approach, we showed that its application could be safely extended to the $\eta$ and $\eta'$ cases, 
which may not be the case for existing calculations. 
Besides, we performed an exact numerical calculation, which is crucial for the $\eta$ and $\eta'$ cases. In contrast, most of the previous 
approaches used approximations, suffering from large systematics.
The current lack of any direct experimental constraint on the double-virtual transition form factor 
was supplied with a very generous estimate based on very general principles in order to avoid as much as possible a strong model-dependence. 
Nonetheless, the introduced uncertainty is well below the experimental one. The achieved predictions represent the most updated results 
and include, for the first time, a systematic error. We confirmed the existing deviation for the $\pi^0\to e^+e^-$ result and 
a slight deviation for the $\eta\to\mu^+\mu^-$, which provides a strong motivation for a future measurement, e.g. there are plans to measure the former 
at NA62, whereas the latter could be measured at LHCb. 
As a result, we studied the new physics scenarios which could provide a reason for such 
discrepancies; these seem to require new light degrees of freedom and some fine-tuning in order to avoid constraints and explain, at the 
same time, both deviations. 
In addition, we discussed the implications of our results for chiral perturbation theory. This is very important as it is the used framework to 
test analogous $K_L$ decays, which provide stringent tests on lepton universality among others. Besides, this is of interest for calculations 
regarding the hyperfine splitting in muonic hydrogen. 
In the future, it would be interesting to perform a similar analysis for the $K_L$ given the available time-like data from single and double 
Dalitz decays. Achieving a precise description, including a careful numerical evaluation, and a reliable systematic error is very important, 
as $K_L\to\bar{\ell}\ell$ decays can place strong constraints on certain new-physics scenarios.
\\

Finally, we calculated the pseudoscalar-pole of the HLbL contribution to $(g_{\mu}-2)$, which was our primary goal in this work. 
Given the current discrepancy among the experimental  
$(g_{\mu}-2)$ extraction and theoretical calculations, planned experiments will 
measure this quantity with improved precision, which urges the theoretical community to improve on the precision of  hadronic 
contributions to this observable. Among others, this requires an error around $10\%$ for the HLbL 
pseudoscalar pole contribution, challenging current theoretical estimates. Such calculation demands, again, a precise 
description of the double-virtual TFFs at low space-like energies, but requires as well an appropriate implementation 
of higher energies, in the region around $1~\textrm{GeV}$. Furthermore, the double-virtual 
behavior is essential, which may be the bottleneck of future dispersive descriptions for the TFF. 
Again, our approach is almost tailor-made for such calculation. The sensitivity of the calculation to 
intermediate energies and the requested precision required the construction of two elements. 
As an important novelty, our approach incorporated, for the first time, an accurate $\eta$ and $\eta'$ description which cannot be neglected 
anymore given the required precision. 
The obtained results provide, for the first time, a systematic error which is actually the dominant one and provides a step forward 
towards a precise model-independent calculation of the HLbL contribution to $(g_{\mu}-2)$.
The future TFF measurements, specially those regarding double-virtual measurements, will undoubtedly provide very interesting 
results not only for the TFFs but for the $(g_{\mu}-2)$ evaluation as well. Furthermore, the possibility of lattice techniques to access the TFFs and the 
hadronic light-by-light tensor will provide valuable inputs for this calculation.

\appendix

\chapter{Definitions and conventions}
\label{app:conv}

\section{Conventions}

We follow the conventions from Peskin and Schroeder's book~\cite{Peskin:1995ev}. This means, among others, to use units in which $\hbar=c=1$ and the 
following conventions for the (diagonal) metric $g^{\mu\nu} $and antisymmetric tensor $\epsilon^{\mu\nu\rho\sigma}$
\begin{equation}
 g^{\mu\nu} = \operatorname{diag}(+1,-1,-1,-1), \quad \epsilon^{0123}= -\epsilon_{0123} = +1.
\end{equation}
Consequently, for time-like quantities $q^2  >0$, whereas  for space-like quantities, $q^2 <0$, which is often noted in capital 
letters as $Q^2\equiv-q^2>0$. Four vectors are often noted as $q=\left(q^0,\boldsymbol{q}\right)$, with $\boldsymbol{q}$ denoting a 
space-component. The slashed notation, $\gamma^{\mu}k^{\nu}g_{\mu\nu}\equiv\slashed{k}$ with $\gamma^{\mu}$ a Dirac matrix is employed.

\section{Feynman rules and spinors}

We make use of the Feynman rules following from the QED lagrangian 
\begin{equation}
\mathcal{L_{\textrm{QED}}} = \overline{\psi}(i\slashed{D} -m)\psi - \frac{1}{4}(F_{\mu\nu})^2, \qquad D_{\mu} = \partial_{\mu} - i \mathcal{Q}A_{\mu}, 
\end{equation}
which can be read from the diagrams below. The figures are to be read from left to right; momentum $p$ flows from left to right; the dot denotes the vertex to which the lines 
attach; the fermion arrow gives the fermion number flow; \\
\newline
\includegraphics[width=1.03\textwidth]{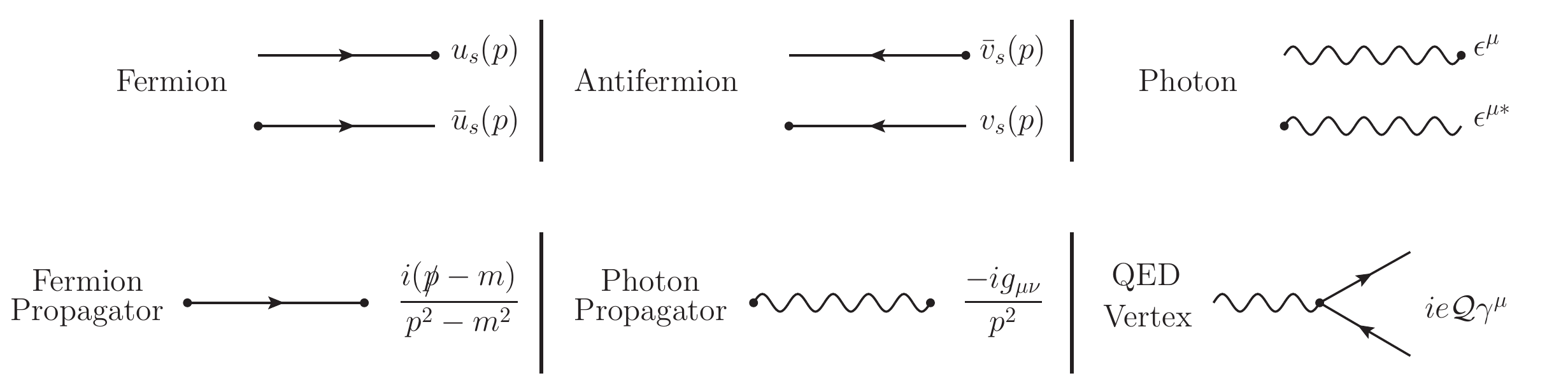}
\newline
$u_s(p)$ and $v_s(p)$ are Dirac spinors fulfilling Dirac equation
\begin{equation}
       \bar{u}_s(p)(\slashed{p}-m) =    (\slashed{p}-m)u_s(p) = 0, \quad        \bar{v}_s(p)(\slashed{p}+m) =    (\slashed{p}+m)v_s(p) = 0.
\end{equation}
For convenience, we also employ along this work the (shorter) notation $u_{p,s}\equiv u_s(p)$ and similar for $v_s(p)$. 
The $\gamma^{\mu}$ and $\gamma_5\equiv i\gamma^0\gamma^1\gamma^2\gamma^3$ matrices are defined in the Weyl or chiral basis
\begin{equation}
\gamma^{\mu} =    \begin{pmatrix}   0  & \sigma^{\mu} \\ \bar{\sigma}^{\mu} & 0  \end{pmatrix} , \quad
\gamma_{5} =    \begin{pmatrix}   -\mathds{1}  & 0 \\ 0 & \mathds{1}  \end{pmatrix} , \quad 
\sigma^{\mu} = (\mathds{1},\boldsymbol{\sigma}), \quad \bar{\sigma}^{\mu} = (\mathds{1},-\boldsymbol{\sigma})
\end{equation}
with $\boldsymbol{\sigma}$ referring to the Pauli matrices 
\begin{equation}
\sigma^1 =    \begin{pmatrix}   0  & 1 \\ 1 & 0  \end{pmatrix} , \quad
\sigma^2 =    \begin{pmatrix}   0  & -i \\ i & 0  \end{pmatrix} , \quad 
\sigma^3 =  \begin{pmatrix}   1  & 0 \\ 0 & -1  \end{pmatrix}.
\end{equation}

\section{$S$-matrix, cross sections and decay rates}
\label{sec:crossdecay}

The amplitude for some particular process is given in terms of the $S$-matrix element, 
\begin{equation}
  \bra{p_1,p_2,...} S \ket{ p_{\mathcal{A}}p_{\mathcal{B}} } \equiv  i  \mathcal{M}(p_{\mathcal{A}}p_{\mathcal{B}}\to \{ p_f \})   (2\pi)^4\delta^{(4)} 
  (p_{\mathcal{A}} + p_{\mathcal{B}} - {\scriptsize{\sum}}~p_f)
\end{equation}
where the amplitude for the process, $\mathcal{M}$ for short, is calculated from the Feynman rules. $\ket{p_{\mathcal{A,B}}}$ denotes the initial 
asymptotic states whereas $p_f$ denote the final ones. Cross sections can be obtained then as 
\begin{multline}
d\sigma = \frac{1}{2E_\mathcal{A}2E_\mathcal{B} |v_\mathcal{A} - v_\mathcal{B}| } \bigg(  \prod_f \frac{d^3p_f}{(2\pi)^3} \frac{1}{2E_f} \bigg) 
  \left| \mathcal{M}(p_{\mathcal{A}}p_{\mathcal{B}}\to \{ p_f \}) \right|^2  \\  \times  (2\pi)^4\delta^{(4)}  (p_{\mathcal{A}} + p_{\mathcal{B}} - {\scriptsize{\sum}}~p_f),
\end{multline}
with $v_\mathcal{A,B}$ the initial particles velocity. Decay rates are expressed as
\begin{equation}
d\Gamma = \frac{1}{2m_\mathcal{A}} \bigg(  \prod_f \frac{d^3p_f}{(2\pi)^3} \frac{1}{2E_f} \bigg) 
  \left| \mathcal{M}(m_{\mathcal{A}}\to \{ p_f \}) \right|^2  \\  (2\pi)^4\delta^{(4)}  (p_{\mathcal{A}}  - {\scriptsize{\sum}}~p_f),
\end{equation}
with $m_{\mathcal{A}}$ the initial particle mass. For the particular case of two-body decays with equal masses, $m_f$, reads
\begin{equation}
d\Gamma = d\Omega \frac{\beta}{64\pi^2m_{\mathcal{A}}}  \left| \mathcal{M}(m_{\mathcal{A}}\to \{ p_f \}) \right|^2, 
\qquad \beta=\sqrt{1-4m_f^2/m_{\mathcal{A}}^2}.
\end{equation} 
Note that for indistinguishable particles in the final state (i.e. $\gamma\gamma$) an extra $1/2$ factor appears.

%

\chapter{Supplementary material}

\section{Formulae for the $g_{VP\gamma}$ couplings}
\label{sec:AppGvp}

%
%

Proceeding in the lines of~\cite{Ball:1995zv,Escribano:2005qq} and including the OZI-violating term $\Lambda_3$ appearing in the anomalous QCD sector~\cite{Feldmann:1999uf}, 
and $K_2$ we obtain that 
\begin{align}
\frac{g_{\rho\eta\gamma}}{m_{\rho}} &= 
   \frac{\sqrt{6}}{8\pi^2f_{\rho}}\frac{1}{c_{\theta_8-\theta_{0}}}
   \left( \frac{c_{\theta_{0}}}{F_8}(1\!+\!K_2\mathring{M}_{\pi}^2) -\frac{\sqrt{2}s_{\theta_8}}{F_{0}}(1\!+\!K_2\mathring{M}_{\pi}^2\!+\!\Lambda_3) \right), \\
\frac{g_{\rho\eta'\gamma}}{m_{\rho}} &= 
   \frac{\sqrt{6}}{8\pi^2f_{\rho}}\frac{1}{c_{\theta_8-\theta_{0}}}
   \left( \frac{s_{\theta_{0}}}{F_8}(1\!+\!K_2M_{\pi}^2) +\frac{\sqrt{2}c_{\theta_8}}{F_{0}}(1\!+\!K_2M_{\pi}^2\!+\!\Lambda_3) \right), \\
\frac{g_{\omega\eta\gamma}}{m_{\omega}} &= 
   \left( \frac{c_{\theta_{0}} \! \! \left[ c_{\theta_V}\!(1\!+\!\delta^8_a)\!-\!\frac{s_{\theta_V}}{\sqrt{2}}(1\!+\!\delta^8_b) \right]}{4\pi^2f_{\omega}F_8c_{\theta_8-\theta_{0}}} - \frac{s_{\theta_8}\!\!\left[s_{\theta_V}\!(1\!+\!\delta^0_a) \! + \!c_{\theta_V}\delta^0_b\right]}{4\pi^2f_{\omega}F_{0}c_{\theta_8-\theta_{0}}}\right),\\
\frac{g_{\omega\eta'\gamma}}{m_{\omega}} &= 
   \left( \frac{s_{\theta_{0}} \! \! \left[ c_{\theta_V}\!(1\!+\!\delta^8_a)\!-\!\frac{s_{\theta_V}}{\sqrt{2}}(1\!+\!\delta^8_b) \right]}{4\pi^2f_{\omega}F_8c_{\theta_8-\theta_{0}}} + \frac{c_{\theta_8}\!\!\left[s_{\theta_V}\!(1\!+\!\delta^0_a) \! + \!c_{\theta_V}\delta^0_b\right]}{4\pi^2f_{\omega}F_{0}c_{\theta_8-\theta_{0}}}\right),\\
\frac{g_{\phi\eta\gamma}}{m_{\phi}} &= 
   \!-\!\left( \frac{c_{\theta_{0}} \! \! \left[ s_{\theta_V}\!(1\!+\!\delta^8_a)\!+\!\frac{c_{\theta_V}}{\sqrt{2}}(1\!+\!\delta^8_b) \right]}{4\pi^2f_{\phi}F_8c_{\theta_8-\theta_{0}}} + \frac{s_{\theta_8}\!\!\left[c_{\theta_V}\!(1\!+\!\delta^0_a) \! - \!s_{\theta_V}\delta^0_b\right]}{4\pi^2f_{\phi}F_{0}c_{\theta_8-\theta_{0}}}\right),\\
\frac{g_{\phi\eta'\gamma}}{m_{\phi}} &= 
   \!-\!\left( \frac{s_{\theta_{0}} \! \! \left[ s_{\theta_V}\!(1\!+\!\delta^8_a)\!+\!\frac{c_{\theta_V}}{\sqrt{2}}(1\!+\!\delta^8_b) \right]}{4\pi^2f_{\phi}F_8c_{\theta_8-\theta_{0}}} - \frac{c_{\theta_8}\!\!\left[c_{\theta_V}\!(1\!+\!\delta^0_a) \! - \!s_{\theta_V}\delta^0_b\right]}{4\pi^2f_{\phi}F_{0}c_{\theta_8-\theta_{0}}}\right),
\end{align}
where $s_{\theta}\equiv \sin\theta$ and $c_{\theta}\equiv \cos\theta$ abbreviations have been employed. 
In the definitions above, $F_{8,0}$ are the decay constants defined in \cref{eq:FP80}. 
Besides, the additional \lcpt NLO corrections are fully introduced in \cref{chap:mixing}, and are encoded in the $\delta$ parameters defined as 
\begin{align}
\delta^8_a &= K_2M_8^2  & \delta^8_b &= K_2(3M_8^2-M_0^2), \\
\delta^0_a &= K_2M_8^2 + \Lambda_3 & \delta^0_b &= K_2M_{80}^2 ,
\end{align}
where the mass parameters above have been introduced  in \cref{eq:m8,eq:m0,eq:m80} and can be defined in terms of the LO $\pi$ and $K$ masses, 
$\mathring{M}_{\pi}^2, \mathring{M}_{K}^2$, that we associate to the physical ones. 
$f_V$ is the vector meson decay constant defined in terms of the matrix element 
$\bra{0} J_{\mu}^V \ket{V}=m_Vf_V\varepsilon_{\mu}$~\cite{Ball:1995zv,Escribano:2005qq}\footnote{$J_{\mu}^{\rho} \equiv J_{\mu}^{3}$, 
$J_{\mu}^{\omega} \equiv \frac{1}{\sqrt{2}}( J_{\mu}^{8}s_{\theta_V} + J_{\mu}^{0}c_{\theta_V} )$ and 
$J_{\mu}^{\phi} \equiv \frac{1}{\sqrt{2}}( J_{\mu}^{8}c_{\theta_V} - J_{\mu}^{0}s_{\theta_V} )$, where $J_{\mu}^a$ are isospin currents as defined in 
\cref{eq:Icurrents}. As an illustration, $\theta_V = \pi-\theta_{ideal}=35.3^{\circ}$ would correspond to 
$J_{\mu}^{\omega} = \frac{1}{\sqrt{2}}(\bar{u}\gamma_{\mu}u + \bar{d}\gamma_{\mu}d)$ and $J_{\mu}^{\phi} = -\bar{s}\gamma_{\mu}s$.} 
with $\theta_V$ the $\omega-\phi$ mixing angle that we take from Ref.~\cite{Escribano:2005qq}, $\theta_V=38.7(2)^{\circ}$ and $m_V$ 
is the vector meson mass. Experimentally, $f_V$ can be related to the vector meson leptonic decay-width,
\begin{equation}
\Gamma_{V\rightarrow e^+e^-} = \frac{4\pi}{3} \alpha^2\frac{f_V^2}{m_V}c_V,
\end{equation} 
where $c_V$ is a charge factor, $c_{\rho,\omega,\phi}=(\frac{1}{\sqrt{2}},\frac{s_{\theta_V}}{\sqrt{6}},\frac{c_{\theta_V}}{\sqrt{6}})$. Taking the values 
from~\cite{Agashe:2014kda}, we find~\cite{Escribano:2005qq}
\begin{equation}
f_{\rho} = 0.221(1)~\textrm{MeV}, \quad f_{\omega} = 0.180(3)~\textrm{MeV}, \quad f_{\phi} = 0.239(4)~\textrm{MeV}.
\end{equation}
The experimental $g_{VP\gamma}$ couplings can be obtained from $V\rightarrow P\gamma$ and $P\rightarrow V\gamma$ processes, which decay-width is given as 
\begin{align}
\Gamma(P\rightarrow V\gamma) = & \ \frac{\alpha}{8}g_{VP\gamma}^2\left(1-\frac{m_V^2}{m_P^2}\right)^3, \\
\Gamma(V\rightarrow P\gamma) = & \  \frac{\alpha}{24}g_{VP\gamma}^2\left(1-\frac{m_P^2}{m_V^2}\right)^3.
\end{align}

\section{Cutcosky rules for additional vector states in \PtoLL}
\label{app:im}

As it was explained in \cref{sec:Cutcosky}, for heavier pseudoscalar states there are additional contributions to the imaginary part beyond the $\gamma\gamma$ one. 
Whereas the $\pi^+\pi^-\gamma$ state, including the resonant contribution, was illustrated in \cref{sec:erroreta} with the aid of a model, the narrow-width vector meson 
contributions can be easily calculated. For this, we only need to note that such contributions are related to a pole in the TFF, corresponding for the 
$\gamma V$ and $VV$ intermediate channels to 
\begin{align}
   \lim_{k^2\to m_V^2}(k^2-m_V^2)F_{P\gamma^*\gamma}(k^2,0) & = \textrm{Res}_{\gamma V}, \\
      \lim_{k_{1,2}^2\to m_V^2}(k_1^2-m_V^2)(k_2^2-m_V^2)F_{P\gamma^*\gamma}(k_1^2,k_2^2)   &= \textrm{Res}_{V V}.
\end{align}
The generalization to additional possible $VV'$ intermediate states is obvious. Accounting for the residues introduced above and following Cutcosky rules, 
we obtain, for the $\gamma V$ intermediate states
\begin{align}
\label{eq:ImGV}
\operatorname{Im}{\mathcal{A}}  = \ & \frac{(-2\pi i)^2}{\pi^2q^2} \int d^4k \frac{(q^2k^2 - (q\cdot k)^2)\textrm{Res}_{\gamma V}}{k^2((p-k)^2-m^2)} \delta(k^2 -m_V^2 )\delta((q-k)^2) \nonumber \\
                        = \ & \frac{ \textrm{Res}_{\gamma V}}{m_P}\frac{1}{m_V^2} \int  \ d\Omega_3 \ dk^0   \frac{\mathbf{k}^3    \delta(k^0- (m_P^2+m_V^2)/(2m_P))    }{ m_V^2 - m_P(k^0 - \beta_{\ell} \mathbf{k} \cos\theta)) }  \nonumber \\
                        = \ & \frac{ \textrm{Res}_{\gamma V}}{4m_V}\left( 1 - \frac{m_V^2}{m_P^2} \right)^2 \int  \ d\Omega_3 \frac{1}{ \beta_{\ell} \cos\theta -1 } \nonumber \\
                        = \ & \frac{ \textrm{Res}_{\gamma V}}{m_V^2}\frac{\pi}{2\beta_{\ell}} \left( 1 - \frac{m_V^2}{m_P^2} \right)^2 \ln \left( \frac{1-\beta_{\ell}}{1+\beta_{\ell}}  \right) \theta(m_P-m_V).
\end{align}
There exist an additional (identical contribution) for the symmetric channel, call it $V\gamma$. Similarly, for the $VV$ intermediate states, and defining $\beta_{V}$ as $\beta_{\ell}$ 
when $m_{\ell} \to m_V$ is replaced, we find
\begin{align}
\label{eq:ImVV}
\operatorname{Im}{\mathcal{A}}  = \ & \frac{(-2\pi i)^2}{\pi^2q^2} \int d^4k \frac{(q^2k^2 - (q\cdot k)^2)  \delta(k^2 -m_V^2 )  \textrm{Res}_{VV}}{k^2(q-k)^2((p-k)^2-m^2)}\delta((q-k)^2-m_V^2) \nonumber \\
                        = \ & \frac{1}{m_P}\frac{\textrm{Res}_{VV}}{m_V^4} \int  \ d\Omega_3 \ dk^0   \frac{\mathbf{k}^3    \delta(k^0- (m_P^2+m_V^2)/(2m_P))  }{ m_V^2 - m_P(k^0 - \beta_{\ell} \mathbf{k} \cos\theta)) } \nonumber \\
                        = \ & \frac{\beta_V^3\textrm{Res}_{VV}}{4m_V^4} \int  \ d\Omega_3 \frac{1}{ \beta_{\ell}\beta_{V} \cos\theta -\frac{1}{2}(1+\beta_V^2) } \nonumber \\
                        = \ & \frac{\textrm{Res}_{VV}}{m_V^4}\frac{\pi}{2\beta_{\ell}} \beta_V^2 \ln \left( \frac{1 + \beta_V^2 -2\beta_{\ell}\beta_V}{1 + \beta_V^2 +2\beta_{\ell}\beta_V}  \right) \theta(m_P-2m_V).
\end{align}

As a particular example, we take a simplified VMD approach where 
\begin{equation}
F_{P\gamma*\gamma^*}(q_1^2,q_2^2) = \frac{m_V^4}{(q_1^2 - m_V^2)(q_2^2 - m_V^2)}, \ \ \textrm{Res}_{\gamma V}=m_V^2,  \ \  \textrm{Res}_{VV}= m_V^4.
\end{equation}
\begin{figure}
  \centering
  \includegraphics[width=0.6\textwidth]{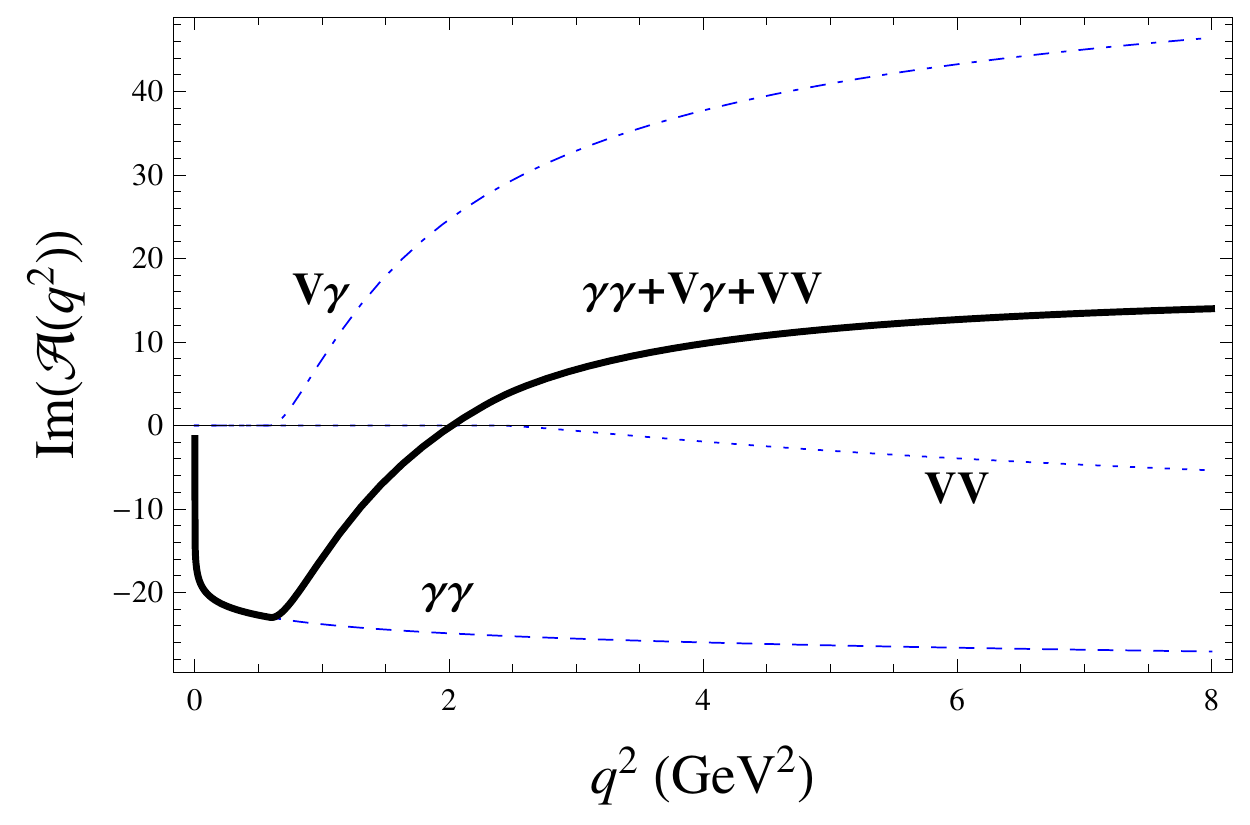}
  \caption{\cref{eq:ImVMD} is shown for $m_V=0.77$~GeV. The different channels open at the threshold values $q^2=0,m_V^2,4m_V^2$, respectively. There is a smooth 
               cancellation as $q^2\rightarrow\infty$.\label{fig:ImPart}}
\end{figure}
Taking into account all the channels, we obtain for the imaginary part
\begin{align}
\label{eq:ImVMD}
  \operatorname{Im}{\mathcal{A}}(q^2) = \ &  \frac{\pi}{2\beta_{\ell}} \ln \left( \frac{1-\beta_{\ell}}{1+\beta_{\ell}}  \right)  -\frac{\pi}{\beta_{\ell}} \left( 1 - \frac{m_V^2}{q^2} \right)^2 \ln \left( \frac{1-\beta_{\ell}}{1+\beta_{\ell}}  \right) \theta(q^2-m_V^2) \nonumber \\
 & +\frac{\pi}{2\beta_{\ell}} \beta_V^2 \ln \left( \frac{1 + \beta_V^2 -2\beta_{\ell}\beta_V}{1 + \beta_V^2 +2\beta_{\ell}\beta_V}  \right) \theta(q^2-4m_V^2).
\end{align}
The shape for the imaginary part is illustrated in \cref{fig:ImPart} as a function of the pseudoscalar mass for the individual and total contributions using $m_V=0.77$~GeV. 
The resulting function approaches $0$ asymptotically as it should, since the vector channels provide a finite result for the loop integral \cref{eq:loopamp}.

\section{Fierz transformations}
\label{sec:fierz}

Given the quantum numbers of the system, $J^{PC}=0^{-+}$, any new contribution to \PtoLL decays necessarily results from an effective 
$(\overline{q}\Gamma q) (\bar{\ell}\Gamma\ell)$ interaction 
where $\Gamma= \Gamma_{P,A}\equiv i\gamma_5,\gamma^{\mu}\gamma_5$ and, again, $q=u,d,s$ and $\ell=e,\mu$. 
Note that this does not necessarily implies that such term arises from an UV completion featuring an intermediate axial or pseudoscalar field as calculated in \cref{sec:np};  
it could arise as well from an effective leptoquark-like interaction $(\bar{q}\tilde{\Gamma}\ell)(\bar{\ell}\tilde{\Gamma}q) $ where $\tilde{\Gamma}\neq \Gamma_{P,V}$.
Still, such term can be Fierz rearranged, 
this means, expressed as~\cite{Soni:1974aw,Nishi:2004st}  
\begin{equation}
\label{eq:fierz1}
(\bar{q}\tilde{\Gamma}_i\ell)(\bar{\ell}\tilde{\Gamma}_iq) = \sum_j \lambda_{ij}(\bar{q}\Gamma_l q) (\bar{\ell}\Gamma_j\ell),
\end{equation}
\begin{equation}
\label{eq:fierz2}
   \Gamma_S = 1 \quad \Gamma_P = i\gamma_5 \quad  \Gamma_V = \gamma^{\mu} \quad  \Gamma_A = \gamma_5\gamma^{\mu} \quad  \Gamma_T = \sigma^{\mu\nu}.
\end{equation}
Then, only the relevant effective pseudoscalar and axial interactions do contribute to the process, which can be obtained using\footnote{Note an extra sign arising 
from the anticommuting nature of the spinor fields $q,\ell$. This should be removed if dealing with numeric quantities such as the spinors $u(v)_{s,p}$}
\begin{equation}
\label{eq:fierz3}
 \lambda_{SA}=\frac{1}{4} \quad \lambda_{VA}=\frac{1}{2} \quad \lambda_{TA}=0 \quad \lambda_{AA}=\frac{1}{2} \quad \lambda_{PA}=\frac{1}{4},
\end{equation}
\begin{equation}
\label{eq:fierz4}
\lambda_{SP}=\frac{1}{4} \quad \lambda_{VP}=-1 \quad \lambda_{TP}=3 \quad \lambda_{AP}=1 \quad \lambda_{PP}=-\frac{1}{4}.
\end{equation}
Consequently, any leptoquark contribution can be obtained from the results given in \cref{eq:NP,eq:NPA,eq:NPP} using \cref{eq:fierz1,eq:fierz2,eq:fierz3,eq:fierz4}.

\bibliographystyle{apsrev4-1}
\bibliography{test}

%
 
\end{document}